\documentclass[12pt,english]{article}
\usepackage[T1]{fontenc}
\usepackage[latin9]{inputenc}
\usepackage{geometry}
\usepackage{babel}
\usepackage{array}
\usepackage{mathrsfs}
\usepackage{multirow}
\usepackage{amsmath}
\usepackage{amsthm}
\usepackage{amssymb}
\usepackage{setspace}
\usepackage[round]{natbib}
\usepackage{graphicx}
\usepackage{comment}

\usepackage{chngcntr}
\usepackage{apptools}
\AtAppendix{\counterwithin{lemma}{section}}
	
\usepackage[unicode=true,pdfusetitle,
 bookmarks=true,bookmarksnumbered=false,bookmarksopen=false,
 breaklinks=false,pdfborder={0 0 1},backref=false,colorlinks=true, linkcolor = green, citecolor = green]
 {hyperref}

\makeatletter

\providecommand{\tabularnewline}{\\}

\theoremstyle{plain}
\newtheorem{assumption}{\protect\assumptionname}
\theoremstyle{remark}
\newtheorem{remark}{\protect\remarkname}
\theoremstyle{plain}

\theoremstyle{plain}
\newtheorem{lemma}{\protect\lemmaname}
\theoremstyle{plain}
\newtheorem{theorem}{\protect\theoremname}

\makeatother

\providecommand{\assumptionname}{Assumption}
\providecommand{\lemmaname}{Lemma}
\providecommand{\propositionname}{Proposition}
\providecommand{\remarkname}{Remark}
\providecommand{\theoremname}{Theorem}

\usepackage{authblk}

\title{Estimating High Dimensional Monotone Index Models\\  by Iterative Convex Optimization\footnote{ We are grateful to conference participants at the 
 BC/BU 2020 Econometrics Workshop, the 2019 Midwestern Econometrics Study group, the 2021 NASM of the Econometric Society, 2022 CIRAQ Econometrics conference, 2022 ISNPS conference, 2022 Advanced Methods Conference at TSE, and seminar participants from Georgetown, UC Berkeley, UC Louvain, UC Riverside, University of Bristol, UVA, University of Warwick and Yale for helpful comments.}}

\author[1]{Shakeeb Khan}
\author[1]{Xiaoying Lan}
\author[2]{Elie Tamer}
\author[1]{Qingsong Yao}

\affil[1]{Dept. of Economics, Boston College \authorcr
  \{ \tt shakeeb.khan Xiaoyang.lan Qingsong.yao\}@bc.edu}
\affil[2]{Dept. of Economics, Harvard University \authorcr
  \{\tt elietamer\}@fas.harvard.edu}

\date{First Version: 6/2021, This Version: \today}
\begin{document}
\maketitle
\vskip -.5in
{\footnotesize \singlespace
\begin{abstract}
 In this paper we propose   new approaches to estimating large dimensional  monotone index models. This class of models has been  popular in the applied and theoretical econometrics literatures as it includes discrete choice, nonparametric transformation, and duration models. A main  advantage of our approach    is computational. For instance, rank estimation procedures such as those proposed in \cite{han1987non} and \cite{cavanagh1998rank} that optimize a nonsmooth, non convex objective function are difficult to use with more than a few regressors and so limits their use in with economic data sets. For   such monotone index models with increasing dimension, we propose to use a new  class of estimators based on {\em batched gradient descent (BGD) } involving nonparametric methods such as kernel estimation or  sieve estimation,  and study their asymptotic properties. The BGD algorithm uses an iterative procedure where the key step exploits a strictly convex objective function, resulting in computational advantages.  A contribution of our approach is that our model is large dimensional and semiparametric and so does not require the use of parametric distributional assumptions.
\end{abstract}
}

\noindent
{\small 

 }

\


\thispagestyle{empty}

\newpage

\setcounter{page}{1}

\setlength{\abovedisplayshortskip}{5pt}
\setlength{\belowdisplayshortskip}{5pt} 
\setlength{\abovedisplayskip}{5pt}
\setlength{\belowdisplayskip}{5pt}

\setcounter{equation}{0}
\section{Introduction}
Monotone index models have received  a great deal of attention in both the theoretical
and applied econometrics literature, as many economic variables of interest are of a limited or qualitative nature.
A leading special case in this class is the binary choice model
which is usually represented by some variation of the following equation:
\begin{equation} y_i=I[x_i'\beta^\ast_e -u_i \geq 0 ] \end{equation}
where $I[\cdot]$ is the usual indicator function, $y_i$ is the observed response variable, taking the values
 0 or 1 and  $x_i$ is an observed $p$ dimensional vector of
covariates which effect  the behavior of $y_i$. Both the scalar disturbance term $u_i$ with distribution function denoted by $G(\cdot)$, and the $p-$ dimensional  vector
$\beta_e^\ast$ are unobserved, the latter often being the parameter estimated from a random sample
$(y_i,x_i') \ \ i=1,2,...n$.

The disturbance term $u_i$ is restricted in ways that
ensure identification of $\beta_e^\ast$. Parametric restrictions
specify the distribution of $u_i$ up to a finite dimensional
parameter and assume that $u_i$ distributed independently of the
covariates $x_i$. Under such a restriction, $\beta_e^\ast$
 can be estimated (up to scale) using maximum likelihood or nonlinear least squares.  Estimators that are robust to these parametric distributional assumptions have been proposed and analyzed resulting in a variety of estimation
procedures for $\beta_e^\ast$. 



An important class of semiparametric restrictions used in the literature
were based  on independence/index restrictions.  Estimation procedures
under this restriction include those proposed by \cite{han1987non},  \cite{ichimura1993semiparametric}, \cite{klein1993efficient}. These cover but are not limited to the above binary response model. 
This class of index models have a robustness advantage over parametric approaches, but 
estimators within this class are difficult to 
compute\footnote{Other estimation of index models includes \citep*{stoker1986consistent,powell1989semiparametric}. While these are relatively easy to compute, such derivative based estimators cannot be applied unless all components of $x_i$ are continuously distributed.} due to nonconvexity and in some cases also nonsmoothness of their respective objective functions. Furthermore the difficulty increases with the dimension of $x_i$.
Recent work which is motivated by computational concerns is \citet*{ahn2018simple}. However, their two step procedure involves a fully nonparametric estimator in the first stage, so is also not suitable for models with a large number of regressors.

A  related drawback of all these procedures is that they are designed to estimate parameters in models
of a small and {\em fixed} dimension. A relatively recent and thriving literature in econometrics and machine learning is recognizing the many advantages
of allowing for large dimensional models or models with a large set of controls. 
This class is a special case of models that  consider the situation when the dimension of $x_i$ is large, and this is now often modeled with 
its dimension increasing with the sample size. Due primarily to its empirical relevance, there has been a burgeoning literature on estimation and inference in certain econometric and statistics models with a large number of regressors or a large number of moment conditions. For a surevey of examples in economics and finance, see \cite{fanlvqi2011}. Recent papers include 
 \cite{neweywind2009},
 \cite{chernozhukov2017central},\cite{bellonietal2018},  \cite{cattaneoetal2018a}, \cite{cattaneoetal2018b},

Related to our work is the recent literature on estimating large dimensional binary choice or monotone index models  in \cite{sur2019modern} and \cite{fan2020rank}.
\cite{sur2019modern} considers inference in a large dimensional logit model, relying on the logistic distribution of the disturbance term where it is shown that $\chi^2$ asymptotic approximations of the LR statistic are suspect when the dimension of $x$ is large. 
\citet*{fan2020rank} on the other hand  estimate parameters by optimizing the objective function introduced in \cite{han1987non}, but with the number parameters increasing with the sample size. Optimizing these rank based objective functions is unfortunately hard  even with recent developments in algorithms and search methods for optimizing non smooth and/or non convex objective
functions. See for example important  recent work based on mixed integer programming (MIP) as in, e.g. \cite{fan2020rank} and \cite{ShinTodorov2020}.


Therefore, in light of the drawbacks in the existing literature, this paper proposes a new estimation procedure that is amenable to easier computattion.
Specifically we aim to construct a computationally feasible  estimator for a semiparametric binary choice and monotone index models with {\em increasing} dimension based on a convex objective function and then establish its asymptotic properties.
As we will discuss in detail in the next section, our algorithm  uses an iterative estimator based on a batched gradient descent (BGD) method, 
and  we show how to  use nonparametric methods  to approximate the distribution in each stage of the iteration.
One is the method of  sieves\footnote{ See \cite{chen2007large} who pioneered the use of sieve methods in econometrics.}, and the other 
is kernel  regression. 

\ \ \ 

\baselineskip = .75\baselineskip
{\bf Notation:}

Throughout the rest of this paper, to facilitate the description and properties of estimation procedures we will be using the following notation.
For any real sequences $\left\{ a_{n}\right\} _{n=1}^{\infty}$ and
$\left\{ b_{n}\right\} _{n=1}^{\infty}$, we write $a_{n}=o\left(b_{n}\right)$
if $\lim\sup_{n\rightarrow\infty}\left|a_{n}/b_{n}\right|=0$, $a_{n}=O\left(b_{n}\right)$
if $\lim\sup_{n\rightarrow\infty}\left|a_{n}/b_{n}\right|<\infty$,
and $a_{n}\sim b_{n}$ if both $a_{n}=O\left(b_{n}\right)$ and $b_{n}=O\left(a_{n}\right)$.
For any random sequences $\left\{ a_{n}\right\} _{n=1}^{\infty}$
and $\left\{ b_{n}\right\} _{n=1}^{\infty}$, we write $a_{n}=O_{p}\left(b_{n}\right)$
if for any $0<\tau<1$ there are $N$ and $C>0$ such that $P\left\{ \left|a_{n}/b_{n}\right|>C\right\} <\tau$
holds for all $n\geq N$, we write $a_{n}=o_{p}\left(b_{n}\right)$
if for any $C>0$, $\lim_{n\rightarrow\infty}P\left\{ \left|a_{n}/b_{n}\right|>C\right\} \rightarrow0$.
For any Borel sets $A\subseteq\mathbb{R}^{k}$, denote its Lebesgue
measure as $m\left(A\right)$. For any symmetric matrix $A$, we write
$A\succ0$ if $A$ is positive definite, and $A\succeq0$ if $A$
is positive semi-definite. For any symmetric matrices $A$ and $B$,
we write $A\succ B$ if $A-B\succ0$ and $A\succeq B$ if $A-B\succeq0$.
For any matrix $A$, we denote $\sigma\left(A\right)$ as its singular
value, and denote $\overline{\sigma}\left(A\right)$ and $\underline{\sigma}\left(A\right)$
as its largest and smallest singular value. For any symmetric matrix
$A$, we denote $\lambda\left(A\right)$ as its eigenvalue, and denote
$\overline{\lambda}\left(A\right)$ and $\underline{\lambda}\left(A\right)$
as its largest and smallest eigenvalue. For any vector $\boldsymbol{x}=\left(x_{1},\cdots,x_{p}\right)^{\mathrm{T}}$,
we denote its Euclidean norm as $\left\Vert \boldsymbol{x}\right\Vert =\sqrt{\sum_{i=1}^{p}x_{i}^{2}}$.
For any matrices $A=\left(a_{ij}\right)_{n\times m}$, we denote $\left\Vert A\right\Vert =\sqrt{\sum_{i=1}^{n}\sum_{j=1}^{m}a_{ij}^{2}}$.
Note that when $A$ is positive semi-definite, there holds $\left\Vert A\boldsymbol{x}\right\Vert \leq\overline{\lambda}\left(A\right)\cdot\left\Vert \boldsymbol{x}\right\Vert $;
for general square matrix $A$, there holds $\left\Vert A\boldsymbol{x}\right\Vert \leq\overline{\sigma}\left(A\right)\cdot\left\Vert \boldsymbol{x}\right\Vert $.
Finally, for any function $f\left(\boldsymbol{x}\right)$ with domain
$D$, define $\left\Vert f\right\Vert _{\infty}=\sup_{\boldsymbol{x}\in D}f\left(\boldsymbol{x}\right)$.

\baselineskip = 1.333\baselineskip

\section{\label{section2}The  BGD Estimator }

To provide some intuition for our semiparametric estimators that will be introduced
in the following sections, in this section we consider a simplified version
of the model where the cumulative distribution function $G\left(\cdot\right)$
is completely known. Under such setup, we explore the \textit{batch gradient descent estimator} (BGD estimator)
of $\boldsymbol{\beta}_{e}^{\star}$ when its dimensionality $p$ may increase, which is also important on its own right.
Throughout the following analysis we assume that the data set satisfies the following
assumption.
\begin{assumption}
\label{assu1}An i.i.d. data set $\mathscr{D}_{n}=\left\{ \left(\mathbf{X}_{e,i},y_{i}\right)\right\} _{i=1}^{n}$
of sample size $n$ is observed, where $y_{i}$ is generated \footnote{Here we are decomposing the vector $\mathbf{X}_{e,i}$ into a scalar component 
$X_{0,i}$ and the vector $\mathbf{X}_i$, and decomposing the vector of parameters $\boldsymbol{\beta}^{\star}_e$ into the scalar term $\beta_0^\star$ and the vector $\boldsymbol{\beta}^{\star}$. As we will see this is done for notational convenience when imposing scale normalizations. } by $
	y_i=I\left(X_{0,i}\beta_{0}^{\star}+\mathbf{X}_i^{\mathrm{T}}\boldsymbol{\beta}^{\star}-u_i>0\right)$ with unobserved shock $u_{i}$ that is independent
of $\mathbf{X}_{e,i}$ and has CDF $G\left(\cdot\right)$.
\end{assumption}
Given any loss function $\ell_{G}\left(\boldsymbol{\beta}_{e},\mathbf{X}_{e},y\right)$
that depends on $G$ and is a.s. differentiable with respect to $\boldsymbol{\beta}_{e}\in\mathcal{B}_{e}$,
the BGD estimator of $\boldsymbol{\beta}_{e}^{\star}$ is based on
the following iteration,
\begin{equation}
\boldsymbol{\beta}_{e,k+1}=\boldsymbol{\beta}_{e,k}-\frac{\delta_{k}}{n}\sum_{i=1}^{n}\partial\ell_{G}\left(\boldsymbol{\beta}_{e,k},\mathbf{X}_{e,i},y_{i}\right)/\partial\boldsymbol{\beta}_{e},\label{known_loss}
\end{equation}
where $\delta_{k}>0$ is the learning rate. Note that $n^{-1}\sum_{i=1}^{n}\partial\ell_{G}\left(\boldsymbol{\beta}_{e},\mathbf{X}_{e,i},y_{i}\right)/\partial\boldsymbol{\beta}_{e}$
constitutes a sample analogue of the derivative $\partial\mathbb{E}\left[\ell_{G}\left(\boldsymbol{\beta}_{e},\mathbf{X}_{e},y\right)\right]/\partial\boldsymbol{\beta}_{e}$. 
Unlike
the stochastic gradient descent (SGD) algorithm, in the  BGD algorithm, in each round of update we evaluate the
derivative of the loss function over all data points. This increases
the computational burden but provides a more accurate estimator for
the derivative of the expected loss function. Given the initial guess
of the parameter, $\boldsymbol{\beta}_{e,1}$, we iterate based on
(\ref{known_loss}) until some terminating conditions are reached. 

In this paper, we consider the following loss function 
\begin{equation}
\ell_{G}\left(\boldsymbol{\beta}_{e},\mathbf{X}_{e},y\right)=\int_{-A}^{\mathbf{X}_{e}^{\mathrm{T}}\boldsymbol{\beta}_{e}}G\left(z\right)dz-y\mathbf{X}_{e}^{\mathrm{T}}\boldsymbol{\beta}_{e},\label{loss_function}
\end{equation}
for some sufficiently large positive constant $A$. The loss function (\ref{loss_function}) was also considered in \citet{agarwal2014least}
and has many nice properties. For instance, under some mild conditions, we can show that
\begin{align*}
\frac{\partial\mathbb{E}\left(\ell_{G}\left(\boldsymbol{\beta}_{e}^{\star},\mathbf{X}_{e},y\right)\right)}{\partial\boldsymbol{\beta}_{e}} & =\mathbb{E}\left\{ \left(G\left(\mathbf{X}_{e}^{\mathrm{T}}\boldsymbol{\beta}_{e}^{\star}\right)-y\right)\mathbf{X}_{e}\right\} \\
 & =\mathbb{E}\left\{ \left(G\left(\mathbf{X}_{e}^{\mathrm{T}}\boldsymbol{\beta}_{e}^{\star}\right)-\mathbb{E}\left(\left.y\right|\mathbf{X}_{e}\right)\right)\mathbf{X}_{e}\right\} =0,
\end{align*}
and
\[
\frac{\partial^{2}\mathbb{E}\left(\ell_{G}\left(\boldsymbol{\beta}_{e},\mathbf{X}_{e},y\right)\right)}{\partial\boldsymbol{\beta}_{e}\partial\boldsymbol{\beta}_{e}^{\mathrm{T}}}=\mathbb{E}\left\{ G^{\prime}\left(\boldsymbol{\mathbf{X}}_{e}^{\mathrm{T}}\boldsymbol{\beta}_{e}\right)\mathbf{X}_{e}\mathbf{X}_{e}^{\mathrm{T}}\right\} \succ0,\forall\boldsymbol{\beta}_{e}\in\mathcal{B}_{e}.
\]
So $\boldsymbol{\beta}_{e}^{\star}$ uniquely minimizes $\mathbb{E}\ell_{G}\left(\boldsymbol{\beta}_{e},\mathbf{X}_{e},y\right)$
over $\mathcal{B}_{e}$. Another desirable property of the loss function (\ref{loss_function})
is that the derivative of (\ref{loss_function}) with respect to $\boldsymbol{\beta}_{e}$,
which is $\left(G\left(\mathbf{X}_{e}^{\mathrm{T}}\boldsymbol{\beta}_{e}\right)-y\right)\mathbf{X}_{e}$,
depends only on $G\left(\cdot\right)$ instead of on its derivatives.
So when we conduct a  semiparametric iteration in the
following sections, we only need to nonparametrically approximate $G\left(\cdot\right)$,
which is generally more robust compared with approximating its derivatives.
Based on loss function (\ref{loss_function}), the BGD estimator is
obtained based on the following iteration 
\begin{equation}
\boldsymbol{\beta}_{e,k+1}=\boldsymbol{\beta}_{e,k}-\frac{\delta_{k}}{n}\sum_{i=1}^{n}\left(G\left(\mathbf{X}_{e,i}^{\mathrm{T}}\boldsymbol{\beta}_{e,k}\right)-y_{i}\right)\mathbf{X}_{e,i}.\label{known_G}
\end{equation}

\ \ 

\begin{remark}
Key to the above approach is the construction of a convex objective function that facilitates computation even with high dimensions. This transformed convex objective works for any  monotone model.  In particular, for any model of the form $y_i =G(x_i'\beta) + \epsilon$ with $E[\epsilon_i|x_i]=0$ and monotone $G(.)$, a similar {\it convex} criterion as in (\ref{loss_function}) can be used for inference on $\beta.$ 
\end{remark}

We now describe the asymptotic properties of $\boldsymbol{\beta}_{e,k}$.
We first make the following assumption. 
\begin{assumption}
\label{assump:2}(i) $\mathcal{X}_{e}=\left[-1,1\right]^{p+1}$; (ii)
$\mathcal{B}_{e}$ is convex, and there exists some constant $B_{0}>0$
such that for any $\boldsymbol{\beta}_{e}\in\mathcal{B}_{e}$, $\left|\beta_{j}\right|\leq B_{0}$
for any $0\leq j\leq p$; (iii) there exists integer $\upsilon_{G}$
such that $G$ has up to $\upsilon_{G}$-th bounded derivatives; (iv)
Define $M_{n}\left(\boldsymbol{\beta}_{e}\right)=\frac{1}{n}\sum_{i=1}^{n}G^{\prime}\left(\mathbf{X}_{e,i}^{\mathrm{T}}\boldsymbol{\beta}_{e}\right)\mathbf{X}_{e,i}\mathbf{X}_{e,i}^{\mathrm{T}}$
and $M\left(\boldsymbol{\beta}_{e}\right)=\mathbb{E}[M_{n}\left(\boldsymbol{\beta}_{e}\right)]$.
For any $\boldsymbol{\beta}_{e}\in\mathcal{B}_{e}$, there holds $0<\underline{\lambda}_{e}\leq\underline{\lambda}\left(M\left(\boldsymbol{\beta}_{e}\right)\right)\leq\overline{\lambda}\left(M\left(\boldsymbol{\beta}_{e}\right)\right)\leq\overline{\lambda}_{e}<\infty$. 
\end{assumption}
\begin{remark}
\autoref{assump:2}(i) and \autoref{assump:2}(ii) are convenient
normalizations that facilitate the assessment of our model. 
Note that to ensure that $\boldsymbol{\beta}_{e,k}$ falls into a
compact set for each $k$, some form of truncation on $\boldsymbol{\beta}_{e,k+1}$
in (\ref{known_G}) is needed. While according to our results below, as long as $\mathcal{B}_{e}$
is sufficiently large, it can be
shown that $\boldsymbol{\beta}_{e,k}$ will fall into $\mathcal{B}_{e}$
for all $k$ with probability going to 1. We then assume that
$\boldsymbol{\beta}_{e,k}\in\mathcal{B}_{e}$ for all $k$.  
\autoref{assump:2}(iii) imposes some smoothness conditions on $G$, where the requirement on $\upsilon_G$ will be stated in the following propositions and theorems. 
\autoref{assump:2}(iv) requires that the eigenvalue of $M_{n}\left(\boldsymbol{\beta}_{e}\right)$
is bounded from both below and above uniformly over $\mathcal{B}_{e}$.

\end{remark}
For any $\boldsymbol{\beta}_{e}\in\mathcal{B}_{e}$, define $\Delta\boldsymbol{\beta}_{e}=\boldsymbol{\beta}_{e}-\boldsymbol{\beta}_{e}^{\star}$.
Also define $\varepsilon_{i}=y_{i}-G\left(\mathbf{X}_{e,i}^{\mathrm{T}}\boldsymbol{\beta}_{e}^{\star}\right)$,
where $\mathbb{E}\left[\left.\varepsilon_{i}\right|\mathbf{X}_{e,i}\right]=0$.
When \autoref{assu1} and \autoref{assump:2} hold, we have the
following result.
\begin{theorem}
\label{prop:Known_G}Suppose that \autoref{assu1} and \autoref{assump:2}
hold with $\upsilon_{G}=3$, 
that $p^{5}\left(\log p\right)^{2}n^{-1}\rightarrow0$, that the learning rate is chosen such that  $\delta_{k}=\delta\leq2/\left(3\overline{\lambda}_{e}\right)$, and that $\boldsymbol{\beta}_e$ is updated under (\ref{known_G}). We have
that

(i) Define 
\[
k_{1,n}^{BGD}=\frac{\log\left\Vert \Delta\boldsymbol{\beta}_{e,1}\right\Vert +\frac{1}{2}\log\left(n/\left(p\log p\right)\right)}{-\log\left(1-\underline{\lambda}_{e}\delta/2\right)},
\]
we then have
\[
\sup_{k\geq k_{1,n}^{BGD}+1}\left\Vert \Delta\boldsymbol{\beta}_{e,k}\right\Vert =O_{p}\left(\sqrt{p\left(\log p\right)/n}\right);
\]

(ii) Define $k_{2,n}^{BGD}$ such that $\left(1-\underline{\lambda}_e\delta\right)^{k_{2,n}^{BGD}}\sqrt{p\log p}\rightarrow 0$, we have
\[
\sup_{k\geq k_{2,n}^{BGD}+1}\left\Vert \Delta\boldsymbol{\beta}_{e,k+k_{1,n}^{BGD}}-M^{-1}\left(\boldsymbol{\beta}_{e}^{\star}\right)\frac{1}{n}\sum_{i=1}^{n}\varepsilon_{i}\mathbf{X}_{e,i}\right\Vert =o_{p}\left(1/\sqrt{n}\right);
\]

(iii) For any $k\geq k_{1,n}^{BGD} + k_{2,n}^{BGD} + 1$, define
$\widehat{\boldsymbol{\beta}}_{e}=\widehat{\boldsymbol{\beta}}_{k}$.
Also define 
\[
\Sigma_{1}^{\star}=M^{-1}\left(\boldsymbol{\beta}_{e}^{\star}\right)\mathbb{E}\left[G_{i}^{\star}\left(1-G_{i}^{\star}\right)\boldsymbol{\mathbf{X}}_{e,i}\boldsymbol{\mathbf{X}}_{e,i}^{\mathrm{T}}\right]M^{-1}\left(\boldsymbol{\beta}_{e}^{\star}\right),
\]
and
\[
\widehat{\Sigma}_{1,n}=M_n^{-1}\left(\widehat{\boldsymbol{\beta}}_{e}\right)\left\{ \frac{1}{n}\sum_{i=1}^{n}\widehat{G}_{i}\left(1-\widehat{G}_{i}\right)\boldsymbol{\mathbf{X}}_{e,i}\boldsymbol{\mathbf{X}}_{e,i}^{\mathrm{T}}\right\} M_n^{-1}\left(\widehat{\boldsymbol{\beta}}_{e}\right),
\]
where $G_{i}^{\star}=G\left(\boldsymbol{\mathbf{X}}_{e,i}^{\mathrm{T}}\boldsymbol{\beta}_{e}^{\star}\right)$
and $\widehat{G}_{i}=G\left(\boldsymbol{\mathbf{X}}_{e,i}^{\mathrm{T}}\widehat{\boldsymbol{\beta}}_{e}\right)$. Suppose further that $\mathbb{E}\left(\mathbf{X}_{e,i}\mathbf{X}_{e,i}^{\mathrm{T}}\right)$  has uniformly (with respect to $p$) upper bounded eigenvalues, 
there holds 
\[
\left\Vert \widehat{\Sigma}_{1,n}-\Sigma_{1}^{\star}\right\Vert \rightarrow_{p} 0.
\]

(iv) For any $p+1$ vector $\rho$ such that $\lim_{n\rightarrow\infty}\left\Vert \rho\right\Vert <\infty$,
$\lim_{n\rightarrow\infty}\rho^{\mathrm{T}}\Sigma_{1}^{\star}\rho=\sigma^{2}\left(\rho\right)$,
and that $\rho^{\mathrm{T}}M^{-1}\left(\boldsymbol{\beta}_{e}^{\star}\right)\frac{1}{\sqrt{n}}\sum_{i=1}^{n}\varepsilon_{i}\mathbf{X}_{e,i}\rightarrow_{d}N\left(0,\sigma^{2}\left(\rho\right)\right)$,
we have that 
\[
\rho^{\mathrm{T}}\Delta\widehat{\boldsymbol{\beta}}_{e}/\sqrt{\widehat{\sigma}^2\left(\rho\right)/n}\rightarrow_{d}N\left(0,1\right),
\]
where $\widehat{\sigma}^{2}\left(\rho\right)=\rho^{\mathrm{T}}\widehat{\Sigma}_{1,n}\rho$. 
\end{theorem}
\begin{proof}[Proof of \autoref{prop:Known_G}]
See \autoref{appendixB}. 
\end{proof}
When $p$ is fixed,  \autoref{prop:Known_G}(i)
implies that 
$
\sup_{k\geq k_{1,n}^{BGD}+1}\left\Vert \Delta\boldsymbol{\beta}_{e,k}\right\Vert =O_{p}\left(1/\sqrt{n}\right),
$
and  \autoref{prop:Known_G}(ii) implies that for $k$ sufficiently
large, the BGD estimator is an asymptotically linear estimator, so
there holds 
$
\sqrt{n}\Delta\boldsymbol{\beta}_{e,k+k_{1,n}^{BGD}}\rightarrow_{d}N\left(0,\Sigma_{1}^{\star}\right)
$ 
by the central limit theorem. The asymptotic
variance can be estimated based on   \autoref{prop:Known_G}(iii).
The number of iterations required to obtain root-$n$ consistency, $k_{1,n}^{BGD}$,
is determined by many factors including the sample size $n$, the
distance between the true parameter and the initial guess $||\Delta\boldsymbol{\beta}_{e,1}||$,
as well as the lower bound of the eigenvalues of $M_{n}\left(\boldsymbol{\beta}_{e}\right)$.
In general, $k_{1,n}^{BGD}$ is of order $O\left(\log n\right)$, but
in practice when we apply the above algorithm, 
the specific number of iteration
 is difficult to determine. For detailed discussion of the number
of iterations, see  \autoref{rem4} at the end of Section \ref{section4}. The
inference on $\boldsymbol{\beta}_{e}^{\star}$ based on the BGD estimator
is given by  \autoref{prop:Known_G}(iv). Note that for any
given vector $\rho$, we require that $\frac{1}{\sqrt{n}}\rho^{\mathrm{T}}M^{-1}\left(\boldsymbol{\beta}_{e}^{\star}\right)\sum_{i=1}^{n}\varepsilon_{i}\mathbf{X}_{e,i}$
is asymptotically normally distributed. An alternative approach is
to apply the high-dimensional central limit theorem to $\frac{1}{n}\sum_{i=1}^{n}M^{-1}\left(\boldsymbol{\beta}_{e}^{\star}\right) \mathbf{X}_{e,i}\varepsilon_{i}$  \citep[e.g.,][]{chernozhukov2017central}. 

Before we conclude this section and move to semiparametric estimation, we further comment on \autoref{prop:Known_G}.
Different from the stochastic gradient descent  algorithm \citep[e.g.,][]{toulis2017asymptotic}, we show in \autoref{prop:Known_G}
that the learning rate $\delta_{k}$ can be selected as a sufficiently
small constant. Indeed, in the following results, we show that $\delta_{k}$
can decay to zero at any rate as long as $\sum_{k=1}^{\infty}\delta_{k}=\infty$
holds, and the choice of $\delta_{k}$ will not change the asymptotic
results displayed in \autoref{prop:Known_G}. In particular,
we have the following proposition.
\begin{theorem}
\label{prop2}Suppose that all the conditions in  \autoref{prop:Known_G}
hold and that $\boldsymbol{\beta}_e$ is updated under (\ref{known_G}). For any sequence of tuning parameters $\left\{ \delta_{k}\right\} _{k=1}^{\infty}$
satisfying $\delta_{k}\geq0$, $\delta_{k}\rightarrow0$, $\limsup_{k\rightarrow\infty}\delta_{k-1}/\delta_{k}<\infty$,
and $\sum_{k=1}^{\infty}\delta_{k}=\infty$, we have that

(i) Define $\widetilde{k}_{1,n}^{BGD}$ such that 
$
\sum_{k=1}^{\widetilde{k}_{1,n}^{BGD}}\delta_{k}\geq\underline{\lambda}_{e}^{-1}\left\{ \log\left(n/p\left(\log p\right)\right)+2\log\left\Vert \Delta\boldsymbol{\beta}_{e,1}\right\Vert \right\} ,
$
 and that 
$
\sup_{k\geq\widetilde{k}_{1,n}^{BGD}+1}\delta_{k}\leq2/\underline{\lambda}_{e},
$
then  there  holds
\[
\sup_{k\geq \widetilde{k}_{1,n}^{BGD}+1}\left\Vert \Delta\boldsymbol{\beta}_{e,k}\right\Vert =O_{p}\left(\sqrt{p\left(\log p\right)/n}\right);
\]

(ii) Define $\widetilde{k}_{2,n}^{BGD}$ such that $\sum_{k=\widetilde{k}_{1,n}^{BGD}+1}^{k=\widetilde{k}_{2,n}^{BGD}}\delta_k /\log p \rightarrow \infty$, then we have that 
\[
\sup_{k\geq \widetilde{k}_{2,n}^{BGD}+1}\left\Vert \Delta\boldsymbol{\beta}_{e,k+\widetilde{k}_{1,n}^{BGD}}-M^{-1}\left(\boldsymbol{\beta}_{e}^{\star}\right)\frac{1}{n}\sum_{i=1}^{n}\varepsilon_{i}\mathbf{X}_{e,i}\right\Vert =o_{p}\left(1/\sqrt{n}\right);
\]

(iii) For any  $k\geq \widetilde{k}_{1,n}^{BGD} + \widetilde{k}_{2,n}^{BGD} + 1$, define
$\widehat{\boldsymbol{\beta}}_{e}=\widehat{\boldsymbol{\beta}}_{k}$.
We have that  \autoref{prop:Known_G}(iii) and (iv) hold.
\end{theorem}
\begin{proof}[Proof of \autoref{prop2}]
See \autoref{appendixB}. 
\end{proof}
\autoref{prop2} shows that the choice of the learning rate basically
does not affect the convergence rate as well as the asymptotic distribution
of the BGD estimators. The main advantage of using a sequence of decaying
learning rates is that we do not need to choose the constant $\delta$
as required in \autoref{prop:Known_G}, since for $k$ sufficiently
large, $\delta_{k}\leq2/\left(3\overline{\lambda}_{e}\right)$ will
automatically hold. However, the disadvantage of using
decaying learning rates is that such procedure takes much longer time
to converge because the magnitude of the update in the $k$-th round
decreases as $k$ increases. For
instance, suppose that we choose $\delta_{k}\sim k^{-\upsilon}$ for some
$0\leq\upsilon<1$, we have that $\sum_{j=1}^{k}\delta_{j}\sim k^{1-\upsilon}$.
Then to ensure that $\sum_{j=1}^{\widetilde{k}_{1,n}^{BGD}}\delta_{j}\geq\underline{\lambda}_{e}^{-1}\left(\log n+2\log\left\Vert \Delta\boldsymbol{\beta}_{e,1}\right\Vert \right)$,
we need $\widetilde{k}_{1,n}^{BGD}\sim\left(\log n\right)^{\frac{1}{1-\upsilon}}$. Obviously,
setting $\upsilon=0$ leads to $k\sim\log n$, which corresponds to the
requirement in  \autoref{prop:Known_G}(i); when $\upsilon>0$,
we can see that more rounds of iteration is needed compared with required
in  \autoref{prop:Known_G}(i). 

\section{\label{section3}Semiparametric BGD Estimation}

In the previous section, we focused on 
iterative estimators based on the BGD algorithm for the parametric binary choice
models. We show that when the CDF of the error term is known, the
iterative estimators based on the BGD algorithm are  consistent
and attain asymptotic normality under mild conditions. However, having
prior knowledge of the form of $G$ is generally too strong an assumption.
In most applications, the source of the individual shock
$u$ in  \autoref{assu1}  is difficult to justify, which makes
it quite difficult, if not completely impossible, to know the exact
expression of $G$. In this scenario, the algorithm proposed in the
previous section is infeasible. 
To overcome such problem, this section generalizes the BGD estimator
proposed in Section \ref{section2} to the semiparametric setting where $G$ is unknown. 

In this setup, to ensure identification we set $\beta_{0}^{\star}$
to be 1, so our estimation target is $\boldsymbol{\beta}^{\star}$.
To simplify our notation, we denote the space of $\mathbf{X}$ as
$\mathcal{X}$, and the corresponding parameter space of $\boldsymbol{\beta}$
as $\mathcal{B}$. Suppose that an initial guess for \textbf{$\boldsymbol{\beta}^{\star}$
}is  given by $\boldsymbol{\beta}_{1}$.
In the $k$-th round of iteration, to update $\boldsymbol{\beta}$
based on the BGD algorithm, we require the knowledge of $G$ as in
Section \ref{section2}, which is infeasible when $G$ is unknown. A natural idea
is that  we can construct an estimator for
$G$ based on the index constructed from the updated parameter in the previous round. More intuitively,
suppose for a moment that in the $k$-th round of iteration, $\boldsymbol{\beta}_{k}$
happens to be identical to the unknown true parameter $\boldsymbol{\beta}^{\star}$,
then we have that
$
G\left(z\right) =\mathbb{E}\left[\left.y\right|X_{0}+\boldsymbol{\mathbf{X}}^{\mathrm{T}}\boldsymbol{\beta}^{\star}=z\right]=\mathbb{E}\left[\left.y\right|X_{0}+\boldsymbol{\mathbf{X}}^{\mathrm{T}}\boldsymbol{\beta}_{k}=z\right]
$
for any $z\in R$. 

This motivates semiparametric estimation by using nonparametric methods to estimate $G\left(\cdot\right)$.
We consider kernel estimation and the method of sieves in each of the following subsections.

\subsection{The KBGD Estimator}

In this section we consider tkernel estimation to estimate $G\left(\cdot\right)$.
The Nadaraya-Watson kernel
estimator of $G\left(\cdot\right)$ is of the form
\begin{equation}
\widehat{G}\left(\left.z\right|\boldsymbol{\beta}_{k}\right)=\frac{\sum_{j=1}^{n}K_{h_{n}}\left(z-X_{0,j}-\boldsymbol{\mathbf{X}}_{j}^{\mathrm{T}}\boldsymbol{\beta}_{k}\right)y_{j}}{\sum_{j=1}^{n}K_{h_{n}}\left(z-X_{0,j}-\boldsymbol{\mathbf{X}}_{j}^{\mathrm{T}}\boldsymbol{\beta}_{k}\right)},z\in R,\label{kernel_estimator}
\end{equation}
where $K_{h}\left(\cdot\right)=h^{-1}K\left(\cdot/h\right)$, $K\left(\cdot\right)$
is some kernel function, and $h_{n}$ is some bandwidth parameter
depending on $n$. Given the estimated CDF $\widehat{G}\left(\left.\cdot\right|\boldsymbol{\beta}_{k}\right)$,
we can update the parameter as if it were the true CDF $G\left(\cdot\right)$.
In particular, $\boldsymbol{\beta}_{k}$ is updated as 
\begin{equation}
\boldsymbol{\beta}_{k+1}=\boldsymbol{\beta}_{k}-\frac{\delta_{k}}{n}\sum_{i=1}^{n}\left(\widehat{G}\left(\left.X_{0,i}+\mathbf{X}_{i}^{\mathrm{\mathrm{T}}}\boldsymbol{\beta}_{k}\right|\boldsymbol{\beta}_{k}\right)-y_{i}\right)\mathbf{X}_{i}.\label{update}
\end{equation}
Keep updating $\boldsymbol{\beta}_{k}$
based on (\ref{kernel_estimator}) and (\ref{update}), until some
terminating conditions are reached. The resulting estimator is labeled as
the \textit{kernel-based batch gradient descent estimator} (KBGD estimator).
\begin{remark}
In essence, the KBGD estimator can not be classified as a BGD estimator
based on a semiparametric loss function. In the semiparametric setup,
given any loss function $\ell_G\left(\boldsymbol{\beta},\mathbf{X}_{e},y\right)$
(quadratic distance in \citet{ichimura1993semiparametric}, log-likelihood
in \citet{klein1993efficient}, or loss function given in (\ref{loss_function}))
with unknown function $G$, it's a common practice to replace $G$
with its nonparametric estimator $\widehat{G}$ and then minimize
(or maximize) the resulting loss function to obtain the 
estimator of $\boldsymbol{\beta}$. Note that under the single-index framework, $\widehat{G}$ usually
involves the unknown parameter $\boldsymbol{\beta}$, which is 
of the form $\widehat{G}\left(\cdot\right)=\widehat{G}\left(\left.\cdot\right|\boldsymbol{\beta}\right)$.
In this scenario, the BGD estimator is constructed by the following
iteration
\[
\boldsymbol{\beta}_{k+1}^{BGD}=\boldsymbol{\beta}_{k}^{BGD}-\frac{\delta_{k}}{n}\sum_{i=1}^{n}\frac{\partial\ell_{\widehat{G}\left(\left.\cdot\right|\boldsymbol{\beta}_{k}^{BGD}\right)}\left(\boldsymbol{\beta}_{k}^{BGD},\mathbf{X}_{e,i},y_{i}\right)}{\partial\boldsymbol{\beta}},
\]
where $\partial\ell_{\widehat{G}\left(\left.\cdot\right|\boldsymbol{\beta}_{k}^{BGD}\right)}\left(\boldsymbol{\beta}_{k}^{BGD},\mathbf{X}_{e,i},y_{i}\right)/\partial\boldsymbol{\beta}$
 involves $\partial\widehat{G}\left(\left.\cdot\right|\boldsymbol{\beta}_k\right)/\partial\boldsymbol{\beta}$,
a complicated functions of $\boldsymbol{\beta}_k$. In particular, the BGD estimator under loss function (\ref{loss_function}) is given by
\[
\boldsymbol{\beta}_{k+1}=\boldsymbol{\beta}_{k}-\frac{\delta_{k}}{n}\sum_{i=1}^{n}\left(\widehat{G}\left(\left.X_{0,i}+\mathbf{X}_{i}^{\mathrm{\mathrm{T}}}\boldsymbol{\beta}_{k}\right|\boldsymbol{\beta}_{k}\right) + \int_{-\infty}^{X_{0,i}+\mathbf{X}_i^{\mathrm{T}}\boldsymbol{\beta}_k}\frac{\partial\widehat{G}\left(\left.z\right|\boldsymbol{\beta}_k\right)}{\partial \boldsymbol{\beta}}dz-y_{i}\right)\mathbf{X}_{i}.
\]
Obviously, an additional term  is introduced compared with (\ref{update}). On the contrary,
during the construction (\ref{update}), we take $G$ as given
when taking the first order derivative of the loss function and then replace the unknown $G$
with its non-parametric estimator in the derivative. More specifically,
the KBGD estimator is updated as follows
\[
\boldsymbol{\beta}_{k+1}=\boldsymbol{\beta}_{k}-\frac{\delta_{k}}{n}\sum_{i=1}^{n}\left.\frac{\partial\ell_G\left(\boldsymbol{\beta}_{k},\mathbf{X}_{e,i},y_{i}\right)}{\partial\boldsymbol{\beta}}\right|_{G\left(\cdot\right)=\widehat{G}\left(\left.\cdot\right|\boldsymbol{\beta}_{k}\right)},
\]
so additional terms involving $\partial\widehat{G}\left(\left.\cdot\right|\boldsymbol{\beta}_k\right)/\partial\boldsymbol{\beta}$ are avoided. Finally, as we discussed in Section \ref{section2}, the derivative of loss function
(\ref{loss_function}) with respect to $\boldsymbol{\beta}$  
depends only on $G$, so we also avoid approximating the derivative of $G$, which has poorer finite-sample performance compared with approximating $G$. Such update also ensures contraction map under some conditions, see \autoref{assu:5}.
\end{remark}
For any fixed $z$ and $\boldsymbol{\beta}$, under mild
conditions  there holds
$
\widehat{G}\left(\left.z\right|\boldsymbol{\beta}\right)\rightarrow_{p}\mathbb{E}\left[\left.y\right|X_{0}+\mathbf{X}^{\mathrm{T}}\boldsymbol{\beta}=z\right].
$
Denote such limit as $L\left(z,\boldsymbol{\beta}\right)$. Obviously,
$L\left(z,\boldsymbol{\beta}^{\star}\right)=G\left(z\right)$ holds
for any $z\in\mathbb{R}$. Before we move to a formal description of
the statistical properties of the KBGD estimator based on (\ref{update}), we first provide
some further discussion on $L\left(z,\boldsymbol{\beta}\right)$.
For simplicity, in the following we only focus on the case where all the covariates
are continuous which permit continuous joint density function. We leave further
discussion of the case where some covariates are discrete to 
\autoref{rem5}. We point  that when there are discrete covariates, our algorithm can be directly applied without any modification, although some further assumptions will be required. 

When all the covariates are continuous, denote the joint density of $\mathbf{X}_{e}$ and
$\mathbf{X}$ as $f_{e}\left(\mathbf{X}_{e}\right)=f_{e}\left(X_{0},\mathbf{X}\right)$
 and $f\left(\mathbf{X}\right)=\int f_{e}\left(X_{0},\mathbf{X}\right)dX_{0}$,
respectively. Denote $z\left(\mathbf{X}_{e},\boldsymbol{\beta}\right)=X_{0}+\mathbf{X}^{\mathrm{T}}\boldsymbol{\beta}$.
Also denote $f_{\mathbf{X},z}\left(\left.\mathbf{X},z\right|\boldsymbol{\beta}\right)$
as the joint density of $\mathbf{X}$ and $z\left(\mathbf{X}_{e},\boldsymbol{\beta}\right)$
given $\boldsymbol{\beta}$. Note that for any $\boldsymbol{x}$ and
$z$, 
\begin{align*}
P\left[\mathbf{X}\leq\boldsymbol{x},z\left(\mathbf{X}_{e},\boldsymbol{\beta}\right)\leq z\right] & =\int_{\widetilde{\mathbf{X}}\leq\boldsymbol{x},\widetilde{X}_{0}+\widetilde{\mathbf{X}}^{\mathrm{T}}\boldsymbol{\beta}\leq z}f_{e}\left(\widetilde{X}_{0},\widetilde{\mathbf{X}}\right)d\widetilde{X}_{0}d\widetilde{\mathbf{X}}\\
 & =\int_{\widetilde{\mathbf{X}}\leq\boldsymbol{x}}\left[\int_{\widetilde{X}_{0}\leq z-\widetilde{\mathbf{X}}^{\mathrm{T}}\boldsymbol{\beta}}f_{e}\left(\widetilde{X}_{0},\widetilde{\mathbf{X}}\right)d\widetilde{X}_{0}\right]d\widetilde{\mathbf{X}}.
\end{align*}
This implies that the joint density of $\mathbf{X}$ and $z\left(\mathbf{X}_{e},\boldsymbol{\beta}\right)$
given $\boldsymbol{\beta}$ is given by 
\begin{equation}
f_{\mathbf{X},z}\left(\left.\mathbf{X},z\right|\boldsymbol{\beta}\right)=f_{e}\left(z-\mathbf{X}^{\mathrm{T}}\boldsymbol{\beta},\mathbf{X}\right),\label{eq:joint_pdf_X_Z}
\end{equation}
and the marginal density of $z\left(\mathbf{X}_{e},\boldsymbol{\beta}\right)$
is given by 
\begin{equation}
f_{z}\left(\left.z\right|\boldsymbol{\beta}\right)=\int_{\mathcal{X}}f_{\mathbf{X},z}\left(\left.\mathbf{X},z\right|\boldsymbol{\beta}\right)d\mathbf{X} = \int_{\mathcal{X}}f_{e}\left(z-\mathbf{X}^{\mathrm{T}}\boldsymbol{\beta},\mathbf{X}\right)d\mathbf{X}.\label{eq:marginal_pdf_Z}
\end{equation}
Define $f_{\mathbf{X}|z}\left(\left.\mathbf{X}\right|z,\boldsymbol{\beta}\right)=f_{\mathbf{X},z}\left(\left.\mathbf{X},z\right|\boldsymbol{\beta}\right)/f_{z}\left(\left.z\right|\boldsymbol{\beta}\right)$
as the conditional density of $\boldsymbol{\mathbf{X}}$ given $z$
and $\boldsymbol{\beta}$, we have that
\begin{align}
L\left(z,\boldsymbol{\beta}\right) & =\mathbb{E}\left(\left.G\left(z-\boldsymbol{\mathbf{X}}^{\mathrm{T}}\Delta\boldsymbol{\beta}\right)\right|z\left(\boldsymbol{\mathbf{X}}_{e},\boldsymbol{\beta}\right)=z\right)\nonumber \\
 & =\int_{\mathcal{X}}G\left(z-\boldsymbol{\mathbf{X}}^{\mathrm{T}}\Delta\boldsymbol{\beta}\right)f_{\mathbf{X}|z}\left(\left.\mathbf{X}\right|z,\boldsymbol{\beta}\right)d\mathbf{X},\label{eq:L}
\end{align}
where $\Delta\boldsymbol{\beta}=\boldsymbol{\beta}-\boldsymbol{\beta}^{\star}$. 

Based on the above notations, now we formally study the asymptotic
properties of the KBGD estimator under increasing dimensions. We first
introduce some further assumptions.
\begin{assumption}
\label{assu:3}The kernel function $K\left(\cdot\right)$ satisfies:
(i) $K$ is bounded and twice continuously differentiable with bounded
first and second derivatives, and the second derivative satisfies
Lipschitz condition on the whole real line; (ii) $\int K\left(s\right)ds=1$;
(iii) there exists positive integer $\upsilon_{K}$ such that $\int s^{\upsilon}K\left(s\right)du=0$
for $1\leq\upsilon\leq\upsilon_{K}-1$ and $\int u^{\upsilon_{K}}K\left(u\right)du\neq0$;
(iv) $K\left(s\right)=0$ for $\left|s\right|>1$. 
\end{assumption}
\begin{assumption}
\label{assu:4}(i) There exists some constant $\zeta>1$ such that
$\zeta^{-1}\leq f_{e}\left(\mathbf{X}_{e}\right)\leq\zeta$ holds
for all $\mathbf{X}_{e}\in\mathcal{X}_{e}$; (ii) there exists positive
integer $\upsilon_{f}$ such that $f_{e}\left(\mathbf{X}_{e}\right)$
has bounded up to $\upsilon_{f}$-th derivatives.
\end{assumption}
\begin{remark}
\autoref{assu:4}(i) together with \autoref{assump:2}(i)
is a commonly-used assumption in the machine learning literature 
\citep[e.g.,][]{wager2018estimation}. It basically requires that the joint density
of $\mathbf{X}_{e}$ is uniformly bounded from both above and below
over $\mathcal{X}_{e}$, so the density does not degenerate over $\mathcal{X}_{e}$.
\autoref{assu:4}(i) basically allows us to construct a subset of 
$\mathcal{X}_{e}$ such that $f_z\left(z\left(\mathbf{X}_{e},\boldsymbol{\beta}\right)|\boldsymbol{\beta}\right)$
is uniformly lowered bounded from zero over such subset. 
\end{remark}
The following lemma will be useful in the proof of our theorem. 
\begin{lemma}
\label{lem:3.1}Suppose that \autoref{assu1}, \autoref{assump:2}(i)-(iii),
\autoref{assu:3}, and \autoref{assu:4} hold with $\upsilon_{G}=3$,
$\upsilon_{K}=2$, and $\upsilon_{f}=3$. Define $\psi\left(n,p,h\right)=h^{-1}\sqrt{\log\left(pnh^{-1}\right)/n}+h^{2}.$ 
If $h_{n}\rightarrow0$ and $p^{\frac{5p+1}{2\left(p+1\right)}}\psi^{\frac{1}{p+1}}\left(n,p,h_{n}\right)\rightarrow0$
further hold, we have that 
\begin{align*}
\sup_{\boldsymbol{\beta}\in\mathcal{B}}\left\Vert \frac{1}{n}\sum_{i=1}^{n}\widehat{G}\left(\left.z\left(\boldsymbol{\mathbf{X}}_{e,i},\boldsymbol{\beta}\right)\right|\boldsymbol{\beta}\right)\boldsymbol{\mathbf{X}}_{i}-\mathbb{E}\left[L\left(z\left(\boldsymbol{\mathbf{X}}_{e,i},\boldsymbol{\beta}\right),\boldsymbol{\beta}\right)\boldsymbol{\mathbf{X}}_{i}\right]\right\Vert = & O_{p}\left(p^{\frac{5p+1}{2\left(p+1\right)}}\psi^{\frac{1}{p+1}}\left(n,p,h_{n}\right)\right).
\end{align*}
\end{lemma}
\begin{proof}[Proof of \autoref{lem:3.1}]
See \autoref{appendixA}.
\end{proof}

\label{rem3.2} \autoref{lem:3.1} implies that $\frac{1}{n}\sum_{i=1}^{n}\widehat{G}\left(\left.Z\left(\mathbf{X}_{e,i},\boldsymbol{\beta}\right)\right|\boldsymbol{\beta}\right)\boldsymbol{\mathbf{X}}_{i}$
will be closer to $\mathbb{E}\left[L\left(z\left(\boldsymbol{\mathbf{X}}_{e,i},\boldsymbol{\beta}\right),\boldsymbol{\beta}\right)\boldsymbol{\mathbf{X}}_{i}\right]$
uniformly with respect to $\boldsymbol{\beta}$ as $n$ increases.
Note that such uniform convergence
results  are free of trimming; we do not need
to trim $\boldsymbol{\mathbf{X}}_{e,i}$ even when the density of
$z\left(\boldsymbol{\mathbf{X}}_{e,i},\boldsymbol{\beta}\right)$
is small. So even when $\widehat{G}\left(\left.z\left(\boldsymbol{\mathbf{X}}_{e,i},\boldsymbol{\beta}\right)\right|\boldsymbol{\beta}\right)$
is a poor estimator for $L\left(z\left(\boldsymbol{\mathbf{X}}_{e,i},\boldsymbol{\beta}\right),\boldsymbol{\beta}\right)$
for some $\boldsymbol{\mathbf{X}}_{e,i}$ and $\boldsymbol{\beta}$,
our results are still valid. While on the same time, the cost of not
conducting any trimming is that our guaranteed convergence rate depends
heavily on the dimensionality. As is required in  \autoref{lem:3.1},
the dimension $p$ must satisfy $p^{\frac{5p+1}{2\left(p+1\right)}}\psi^{\frac{1}{p+1}}\left(n,p, h_{n}\right)\rightarrow0$.
Suppose that $p/n\rightarrow0$ and we choose $h_{n}=\left(\left(\log n\right)/n\right)^{1/6}$,
we have that $\psi\left(n,p,h_{n}\right)\sim\left(\left(\log n\right)/n\right)^{1/3}$.
This implies that when $p$ is fixed, the convergence rate in 
\autoref{lem:3.1} is $\left(\left(\log n\right)/n\right)^{1/3\left(p+1\right)}$.
When $p$ increases with $n$, the dimension $p$ should satisfy 
 $p\log p=O\left(\log n\right)$, implying that  $p$ is allowed
to increase only mildly with $n$. The restriction on $p$ basically
comes from the fact that as $\mathbf{X}_{e,i}$ moves towards the
boundary of $\mathcal{X}_{e}$, the density of random variable $z\left(\mathbf{X}_{e,i},\boldsymbol{\beta}\right)$
decreases faster towards zero given a larger $p$, which makes the
convergence rate sensitive to the increase of $p$.

For notational simplicity, in the following we denote $z\left(\mathbf{X}_{e,i},\boldsymbol{\beta}_{k}\right)$
and $z\left(\mathbf{X}_{e,i},\boldsymbol{\beta}^{\star}\right)$
as $z_{i,k}$ and $z_{i}^{\star}$. Based on the results in 
\autoref{lem:3.1}, we have that under all conditions as imposed in \autoref{lem:3.1}, there holds 
\begin{align}\label{kernel_representation}
\boldsymbol{\beta}_{k+1} & =\boldsymbol{\beta}_{k}-\delta_{k}\mathbb{E}\left[\left(L\left(z_{i,k},\boldsymbol{\beta}_{k}\right)-G\left(z_{i}^{\star}\right)\right)\cdot\mathbf{X}_{i}\right]+\delta_{k}\cdot\left(\text{small order terms}\right).
\end{align}
Note that $z_{i,k}=z_{i}^{\star}+\mathbf{X}_{i}^{\mathrm{T}}\Delta\boldsymbol{\beta}_{k}$
and $L\left(z_{i,k},\boldsymbol{\beta}_{k}\right)=\int_{\mathcal{X}}G\left(z_{i,k}-\mathbf{X}^{\mathrm{T}}\Delta\boldsymbol{\beta}_{k}\right)f_{\mathbf{X}|z}\left(\left.\mathbf{X}\right|z_{i,k},\boldsymbol{\beta}_{k}\right)d\mathbf{X}$,
so $\left(L\left(z_{i,k},\boldsymbol{\beta}_{k}\right)-G\left(z_{i}^{\star}\right)\right)\cdot\mathbf{X}_{i}$
equals to 
\begin{align}\label{kernel_representation_mean_value}
 & \left\{ \int_{\mathcal{X}}\left[G\left(z_{i}^{\star}+\mathbf{X}_{i}^{\mathrm{T}}\Delta\boldsymbol{\beta}_{k}-\mathbf{X}^{\mathrm{T}}\Delta\boldsymbol{\beta}_{k}\right)-G\left(z_{i}^{\star}\right)\right]f_{\mathbf{X}|z}\left(\left.\mathbf{X}\right|z_{i,k},\boldsymbol{\beta}_{k}\right)d\mathbf{X}\right\} \cdot\mathbf{X}_{i} \nonumber \\ 
 & =\int_{0}^{1}\int_{\mathcal{X}}\left[G^{\prime}\left(z_{i}^{\star}+t\left(\mathbf{X}_{i}-\mathbf{X}\right)^{\mathrm{T}}\Delta\boldsymbol{\beta}_{k}\right)f_{\mathbf{X}|z}\left(\left.\mathbf{X}\right|z_{i,k},\boldsymbol{\beta}_{k}\right)\left(\mathbf{X}_{i}\boldsymbol{\mathbf{X}}_{i}^{\mathrm{T}}-\mathbf{X}_{i}\mathbf{X}^{\mathrm{T}}\right)\right]\Delta\boldsymbol{\beta}_{k}d\mathbf{X}dt,
\end{align}
where the integration is understood to be element-wise. To further
simplify our notation, define 
\[
W\left(\boldsymbol{\mathbf{X}}_{e},\widetilde{\boldsymbol{\mathbf{X}}}_{e},\boldsymbol{\beta}\right)=G^{\prime}\left(z\left(\boldsymbol{\mathbf{X}}_{e},\boldsymbol{\beta}^{\star}\right)+\left(\boldsymbol{\mathbf{X}}-\widetilde{\boldsymbol{\mathbf{X}}}\right)^{\mathrm{T}}\Delta\boldsymbol{\beta}\right)f_{\boldsymbol{X}|z}\left(\left.\widetilde{\boldsymbol{\mathbf{X}}},\right|z\left(\boldsymbol{\mathbf{X}}_{e},\boldsymbol{\beta}\right),\boldsymbol{\beta}\right),
\]
\[
V\left(\boldsymbol{\mathbf{X}}_{e},\widetilde{\boldsymbol{\mathbf{X}}}_{e},\boldsymbol{\beta}\right)=\left(\boldsymbol{\boldsymbol{\mathbf{X}}}\boldsymbol{\boldsymbol{\mathbf{X}}}^{\mathrm{T}}-\boldsymbol{\boldsymbol{\mathbf{X}}}\widetilde{\boldsymbol{\boldsymbol{\mathbf{X}}}}^{\mathrm{T}}\right)W\left(\boldsymbol{\mathbf{X}}_{e},\widetilde{\boldsymbol{\mathbf{X}}}_{e},\boldsymbol{\beta}\right),
\]
and 
\[
\varLambda\left(\boldsymbol{\beta}\right)=\mathbb{E}\left[\int_{\mathcal{X}}V\left(\boldsymbol{\mathbf{X}}_{e,i},\mathbf{X}_{e},\boldsymbol{\beta}\right)d\mathbf{X}\right],
\]
we have that 
\[
\mathbb{E}\left[\left(L\left(z_{i,k},\boldsymbol{\beta}_{k}\right)-G\left(z_{i}^{\star}\right)\right)\cdot\mathbf{X}_{i}\right]=\int_{0}^{1}\varLambda\left(\boldsymbol{\beta}^{\star}+t\Delta\boldsymbol{\beta}_{k}\right)\Delta\boldsymbol{\beta}_{k}dt,
\]
which indicates that 
\[
\Delta\boldsymbol{\beta}_{k+1}=\left\{ \int_{0}^{1}\left(I_{p}-\delta_{k}\varLambda\left(\boldsymbol{\beta}^{\star}+t\Delta\boldsymbol{\beta}_{k}\right)\right)dt\right\} \Delta\boldsymbol{\beta}_{k}+\delta_{k}\cdot\left(\text{small order terms}\right).
\]
To ensure that with probability going to 1 the above iteration shrinks
$\left\Vert \Delta\boldsymbol{\beta}_{k}\right\Vert $, we make the
following assumption.
\begin{assumption}
\label{assu:5}There hold 
\[
\sup_{\boldsymbol{\beta}\in\mathcal{B}}\overline{\lambda}\left(\varLambda\left(\boldsymbol{\beta}\right)+\varLambda^{\mathrm{T}}\left(\boldsymbol{\beta}\right)\right)\leq\overline{\lambda}_{\varLambda}<\infty,
\]
and 
\[
\inf_{\boldsymbol{\beta}\in\mathcal{B}}\underline{\lambda}\left(\varLambda\left(\boldsymbol{\beta}\right)+\varLambda^{\mathrm{T}}\left(\boldsymbol{\beta}\right)\right)\geq\underline{\lambda}_{\varLambda}>0.
\]
\end{assumption}

Based on the above assumptions, we have the following result.
\begin{theorem}
\label{thm:3.1}Suppose that \autoref{assu1}, \autoref{assump:2}(i)--(iii), \autoref{assu:3}--\autoref{assu:5} hold with $\upsilon_{G}=3$, $\upsilon_{K}=2$,
and  $\upsilon_{f}=3$,  $\delta_{k}=\delta$ such that $\delta<\min\left\{ 1/\left(2\underline{\lambda}_{\varLambda}\right),1/\left(4p^{2}\left\Vert G^{\prime}\right\Vert _{\infty}\right)\right\} $,  and that $\boldsymbol{\beta}$ is updated under (\ref{kernel_estimator})
and (\ref{update}). 
Define 
\[
k_{1,n}^{KBGD}=\frac{\log\left(\left\Vert \Delta\boldsymbol{\beta}_{1}\right\Vert \right)-\log\left(p^{\frac{5p+1}{2\left(p+1\right)}}\psi^{\frac{1}{p+1}}\left(n,p,h_{n}\right)\right)}{-\log\left(1-\delta\underline{\lambda}_{\varLambda}/4\right)}.
\]
Then if $h_{n}\rightarrow0$ and $p^{\frac{5p+1}{2\left(p+1\right)}}\psi^{\frac{1}{p+1}}\left(n,p,h_{n}\right)\rightarrow0$
hold, we have that 
\[
\sup_{k\geq k_{1,n}^{KBGD}+1}\left\Vert \Delta\boldsymbol{\beta}_{k}\right\Vert =O_{p}\left(p^{\frac{5p+1}{2\left(p+1\right)}}\psi^{\frac{1}{p+1}}\left(n,p,h_{n}\right)\right).
\]
In particular, if $h_{n}$ is chosen such that $h_{n}=\left(\left(\log n\right)/n\right)^{1/6}$,
then 
\[
\sup_{k\geq k_{1,n}^{KBGD}+1}\left\Vert \Delta\boldsymbol{\beta}_{k}\right\Vert =O_{p}\left(p^{\frac{5p+1}{2\left(p+1\right)}}\left(\frac{\log n}{n}\right)^{\frac{1}{3p+3}}\right).
\]
\end{theorem}

\begin{proof}[Proof of \autoref{thm:3.1}] See \autoref{appendixB}. 
\end{proof}
\autoref{thm:3.1} implies that the iterative estimator based
on (\ref{kernel_estimator}) and (\ref{update}) is consistent under
increasing dimensions, no matter whether the starting point is close
to the unknown true parameter or not. However, the convergence speed heavily depends on the dimensionality
of the problem, $p$, even when $p$ is fixed. This is not ideal under
our single-index setup but is not surprising since our algorithm does
not involve any trimming procedure as we have discussed in 
\autoref{rem3.2}.

We proceed to establish the asymptotic normality of
the KBGD estimator. Due to technical difficulties, throughout the following analysis in this section we only consider the case where 
$p$ is fixed. As we can see in  \autoref{thm:3.1}, even in the
case of fixed dimensionality, the guaranteed convergence rate of the KBGD estimator
based on (\ref{kernel_estimator}) and (\ref{update}) is at best
$\left(\left(\log n\right)/n\right)^{\frac{1}{3p+3}}$, which still
depends on $p$. To obtain asymptotic normality, we need to slightly
modify our algorithm to get rid of the dependence on dimensionality. In particular,
we introduce trimming to our algorithm. When updating the parameter,
we only use observations that fall into a pre-selected region as did
in \citet{ichimura1993semiparametric}. In particular, the algorithm is modified as, 
\begin{equation}
\boldsymbol{\beta}_{k+1}=\boldsymbol{\beta}_{k}-\frac{\delta_{k}}{n}\sum_{i=1}^{n}I_{i}^{\phi}\cdot\left(\widehat{G}\left(\left.z_{i,k}\right|\boldsymbol{\beta}_{k}\right)-y_{i}\right)\boldsymbol{\mathbf{X}}_{i},\label{eq:truncation_update}
\end{equation}
where $\widehat{G}\left(\left.z_{i,k}\right|\boldsymbol{\beta}_{k}\right)=\widehat{G}\left(\left.z\left(\boldsymbol{\mathbf{X}}_{e,i},\boldsymbol{\beta}_{k}\right)\right|\boldsymbol{\beta}_{k}\right)$
is defined in (\ref{kernel_estimator}), $I_{i}^{\phi}=I\left(\boldsymbol{\mathbf{X}}_{e,i}\in\mathcal{X}_{e}^{\phi}\right)$,
and $\mathcal{X}_{e}^{\phi}$ is a subset of $\mathcal{X}_{e}$
given by
\begin{equation}
\mathcal{X}_{e}^{\phi}=\left\{ \boldsymbol{\mathbf{X}}_{e}\in\mathcal{X}_{e}:\left|X_{j}\right|\leq1-\phi,0\leq j\leq p\right\} \label{eq:truncation_seet}
\end{equation}
for some $\phi>0$ whose value will be determined later. Different
from (\ref{update}), the update of $\boldsymbol{\beta}_{k}$ based
on (\ref{eq:truncation_update}) uses only a subset of the whole sample
for which the covariate vector $\boldsymbol{\mathbf{X}}_{e,i}$ falls
into $\mathcal{X}_{e}^{\phi}$. The reason why we choose
the trimming set as in (\ref{eq:truncation_seet}) is that, as we
show in the \autoref{appendixA}, for any $0<\phi<1$, there holds 
$
\inf_{\left(\mathbf{X}_{e},\boldsymbol{\beta}\right)\in\mathcal{X}_{e}^{\phi}\times\mathcal{B}}f_{z}\left(\left.z\left(\mathbf{X}_{e},\boldsymbol{\beta}\right)\right|\boldsymbol{\beta}\right)\geq C\phi^{p}p^{-p}
$
for some constant $C>0$ that depends on $\phi$. When $p$ and $\phi$
are both fixed, $f_{z}\left(\left.z\left(\boldsymbol{\mathbf{X}}_{e},\boldsymbol{\beta}\right)\right|\boldsymbol{\beta}\right)$
is uniformly lower bounded from zero for any combination $\left(\mathbf{X}_{e},\boldsymbol{\beta}\right)\in\mathcal{X}_{e}^{\phi}\times\mathcal{B}$,
so the uniform estimation accuracy of $L\left(z\left(\boldsymbol{\mathbf{X}}_{e,i},\boldsymbol{\beta}\right),\boldsymbol{\beta}\right)$
over $\boldsymbol{\mathbf{X}}_{e,i}$ and $\boldsymbol{\beta}$ will
be improved. Note that trimming will cause some efficiency loss by
dropping some observations,
but such loss can be controlled to be small if we choose $\phi$ to
be close to zero. We also point  that trimming is
only applied to the update of the parameter; when nonparametrically
estimating  $G$, we still use all the data
points.

To simplify our following notation, given the trimming parameter $\phi$,
we denote $I^{\phi} \cdot \mathbf{X}$ as $\boldsymbol{\mathbf{X}}^{\phi}$.
We also define
\[
\varLambda_{\phi}\left(\boldsymbol{\beta}\right)=\mathbb{E}\left[I_{i}^{\phi}\cdot\int_{\mathcal{X}}V\left(\boldsymbol{\mathbf{X}}_{e,i},\boldsymbol{\mathbf{X}}_{e},\boldsymbol{\beta}\right)d\boldsymbol{\mathbf{X}}\right].
\]
The following theorem provides a counterpart to the results in 
\autoref{thm:3.1}. 
\begin{theorem}
\label{thm:3.2}Suppose that all the assumptions and conditions
on $\upsilon_{G}$, $\upsilon_{K}$, and $\upsilon_{f}$ in 
\autoref{thm:3.1} hold. Suppose moreover that  $h_{n}\rightarrow0$, $\delta_{k}=\delta<\min\left\{ 1/\left(2\underline{\lambda}_{\varLambda}\right),1/\left(4p^{2}\left\Vert G^{\prime}\right\Vert _{\infty}\right)\right\} $, 
$\phi<\delta\underline{\lambda}_{\varLambda}/\left(16p^{2}\left\Vert G^{\prime}\right\Vert _{\infty}\zeta\right)$, and that $\boldsymbol{\beta}$ is updated under (\ref{kernel_estimator})
and (\ref{eq:truncation_update}). Define 
\[
\widetilde{k}_{1,n}^{KBGD}=\frac{\log\left(\left\Vert \Delta\boldsymbol{\beta}_{1}\right\Vert \right)-\log\left(\psi\left(n,p,h_{n}\right)\right)}{-\log\left(1-\delta\underline{\lambda}_{\varLambda}/8\right)},
\]
then there holds 
\[
\sup_{k\geq\widetilde{k}_{1,n}^{KBGD}+1}\left\Vert \Delta\boldsymbol{\beta}_{k}\right\Vert =O_{p}\left(\psi\left(n,p,h_{n}\right)\right).
\]
\end{theorem}
\begin{proof}[Proof of \autoref{thm:3.2}]
	See \autoref{appendixB}.
\end{proof}

Note that when $p$ is fixed, $\psi\left(n,p,h_{n}\right)$
 no longer depends on $p$ asymptotically. The improvement over the convergence
rate basically comes from the improvement of the uniform convergence
rate of the kernel estimator due to trimming. Also note that under
trimming, the minimum number of iteration in  \autoref{thm:3.1}(i),
$\widetilde{k}_{1,n}^{KBGD}$, is of order $\log n$ as long as $nh_{n}\rightarrow\infty$.
This implies that under trimming, a faster convergence rate is guaranteed
with the minimum number of iterations being of the same magnitude
as that of the estimator without trimming. 

We now proceed to establish the asymptotic normality of $\boldsymbol{\beta}_{k}$.
Define 

\[
\boldsymbol{\xi}_{n}^{\phi}=\frac{1}{n}\sum_{i=1}^{n}\left(\widehat{G}\left(\left.z_{i}^{\star}\right|\boldsymbol{\beta}^{\star}\right)-y_{i}\right)\mathbf{X}_{i}^{\phi}.
\]
We note that 
\begin{align}
\Delta\boldsymbol{\beta}_{k+1} & =\Delta\boldsymbol{\beta}_{k}-\frac{\delta_{k}}{n}\sum_{i=1}^{n}\left(\widehat{G}\left(\left.z_{i,k}\right|\boldsymbol{\beta}_{k}\right)-y_{i}\right)\mathbf{X}_{i}^{\phi},\nonumber \\
 & =\Delta\boldsymbol{\beta}_{k}-\frac{\delta_{k}}{n}\sum_{i=1}^{n}\left(\widehat{G}\left(\left.z_{i,k}\right|\boldsymbol{\beta}_{k}\right)-\widehat{G}\left(\left.z_{i}^{\star}\right|\boldsymbol{\beta}^{\star}\right)\right)\mathbf{X}_{i}^{\phi}-\delta_{k}\boldsymbol{\xi}_{n}^{\phi}\nonumber \\
 & =\int_{0}^{1}\left\{ I_{p}-\frac{\delta_{k}}{n}\sum_{i=1}^{n}\left[\mathbf{X}_{i}^{\phi}\left.\frac{\partial\widehat{G}\left(\left.z\left(\boldsymbol{X}_{e,i},\boldsymbol{\beta}\right)\right|\boldsymbol{\beta}\right)}{\partial\boldsymbol{\beta}^{\mathrm{T}}}\right|_{\boldsymbol{\beta}=\boldsymbol{\beta}^{\star}+t\Delta\boldsymbol{\beta}_{k}}\right]\right\} dt\Delta\boldsymbol{\beta}_{k}-\delta_{k}\boldsymbol{\xi}_{n}^{\phi},\label{eq:gradient_update}
\end{align}
where the integration is understood to be element-wise. To understand
the properties of the above algorithm, we need the following lemmas.
\begin{lemma}
\label{lem3.2}Suppose that all the assumptions in  \autoref{thm:3.1}
hold with $\upsilon_{G}=4$, $\upsilon_{K}=3$, and $\upsilon_{f}=4$.
For any sequence of subset $\left\{ \mathcal{B}_{n}\right\} _{n=1}^{\infty}$
with $\mathcal{B}_{n}\subseteq\mathcal{B}$, we have that 
\[
\sup_{\boldsymbol{\beta}\in\mathcal{B}_{n}}\left\Vert \frac{1}{n}\sum_{i=1}^{n}\mathbf{X}_{i}^{\phi}\frac{\partial\widehat{G}\left(\left.z\left(\mathbf{X}_{e,i},\boldsymbol{\beta}\right)\right|\boldsymbol{\beta}\right)}{\partial\boldsymbol{\beta}^{\mathrm{T}}}-\varLambda_{\phi}\left(\boldsymbol{\beta}\right)\right\Vert =O_{p}\left(h_{n}^{-2}\sqrt{\left(\log\left(nh_{n}^{-1}\right)\right)/n}+h_{n}^{3}+\sup_{\boldsymbol{\beta}\in\mathcal{B}_{n}}\left\Vert \Delta\boldsymbol{\beta}\right\Vert \right).
\]
\end{lemma}
\begin{proof}[Proof of \autoref{lem3.2}]
See \autoref{appendixA}.
\end{proof}
\begin{lemma}
\label{lem3.3}Suppose that all the assumptions in \autoref{thm:3.1}
hold with $\upsilon_{G}=4$, $\upsilon_{K}=3$, and $\upsilon_{f}=4$.
If $h_{n}$ is chosen such that $h_{n}^{6}n\rightarrow0$, we have
that $\sqrt{n}\boldsymbol{\xi}_{n}^{\phi}\rightarrow_{d}N\left(0,\Sigma_{\boldsymbol{\xi}}^{\phi}\right)$,
where 
\[
\Sigma_{\boldsymbol{\xi}}^{\phi}=\mathbb{E}\left[\left(1-G\left(z_{i}^{\star}\right)\right)G\left(z_{i}^{\star}\right)\left(\mathbf{X}_{i}^{\phi}-\mathbb{E}\left(\left.\mathbf{X}_{i}^{\phi}\right|z_{i}^{\star}\right)\right)\left(\mathbf{X}_{i}^{\phi}-\mathbb{E}\left(\left.\mathbf{X}_{i}^{\phi}\right|z_{i}^{\star}\right)\right)^{\mathrm{T}}\right].
\]
\end{lemma}
\begin{proof}[Proof of \autoref{lem3.3}]
See \autoref{appendixA}.
\end{proof}
Now we are in a position to illustrate the results of the asymptotic
normality of our KBGD estimator.
\begin{theorem}
\label{thm:3.3}Suppose that all the assumptions in \autoref{thm:3.1}
hold with $\upsilon_{G}=4$, $\upsilon_{K}=3$, and $\upsilon_{f}=4$.
Suppose moreover that 
$\delta_{k}=\delta<\min\left\{ 1/\left(2\underline{\lambda}_{\varLambda}\right),1/\left(4p^{2}\left\Vert G^{\prime}\right\Vert _{\infty}\right)\right\} $, $\phi<\delta \underline{\lambda}_{\varLambda}/\left(16p^{2}\left\Vert G^{\prime}\right\Vert _{\infty}\zeta\right)$, 
$h_{n}$ is chosen such that $nh_{n}^{6}\rightarrow0$ and $h_{n}^{4}n/\left(\log n\right)^{2}\rightarrow\infty$,
and that  $\boldsymbol{\beta}$ is updated under (\ref{kernel_estimator})
and (\ref{eq:truncation_update}). Then 

(i) There holds
\[
\sup_{k\geq\widetilde{k}_{1,n}^{KBGD}+k_{2,n}^{KBGD}+1}\left\Vert \Delta\boldsymbol{\beta}_{k}\right\Vert =O_{p}\left(n^{-1/2}\right),
\]
where  $k_{2,n}^{KBGD}$ is given by 
\[
k_{2,n}^{KBGD}=\frac{\log\left(n^{1/2}\right)+\log\left(\psi\left(n,p,h_{n}\right)\right)}{-\log\left(1-\delta\underline{\lambda}_{\varLambda}/16\right)};
\]

(ii) Define $\widehat{\boldsymbol{\beta}}=\widehat{\boldsymbol{\beta}}_{k}$
for any $k -\widetilde{k}_{1,n}^{KBGD}-k_{2,n}^{KGBD}  \rightarrow\infty$, we have that 
\[
\sqrt{n}\left(\widehat{\boldsymbol{\beta}}-\boldsymbol{\beta}^{\star}\right)\rightarrow N\left(0,\Sigma_{\boldsymbol{\beta}}^{\phi}\right),
\]
where $\Sigma_{\boldsymbol{\beta}}^{\phi}=\varLambda_{\phi}^{-1}\left(\boldsymbol{\beta}^{\star}\right)\Sigma_{\boldsymbol{\xi}}^{\phi}\left(\varLambda_{\phi}^{-1}\left(\boldsymbol{\beta}^{\star}\right)\right)^{\mathrm{T}}$. 
\end{theorem}
\begin{proof}[Proof of \autoref{thm:3.3}]
See \autoref{appendixB}.
\end{proof}
We introduce the estimator for the variance matrix, based on which the confidence interval of $\boldsymbol{\beta}^{\star}$ can be then constructed. 
\begin{theorem}\label{thm:3.4}
Suppose that all the assumptions and conditions in \autoref{thm:3.3}
hold. Suppose also that $\widehat{\boldsymbol{\beta}}$ is defined
as in \autoref{thm:3.3}. Define 
\[
\widehat{\Sigma}_{\boldsymbol{\xi}}^{\phi}=\frac{1}{n}\sum_{i=1}^{n}\left(\widehat{G}_{i}\left(1-\widehat{G}_{i}\right)\left(\mathbf{X}_{i}^{\phi}-\widehat{\mathbb{E}}\left(\left.\mathbf{X}_{i}^{\phi}\right|\widehat{z}_{i}\right)\right)\left(\mathbf{X}_{i}^{\phi}-\widehat{\mathbb{E}}\left(\left.\mathbf{X}_{i}^{\phi}\right|\widehat{z}_{i}\right)\right)^{\mathrm{T}}\right),
\]
and 
\[
\widehat{\varLambda}_{\phi}\left(\widehat{\boldsymbol{\beta}}\right)=\frac{1}{n}\sum_{i=1}^{n}\mathbf{X}_{i}^{\phi}\frac{\partial\widehat{G}\left(\left.z\left(\mathbf{X}_{e,i},\widehat{\boldsymbol{\beta}}\right)\right|\widehat{\boldsymbol{\beta}}\right)}{\partial\boldsymbol{\beta}^{\mathrm{T}}},
\]
where 
\[
\widehat{G}_{i}=\frac{\sum_{j=1}^{n}K_{h_{n}}\left(\widehat{z}_{i}-\widehat{z}_{j}\right)y_{j}}{\sum_{j=1}^{n}K_{h_{n}}\left(\widehat{z}_{i}-\widehat{z}_{j}\right)},\ \widehat{\mathbb{E}}\left(\left.\mathbf{X}_{i}^{\phi}\right|\widehat{z}_{i}\right)=\frac{\sum_{j=1}^{n}K_{h_{n}}\left(\widehat{z}_{i}-\widehat{z}_{j}\right)\mathbf{X}_{j}^{\phi}}{\sum_{j=1}^{n}K_{h_{n}}\left(\widehat{z}_{i}-\widehat{z}_{j}\right)},
\]
and $\widehat{z}_{i}=X_{0,i}+\mathbf{X}_{i}^{\mathrm{T}}\widehat{\boldsymbol{\beta}}$.
Then we have that 
\[
\left\Vert \widehat{\varLambda}_{\phi}^{-1}\left(\widehat{\boldsymbol{\beta}}\right)\widehat{\Sigma}_{\boldsymbol{\xi}}^{\phi}\left(\widehat{\varLambda}_{\phi}^{-1}\left(\widehat{\boldsymbol{\beta}}\right)\right)^{\mathrm{T}}-\Sigma_{\boldsymbol{\beta}}^{\phi}\right\Vert \rightarrow_{p}0.
\]
\end{theorem}
\begin{proof}[Proof of \autoref{thm:3.4}]
See \autoref{appendixB}.
\end{proof}
We finally provide some remarks for the KBGD estimators.
\begin{remark}
\label{rem4}We first provide some remarks on the implementation
of our KBGD estimator. The KBGD estimator might be sensitive to the
data magnitude. So when implementing such an estimator, we recommend
first standardizing the data so that each covariate has zero mean
and unit variance. Note that when constructing the KBGD estimator,
we normalize the coefficient of $X_{0,i}$ to 1, indicating that the
coefficients of $\mathbf{X}_{e,i}$ can not all be zeros. So we need
to test whether at least one covariate affects the conditional probability
of $y_{i}=1$. One option is to run a Logit or Probit regression and
test whether all the coefficients are equal to zero. 

When applying our algorithm, it is also crucial to determine the learning rate $\delta$, bandwidth of kernel estimator $h_{n}$, and terminating
conditions of the algorithm. In \autoref{thm:3.3}, the tuning
parameter $\delta$ is required to be smaller than $1/\left(2\underline{\lambda}_{\varLambda}\right)$
and $1/\left(4p^{2}\left\Vert G^{\prime}\right\Vert _{\infty}\right)$,
neither of which is known. So we recommend setting $\delta$ to be
1 in the first place, and gradually shrink it if the iteration does
not converge. For the choice of the bandwidth $h_{n}$, \autoref{thm:3.3}
requires that $h_{n}$ is chosen such that $nh_{n}^{6}\rightarrow0$
and $nh_{n}^{4}/\left(\log n\right)^{2}\rightarrow\infty$. As a rule
of thumb, we recommend choosing $h_{n}=C\cdot n^{-1/5}$. For the
choice of the constant $C$, we can choose $C=C_{k}=\text{std}\left(z_{i,k}\right)$
for the $k$-th round of iteration and $C=\text{std}\left(\widehat{z}_{i}\right)$
when estimating the variance $\Sigma_{\boldsymbol{\beta}}^{\phi}$.
We finally discuss the terminating conditions. As we show in 
\autoref{thm:3.3}, to obtain root-$n$ consistency and asymptotic normality,
the iteration number is required to be only of order $\log\left(n\right)$.
However, such rule can not be directly applied to determine the number
of iterations since the initial distance $\left\Vert \Delta\boldsymbol{\beta}_{1}\right\Vert $
as well as the lower bounded on the eigenvalues $\underline{\lambda}_{\varLambda}$
are both unknown. We recommend the terminating condition
$\max_{1\leq j\leq p}|\widehat{\beta}_{j,k+1}-\widehat{\beta}_{j,k}|<\varrho$
for some predetermined tolerance $\varrho$. During the simulation,
we choose $\varrho=10^{-5}$. Note that in many cases, $\max_{1\leq j\leq p}|\widehat{\beta}_{j,k+1}-\widehat{\beta}_{j,k}|$
may not be monotonically decreasing with $k$; in some extreme cases,
$\max_{1\leq j\leq p}|\widehat{\beta}_{j,k+1}-\widehat{\beta}_{j,k}|$
may even be oscillating and does not shrink to zero. On these condition,
we recommend decreasing $\delta$ or choosing $h_{n}=C\cdot n^{-1/5}$
with $C=1$ when iterating. If the maximum distance still oscillates,
we recommend stop iteration when the maximum distance achieves its
minimum value.
\end{remark}
\begin{remark}
\label{rem5}Our previous discussion has be confined to the case where
all the covariates are continuously distributed, while our algorithm
can be directly applied to the case where there are discrete covariates
without any modifications. The basic reason is that, in contrast to  the
average derivative approach \citep{stoker1986consistent,powell1989semiparametric} that uses the differentiation with respect to covariates,
the KBGD estimator performs differentiation with respect to the parameters,
so it does not impose requirements on the continuity of the covariates.
It should be noted that we do require at least one continuous covariate
to guarantee identification of the parameters. For simplicity, we
recommend choosing a continuous covariate as the standardization covariate
$X_{0}$. Finally, we point out that stronger assumption should be
imposed to make our results valid when there are discrete covariates.
In particular, suppose that $\mathbf{X}_{e}=\left(\mathbf{X}_{c}^{\mathrm{T}},\mathbf{X}_{d}^{\mathrm{T}}\right)^{\mathrm{T}}$,
where $\mathbf{X}_{c}$ is the collection of all the continuous covariates,
whereas $\mathbf{X}_{d}$ is the collection of all the discrete covariates.
Also denote the density function of $\mathbf{X}_{c}$ conditional
on $\mathbf{X}_{d}$ as $f_{\mathbf{X}_{c}|\mathbf{X}_{d}}\left(\mathbf{X}_{c}|\mathbf{X}_{d}\right)$.
Then we require that all the conditions imposed on the $f_{e}\left(\mathbf{X}_{e}\right)$
hold for $f_{\mathbf{X}_{c}|\mathbf{X}_{d}}\left(\mathbf{X}_{c}|\mathbf{X}_{d}\right)$
for any realizations of $\mathbf{X}_{d}$. 
\end{remark}

\subsection{\label{section4} The SBGD Estimator}

	In the previous section, we introduced the  KBGD
	algorithm, where the update of the parameter is based on a BGD-type procedure while the unknown CDF is replaced with
	its Nadaraya-Watson kernel estimator constructed by the initial parameter. In this section, we consider
	an alternative nonparametric approximation for the unknown CDF based
	on the method of sieves. Given a set of basis functions $\{r_{j}\left(z\right)\}_{j=0}^{\infty}$
	that is complete in $C\left(\mathbb{R}\right)$ space, any smooth
	CDF $G$ can be represented by $G\left(z\right)=\sum_{j=0}^{\infty}\pi_{j}^{\star}r_{j}\left(z\right)$
	for any $z\in R$, where $\{\pi_{j}^{\star}\}_{j=0}^{\infty}$ is
	the unknown coefficients of the basis functions. In practice, to make
	our algorithm tractable, we truncate the sequence of the basis functions and only use the first $q+1$ basis functions for
	approximation, where $q$ increases with sample size $n$ at some
	rate. To approximate $G$, it then remains to provide an estimator for the unknown coefficients
	of the basis functions $\{\pi_{j}^{\star}\}_{j=0}^{q}$. Our estimation
	procedure for $\{\pi_{j}^{\star}\}_{j=0}^{q}$ shares similar intuition
	as the one that motivates the Nadaraya-Watson kernel estimator in
	the previous section. In particular, suppose for a moment that in
	the $k$-th round of update, we start with $\boldsymbol{\beta}_{k}$, which	happens to be identical to the unknown true  parameter $\boldsymbol{\beta}^{\star}$.
	In this case, define $\boldsymbol{r}_{q}(z)=\left(r_{0}\left(z\right),\cdots,r_{q}\left(z\right)\right)^{\mathrm{T}}$
	 and $\boldsymbol{\pi}_{q}^{\star}=\left(\pi_{1}^{\star},\cdots,\pi_{q}^{\star}\right)^{\mathrm{T}}$, we have that 
	\[
	y_{i}=G\left(z_{i,k}\right)+\varepsilon_{i}\approx\boldsymbol{r}_{q}^{\mathrm{T}}\left(z_{i,k}\right)\boldsymbol{\pi}_{q}^{\star}+\varepsilon_{i},
	\]
		where recall that $z_{i,k}=X_{0,i}+\mathbf{X}_{i}^{\mathrm{T}}\boldsymbol{\beta}_{k}$. The above relationship motivates the following OLS estimator for the sieve coefficients
	\begin{equation}
		\widehat{\boldsymbol{\pi}}_{q,n,k}=\left(\sum_{i=1}^{n}\boldsymbol{r}_{q}\left(z_{i,k}\right)\boldsymbol{r}_{q}^{\mathrm{T}}\left(z_{i,k}\right)\right)^{-1}\left(\sum_{i=1}^{n}\boldsymbol{r}_{q}\left(z_{i,k}\right)y_{i}\right).\label{OLS_update}
	\end{equation}
	Given the estimator of the sieve coefficients $\widehat{\boldsymbol{\pi}}_{q,n,k}$,
	the unknown CDF $G$ in the $k$-th round of update is approximated
	by 
	\begin{equation}
		\widehat{G}\left(\left.z\right|\boldsymbol{\beta}_{k}\right)=\boldsymbol{r}_{q}^{\mathrm{T}}\left(z\right)\widehat{\boldsymbol{\pi}}_{n,q,k},\ -\infty<z<\infty.\label{sivev_estimator}
	\end{equation}
	Based on the estimated CDF $\widehat{G}\left(\left.z\right|\boldsymbol{\beta}_{k}\right)$, the update of the parameter can be carried out based on (\ref{update}).
	We iterate sequentially based on (\ref{OLS_update}), (\ref{sivev_estimator})
	and (\ref{update}) until some terminating conditions are satisfied.
	The resulting estimator is then labeled as the \textit{sieve-based batch gradient
	descent estimator} (SBGD  estimator).

\begin{remark}\label{rem6}
		In the above SBGD procedure, we update the sieve parameter based on the
		OLS-type estimation. An alternative procedure can be based on the flexible
		Logit regression proposed by \citet{hirano2003efficient}. The advantage
		of using flexible Logit regression is that the estimated CDF $\widehat{G}\left(\left.z\right|\boldsymbol{\beta}_{k}\right)$
		always falls between 0 and 1 for all $z$, which makes the update more stable. While
		the disadvantage of such update is that the flexible Logit regression
		is based on MLE, which does not allow for an analytical solution. Using
		numerical optimization to solve for the sieve coefficients in each
		round of update will add to additional computational burdens.

\end{remark}
\begin{remark}\label{rem7}

		Compared with the KBGD algorithm, the SBGD procedure has at least
		two advantages. On the one side, the sieve-based approximation for
		the unknown CDF is global and guarantees uniform approximation error
		rate. This allows us to update the parameter without performing any
		form of trimming as we did for the KBGD estimator. Moreover, this allows us to  develop the asymptotic distribution of the SBGD estimator for the case of increasing dimensionality. On the otherhand,
		the KBGD procedure relies on the kernel estimation of CDF $G$ at
		$n$ data points, whose computational complexity of each update is
		of order $O\left(n^{2}\right)$. While the most time-consuming part
		of the SBGD procedure is the OLS procedure (\ref{OLS_update}), whose
		computational complexity is of order $O\left(nq^{2}+q^{3}\right)$.
		When $q/\sqrt{n}\rightarrow0$, the computational burden of SBGD estimator
		will be substantially lower than that of KBGD estimator.

\end{remark}

	Define $R_{q}\left(z\right)=G\left(z\right)-\boldsymbol{r}^{\mathrm{T}}\left(z\right)\boldsymbol{\pi}_{q}^{\star}$,
	$\Gamma_{q,n}\left(\boldsymbol{\beta}\right)=\frac{1}{n}\sum_{i=1}^{n}\boldsymbol{r}_{q}\left(X_{0,i}+\mathbf{X}_{i}^{\mathrm{T}}\boldsymbol{\beta}\right)\boldsymbol{r}_{q}^{\mathrm{T}}\left(X_{0,i}+\mathbf{X}_{i}^{\mathrm{T}}\boldsymbol{\beta}\right)$,
	$\Gamma_{q,n,k}=\Gamma_{q,n}\left(\boldsymbol{\beta}_{k}\right)$,
	and $\mathfrak{X}_{q,n}\left(z,\boldsymbol{\beta}\right)=\frac{1}{n}\sum_{i=1}^{n}\left(\boldsymbol{r}_{q}^{\mathrm{T}}\left(X_{0,i}+\mathbf{X}_{i}^{\mathrm{T}}\boldsymbol{\beta}\right)\Gamma_{q,n}^{-1}\left(\boldsymbol{\beta}\right)\boldsymbol{r}_{q}\left(z\right)\mathbf{X}_{i}\right).$
	Through tedious algebra, we can show that the SBGD procedure has the
	following representation, 
	\begin{align}
		\boldsymbol{\beta}_{k+1} & =\boldsymbol{\beta}_{k}-\frac{\delta_{k}}{n}\sum_{i=1}^{n}\left(\mathbf{X}_{i}-\mathfrak{X}_{q,n}\left(z_{i,k},\boldsymbol{\beta}_{k}\right)\right)\left(G\left(z_{i,k}\right)-G\left(z_{i}^{\star}\right)\right)\nonumber \\
		& -\frac{\delta_{k}}{n}\sum_{i=1}^{n}\mathbf{X}_{i}\boldsymbol{r}_{q}^{\mathrm{T}}\left(z_{i,k}\right)\Gamma_{q,n,k}^{-1}\left(\frac{1}{n}\sum_{j=1}^{n}\boldsymbol{r}_{q}\left(z_{j,k}\right)R_{q}\left(z_{j,k}\right)+\frac{1}{n}\sum_{i=1}^{n}\boldsymbol{r}_{q}\left(z_{j,k}\right)\varepsilon_{j}\right)\nonumber \\
		& +\frac{\delta_{k}}{n}\sum_{i=1}^{n}\left(R_{q}\left(z_{i,k}\right)\mathbf{X}_{i}+\varepsilon_{i}\mathbf{X}_{i}\right),\label{SBGD_expression}
	\end{align}
	where recall that 
 $z_{i}^{\star}=X_{0,i}+\mathbf{X}_{i}^{\mathrm{T}}\boldsymbol{\beta}^{\star}$. To study the properties of the above procedure, we introduce some
	additional assumptions. 
\begin{assumption}
	\label{assu:6}(i) There holds $\max_{0\leq j\leq q}\left\Vert r_{j}\right\Vert _{\infty}\leq D_{q,0}$,
	$\max_{0\leq j\leq q}\left\Vert r_{j}^{\prime}\right\Vert _{\infty}\leq D_{q,1}$,
	and $\max_{0\leq j\leq q}\left\Vert r_{j}^{\prime\prime}\right\Vert _{\infty}\leq D_{q,2}$;
	(ii) Define $\Gamma_{q}\left(\boldsymbol{\beta}\right)=\mathbb{E}\left(\boldsymbol{r}_{q}\left(X_{0}+\mathbf{X}^{\mathrm{T}}\boldsymbol{\beta}\right)\boldsymbol{r}_{q}^{\mathrm{T}}\left(X_{0}+\mathbf{X}^{\mathrm{T}}\boldsymbol{\beta}\right)\right),$
	there hold $\inf_{\boldsymbol{\beta}\in\mathcal{B}}\underline{\lambda}\left(\Gamma_{q}\left(\boldsymbol{\beta}\right)\right)\geq\underline{\lambda}_{\Gamma}>0$
	and $\sup_{\boldsymbol{\beta}\in\mathcal{B}}\overline{\lambda}\left(\Gamma_{q}\left(\boldsymbol{\beta}\right)\right)\leq\overline{\lambda}_{\Gamma}<\infty$
	for all $q$; (iii) There hold $\sup_{z\in R} \left|G\left(z\right) - \boldsymbol{r}^{\mathrm{T}}\left(z\right)\boldsymbol{\pi}_q^{\star}\right| \leq \mathcal{E}_{q,0}$ and $\sup_{z\in R} \left|G^{\prime}\left(z\right) - \left(\boldsymbol{r}^{\prime }\left(z\right)\right)^\mathrm{T}\boldsymbol{\pi}_q^{\star}\right| \leq \mathcal{E}_{q,1}$, where $\boldsymbol{r}^{\prime}(z) = \left(r^{\prime}_0(z), \cdots, r^{\prime}_q(z)\right)^{\mathrm{T}}$. 
\end{assumption}
For any $-\infty<z<\infty$, define the population counterpart of
$\mathfrak{X}_{q,n}\left(z,\boldsymbol{\beta}\right)$ as 

	\[
	\mathfrak{X}_{q}\left(z,\boldsymbol{\beta}\right)=\mathbb{E}\left(\boldsymbol{r}_{q}^{\mathrm{T}}\left(z\left(\mathbf{X}_{e},\boldsymbol{\beta}\right)\right)\Gamma_{q}^{-1}\left(\boldsymbol{\beta}\right)\boldsymbol{r}_{q}\left(z\right)\mathbf{X}\right).
	\]
	Then we have the following lemma.

\begin{lemma}
	\label{lem4.1}Define 
$
	\chi_{1,n}=\sqrt{pq^{2}D_{q,0}^{4}\log\left(pqD_{q,0}D_{q,1}n\right)/n},
$ 
	and 
$
	\chi_{2,n}= \sqrt{p}qD_{q,0}^{2}\left(\chi_{1,n}+ \mathcal{E}_{q,0}\right).
$ 
	Suppose that \autoref{assu1},  \autoref{assump:2}(i)-(iii),
	and \autoref{assu:6} hold, and moreover, $\upsilon_G\geq 1$ and the combination of $p$, $q$ and $\upsilon_{G}$ guarantees that   $\chi_{1,n}\rightarrow0$
	as $n\rightarrow\infty$.  Then the following holds,
	\[
	\boldsymbol{\beta}_{k+1}=\boldsymbol{\beta}_{k}-\delta_{k}\mathbb{E}\left[\left(\mathbf{X}-\mathfrak{X}_{q}\left(z\left(\mathbf{X}_{e},\boldsymbol{\beta}_{k}\right),\boldsymbol{\beta}_{k}\right)\right)\left(G\left(z\left(\mathbf{X}_{e},\boldsymbol{\beta}_{k}\right)\right)-G\left(z\left(\mathbf{X}_{e},\boldsymbol{\beta}^{\star}\right)\right)\right)\right]+\delta_{k}\mathfrak{R}_{n,k},
	\]
	where $\sup_{k\geq1}\left\Vert \mathfrak{R}_{n,k}\right\Vert =O_{p}\left(\chi_{2,n}\right)$.
\end{lemma}

\begin{proof}[Proof of \autoref{lem4.1}]
	See \autoref{appendixA}.
\end{proof}
 Obviously, \autoref{lem4.1} provides a parallel result  to (\ref{kernel_representation}). In particular, define 
\[
\Psi_{q}\left(t,\boldsymbol{\beta}\right)=\mathbb{E}\left[G^{\prime}\left(z\left(\mathbf{X}_{e},\boldsymbol{\beta}^{\star}\right)+t\mathbf{X}^{\mathrm{T}}\Delta\boldsymbol{\beta}\right)\left(\mathbf{X}\mathbf{X}^{\mathrm{T}}-\mathfrak{X}_{q}\left(z\left(\mathbf{X}_{e},\boldsymbol{\beta}\right),\boldsymbol{\beta}\right)\mathbf{X}^{\mathrm{T}}\right)\right],
\]
under all the conditions imposed in \autoref{lem4.1}, we have that 
\begin{align}
	\Delta\boldsymbol{\beta}_{k+1} & =\left\{ \int_{0}^{1}\left(I_{p}-\delta_{k}\Psi_{q}\left(t,\boldsymbol{\beta}_{k}\right)\right)dt\right\} \Delta\boldsymbol{\beta}_{k}+\delta_{k}\mathfrak{R}_{n,k}.\label{contraction_sieve}
\end{align}
Obviously,  (\ref{contraction_sieve}) is also a parallel result to (\ref{kernel_representation_mean_value}). As a result, to ensure that (\ref{contraction_sieve})
actually constitutes a contraction for $\left\Vert \Delta\boldsymbol{\beta}_{k}\right\Vert $,
we impose the following assumption that is similar to \autoref{assu:5}. 

\begin{assumption}
	\label{assu:7}For any $q\geq0$, there hold
	\[
	\inf_{0\leq t\leq1,\boldsymbol{\beta}\in\mathcal{B}}\underline{\lambda}\left(\Psi_{q}\left(t,\boldsymbol{\beta}\right)+\Psi_{q}^{\mathrm{T}}\left(t,\boldsymbol{\beta}\right)\right)\geq\underline{\lambda}_{\Psi}>0,
	\]
	\[
	\sup_{0\leq t\leq1,\boldsymbol{\beta}\in\mathcal{B}}\underline{\lambda}\left(\Psi_{q}\left(t,\boldsymbol{\beta}\right)+\Psi_{q}^{\mathrm{T}}\left(t,\boldsymbol{\beta}\right)\right)\geq\overline{\lambda}_{\Psi}<\infty.
	\]
\end{assumption}
Based on the above assumptions, we have the following result.
\begin{theorem}

		\label{thm4.1}Suppose that  \autoref{assu1}, \autoref{assump:2}(i)-(iii),
\autoref{assu:6} and \autoref{assu:7} hold, $\upsilon_G\geq 1$, and the combination of $p$, $q$ and $\upsilon_{G}$ guarantees that  
		$\chi_{1,n}\rightarrow0$ as $n\rightarrow \infty$. Suppose moreover that the learning rate is chosen such that $\delta_{k}=\delta$
		with
		$
		0<\delta<\min\left\{ 1/\left(2\underline{\lambda}_{\Psi}\right),\underline{\lambda}_{\Psi}/\left(2\left\Vert G^{\prime}\right\Vert _{\infty}^{2}p^{2}\left\{ 1+\underline{\lambda}_{\Gamma}^{-1}qD_{q,0}^{2}\right\} ^{2}\right)\right\}
		$,  and that $\boldsymbol{\beta}$ is updated based on  (\ref{OLS_update}), (\ref{sivev_estimator})
		and (\ref{update}).
		Define 
		\[
		k_{1,n}^{SBGD}=\frac{\log\left(\left\Vert \Delta\boldsymbol{\beta}_{1}\right\Vert \right)-\log\left(\chi_{2,n}\right)}{-\log\left(1-\underline{\lambda}_{\Psi}\delta/4\right)},
		\]
		then we have that 
		\[
		\sup_{k\geq k_{1,n}^{SBGD}+1}\left\Vert \Delta\boldsymbol{\beta}_{k}\right\Vert =O_{p}\left(\chi_{2,n}\right).
		\]

\end{theorem}
\begin{proof}[Proof of \autoref{thm4.1}]
	See \autoref{appendixB}. 
\end{proof}
According to \autoref{thm4.1}, when $\chi_{2,n} \rightarrow 0$ as $n\rightarrow\infty$, the SBGD estimator is consistent as long as the number of updates exceeds $k_{1,n}^{SBGD}$. Based on such consistent estimator, we are ready to establish the asymptotic normality of our SBGD
estimator. Apply the mean value theorem to (\ref{SBGD_expression}), we have that 
\begin{align*}
	\Delta\boldsymbol{\beta}_{k+1} & = \left\{I_p-\delta_{k} \int^{1}_0\frac{1}{n}\sum_{i=1}^{n}G^{\prime}\left(z_i^{\star} + t\mathbf{X}_i^{\mathrm{T}}\Delta\boldsymbol{\beta}_k\right)\left(\mathbf{X}_{i}\mathbf{X}_{i}^{\mathrm{T}}-\mathfrak{X}_{q,n}\left(z_{i,k},\boldsymbol{\beta}_{k}\right)\mathbf{X}_{i}^{\mathrm{T}}\right)dt\right\}\Delta\boldsymbol{\beta}_{k} \\
	& -\frac{\delta_{k}}{n}\sum_{i=1}^{n}\mathbf{X}_{i}\boldsymbol{r}_{q}^{\mathrm{T}}\left(z_{i,k}\right)\Gamma_{q,n,k}^{-1}\left(\frac{1}{n}\sum_{j=1}^{n}\boldsymbol{r}_{q}\left(z_{j,k}\right)R_{q}\left(z_{j,k}\right)+\frac{1}{n}\sum_{i=1}^{n}\boldsymbol{r}_{q}\left(z_{j,k}\right)\varepsilon_{j}\right) \\
	& +\frac{\delta_{k}}{n}\sum_{i=1}^{n}\left(R_{q}\left(z_{i,k}\right)\mathbf{X}_{i}+\varepsilon_{i}\mathbf{X}_{i}\right).
\end{align*}
Define $\Psi_{q}^{\star}=\mathbb{E}\left[G^{\prime}\left(z\left(\mathbf{X}_{e},\boldsymbol{\beta}^{\star}\right)\right)\left(\mathbf{X}\mathbf{X}^{\mathrm{T}}-\mathfrak{X}_{q}\left(z\left(\mathbf{X}_{e},\boldsymbol{\beta}^{\star}\right),\boldsymbol{\beta}^{\star}\right)\mathbf{X}^{\mathrm{T}}\right)\right]$
and $\mathfrak{V}_{q}=\mathbb{E}\left(\mathbf{X}_{i}\boldsymbol{r}_{q}^{\mathrm{T}}\left(z_{i}^{\star}\right)\Gamma_{q}^{-1}\left(\boldsymbol{\beta}^{\star}\right)\right)$. Similar to \autoref{lem3.2} and \autoref{lem3.3}, 
we provide two additional lemmas that are useful to understand the above
algorithm.
\begin{lemma}
	\label{lem4.2}Suppose that  \autoref{assu1}, \autoref{assump:2}(i)-(iii),
	and \autoref{assu:6} hold, $\upsilon_G\geq 2$ and the combination of $p$, $q$ and $\upsilon_{G}$ guarantees that  
	$\chi_{1,n}\rightarrow0$ as $n\rightarrow \infty$. Then for any sequence $\left\{ \mathcal{B}_{n}\right\} _{n=1}^{\infty}$
	with \textup{$\mathcal{B}_{n}\subseteq\mathcal{B}$} we have that
	\begin{align*}
		& \sup_{0\leq t\leq1,\boldsymbol{\beta}\in\mathcal{B}_{n}}\left\Vert \frac{1}{n}\sum_{i=1}^{n}G^{\prime}\left(z_{i}^{\star}+t\mathbf{X}_{i}^{\mathrm{T}}\Delta\boldsymbol{\beta}\right)\left(\mathbf{X}_{i}\mathbf{X}_{i}^{\mathrm{T}}-\mathfrak{X}_{q,n}\left(z\left(\mathbf{X}_{e,i},\boldsymbol{\beta}\right),\boldsymbol{\beta}\right)\mathbf{X}_{i}^{\mathrm{T}}\right)-\Psi_{q}^{\star}\right\Vert \\
		& =O_{p}\left(pqD_{q,0}^{2}\chi_{1,n} + \sqrt{p^3}q^{2}D_{q,0}^{3}D_{q,1}\sup_{\boldsymbol{\beta}\in\mathcal{B}_{n}}\left\Vert \Delta\boldsymbol{\beta}\right\Vert \right).
	\end{align*}
\end{lemma}

\begin{proof}[Proof of \autoref{lem4.2}]
	See \autoref{appendixA}.
\end{proof}
\begin{lemma}
	\label{lem4.3}Suppose that  \autoref{assu1}, \autoref{assump:2}(i)-(iii), \autoref{assu:6}, and \autoref{assu:7}  hold, and the combination of $p$, $q$ and $\upsilon_{G}$ guarantees that  
	$\chi_{1,n}\rightarrow0$ as $n\rightarrow \infty$. Define $\boldsymbol{r}_{q,i,k}=\boldsymbol{r}_{q}\left(z_{i,k}\right)$,  and 
	$R_{q,i,k}=R_q\left(z_{i,k}\right)$.
	Also define 
	\[
	\chi_{3,n}=\sqrt{p^{2}qD_{q,1}^{2}\log\left(pqD_{q,2}n\right)/n},
	\]
	then we have that 
	\begin{align*}
		\sup_{k\geq k_{1,n}^{SBGD}+1} & \left\Vert \frac{1}{n}\sum_{i=1}^{n}\mathbf{X}_{i}\boldsymbol{r}_{q,i,k}^{\mathrm{T}}\Gamma_{q,n,k}^{-1}\left(\frac{1}{n}\sum_{j=1}^{n}\boldsymbol{r}_{q,j,k}R_{q,j,k}+\frac{1}{n}\sum_{j=1}^{n}\boldsymbol{r}_{q,j,k}\varepsilon_{j}\right)+\right.\\
		&
		\left.\frac{1}{n}\sum_{i=1}^{n}R_{q}\left(z_{i,k}\right)\mathbf{X}_{i} -\frac{1}{n}\sum_{i=1}^{n}\mathfrak{X}_{q}\left(z_{i}^{\star},\boldsymbol{\beta}^{\star}\right)\varepsilon_{j}\right\Vert =O_{p}\left(\chi_{4,n}\right),
	\end{align*}
	where $\chi_{4,n}=\sqrt{p}qD_{q,0}^{2}\mathcal{E}_{q,0}+\sqrt{pq}D_{q,0}\chi_{2,n}\chi_{3,n}+\chi_{2,n}\sqrt{p^2q^4D_{q,0}^{6}D_{q,1}^2\left(\log q\right)/n}$. 
\end{lemma}

\begin{proof}[Proof of \autoref{lem4.3}]
	See \autoref{appendixA}.
\end{proof}

Based on the above two lemmas, we are now ready to study the asymptotic distribution of the SBGD estimator.
\begin{theorem}
	\label{thm4.2}Suppose that  \autoref{assu1}, \autoref{assump:2}(i)-(iii),
	\autoref{assu:6} and  \autoref{assu:7} hold, $\upsilon_G\geq 2$, the combination of $p$, $q$ and $\upsilon_{G}$ guarantees that  
	$\chi_{1,n}\rightarrow0$ as $n\rightarrow \infty$,  and that $\boldsymbol{\beta}$ is updated based on  (\ref{OLS_update}), (\ref{sivev_estimator})
	and (\ref{update}). We have
	that
	
	(i) There holds
	\[
	\Delta\boldsymbol{\beta}_{k+1}=\left(I_{p}-\delta\Psi_{q}^{\star}\right)\Delta\boldsymbol{\beta}_{k}+\frac{\delta}{n}\sum_{i=1}^{n}\left(\mathbf{X}_{i} - \mathfrak{X}_{q}\left(z_{i}^{\star},\boldsymbol{\beta}^{\star}\right) \right)\varepsilon_{i}+\widetilde{\mathfrak{R}}_{n,k},
	\]
	where $\sup_{k\geq k_{1,n}^{SBGD}+1}\left\Vert \widetilde{\mathfrak{R}}_{n,k}\right\Vert =O_{p}\left(\chi_{5,n}\right)$
	with 
	\[
	\chi_{5,n}=\sqrt{p}qD_{q,0}^{2}\left(p+qD_{q,0}D_{q,1}\right)\chi_{2,n}^{2}+\chi_{4,n};
	\]
	
	(ii) Define $\widehat{\boldsymbol{\beta}}=\boldsymbol{\beta}_{k+k_{1,n}^{SBGD}+k_{2,n}^{SBGD}+1}$
	with 
	\[
	k_{2,n}^{SBGD}=\frac{-\log\chi_{2,n}+\log\sqrt{n}}{-\log\left(1-\underline{\lambda}_{\Psi}\delta/4\right)},
	\]
	and any $k\geq1$. If the combination of $p$, $q$ and $\upsilon_{G}$ further guarantees that   $\sqrt{n}\chi_{5,n}\rightarrow0$ as $n\rightarrow\infty$,
	we have that
	\[
	\sqrt{n}\left(\widehat{\boldsymbol{\beta}}-\boldsymbol{\beta}^{\star}\right)=\Psi_{q}^{\star-1}\frac{1}{\sqrt{n}}\sum_{i=1}^{n}\left( \mathbf{X}_{i} - \mathfrak{X}_{q}\left(z_{i}^{\star},\boldsymbol{\beta}^{\star}\right) \right)\varepsilon_{i}+o_{p}\left(n^{-\frac{1}{2}}\right).
	\]
	Then for any $p\times1$ vector $\rho$ such that $\left\Vert \rho\right\Vert <\infty$
	and $\frac{1}{\sqrt{n}}\sum_{i=1}^{n}\rho^{\mathrm{T}}\Psi_{q}^{\star-1}\left(\mathbf{X}_{i} - \mathfrak{X}_{q}\left(z_{i}^{\star},\boldsymbol{\beta}^{\star}\right) \right)\varepsilon_{i}\rightarrow_{d}N\left(0,\sigma_{S}^{2}\left(\rho\right)\right)$
	with 
	\[
	\sigma_{S}^{2}\left(\rho\right)=\lim_{n\rightarrow\infty}\rho^{\mathrm{T}}\Psi_{q}^{\star-1}\mathbb{E}\left\{ G\left(z_{i}^{\star}\right)\left(1-G\left(z_{i}^{\star}\right)\right)\left(\mathbf{X}_{i} - \mathfrak{X}_{q}\left(z_{i}^{\star},\boldsymbol{\beta}^{\star}\right) \right)\left( \mathbf{X}_{i} - \mathfrak{X}_{q}\left(z_{i}^{\star},\boldsymbol{\beta}^{\star}\right) \right)^{\mathrm{T}}\right\} \left(\Psi_{q}^{\star-1}\right)^{\mathrm{T}}\rho,
	\]
	there holds \[\sqrt{n}\rho^{\mathrm{T}}\left(\widehat{\boldsymbol{\beta}}-\boldsymbol{\beta}^{\star}\right)\rightarrow_{d}N\left(0,\sigma_{S}^{2}\left(\rho\right)\right)\].
\end{theorem}
\begin{proof}[Proof of \autoref{thm4.2}]
	See \autoref{appendixB}.
\end{proof}

We now provide the estimator for the variance. 

\begin{theorem}\label{thm4.3}
	Suppose that all the conditions listed in  \autoref{thm4.2} hold  and $pq^{2}D_{q,0}^{4}\mathcal{E}_{q,1}\rightarrow 0$ as $n\rightarrow0$. Let
	$\widehat{\boldsymbol{\beta}}$ be as defined as in  \autoref{thm4.2}.
	Define $\widehat{\boldsymbol{r}}_{q,i}=\boldsymbol{r}_{q}\left(z\left(\mathbf{X}_{e,i},\widehat{\boldsymbol{\beta}}\right)\right)$,
	$\widehat{\boldsymbol{r}}_{q,i}^{\prime}=\boldsymbol{r}_{q}^{\prime}\left(z\left(\mathbf{X}_{e,i},\widehat{\boldsymbol{\beta}}\right)\right)$,
	$\widehat{\boldsymbol{\pi}}_{q}=\left(\sum_{i=1}^{n}\widehat{\boldsymbol{r}}_{q,i}\widehat{\boldsymbol{r}}_{q,i}^{\mathrm{T}}\right)^{-1}\left(\sum_{i=1}^{n}\widehat{\boldsymbol{r}}_{q,i}y_{i}\right),$
	$\widehat{G}_{i}=\widehat{\boldsymbol{r}}_{q,i}^{\mathrm{T}}\widehat{\boldsymbol{\pi}},\ \widehat{G}_{i}^{\prime}=\widehat{\boldsymbol{r}}_{q,i}^{\prime\mathrm{T}}\widehat{\boldsymbol{\pi}}_{q},$
	$\widehat{\Psi}_{q,i}^{\star}=\frac{1}{n}\sum_{i=1}^{n}\widehat{G}_{i}^{\prime}\cdot\left(\mathbf{X}_{i}\mathbf{X}_{i}^{\mathrm{T}}-\mathfrak{X}_{q,n}\left(\widehat{z}_{i},\widehat{\boldsymbol{\beta}}\right)\mathbf{X}_{i}^{\mathrm{T}}\right)$, $\widehat{\mathfrak{X}}_{q,i}=\frac{1}{n}\sum_{j=1}^{n}\mathbf{X}_{j}\widehat{\boldsymbol{r}}_{q,j}^{\mathrm{T}}\Gamma_{q,n}^{-1}\left(\widehat{\boldsymbol{\beta}}\right)\widehat{\boldsymbol{r}}_{q,i},$
	and 
	\[
	\widehat{\sigma}_{S}^{2}\left(\rho\right)=\rho^{\mathrm{T}}\widehat{\Psi}_{q}^{\star-1}\frac{1}{n}\sum_{i=1}^{n}\left\{ \widehat{G}_{i}\left(1-\widehat{G}_{i}\right)\left(\mathbf{X}_{i} - \widehat{\mathfrak{X}}_{q,i} \right)\left( \mathbf{X}_{i} - \widehat{\mathfrak{X}}_{q,i} \right)^{\mathrm{T}}\right\} \left(\widehat{\Psi}_{q}^{\star-1}\right)^{\mathrm{T}}\rho,
	\]
	Then for any $p\times1$ vector $\rho$ such that $\left\Vert \rho\right\Vert <\infty$,
	there holds 
	\[
	\left|\widehat{\sigma}_{S}^{2}\left(\rho\right)-\sigma_{S}^{2}\left(\rho\right)\right|\rightarrow_{p}0.
	\]
\end{theorem}

\begin{proof}[Proof of \autoref{thm4.3}]
	See \autoref{appendixB}.
\end{proof}

We finally provide some remarks on the empirical applications of the SBGD estimator.

\begin{remark}\label{rem8}

For the choice of sieve functions, we can use polynomial series for the case where the error term $u_i$ has bounded support and Hermite polynomials for the case where $u_i$ has unbounded support. Note that when using polynomial series $\left\{1, z, z^2, \cdots, z^q\right\}$, the correlation between the sieve functions increases as the approximation order $q$ increases, which may lead to a violation of \autoref{assu:6}(ii). To improve the finite sample performance of our method, we recommend using Chebyshev or Legendre polynomials. Moreover, in the case where $u_i$ has unbounded support, following \citet{bierens2014consistency}, we recommend first conducting the following transformation $G\left(z\right)$ = $\widetilde{G}\left(T\left(z\right)\right)$, where $T:R \mapsto [-1,1]$ is a differentiable function, and then using standard Chebyshev or Legendre polynomials to approximate $\widetilde{G}$. For example, in our following simulations and empirical applications in Section \ref{section6}, we use $T\left(z\right) = 2\pi^{-1}\arctan\left(z\right)$. For the uniform error bound of truncated Legendre polynomials, see \citet{wang2012convergence}.
\end{remark}

\section{\label{section4}Monte Carlo Experiments}

This section conducts Monte Carlo simulations to study the performance
of our KBGD and SBGD estimators. We focus on two aspects of our estimators.
First we study the finite-sample properties of the KBGD
estimator, including the bias and the root mean squared error (RMSE).
Let the $j$-th argument of the true parameter be $\beta_{j}^{\star}$,
and the simulation is repeated $R$ times, where its estimator in
the $r$-th round of simulation is $\widehat{\beta}_{j}^{r}$, then
the bias and RMSE are respectively given by $\text{Bias}=|\frac{1}{R}\sum_{r=1}^{R}(\widehat{\beta}_{j}^{r}-\beta_{j}^{\star})|$
and $\text{RMSE}=\sqrt{\sum_{r=1}^{R}(\widehat{\beta}_{j}^{r}-\beta_{j}^{\star})^{2}/R}.$
We also investigate whether the confidence interval based on the asymptotic
distribution has good coverage rate. We consider nominal coverage
rate $\alpha=0.95$, so the confidence interval for $\beta_{j}^{\star}$
in the $r$-th round of repetition is given by $CI_{j}^{r}=[\widehat{\beta}_{j}^{r}-1.96\cdot\widehat{\text{std}}_{j}^{r},\widehat{\beta}_{j}^{r}+1.96\cdot\widehat{\text{std}}_{j}^{r}]$,
where $\widehat{\text{std}}_{j}^{r}$ is the estimated standard deviation
of $\widehat{\beta}_{j}^{r}$. The actual coverage rate is then given
by $CR=\frac{1}{R}\sum_{r=1}^{R}I(\beta_{j}^{\star}\in CI_{j}^{r}).$

We are also interested in how sensitive our estimators are to the
initial guess of the true parameter. In each repetition of our simulation,
we consider three different initial guesses: the true parameter vector,
the parameter vector estimated based on the Logit regression, and
the parameter with all elements being zeros. If the estimation results
starting from different initial guesses are close or even identical
to each other, the estimation methods are insensitive to the initial
guesses and thus are robust in terms of computation. Denote $\widehat{\boldsymbol{\beta}}_{T}^{r}$,
$\widehat{\boldsymbol{\beta}}_{L}^{r}$, and $\widehat{\boldsymbol{\beta}}_{Z}^{r}$
as the estimators with starting points being true parameter, Logit
estimator, and vector of zeros. We use $S_{L}=\sqrt{\frac{1}{R}\sum_{i=1}^{n}||\widehat{\boldsymbol{\beta}}_{L}^{r}-\widehat{\boldsymbol{\beta}}_{T}^{r}||^{2}}$
and $S_{Z}=\sqrt{\frac{1}{R}\sum_{i=1}^{n}||\widehat{\boldsymbol{\beta}}_{Z}^{r}-\widehat{\boldsymbol{\beta}}_{T}^{r}||^{2}}$
as the measurement of the sensitivity. 

We consider data generating process $y_{i}=I(X_{0,i}+\beta_{1}^{\star}X_{1,i}\cdots+\beta_{10}^{\star}X_{10,i}-u_{i}>0),i=1,2,\cdots,n,$
where data are i.i.d over $i$, and $X_{0,i},X_{1,i},\cdots,X_{10,i},u_{i}$
are also independent. We set $\boldsymbol{\beta}^{\star}=(1,0.5,-0.5,1,-1,2,-2,4,-4,1.5,-1.5)^{\mathrm{T}}$,
$X_{j,i}\sim N\left(0,1\right)$ for $0\leq j\leq8$, $X_{9,i}\sim\text{Bernoulli}\left(1/2\right)$,
$X_{10,i}\sim\text{Poisson}\left(2\right)$, and $u_{i}\sim Cauchy$.
We consider two sample sizes $n=2500$ and $5000$. Finally, for finite-sample
performance, we repeat the simulation 500 times; for sensitivity analysis,
we repeat 100 times. 

\begin{table}
\begin{centering}
\caption{\label{tab1}Finite Sample Performance of KBGD and SBGD Estimators}
\begin{tabular}{>{\centering}p{0.6cm}>{\centering}p{0.9cm}>{\centering}p{0.9cm}>{\centering}p{0.9cm}>{\centering}p{0.9cm}>{\centering}p{0.9cm}>{\centering}p{0.9cm}c>{\centering}p{0.9cm}>{\centering}p{0.9cm}>{\centering}p{0.9cm}>{\centering}p{0.9cm}>{\centering}p{0.9cm}>{\centering}p{0.9cm}}
\hline 
 & \multicolumn{2}{c}{Bias} & \multicolumn{2}{c}{RMSE} & \multicolumn{2}{c}{CR} &  & \multicolumn{2}{c}{Bias} & \multicolumn{2}{c}{RMSE} & \multicolumn{2}{c}{CR}\tabularnewline
\hline 
 & K{\small{}BGD} & S{\small{}BGD} & K{\small{}BGD} & S{\small{}BGD} & K{\small{}BGD} & S{\small{}BGD} &  & K{\small{}BGD} & S{\small{}BGD} & K{\small{}BGD} & S{\small{}BGD} & K{\small{}BGD} & S{\small{}BGD}\tabularnewline
\hline 
\multicolumn{7}{c}{$n=2500$} &  & \multicolumn{6}{c}{$n=5000$}\tabularnewline
\hline 
$\beta_{1}$ & 0.0024 & 0.0031 & 0.1193 & 0.1240 & 0.9600 & 0.9680 &  & 0.0047 & 0.0005 & 0.0844 & 0.0867 & 0.9500 & 0.9600\tabularnewline
$\beta_{2}$ & 0.0002 & 0.0055 & 0.1255 & 0.1336 & 0.9480 & 0.9500 &  & 0.0031 & 0.0074 & 0.0846 & 0.0878 & 0.9520 & 0.9540\tabularnewline
$\beta_{3}$ & 0.0136 & 0.0260 & 0.1544 & 0.1791 & 0.9480 & 0.9460 &  & 0.0004 & 0.0074 & 0.1053 & 0.1112 & 0.9320 & 0.9320\tabularnewline
$\beta_{4}$ & 0.0093 & 0.0213 & 0.1551 & 0.1706 & 0.9500 & 0.9440 &  & 0.0012 & 0.0095 & 0.1035 & 0.1117 & 0.9600 & 0.9500\tabularnewline
$\beta_{5}$ & 0.0257 & 0.0482 & 0.2511 & 0.2968 & 0.9540 & 0.9400 &  & 0.0007 & 0.0168 & 0.1648 & 0.1889 & 0.9400 & 0.9480\tabularnewline
$\beta_{6}$ & 0.0236 & 0.0477 & 0.2502 & 0.2860 & 0.9480 & 0.9580 &  & 0.0121 & 0.0269 & 0.1723 & 0.1931 & 0.9540 & 0.9360\tabularnewline
$\beta_{7}$ & 0.0500 & 0.0964 & 0.4513 & 0.5416 & 0.9640 & 0.9420 &  & 0.0051 & 0.0352 & 0.3083 & 0.3525 & 0.9440 & 0.9420\tabularnewline
$\beta_{8}$ & 0.0447 & 0.0920 & 0.4662 & 0.5441 & 0.9360 & 0.9520 &  & 0.0098 & 0.0394 & 0.3121 & 0.3477 & 0.9420 & 0.9440\tabularnewline
$\beta_{9}$ & 0.0242 & 0.0454 & 0.2921 & 0.3303 & 0.9480 & 0.9500 &  & 0.0072 & 0.0048 & 0.1840 & 0.1909 & 0.9540 & 0.9560\tabularnewline
$\beta_{10}$ & 0.0168 & 0.0338 & 0.1881 & 0.2223 & 0.9520 & 0.9440 &  & 0.0030 & 0.0147 & 0.1247 & 0.1402 & 0.9440 & 0.9380\tabularnewline
\hline 
\end{tabular}
\par\end{centering}
{\footnotesize{}NOTE: For KBGD estimator, we use fourth-order Epanechinikov
kernel to construct the Nadaraya-Watson estimator. We choose $\delta=1$.
In each round of iteration, the bandwidth $h_{n}$ is chosen as $h_{n}=\sigma_{\widehat{z}}\cdot n^{-1/5}$,
where $n$ is sample size, $\sigma_{\widehat{z}}$ is the standard
deviation of $z_{i,k}$, and $z_{i,k}=X_{0,i}+\mathbf{X}_{i}^{\mathrm{T}}\boldsymbol{\beta}_{k}$.
For SBGD estimator, we choose $q=9$ and use Legendre polynomials
with transformation discussed in \autoref{rem8}. For both estimators, the
stopping rule is either $\max_{1\leq j\leq p}|\widehat{\beta}_{j,k+1}-\widehat{\beta}_{j,k}|<10^{-5}$
or $k\geq20000$. The above also applies to our empirical analysis
in Section \ref{section6}. Trimming is ignored during all the simulations.
Due to the outliers of the simulation, we trim out the lower and upper
2\% simulation results and calculate the bias and RMSE. }{\footnotesize\par}
\end{table}

\autoref{tab1} reports the finite-sample properties of our estimators.
It can be seen that our estimators works well in finite sample cases.
Both estimators have small bias, whose RMSE decrease with sample size.
Moreover, the confidence interval constructed based on the asymptotic
variance and normal approximation has actual coverage rate that is
quite close to the nominal rate $0.95$.

\begin{table}
\begin{centering}
\caption{\label{tab2}Sensitivity of KBGD and SBGD Estimators: Fixed Coefficients}
\begin{tabular}{c>{\centering}p{1.3cm}>{\centering}p{1.2cm}>{\centering}p{1.2cm}c>{\centering}p{1.2cm}>{\centering}p{1.2cm}>{\centering}p{1.2cm}}
\hline 
 &  & \multicolumn{2}{c}{Sensitivity} &  & \multicolumn{3}{c}{Running Time}\tabularnewline
\hline 
 & Method & $S_{L}$ & $S_{Z}$ &  & True & Logit & Zeros\tabularnewline
\hline 
\multirow{2}{*}{$n=2500$} & KBGD & 0.0242 & 0.0198 &  & 113.21 & 79.120 & 158.91\tabularnewline
 & SBGD & 0.0175 & 0.0259 &  & 0.9504 & 0.9482 & 1.1587\tabularnewline
\hline 
\multirow{2}{*}{$n=5000$} & KBGD & 0.0241 & 0.0175 &  & 157.48 & 87.954 & 230.07\tabularnewline
 & SBGD & 0.0189 & 0.0282 &  & 1.4644 & 1.4722 & 1.9074\tabularnewline
\hline 
\end{tabular}
\par\end{centering}
{\footnotesize{}NOTE: The running time is in seconds. Due to the outliers
of the simulation, we trim out the lower and upper 2\% simulation
results and calculate the corresponding results. The above also applies
to  \autoref{tab3}.}{\footnotesize\par}
\end{table}

\autoref{tab2} reports the sensitivity of our estimators to the starting
points. We can see that for both estimators, $S_{L}$ and $S_{Z}$
are close to zero, indicating that the resulting estimators starting
from Logit estimator or zeros are almost identical to the ones starting
from the unknown true parameter. Such a result demonstrates that our
algorithms are robust to different initial guesses. We also find that
compared with SBGD, our KBGD estimator takes much longer time to converge. 

\begin{table}
\centering{}\caption{\label{tab3}Sensitivity of KBGD and SBGD Estimators: Random Coefficients}
\begin{tabular}{c>{\centering}p{1.3cm}>{\centering}p{1.2cm}>{\centering}p{1.2cm}c>{\centering}p{1.2cm}>{\centering}p{1.2cm}>{\centering}p{1.2cm}}
\hline 
 &  & \multicolumn{2}{c}{Sensitivity} &  & \multicolumn{3}{c}{Running Time}\tabularnewline
\hline 
 & Method & $S_{L}$ & $S_{Z}$ &  & True & Logit & Zeros\tabularnewline
\hline 
\multirow{2}{*}{$n=2500$} & KBGD & 0.0270 & 0.0214 &  & 122.00 & 74.433 & 166.94\tabularnewline
 & SBGD & 0.0123 & 0.0246 &  & 1.0132 & 0.8252 & 1.2044\tabularnewline
\hline 
\multirow{2}{*}{$n=5000$} & KBGD & 0.0234 & 0.0232 &  & 163.74 & 91.449 & 247.49\tabularnewline
 & SBGD & 0.0077 & 0.0234 &  & 1.5529 & 1.4377 & 1.9217\tabularnewline
\hline 
\end{tabular}
\end{table}

The robustness of our algorithm might be sensitive to the setups of
coefficients. To check whether this is the case, instead of using
the fixed parameters specified before, in each round of simulation
we randomly draw true parameter $\boldsymbol{\beta}^{\star}$ as follows
$\beta_{1}^{\star},\beta_{2}^{\star},\beta_{9}^{\star},\beta_{10}^{\star}\sim N\left(0,1\right)$,
$\beta_{3}^{\star},\beta_{4}^{\star},\beta_{5}^{\star},\beta_{6}^{\star}\sim2N\left(0,1\right)$,
and $\beta_{7}^{\star},\beta_{8}^{\star}\sim4N\left(0,1\right)$.
The simulation results are reported in \autoref{tab3}. We can see
that the results are similar to those under fixed parameters, indicating
that our algorithm is robust to initial point under different parameter
setups.

\section{\label{section6}Empirical Application}

	As an empirical illustration of our new methods, this section applies
	our KBGD and SBGD estimation procedures to study how education affects
	the risk aversion. In the existing researches, it's extensively documented
	that, on the individual level, risk aversion is significantly correlated
	with the level of education, although the directions of correlation
	are mixed, see \citet{outreville2015relationship} for a comprehensive review. In this study, we investigate
	how educational background of the family affects the risk aversion
	of the household as well as household-level investing behaviors. We
	use the national survey data from 2019 China Household Financial Survey
	Project (CHFS) \citep{gan2014data}, which provides household-level
	information over demographics, asset and debt, income and consumption,
	social security and insurance, and various household's subjective
	preferences. The dependent variable we are interested in is the degree
	of risk aversion of the household. In particular, $y_{i}$ is constructed
	to take value of $0$ if the $i$-th household is completely against
	any form of risks and thus is described as being extremely risk averse;
	it takes value of $1$ if the family is willing to bear some form
	of risks when making investments. We study how the probability of
	$y_{i}=1$ is affected by a set of factors based on the binary choice
	model. The key factor that we are particularly interested in is the
	educational backgrounds, which is defined as the year of education
	of the head of the household. We also consider a set of other control
	variables including gender, ethnicity, health conditions, marital
	status, region of residence, economic knowledge, total income and
	total asset, whose impacts on the risk aversion are of interest on
	their own right. See Yao (2023) for detailed discussion on the construction
	of the data sets. 
	
	Before estimation, we normalize all the continuous variables so that
	the resulting variables all have zero mean and unity variance. To
	provide a comparison to the semiparametric estimation results, we
	first conduct parametric Logit regression and report the normalized
	coefficients in regression (I) in  \autoref{empirical_table}. We
	then conduct KBGD and SBGD estimation and report the estimated coefficients
	of education in (II) and (III). As we can see from \autoref{empirical_table},
	no matter which estimation methods we use, the coefficient of educational
	background is estimated to be positive with significance at $1\%$
	level. This implies that, holding other conditions fixed, on average
	an increase in the year of education of the head in the households
	leads to the increase of willingness to bear risks. Comparing the
	semiparametric estimation results with that of Logit regression, we
	can see that the KBGD and SBGD estimators are close to each other,
	which are both smaller than that of Logit regression, indicating that
	parametric estimation might suffer from model misspecification and
	lead to an overestimation of the impacts of education on risk aversion.
	We finally compare the computation time of each method. We can see that
	both KBGD and SBGD estimators take much longer to converge compared
	with the parametric estimation. Comparatively, the SBGD algorithm
	is significantly faster than the KBGD algorithm, which takes over
	two hours to converge. This result supports the use of SBGD algorithm
	when there are data of large scale. 
	
	\begin{table}
		\caption{\label{empirical_table}Estimation Results}
		
		\begin{centering}
			\begin{tabular}{>{\raggedright}p{4cm}>{\centering}m{2.5cm}>{\centering}m{2.5cm}>{\centering}m{2.5cm}}
				\hline 
				& (I) & (II) & (III)\tabularnewline
				\hline 
				Estd. Coefficients & $2.5543^{***}$
				
				$(0.1070)$ & $2.4832^{***}$
				
				$(0.3638)$ & $2.4647^{***}$
				
				$(0.3239)$\tabularnewline
				Num. of Obs. & $26906$ & $26906$ & $26906$\tabularnewline
				Estimation Methods & Logit & KBGD & SBGD\tabularnewline
				Running Time & 1.4276 & 8573.1 & 40.9941\tabularnewline
				Num. of Iteration & -- & 14996 & 12986\tabularnewline
				\hline 
			\end{tabular}
			\par\end{centering}
		Note: For Logit regression, we report the coefficient of education
		divided by that of total asset. For semiparametric estimation, we
		normalize the coefficient of total asset to be 1. The standard deviations
		are reported in the brackets below the coefficients. $^{***}$ indicates
		significance at $1\%$ level. For both KBGD and SBGD estimators, we
		choose $\delta_{k}=1$. For KBGD estimator, we choose $h_{n}=C\cdot n^{-1/5}$
		with $C=C_{k}=\text{std}(z_{i,k})$, and use the fourth-order Epanechinikov
		kernel. For SBGD estimator, we choose $q=9$ and use Legendre polynomials
		with transformation discussed in \autoref{rem8}. The starting point of iteration
		for both KBGD and SBGD estimators is chosen as the origin point with
		all arguments being 0. The stopping rule is set as $\max_{1\leq j\leq p}|\widehat{\beta}_{j,k+1}-\widehat{\beta}_{j,k}|<\varrho$
		with $\varrho=10^{-5}$. Finally, the running time is in second.
	\end{table}

\pagebreak

\section{Conclusions}\label{conclude}
\setcounter{equation}{0}
In this paper, we proposed  new estimation procedures for binary choice and monotonic index models with increasing dimensions. 
Existing semiparametric estimation procedures for this model cannot be implemented in practice when the number of regressors is large. In contrast, our algorithmic based procedures can be used for many regressor models as it involves convex optimization at each iteration of the procedure. We show this iterative procedure also has desirable asymptotic properties when the number of regressors increases with the sample size in ways that are standard in big data literature.

\newpage{}

\appendix
\section{Lemmas and Proofs}\label{appendixA}

This part provides some lemmas that will be used during the establishment
of our results in the main context. If not otherwise stated, the dimension
$p$ of covariate $\mathbf{X}$ is allowed to increase with sample
size $n$. 

\begin{lemma}
	\label{lemS.1}Consider i.i.d. random variables $\left\{ U_{i}\right\} _{i=1}^{n}$
	on probability space $\left(\Omega,\mathscr{A},P\right)$ and $d_{1}\times d_{2}$
	matrix $A\left(U,\theta\right):\Omega\times\Theta\rightarrow R^{d_{1}\times d_{2}}$
	with $\Theta\subseteq R^{p}$ being compact, $\sup_{U\in\Omega,\theta\in\Theta}\left\Vert A_{s,t}\left(U,\theta\right)\right\Vert \leq D_{A,0}$
	and $\sup_{U\in\Omega}\left\Vert A_{s,t}\left(U,\theta_{1}\right)-A_{s,t}\left(U,\theta_{2}\right)\right\Vert \leq D_{A,1}\left\Vert \theta_{1}-\theta_{2}\right\Vert $
	uniformly for all $1\leq s\leq d_{1}$ and $1\leq t\leq d_{2}$. Then
	there holds 
	\[
	\sup_{\theta\in\Theta}\left\Vert \frac{1}{n}\sum_{i=1}^{n}A\left(U_{i},\theta\right)-\mathbb{E}A\left(U_{i},\theta\right)\right\Vert =O_{p}\left(\sqrt{\frac{pd_{1}d_{2}D_{A,0}^{2}\log\left(d_{1}d_{2}D_{A,1}n\right)}{n}}\right).
	\]
\end{lemma}
\begin{proof}[Proof of \autoref{lemS.1}]
	Note that 
	\begin{align*}
		\sup_{\theta\in\Theta}\left\Vert \frac{1}{n}\sum_{i=1}^{n}A\left(U_{i},\theta\right)-\mathbb{E}A\left(U_{i},\theta\right)\right\Vert  & \leq\max_{1\leq b\leq B}\left\Vert \frac{1}{n}\sum_{i=1}^{n}A\left(U_{i},\theta_{b}\right)-\mathbb{E}A\left(U_{i},\theta_{b}\right)\right\Vert \\
		& +\max_{1\leq b\leq B}\sup_{\left\Vert \theta-\theta_{b}\right\Vert \leq\frac{C}{\sqrt[p]{B}}}\left\Vert \frac{1}{n}\sum_{i=1}^{n}A\left(U_{i},\theta\right)-\frac{1}{n}\sum_{i=1}^{n}A\left(U_{i},\theta_{b}\right)\right\Vert \\
		& +\max_{1\leq b\leq B}\sup_{\left\Vert \theta-\theta_{b}\right\Vert \leq\frac{C}{\sqrt[p]{B}}}\left\Vert \mathbb{E}A\left(U_{i},\theta\right)-\mathbb{E}A\left(U_{i},\theta_{b}\right)\right\Vert .
	\end{align*}
	For the first term, we have that 
	\begin{align*}
		& P\left(\max_{1\leq b\leq B}\left\Vert \frac{1}{n}\sum_{i=1}^{n}A\left(U_{i},\theta_{b}\right)-\mathbb{E}A\left(U_{i},\theta_{b}\right)\right\Vert >\tau\right)\\
		& \leq\sum_{b=1}^{B}P\left(\left\Vert \frac{1}{n}\sum_{i=1}^{n}A\left(U_{i},\theta_{b}\right)-\mathbb{E}A\left(U_{i},\theta_{b}\right)\right\Vert >\tau\right)\\
		& \leq\sum_{b=1}^{B}P\left(\max_{1\leq s\leq d_{1}}\max_{1\leq t\leq d_{2}}\left\Vert \frac{1}{n}\sum_{i=1}^{n}A_{s,t}\left(U_{i},\theta_{b}\right)-\mathbb{E}A_{s,t}\left(U_{i},\theta_{b}\right)\right\Vert >\frac{\tau}{\sqrt{d_{1}d_{2}}}\right)\\
		& \leq\sum_{b=1}^{B}\sum_{s=1}^{d_{1}}\sum_{t=1}^{d_{2}}P\left(\left\Vert \frac{1}{n}\sum_{i=1}^{n}A_{s,t}\left(U_{i},\theta_{b}\right)-\mathbb{E}A_{s,t}\left(U_{i},\theta_{b}\right)\right\Vert >\frac{\tau}{\sqrt{d_{1}d_{2}}}\right)\\
		& \leq\sum_{b=1}^{B}\sum_{s=1}^{d_{1}}\sum_{t=1}^{d_{2}}2\exp\left(-Cn\tau^{2}/\left(d_{1}d_{2}D_{A,0}^{2}\right)\right)=2\exp\left(C\log\left(Bd_{1}d_{2}\right)-Cn\tau^{2}/\left(d_{1}d_{2}D_{A,0}^{2}\right)\right),
	\end{align*}
	indicating that 
	\[
	\max_{1\leq b\leq B}\left\Vert \frac{1}{n}\sum_{i=1}^{n}A\left(U_{i},\theta_{b}\right)-\mathbb{E}A\left(U_{i},\theta_{b}\right)\right\Vert =O_{p}\left(\sqrt{\frac{d_{1}d_{2}D_{A,0}^{2}\log\left(Bd_{1}d_{2}\right)}{n}}\right).
	\]
	On the other side, for the second term we have that 
	\begin{align*}
		& \max_{1\leq b\leq B}\sup_{\left\Vert \theta-\theta_{b}\right\Vert \leq\frac{C}{\sqrt[p]{B}}}\left\Vert \frac{1}{n}\sum_{i=1}^{n}A\left(U_{i},\theta\right)-\frac{1}{n}\sum_{i=1}^{n}A\left(U_{i},\theta_{b}\right)\right\Vert \\
		& \leq\sqrt{d_{1}d_{2}}\max_{1\leq s\leq d_{1}}\max_{1\leq t\leq d_{2}}\sup_{U\in\Omega}\sup_{\left\Vert \theta-\theta_{b}\right\Vert \leq\frac{C}{\sqrt[p]{B}}}\left|A_{s,t}\left(U,\theta\right)-A_{s,t}\left(U,\theta_{b}\right)\right|\leq\frac{\sqrt{d_{1}d_{2}}D_{A,1}}{\sqrt[p]{B}}.
	\end{align*}
	The same bound holds for the third term. Then let $B=\left(\sqrt{n}D_{A,1}\right)^{p}$,
	we finish the proof.
\end{proof}

\begin{lemma}
\label{lem:2}If \autoref{assu1}, \autoref{assump:2}(i)-(iii),
and \autoref{assu:4} hold with $\min\left\{ \upsilon_{G},\upsilon_{f}\right\} \geq2$,
then there exists a constant $C$ that does not depend on $\boldsymbol{\mathbf{X}},z,\boldsymbol{\beta}$
such that the following hold

(i) $\sup_{\boldsymbol{\mathbf{X}},z,\boldsymbol{\beta}}\left|\partial^{s}f_{\boldsymbol{\mathbf{X}},z}\left(\left.\boldsymbol{\mathbf{X}},z\right|\boldsymbol{\beta}\right)/\partial z^{s}\right|\leq C$
for $0\leq s\leq\upsilon_{f}$; 

(ii) $\sup_{z,\boldsymbol{\beta}}\left|\partial^{s}f_{z}\left(\left.z\right|\boldsymbol{\beta}\right)/\partial z^{s}\right|\leq C$
for $0\leq s\leq\upsilon_{f}$; 

(iii) $\sup_{\boldsymbol{\mathbf{X}},z,\boldsymbol{\beta}}\left\Vert \partial f_{\boldsymbol{\mathbf{X}},z}\left(\left.\boldsymbol{\mathbf{X}},z\right|\boldsymbol{\beta}\right)/\partial\boldsymbol{\beta}\right\Vert \leq C\sqrt{p}$;

(iv) $\sup_{\boldsymbol{\mathbf{X}},z,\boldsymbol{\beta}}\left\Vert \partial^{2}f_{\boldsymbol{\mathbf{X}},z}\left(\left.\boldsymbol{\mathbf{X}},z\right|\boldsymbol{\beta}\right)/\partial\boldsymbol{\beta}\partial\boldsymbol{\beta}^{\mathrm{T}}\right\Vert \leq Cp$;

(v) $\left\Vert \partial f_{z}\left(\left.z\right|\boldsymbol{\beta}\right)/\partial\boldsymbol{\beta}\right\Vert \leq C\sqrt{p}$;

(vi) $\left\Vert \partial^{2}f_{z}\left(\left.z\right|\boldsymbol{\beta}\right)/\partial\boldsymbol{\beta}\partial\boldsymbol{\beta}^{\mathrm{T}}\right\Vert \leq Cp$; 

(vii) \textup{$\sup_{z,\boldsymbol{\beta},f_{z}\left(\left.z\right|\boldsymbol{\beta}\right)\neq0}\left|\partial^{s}L\left(z,\boldsymbol{\beta}\right)/\partial z^{s}\right|\leq C$}
for $0\leq s\leq\min\left\{ \upsilon_{G},\upsilon_{f}\right\} $;

(viii) \textup{$\sup_{z,\boldsymbol{\beta},f_{z}\left(\left.z\right|\boldsymbol{\beta}\right)\neq0}\left\Vert \partial L\left(z,\boldsymbol{\beta}\right)/\partial\boldsymbol{\beta}\right\Vert \leq C\sqrt{p}$;}

(ix) \textup{$\sup_{z,\boldsymbol{\beta},f_{z}\left(\left.z\right|\boldsymbol{\beta}\right)\neq0}\left\Vert \partial^{2}L\left(z,\boldsymbol{\beta}\right)/\partial\boldsymbol{\beta}\partial\boldsymbol{\beta}^{\mathrm{T}}\right\Vert \leq Cp$}; 

(x)\textup{ }$\sup_{\boldsymbol{\mathbf{X}}_{e},\boldsymbol{\beta},f_{z}\left(\left.z\left(\boldsymbol{\mathbf{X}}_{e},\boldsymbol{\beta}\right)\right|\boldsymbol{\beta}\right)\neq0}\int_{\mathcal{X}}\left\Vert \partial W\left(\boldsymbol{\mathbf{X}}_{e},\widetilde{\boldsymbol{\mathbf{X}}}_{e},\boldsymbol{\beta}\right)/\partial\boldsymbol{\beta}\right\Vert d\widetilde{\boldsymbol{\mathbf{X}}}\leq C\sqrt{p}$.
\end{lemma}
\begin{proof}
To prove \autoref{lem:2}(i) and \autoref{lem:2}(ii), we note that for any $0\leq s\leq\upsilon_f$, 
\[
\frac{\partial^{s}f_{\mathbf{X},z}\left(\left.\boldsymbol{\mathbf{X}},z\right|\boldsymbol{\beta}\right)}{\partial z^{s}}=\left.\frac{\partial^{s}f_{e}\left(X_{0},\boldsymbol{\mathbf{X}}\right)}{\partial X_{0}^{s}}\right|_{X_{0}=z-\boldsymbol{\mathbf{X}}^{\mathrm{T}}\boldsymbol{\beta}},
\]
and 
\[
\frac{\partial^{s}f_{z}\left(\left.z\right|\boldsymbol{\beta}\right)}{\partial z^{s}}=\int_{\mathcal{X}}\left[\frac{\partial^{s}f_{\mathbf{X},z}\left(\left.\boldsymbol{\mathbf{X}},z\right|\boldsymbol{\beta}\right)}{\partial X_{0}^{s}}\right]d\boldsymbol{\mathbf{X}}.
\]
Since $f_{e}\left(\boldsymbol{\mathbf{X}}_{e}\right)$ has up to $\upsilon_f$-th
bounded derivatives over $\mathcal{X}_{e}$ according to Assumption
\ref{assu:4}(ii) and $X_{j}$ is bounded by 1 for all $1\leq j\leq p$
according to Assumption \ref{assump:2}(i), \autoref{lem:2}(i) and \autoref{lem:2}(ii) hold. 

Similarly, note that 
\[
\frac{\partial f_{\boldsymbol{\mathbf{X}},z}\left(\left.\boldsymbol{\mathbf{X}},z\right|\boldsymbol{\beta}\right)}{\partial\boldsymbol{\beta}}=-\left[\left.\frac{\partial f_{e}\left(X_{0},\boldsymbol{\mathbf{X}}\right)}{\partial X_{0}}\right|_{X_{0}=z-\boldsymbol{\mathbf{X}}^{\mathrm{T}}\boldsymbol{\beta}}\right]\boldsymbol{\mathbf{X}},
\]
\[
\frac{\partial^{2}f_{\boldsymbol{\mathbf{X}},z}\left(\left.\boldsymbol{\mathbf{X}},z\right|\boldsymbol{\beta}\right)}{\partial\boldsymbol{\beta}\partial\boldsymbol{\beta}^{\mathrm{T}}}=\left[\left.\frac{\partial^{2}f_{e}\left(X_{0},\boldsymbol{\mathbf{X}}\right)}{\partial X_{0}^{2}}\right|_{X_{0}=z-\boldsymbol{\mathbf{X}}^{\mathrm{T}}\boldsymbol{\beta}}\right]\boldsymbol{\mathbf{X}}\boldsymbol{\mathbf{X}}^{\mathrm{T}},
\]
\[
\frac{\partial f_{z}\left(\left.z\right|\boldsymbol{\beta}\right)}{\partial\boldsymbol{\beta}}=-\int_{\mathcal{X}}\left[\left.\frac{\partial f_{e}\left(X_{0},\boldsymbol{\mathbf{X}}\right)}{\partial X_{0}}\right|_{X_{0}=z-\boldsymbol{\mathbf{X}}^{\mathrm{T}}\boldsymbol{\beta}}\right]\boldsymbol{\mathbf{X}}d\boldsymbol{\mathbf{X}},
\]
\[
\frac{\partial^{2}f_{z}\left(\left.z\right|\boldsymbol{\beta}\right)}{\partial\boldsymbol{\beta}\partial\boldsymbol{\beta}^{\mathrm{T}}}=\int_{\mathcal{X}}\left[\left.\frac{\partial^{2}f_{e}\left(X_{0},\boldsymbol{\mathbf{X}}\right)}{\partial X_{0}^{2}}\right|_{X_{0}=z-\boldsymbol{\mathbf{X}}^{\mathrm{T}}\boldsymbol{\beta}}\right]\boldsymbol{\mathbf{X}}\boldsymbol{\mathbf{X}}^{\mathrm{T}}d\mathbf{X},
\]
we validate \autoref{lem:2}(iii)-\autoref{lem:2}(vi).

To prove \autoref{lem:2}(vii), note that 
\begin{align*}
\left|\frac{\partial^{s}L\left(z,\boldsymbol{\beta}\right)}{\partial z^{s}}\right| & \leq C\sum_{j=0}^{s}\left|\int_{\mathcal{X}}G^{(j)}\left(z-\boldsymbol{\mathbf{X}}^{\mathrm{T}}\Delta\boldsymbol{\beta}\right)\frac{\partial^{s-j}f_{\boldsymbol{\mathbf{X}}|z}\left(\left.\boldsymbol{\mathbf{X}}\right|z,\boldsymbol{\beta}\right)}{\partial z^{s-j}}d\boldsymbol{\mathbf{X}}\right|\\
 & \leq C\sum_{j=0}^{s}\left\Vert G^{(j)}\right\Vert _{\infty}\cdot\left(\int_{\mathcal{X}}\left|\frac{\partial^{s-j}f_{\boldsymbol{\mathbf{X}}|z}\left(\left.\boldsymbol{\mathbf{X}}\right|z,\boldsymbol{\beta}\right)}{\partial z^{s-j}}\right|d\boldsymbol{\mathbf{X}}\right).
\end{align*}
According to  \autoref{assump:2}(iii), $\left\Vert G^{(j)}\right\Vert _{\infty}$
is bounded for all $0\leq j\leq\upsilon_{G}$. Then it remains to
show that $\int_{\mathcal{X}}\left|\partial^{s-j}f_{\boldsymbol{\mathbf{X}}|z}/\partial z_{\infty}^{s-j}\right|d\boldsymbol{\mathbf{X}}$
is also upper bounded for all $0\leq j\leq\upsilon_{f}$. When $j=s,$
we have that $\int_{\mathcal{X}}\left|\partial^{s-j}f_{\boldsymbol{\mathbf{X}}|z}\left(\left.\boldsymbol{\mathbf{X}}\right|z,\boldsymbol{\beta}\right)/\partial z^{s-j}\right|d\boldsymbol{\mathbf{X}}=1$.
When $j=s-1$, define $\mathbb{X}\left(z,\boldsymbol{\beta}\right)=\left\{ \boldsymbol{\mathbf{X}}:\left(z-\boldsymbol{\mathbf{X}}^{\mathrm{T}}\boldsymbol{\beta},\boldsymbol{\mathbf{X}}\right)\in\mathcal{X}_{e}\right\} $.
We have that 
\begin{align*}
 & \int_{\mathcal{X}}\left|\frac{\partial f_{\boldsymbol{\mathbf{X}}|z}\left(\left.\boldsymbol{\mathbf{X}}\right|z,\boldsymbol{\beta}\right)}{\partial z}\right|d\boldsymbol{\mathbf{X}}\\
 & =\int_{\mathcal{X}}\left|\frac{\partial f_{\boldsymbol{\mathbf{X}},z}\left(\left.\boldsymbol{\mathbf{X}},z\right|\boldsymbol{\beta}\right)/\partial z}{\int_{\mathcal{X}}f_{\boldsymbol{\mathbf{X}},z}\left(\left.\boldsymbol{\mathbf{X}},z\right|\boldsymbol{\beta}\right)d\boldsymbol{\mathbf{X}}}-\frac{f_{\boldsymbol{\mathbf{X}},z}\left(\left.\boldsymbol{\mathbf{X}},z\right|\boldsymbol{\beta}\right)\int_{\mathcal{X}}\left(\partial f_{\boldsymbol{\mathbf{X}},z}\left(\left.\boldsymbol{\mathbf{X}},z\right|\boldsymbol{\beta}\right)/\partial z\right)d\boldsymbol{\mathbf{X}}}{\left(\int_{\mathcal{X}}f_{\boldsymbol{\mathbf{X}},z}\left(\left.\boldsymbol{\mathbf{X}},z\right|\boldsymbol{\beta}\right)d\boldsymbol{\mathbf{X}}\right)^{2}}\right|d\boldsymbol{\mathbf{X}}\\
 & \leq\frac{2\int_{\mathcal{X}}\left|\partial f_{\boldsymbol{\mathbf{X}},z}\left(\left.\boldsymbol{\mathbf{X}},z\right|\boldsymbol{\beta}\right)/\partial z\right|d\boldsymbol{\mathbf{X}}}{\int_{\mathcal{X}}f_{\boldsymbol{\mathbf{X}},z}\left(\left.\boldsymbol{\mathbf{X}},z\right|\boldsymbol{\beta}\right)d\boldsymbol{\mathbf{X}}}\leq\frac{2\left\Vert \partial f_{\boldsymbol{\mathbf{X}},z}/\partial z\right\Vert _{\infty}m\left(\mathbb{X}\left(z,\boldsymbol{\beta}\right)\right)}{\zeta^{-1}m\left(\mathbb{X}\left(z,\boldsymbol{\beta}\right)\right)}\leq C
\end{align*}
according to part (i) of this lemma.
The proof of the case when $j=s-2,\cdots,0$ are similar,
so is omitted.

To prove \autoref{lem:2}(viii), note that 
\begin{align*}
\left\Vert \frac{\partial L\left(z,\boldsymbol{\beta}\right)}{\partial\boldsymbol{\beta}}\right\Vert  & \leq\int_{\mathcal{X}}\left\Vert G^{\prime}\left(z-\boldsymbol{\mathbf{X}}^{\mathrm{T}}\Delta\boldsymbol{\beta}\right)f_{\boldsymbol{\mathbf{X}}|z}\left(\left.\boldsymbol{\mathbf{X}}\right|z,\boldsymbol{\beta}\right)\boldsymbol{\mathbf{X}}\right\Vert d\boldsymbol{\mathbf{X}}\\
 & +\int_{\mathcal{X}}\left\Vert G\left(Z-\boldsymbol{\mathbf{X}}^{\mathrm{T}}\Delta\boldsymbol{\beta}\right)\frac{\partial f_{\boldsymbol{\mathbf{X}}|z}\left(\left.\boldsymbol{\mathbf{X}}\right|z,\boldsymbol{\beta}\right)}{\partial\boldsymbol{\beta}}\right\Vert d\boldsymbol{\mathbf{X}}.
\end{align*}
Obviously, the first term on the RHS is bounded by $\left\Vert G^{\prime}\right\Vert _{\infty}\sqrt{p}$,
and the second term is bounded by $\left\Vert G\right\Vert _{\infty}\int_{\mathcal{X}}\left\Vert \partial f_{\boldsymbol{\mathbf{X}}|z}\left(\left.\boldsymbol{\mathbf{X}}\right|z,\boldsymbol{\beta}\right)/\partial\boldsymbol{\beta}\right\Vert d\boldsymbol{\mathbf{X}}$.
Note that 
\begin{align*}
\int_{\mathcal{X}}\left\Vert \partial f_{\boldsymbol{\mathbf{X}}|z}\left(\left.\boldsymbol{\mathbf{X}}\right|z,\boldsymbol{\beta}\right)/\partial\boldsymbol{\beta}\right\Vert d\boldsymbol{\mathbf{X}} & \leq\frac{2\int_{\mathcal{X}}\left\Vert \partial f_{\boldsymbol{\mathbf{X}},z}\left(\left.\boldsymbol{\mathbf{X}},z\right|\boldsymbol{\beta}\right)/\partial\boldsymbol{\beta}\right\Vert d\boldsymbol{\mathbf{X}}}{\int_{\mathcal{X}}f_{\boldsymbol{\mathbf{X}},z}\left(\left.\boldsymbol{\mathbf{X}},z\right|\boldsymbol{\beta}\right)d\boldsymbol{\mathbf{X}}}\\
 & \leq\frac{2C\sqrt{p}m\left(\mathbb{X}\left(z,\boldsymbol{\beta}\right)\right)}{\zeta^{-1}m\left(\mathbb{X}\left(z,\boldsymbol{\beta}\right)\right)}\leq C\sqrt{p},
\end{align*}
according to part (iii) of this lemma. This proves \autoref{lem:2}(viii). \autoref{lem:2}(ix) can
be similarly proved.

Finally, to show \autoref{lem:2}(x), we note that 
\begin{align*}
 & \int_{\mathcal{X}}\left\Vert \frac{\partial W\left(\boldsymbol{\mathbf{X}}_{e},\widetilde{\boldsymbol{\mathbf{X}}}_{e},\boldsymbol{\beta}\right)}{\partial\boldsymbol{\beta}}\right\Vert d\widetilde{\boldsymbol{X}}\\
 & \leq\int_{\mathcal{X}}\left\Vert G^{\prime\prime}\left(z\left(\boldsymbol{\mathbf{X}}_{e},\boldsymbol{\beta}^{\star}\right)+\left(\boldsymbol{\mathbf{X}}-\widetilde{\boldsymbol{\mathbf{X}}}\right)^{\mathrm{T}}\Delta\boldsymbol{\beta}\right)\left(\boldsymbol{\mathbf{X}}-\widetilde{\mathbf{X}}\right)\right\Vert f_{\boldsymbol{\mathbf{X}}|z}\left(\left.\widetilde{\mathbf{X}}\right|z\left(\boldsymbol{\mathbf{X}}_{e},\boldsymbol{\beta}\right),\boldsymbol{\beta}\right)d\widetilde{\mathbf{X}}\\
 & +\int_{\mathcal{X}}\left|G^{\prime}\left(z\left(\boldsymbol{\mathbf{X}}_{e},\boldsymbol{\beta}^{\star}\right)+\left(\boldsymbol{\mathbf{X}}-\widetilde{\mathbf{X}}\right)^{\mathrm{T}}\Delta\boldsymbol{\beta}\right)\right|\left\Vert \frac{\partial f_{\boldsymbol{\mathbf{X}}|z}\left(\left.\widetilde{\mathbf{X}}\right|z\left(\mathbf{X}_{e},\boldsymbol{\beta}\right),\boldsymbol{\beta}\right)}{\partial\boldsymbol{\beta}}\right\Vert d\widetilde{\mathbf{X}}.
\end{align*}
Obviously, the first term is bounded by $2\sqrt{p}\left\Vert G^{\prime\prime}\right\Vert _{\infty}$,
and the second term is bounded by $\left\Vert G^{\prime}\right\Vert _{\infty}\int_{\mathcal{X}}\left\Vert \partial f_{\boldsymbol{\mathbf{X}}|Z}\left(\left.\widetilde{\mathbf{X}},z\left(\boldsymbol{\mathbf{X}}_{e},\boldsymbol{\beta}\right)\right|\boldsymbol{\beta}\right)/\partial\boldsymbol{\beta}\right\Vert d\widetilde{\mathbf{X}}$.
Note that 
\begin{align*}
\int_{\mathcal{X}}\left\Vert \frac{\partial f_{\mathbf{X}|z}\left(\left.\widetilde{\mathbf{X}}\right|z\left(\boldsymbol{\mathbf{X}}_{e},\boldsymbol{\beta}\right),\boldsymbol{\beta}\right)}{\partial\boldsymbol{\beta}}\right\Vert d\widetilde{\mathbf{X}} & \leq\frac{2\int_{\mathcal{X}}\left\Vert \partial f_{\boldsymbol{\mathbf{X}},z}\left(\left.\widetilde{\mathbf{X}},z\left(\boldsymbol{\mathbf{X}}_{e},\boldsymbol{\beta}\right)\right|\boldsymbol{\beta}\right)/\partial\boldsymbol{\beta}\right\Vert d\widetilde{\mathbf{X}}}{f_{z}\left(\left.z\left(\boldsymbol{\mathbf{X}}_{e},\boldsymbol{\beta}\right)\right|\boldsymbol{\beta}\right)}.
\end{align*}
We can see that 
\begin{align*}
\frac{\partial f_{\boldsymbol{\mathbf{X}},z}\left(\left.\widetilde{\mathbf{X}},z\left(\boldsymbol{\mathbf{X}}_{e},\boldsymbol{\beta}\right)\right|\boldsymbol{\beta}\right)}{\partial\boldsymbol{\beta}} & =\left.\frac{\partial f_{\boldsymbol{\mathbf{X}},z}\left(\left.\widetilde{\mathbf{X}},z\right|\boldsymbol{\beta}\right)}{\partial z}\right|_{z=z\left(\boldsymbol{\mathbf{X}}_{e},\boldsymbol{\beta}\right)}\boldsymbol{\mathbf{X}}\\
 & +\left.\frac{\partial f_{\boldsymbol{\mathbf{X}},z}\left(\left.\widetilde{\mathbf{X}},z\right|\boldsymbol{\beta}\right)}{\partial\boldsymbol{\beta}}\right|_{z=z\left(\boldsymbol{\mathbf{X}}_{e},\boldsymbol{\beta}\right)},
\end{align*}
according to (i) and (ii), we know that $\left\Vert \left.\partial f_{\boldsymbol{\mathbf{X}},z}\left(\left.\widetilde{\mathbf{X}},z\right|\boldsymbol{\beta}\right)/\partial z\right|_{z=z\left(\boldsymbol{\mathbf{X}}_{e},\boldsymbol{\beta}\right)}\right\Vert $
is bounded, and $\left\Vert \left.\partial f_{\boldsymbol{\mathbf{X}},z}\left(\left.\widetilde{\mathbf{X}},z\right|\boldsymbol{\beta}\right)/\partial \boldsymbol{\beta}\right|_{z=z\left(\boldsymbol{\mathbf{X}}_{e},\boldsymbol{\beta}\right)}\right\Vert $
is bounded by $C\sqrt{p}$, so $\left\Vert \partial f_{\boldsymbol{\mathbf{X}},z}\left(\left.\widetilde{\mathbf{X}},z\left(\boldsymbol{\mathbf{X}}_{e},\boldsymbol{\beta}\right)\right|\boldsymbol{\beta}\right)/\partial\boldsymbol{\beta}\right\Vert $
is bounded by $C\sqrt{p}$. So 
\begin{align*}
\frac{\int_{\mathcal{X}}\left\Vert \partial f_{\boldsymbol{\mathbf{X}},z}\left(\left.\widetilde{\mathbf{X}},z\left(\boldsymbol{\mathbf{X}}_{e},\boldsymbol{\beta}\right)\right|\boldsymbol{\beta}\right)/\partial\boldsymbol{\beta}\right\Vert d\widetilde{\mathbf{X}}}{f_{z}\left(\left.z\left(\boldsymbol{\mathbf{X}}_{e},\boldsymbol{\beta}\right)\right|\boldsymbol{\beta}\right)} & \leq\frac{C\sqrt{p}\cdot m\left(\mathbb{X}\left(z\left(\boldsymbol{\mathbf{X}}_{e},\boldsymbol{\beta}\right),\boldsymbol{\beta}\right)\right)}{\zeta^{-1}\cdot m\left(\mathbb{X}\left(z\left(\boldsymbol{\mathbf{X}}_{e},\boldsymbol{\beta}\right),\boldsymbol{\beta}\right)\right)}=C\sqrt{p}.
\end{align*}
This finishes the proof of \autoref{lem:2}(xii). 
\end{proof}
\begin{lemma}
\label{lem3}Suppose that  \autoref{assu1}, \autoref{assump:2}(i)-(iii), \ref{assu:3} and \autoref{assu:4}
hold with $\upsilon_{G}=3$, $\upsilon_{K}=2$, and $\upsilon_{f}=3$.
Define 
\[
A_{n,\cdot}\left(\mathbf{X}_{e},\boldsymbol{\beta}\right)=\frac{1}{nh_{n}}\sum_{j=1}^{n}K\left(\left(z\left(\boldsymbol{\mathbf{X}}_{e},\boldsymbol{\beta}\right)-z\left(\boldsymbol{\mathbf{X}}_{e,j},\boldsymbol{\beta}\right)\right)/h_{n}\right)\cdot\left(\cdot_{j}\right),
\]
where $\cdot$ is $y$ or $1$. Also define $A_{\cdot}\left(\boldsymbol{\mathbf{X}}_{e},\boldsymbol{\beta}\right)=\lim_{n\rightarrow\infty}\mathbb{E}_{\mathscr{D}_{n}}A_{n,\cdot}\left(\boldsymbol{\mathbf{X}}_{e},\boldsymbol{\beta}\right)$, where the expectation $\mathbb{E}_{\mathscr{D}_{n}}$ is taken with respect to the data set $\mathscr{D}_{n}$.
Then

(i) There holds 
\[
\sup_{\left(\boldsymbol{\mathbf{X}}_{e},\boldsymbol{\beta}\right)\in\mathcal{X}_{e}\times\in\mathcal{B}}\left|A_{n,\cdot}\left(\boldsymbol{\mathbf{X}}_{e},\boldsymbol{\beta}\right)-\mathbb{E}_{\mathscr{D}_{n}}A_{n,\cdot}\left(\boldsymbol{\mathbf{X}}_{e},\boldsymbol{\beta}\right)\right|=O_{p}\left(h_{n}^{-1}\sqrt{p\log\left(nph_{n}^{-1}\right)/n}\right);
\]

(ii) There holds
\[
\sup_{\left(\boldsymbol{\mathbf{X}}_{e},\boldsymbol{\beta}\right)\in\mathcal{X}_{e}\times\in\mathcal{B}}\left|\mathbb{E}_{\mathscr{D}_{n}}A_{n,\cdot}\left(\boldsymbol{\mathbf{X}}_{e},\boldsymbol{\beta}\right)-A_{\cdot}\left(\boldsymbol{\mathbf{X}}_{e},\boldsymbol{\beta}\right)\right|=O_{p}\left(h_{n}^{2}\right);
\]

(iii) Define $\psi\left(n,p,h_{n}\right)=h_{n}^{-1}\sqrt{p\log\left(nph_{n}^{-1}\right)/n}+h_{n}^{2}$,
there holds
\[
\sup_{\left(\boldsymbol{\mathbf{X}}_{e},\boldsymbol{\beta}\right)\in\mathcal{X}_{e}\times\in\mathcal{B}}\left|A_{n,\cdot}\left(\boldsymbol{\mathbf{X}}_{e},\boldsymbol{\beta}\right)-A_{\cdot}\left(\boldsymbol{\mathbf{X}}_{e},\boldsymbol{\beta}\right)\right|=O_{p}\left(h_{n}^{-1}\sqrt{p\log\left(nph_{n}^{-1}\right)/n}+h_{n}^{2}\right).
\]
\end{lemma}
\begin{proof}
\autoref{lem3}(i) is a direct result of  \autoref{lemS.1} if we note that \[\left|K\left(\left(z\left(\boldsymbol{\mathbf{X}}_{e},\boldsymbol{\beta}\right)-z\left(\boldsymbol{\mathbf{X}}_{e,j},\boldsymbol{\beta}\right)\right)/h_{n}\right)\cdot\left(\cdot_{j}\right)\right|\leq Ch_h^{-1}\] 
and 
\[\Vert \partial \left(K\left(\left(z\left(\boldsymbol{\mathbf{X}}_{e},\boldsymbol{\beta}\right)-z\left(\boldsymbol{\mathbf{X}}_{e,j},\boldsymbol{\beta}\right)\right)/h_{n}\right)\cdot\left(\cdot_{j}\right)\right)/\partial\boldsymbol{\beta}\Vert \leq C\sqrt{p}h_h^{-2}.\]

To prove \autoref{lem3}(ii), we only need to note that
\begin{align*}
 & \mathbb{E}_{\mathscr{D}_{n}}\left[A_{n,y}\left(\mathbf{X}_{e},\boldsymbol{\beta}\right)\right]\\
 & =\frac{1}{h_{n}}\mathbb{E}_{\mathscr{D}_{n}}\left[K\left(\frac{z\left(\boldsymbol{\mathbf{X}}_{e},\boldsymbol{\beta}\right)-z\left(\boldsymbol{\mathbf{X}}_{e,j},\boldsymbol{\beta}\right)}{h_{n}}\right)y_{j}\right]\\
 & =\frac{1}{h_{n}}\mathbb{E}_{\mathscr{D}_{n}}\left[K\left(\frac{z\left(\boldsymbol{\mathbf{X}}_{e},\boldsymbol{\beta}\right)-z\left(\boldsymbol{\mathbf{X}}_{e,j},\boldsymbol{\beta}\right)}{h_{n}}\right)G\left(z\left(\boldsymbol{\mathbf{X}}_{e,j},\boldsymbol{\beta}\right)-\boldsymbol{\mathbf{X}}_{j}^{\mathrm{T}}\Delta\boldsymbol{\beta}\right)\right]\\
 & =\frac{1}{h_{n}}\int K\left(\frac{z\left(\boldsymbol{\mathbf{X}}_{e},\boldsymbol{\beta}\right)-z}{h_{n}}\right)G\left(z-\boldsymbol{\mathbf{X}}_{j}^{\mathrm{T}}\Delta\boldsymbol{\beta}\right)f_{\boldsymbol{\mathbf{X}},z}\left(\left.\boldsymbol{\mathbf{X}}_{j},z\right|\boldsymbol{\beta}\right)d\boldsymbol{\mathbf{X}}_{j}dz\\
 & =\frac{1}{h_{n}}\int K\left(\frac{z\left(\boldsymbol{\mathbf{X}}_{e},\boldsymbol{\beta}\right)-z}{h_{n}}\right)f_{z}\left(\left.z\right|\boldsymbol{\beta}\right)dz\int_{\mathcal{X}}G\left(z-\boldsymbol{\mathbf{X}}_{j}^{\mathrm{T}}\Delta\boldsymbol{\beta}\right)\frac{f_{\boldsymbol{\mathbf{X}},z}\left(\left.\boldsymbol{\mathbf{X}}_{j},z\right|\boldsymbol{\beta}\right)}{f_{z}\left(\left.z\right|\boldsymbol{\beta}\right)}d\boldsymbol{\mathbf{X}}_{j}\\
 & =\frac{1}{h_{n}}\int K\left(\frac{z\left(\boldsymbol{\mathbf{X}}_{e},\boldsymbol{\beta}\right)-z}{h_{n}}\right)f_{z}\left(\left.z\right|\boldsymbol{\beta}\right)L\left(z,\boldsymbol{\beta}\right)dz\\
 & =\int K\left(z\right)L\left(z\left(\boldsymbol{\mathbf{X}}_{e},\boldsymbol{\beta}\right)-h_{n}z,\boldsymbol{\beta}\right)f_{z}\left(\left.z\left(\boldsymbol{\mathbf{X}}_{e},\boldsymbol{\beta}\right)-h_{n}z\right|\boldsymbol{\beta}\right)dz\\
 & =L\left(z\left(\boldsymbol{\mathbf{X}}_{e},\boldsymbol{\beta}\right)\right)f_{z}\left(\left.z\left(\boldsymbol{\mathbf{X}}_{e},\boldsymbol{\beta}\right)\right|\boldsymbol{\beta}\right)+\frac{h_{n}^{2}}{2}\left[\frac{\partial^{2}L\left(z\left(\boldsymbol{\mathbf{X}}_{e},\boldsymbol{\beta}\right),\boldsymbol{\beta}\right)f_{z}\left(\left.z\left(\boldsymbol{\mathbf{X}}_{e},\boldsymbol{\beta}\right)\right|\boldsymbol{\beta}\right)}{\partial z^{2}}\right]\left[\int K\left(z\right)z^{2}dz\right]\\
 & +\frac{h_{n}^{3}}{6}\left\{ \int K\left(z\right)z^{3}\left[\frac{\partial^{3}L\left(\widetilde{z},\boldsymbol{\beta}\right)f_{z}\left(\left.\widetilde{z}\right|\boldsymbol{\beta}\right)}{\partial z^{3}}\right]dz\right\} ,
\end{align*}
and similarly, 
\begin{align*}
\mathbb{E}_{\mathscr{D}_{n}}\left[A_{n,1}\left(\boldsymbol{\mathbf{X}}_{e},\boldsymbol{\beta}\right)\right] & =\frac{1}{h_{n}}\mathbb{E}_{\mathscr{D}_{n}}\left[K\left(\frac{z\left(\boldsymbol{\mathbf{X}}_{e},\boldsymbol{\beta}\right)-z\left(\boldsymbol{\mathbf{X}}_{e,j},\boldsymbol{\beta}\right)}{h_{n}}\right)\right]\\
 & =\frac{1}{h_{n}}\int\left[K\left(\frac{z\left(\boldsymbol{\mathbf{X}}_{e},\boldsymbol{\beta}\right)-z}{h_{n}}\right)f_{z}\left(\left.z\right|\boldsymbol{\beta}\right)\right]dz\\
 & =\int K\left(z\right)f_{z}\left(\left.z\left(\boldsymbol{\mathbf{X}}_{e},\boldsymbol{\beta}\right)-h_{n}z\right|\boldsymbol{\beta}\right)dz\\
 & =f_{z}\left(\left.z\left(\boldsymbol{\mathbf{X}}_{e},\boldsymbol{\beta}\right)\right|\boldsymbol{\beta}\right)+\frac{h_{n}^{2}}{2}\left[\frac{\partial^{2}f_{z}\left(\left.z\left(\boldsymbol{\mathbf{X}}_{e},\boldsymbol{\beta}\right)\right|\boldsymbol{\beta}\right)}{\partial z^{2}}\right]\left[\int K\left(z\right)z^{2}dz\right]\\
 & +\frac{h_{n}^{3}}{6}\left\{ \int K\left(z\right)z^{3}\left[\frac{\partial^{3}f_{z}\left(\left.\widetilde{z}\right|\boldsymbol{\beta}\right)}{\partial z^{3}}\right]dz\right\} ,
\end{align*}
where $\widetilde{z}$ lies between $z\left(\boldsymbol{\mathbf{X}}_{e},\boldsymbol{\beta}\right)$
and $z$. Note that according to  \autoref{lem:2} (i) and (ii),
$f_{z}\left(\left.z\right|\boldsymbol{\beta}\right)$ and $L\left(z,\boldsymbol{\beta}\right)f_{z}\left(\left.z\right|\boldsymbol{\beta}\right)=\int_{\mathcal{X}}G\left(z-\boldsymbol{\mathbf{X}}^{\mathrm{T}}\Delta\boldsymbol{\beta}\right)f_{\boldsymbol{\mathbf{X}},z}\left(\left.\boldsymbol{\mathbf{X}},z\right|\boldsymbol{\beta}\right)d\boldsymbol{\mathbf{X}}$
both have up to third bounded derivatives with respect to $z$, so
the results hold. 

Finally, \autoref{lem3} (iii) is a combination of \autoref{lem3} (i) and \autoref{lem3} (ii). 
\end{proof}
\begin{lemma}
\label{lem4}Suppose that  \autoref{assu1}, \autoref{assump:2}(i)-(iii), \autoref{assu:3}, and \autoref{assu:4} hold. Given
any positive sequence $\left\{ \phi_{n}\right\} _{n=1}^{\infty}$
satisfying $p\phi_{n}\downarrow0$, define \[\mathcal{X}_{e,n}=\left\{ \boldsymbol{X}_{e}\in\mathcal{X}_{e}:\left|X_{j}\right|\leq 1-\phi_{n},0\leq j\leq p\right\}.\]
Then 

(i) $1-P\left(\mathbf{X}_{e}\in\mathcal{X}_{e,n}\right)=O\left(p\phi_{n}\right)$,
and $\inf_{\left(\boldsymbol{\mathbf{X}}_{e},\boldsymbol{\beta}\right)\in\mathcal{X}_{e,n}\times\mathcal{B}}f_{Z}\left(\left.z\left(\boldsymbol{\mathbf{X}}_{e},\boldsymbol{\beta}\right)\right|\boldsymbol{\beta}\right)\sim\phi_{n}^{p}p^{-p}$;

(ii) If $\psi\left(n,p,h_{n}\right)=o\left(\phi_{n}^{p}p^{-p}\right),$
there holds
\[
\sup_{\left(\boldsymbol{\mathbf{X}}_{e},\boldsymbol{\beta}\right)\in\mathcal{X}_{e,n}\times\mathcal{B}}\left|\widehat{G}\left(\left.z\left(\boldsymbol{\mathbf{X}}_{e},\boldsymbol{\beta}\right)\right|\boldsymbol{\beta}\right)-L\left(z\left(\boldsymbol{\mathbf{X}}_{e},\boldsymbol{\beta}\right),\boldsymbol{\beta}\right)\right|=O_{p}\left(p^{p}\phi_{n}^{-p}\psi\left(n,p,h_{n}\right)\right).
\]
\end{lemma}
\begin{proof}
To prove \autoref{lem4}(i), note that for $p\phi_{n}<1$, 
$
m\left(\mathcal{X}_{e}-\mathcal{X}_{e,n}\right)=1-\left(1-\phi_{n}\right)^{p}\leq p\phi_{n}.
$
So 
$
\int_{\mathcal{X}_{e}-\mathcal{X}_{e,n}}f_{e}\left(\boldsymbol{\mathbf{X}}_{e}\right)d\boldsymbol{\mathbf{X}}_{e}\leq\zeta p\phi_{n}=O\left(p\phi_{n}\right)
$
due to  \autoref{assu:4}(i). To show the lower bound, note
that given any $\boldsymbol{\beta}\in\mathcal{B}$ and $\boldsymbol{\mathbf{X}}_{e}\in\mathcal{X}_{e,n}$, there holds
$
|z\left(\boldsymbol{\mathbf{X}}_{e},\boldsymbol{\beta}\right)-\widetilde{\boldsymbol{\mathbf{X}}}^{\mathrm{T}}\boldsymbol{\beta}-X_{0}| \leq\sum_{j=1}^{p}\left|\beta_{j}\right| |X_{j}-\widetilde{X}_{j}|.
$
This implies that for any $\widetilde{\mathbf{X}}$, $\widetilde{\mathbf{X}}\in\mathbb{X}\left(z\left(\boldsymbol{\boldsymbol{\mathbf{X}}}_{e},\boldsymbol{\beta}\right),\boldsymbol{\beta}\right)$
if 
\[
\widetilde{\mathbf{X}}\in\left\{ \widetilde{\mathbf{X}}\in\left[0,1\right]^{p}:\left(\sup_{\boldsymbol{\beta}\in\mathcal{B}}\left|\beta_{j}\right|\right)\left|X_{j}-\widetilde{X}_{j}\right|\leq\phi_{n}/p\right\} .
\]
Since the above set has Lebesgue measure of order $O\left(\phi_{n}^{p}/p^{p}\right)$,
we have that 
\begin{align*}
 & \inf_{\left(\boldsymbol{\mathbf{X}}_{e},\boldsymbol{\beta}\right)\in\mathcal{X}_{e,n}\times\mathcal{B}}f_{z}\left(\left.z\left(\mathbf{X}_{e},\boldsymbol{\beta}\right)\right|\boldsymbol{\beta}\right)\\
 & \geq\inf_{\left(\boldsymbol{\mathbf{X}}_{e},\boldsymbol{\beta}\right)\in\mathcal{X}_{e,n}\times\mathcal{B}}\int_{\widetilde{\mathbf{X}}\in\mathbb{X}\left(z\left(\boldsymbol{\mathbf{X}}_{e},\boldsymbol{\beta}\right),\boldsymbol{\beta}\right)}f_{e}\left(z\left(\boldsymbol{\mathbf{X}}_{e},\boldsymbol{\beta}\right)-\widetilde{\mathbf{X}}^{\mathrm{T}}\boldsymbol{\beta},\widetilde{\mathbf{X}}\right)d\widetilde{\mathbf{X}}\sim\phi_{n}^{p}/p^{p},
\end{align*}
due to  \autoref{assu:4}(i). This proves \autoref{lem4}(i). 

To prove \autoref{lem4}(ii), note that for any $\boldsymbol{\mathbf{X}}_{e}$ and
$\boldsymbol{\beta}$, we have $\widehat{G}\left(\left.z\left(\boldsymbol{\mathbf{X}}_{e},\boldsymbol{\beta}\right)\right|\boldsymbol{\beta}\right)=A_{n,y}\left(\boldsymbol{\mathbf{X}}_{e},\boldsymbol{\beta}\right)/A_{n,1}\left(\boldsymbol{\mathbf{X}}_{e},\boldsymbol{\beta}\right)$
and $L\left(z\left(\boldsymbol{\mathbf{X}}_{e},\boldsymbol{\beta}\right),\boldsymbol{\beta}\right)=A_{y}\left(\boldsymbol{\mathbf{X}}_{e},\boldsymbol{\beta}\right)/A_{1}\left(\boldsymbol{\mathbf{X}}_{e},\boldsymbol{\beta}\right)$.
So 
\begin{align*}
& \sup_{\left(\boldsymbol{\mathbf{X}}_{e},\boldsymbol{\beta}\right)\in\mathcal{X}_{e,n}\times\mathcal{B}}\left|\widehat{G}\left(\left.z\left(\boldsymbol{\mathbf{X}}_{e},\boldsymbol{\beta}\right)\right|\boldsymbol{\beta}\right)-L\left(z\left(\boldsymbol{\mathbf{X}}_{e},\boldsymbol{\beta}\right),\boldsymbol{\beta}\right)\right|\\
 & \leq\sup_{\left(\boldsymbol{\mathbf{X}}_{e},\boldsymbol{\beta}\right)\in\mathcal{X}_{e,n}\times\mathcal{B}}\frac{\left|A_{n,y}\left(\boldsymbol{\mathbf{X}}_{e},\boldsymbol{\beta}\right)-A_{y}\left(\boldsymbol{\mathbf{X}}_{e},\boldsymbol{\beta}\right)\right|}{A_{n,1}\left(\boldsymbol{\mathbf{X}}_{e},\boldsymbol{\beta}\right)}\\
 & +\sup_{\left(\boldsymbol{\mathbf{X}}_{e},\boldsymbol{\beta}\right)\in\mathcal{X}_{e,n}\times\mathcal{B}}L\left(z\left(\boldsymbol{\mathbf{X}}_{e},\boldsymbol{\beta}\right),\boldsymbol{\beta}\right)\frac{\left|A_{n,1}\left(\boldsymbol{\mathbf{X}},\boldsymbol{\beta}\right)-A_{1}\left(\boldsymbol{\mathbf{X}},\boldsymbol{\beta}\right)\right|}{A_{1}\left(\boldsymbol{\mathbf{X}},\boldsymbol{\beta}\right)}.
\end{align*}
Obviously, since $\psi_{1}\left(n,p,h_{n}\right)=o\left(\phi_{n}^{p}/p^{p}\right)$,
\[
\sup_{\left(\boldsymbol{\mathbf{X}}_{e},\boldsymbol{\beta}\right)\in\mathcal{X}_{e,n}\times\mathcal{B}}\left|A_{n,1}\left(\boldsymbol{\mathbf{X}}_{e},\boldsymbol{\beta}\right)-A_{1}\left(\boldsymbol{\mathbf{X}}_{e},\boldsymbol{\beta}\right)\right|=o_{p}\left(\phi_{n}^{p}/p^{p}\right),
\]
so $\inf_{\left(\boldsymbol{\mathbf{X}}_{e},\boldsymbol{\beta}\right)\in\mathcal{X}_{e,n}\times\mathcal{B}}A_{n,1}^{-1}\left(\boldsymbol{X}_{e},\boldsymbol{\beta}\right)=O_{p}\left(p^{p}\phi_{n}^{-p}\right)$.
Moreover, $L\left(z\left(\mathbf{X}_{e},\boldsymbol{\beta}\right),\boldsymbol{\beta}\right)$
is upper bounded by  \autoref{lem:2}(vii). Then the results hold
according to   \autoref{lem3}.
\end{proof}
\textbf{\medskip{}
}
\begin{flushleft}
\textbf{\medskip{}
Proof of \autoref{lem:3.1}. }
\par\end{flushleft}
\begin{proof}
Note that 
\begin{align}
 & \sup_{\boldsymbol{\beta}\in\mathcal{B}}\left\Vert \frac{1}{n}\sum_{i=1}^{n}\widehat{G}\left(\left.z\left(\mathbf{X}_{e,i},\boldsymbol{\beta}\right)\right|\boldsymbol{\beta}\right)\boldsymbol{\mathbf{X}}_{i}-\mathbb{E}\left[L\left(z\left(\boldsymbol{\mathbf{X}}_{e,i},\boldsymbol{\beta}\right),\boldsymbol{\beta}\right)\boldsymbol{\mathbf{X}}_{i}\right]\right\Vert \nonumber \\
 & \leq\sup_{\boldsymbol{\beta}\in\mathcal{B}}\left\Vert \frac{1}{n}\sum_{i=1}^{n}\left(\widehat{G}\left(\left.z\left(\boldsymbol{\mathbf{X}}_{e,i},\boldsymbol{\beta}\right)\right|\boldsymbol{\beta}\right)\boldsymbol{\mathbf{X}}_{i}-L\left(z\left(\boldsymbol{\mathbf{X}}_{e,i},\boldsymbol{\beta}\right),\boldsymbol{\beta}\right)\right)\boldsymbol{\mathbf{X}}_{i}\right\Vert \label{Ap4}\\
 & +\sup_{\boldsymbol{\beta}\in\mathcal{B}}\left\Vert \frac{1}{n}\sum_{i=1}^{n}L\left(z\left(\boldsymbol{\mathbf{X}}_{e,i},\boldsymbol{\beta}\right),\boldsymbol{\beta}\right)\boldsymbol{\mathbf{X}}_{i}-\mathbb{E}\left[L\left(z\left(\boldsymbol{\mathbf{X}}_{e,i},\boldsymbol{\beta}\right),\boldsymbol{\beta}\right)\boldsymbol{\mathbf{X}}_{i}\right]\right\Vert .\label{Ap5}
\end{align}

Obviously, (\ref{Ap4}) is bounded by 
\begin{align}
 & \sup_{\boldsymbol{\beta}\in\mathcal{B}}\left\Vert \frac{1}{n}\sum_{i=1}^{n}\left(\widehat{G}\left(\left.z\left(\boldsymbol{\mathbf{X}}_{e,i},\boldsymbol{\beta}\right)\right|\boldsymbol{\beta}\right)\boldsymbol{\mathbf{X}}_{i}-L\left(z\left(\boldsymbol{\mathbf{X}}_{e,i},\boldsymbol{\beta}\right),\boldsymbol{\beta}\right)\right)\boldsymbol{\mathbf{X}}_{i}\right\Vert \nonumber \\
 & \leq\frac{1}{n}\sum_{i=1}^{n}\sup_{\boldsymbol{\beta}\in\mathcal{B}}\left\Vert \widehat{G}\left(\left.z\left(\boldsymbol{\mathbf{X}}_{e,i},\boldsymbol{\beta}\right)\right|\boldsymbol{\beta}\right)\boldsymbol{\mathbf{X}}_{i}-L\left(z\left(\boldsymbol{\mathbf{X}}_{e,i},\boldsymbol{\beta}\right),\boldsymbol{\beta}\right)\boldsymbol{\mathbf{X}}_{i}\right\Vert \cdot I_{n,i}\label{Ap6}\\
 & +\frac{1}{n}\sum_{i=1}^{n}\sup_{\boldsymbol{\beta}\in\mathcal{B}}\left\Vert \widehat{G}\left(\left.z\left(\boldsymbol{\mathbf{X}}_{e,i},\boldsymbol{\beta}\right)\right|\boldsymbol{\beta}\right)\boldsymbol{\mathbf{X}}_{i}-L\left(z\left(\boldsymbol{\mathbf{X}}_{e,i},\boldsymbol{\beta}\right),\boldsymbol{\beta}\right)\boldsymbol{\mathbf{X}}_{i}\right\Vert \cdot\left(1-I_{n,i}\right),\label{Ap7}
\end{align}
where $I_{n,i}=I\left(\boldsymbol{\mathbf{X}}_{e,i}\in\mathcal{X}_{e,n}\right)$
and $\mathcal{X}_{e,n}$ is chosen as in  \autoref{lem4}. Note that
(\ref{Ap6}) is bounded by 
\begin{align*}
 & \frac{1}{n}\sum_{i=1}^{n}\sup_{\boldsymbol{\beta}\in\mathcal{B}}\left\Vert \widehat{G}\left(\left.z\left(\boldsymbol{\mathbf{X}}_{e,i},\boldsymbol{\beta}\right)\right|\boldsymbol{\beta}\right)\boldsymbol{\mathbf{X}}_{i}-L\left(z\left(\boldsymbol{\mathbf{X}}_{e,i},\boldsymbol{\beta}\right),\boldsymbol{\beta}\right)\boldsymbol{\mathbf{X}}_{i}\right\Vert \cdot I_{n,i}\\
 & \leq\sup_{\left(\boldsymbol{\mathbf{X}}_{e},\boldsymbol{\beta}\right)\in\mathcal{X}_{e,n}\times\mathcal{B}}\left\Vert \widehat{G}\left(\left.z\left(\boldsymbol{\mathbf{X}}_{e},\boldsymbol{\beta}\right)\right|\boldsymbol{\beta}\right)\boldsymbol{\mathbf{X}}-L\left(Z\left(\boldsymbol{\mathbf{X}}_{e},\boldsymbol{\beta}\right),\boldsymbol{\beta}\right)\boldsymbol{\mathbf{X}}\right\Vert \\
 & =O_{p}\left(p^{p+1/2}\phi_{n}^{-p}\psi_{1}\left(n,p,h_{n}\right)\right),
\end{align*}
according to \autoref{lem4}. For (\ref{Ap7}), we have that 
\begin{align*}
 & \mathbb{E}\frac{1}{n}\sum_{i=1}^{n}\sup_{\boldsymbol{\beta}\in\mathcal{B}}\left\Vert \widehat{G}\left(\left.z\left(\boldsymbol{\mathbf{X}}_{e,i},\boldsymbol{\beta}\right)\right|\boldsymbol{\beta}\right)\boldsymbol{\mathbf{X}}_{i}-L\left(Z\left(\boldsymbol{\mathbf{X}}_{e,i},\boldsymbol{\beta}\right),\boldsymbol{\beta}\right)\boldsymbol{\mathbf{X}}_{i}\right\Vert \cdot\left(1-I_{n,i}\right)\\
 & \leq C\sqrt{p}\mathbb{E}I\left(\boldsymbol{\mathbf{X}}_{e,i}\notin\mathcal{X}_{e,n}\right)=O\left(p^{3/2}\phi_{n}\right),
\end{align*}
according to  \autoref{lem4}(i). Then we have that (\ref{Ap6})
is of order $O_{p}\left(p^{p+1/2}\phi_{n}^{-p}\psi_{1}\left(n,p,h_{n}\right)+p^{3/2}\phi_{n}\right)$. 

Now we go to (\ref{Ap5}). Similar to the above truncation, we have
that 
\begin{align}
 & \sup_{\boldsymbol{\beta}\in\mathcal{B}}\left\Vert \frac{1}{n}\sum_{i=1}^{n}L\left(z\left(\boldsymbol{\mathbf{X}}_{e,i},\boldsymbol{\beta}\right),\boldsymbol{\beta}\right)\boldsymbol{\mathbf{X}}_{i}-\mathbb{E}\left[L\left(z\left(\boldsymbol{\mathbf{X}}_{e,i},\boldsymbol{\beta}\right),\boldsymbol{\beta}\right)\boldsymbol{\mathbf{X}}_{i}\right]\right\Vert \nonumber \\
 & \leq\sup_{\boldsymbol{\beta}\in\mathcal{B}}\left\Vert \frac{1}{n}\sum_{i=1}^{n}L\left(z\left(\boldsymbol{\mathbf{X}}_{e,i},\boldsymbol{\beta}\right),\boldsymbol{\beta}\right)\boldsymbol{\mathbf{X}}_{i}\cdot I_{n,i}-\mathbb{E}\left[L\left(z\left(\boldsymbol{\mathbf{X}}_{e,i},\boldsymbol{\beta}\right),\boldsymbol{\beta}\right)\boldsymbol{\mathbf{X}}_{i}\cdot I_{n,i}\right]\right\Vert \label{Ap8}\\
 & +\sup_{\boldsymbol{\beta}\in\mathcal{B}}\left\Vert \frac{1}{n}\sum_{i=1}^{n}L\left(z\left(\boldsymbol{\mathbf{X}}_{e,i},\boldsymbol{\beta}\right),\boldsymbol{\beta}\right)\boldsymbol{\mathbf{X}}_{i}\cdot\left(1-I_{n,i}\right)-\mathbb{E}\left[L\left(z\left(\boldsymbol{\mathbf{X}}_{e,i},\boldsymbol{\beta}\right),\boldsymbol{\beta}\right)\boldsymbol{\mathbf{X}}_{i}\cdot\left(1-I_{n,i}\right)\right]\right\Vert .\label{Ap9}
\end{align}
Obviously, (\ref{Ap9}) is $O_{p}\left(p^{3/2}\phi_{n}\right)$. For
(\ref{Ap8}), note that $\Vert L\left(z\left(\boldsymbol{\mathbf{X}}_{e,i},\boldsymbol{\beta}\right),\boldsymbol{\beta}\right)X_{j,i}\cdot I_{n,i}\Vert$ is bounded by $C$ and $\partial \Vert L\left(z\left(\boldsymbol{\mathbf{X}}_{e,i},\boldsymbol{\beta}\right),\boldsymbol{\beta}\right)X_{j,i}\cdot I_{n,i}/\partial\boldsymbol{\beta}\Vert$ is bounded by $C\sqrt{p}$ by \autoref{lem:2}(vii) and (viii), we have that (\ref{Ap8}) is of order $O_p\left(\sqrt{p^2n\log(pn)/n}\right)$ using \autoref{lemS.1}.
Then 
\begin{align*}
 & \sup_{\boldsymbol{\beta}\in\mathcal{B}}\left\Vert \frac{1}{n}\sum_{i=1}^{n}L\left(z\left(\mathbf{X}_{e,i},\boldsymbol{\beta}\right),\boldsymbol{\beta}\right)\boldsymbol{\mathbf{X}}_{i}-\mathbb{E}\left[L\left(z\left(\boldsymbol{\mathbf{X}}_{e,i},\boldsymbol{\beta}\right),\boldsymbol{\beta}\right)\boldsymbol{\mathbf{X}}_{i}\right]\right\Vert \\
 & =O_{p}\left(\sqrt{p^{2}\log\left(pn\right)/n}+p^{3/2}\phi_{n}\right).
\end{align*}

Together, we have that
\begin{align*}
 & \sup_{\boldsymbol{\beta}\in\mathcal{B}}\left\Vert \frac{1}{n}\sum_{i=1}^{n}\widehat{G}\left(\left.z\left(\boldsymbol{\mathbf{X}}_{e,i},\boldsymbol{\beta}\right)\right|\boldsymbol{\beta}\right)\boldsymbol{\mathbf{X}}_{i}-\mathbb{E}\left[L\left(z\left(\boldsymbol{\mathbf{X}}_{e,i},\boldsymbol{\beta}\right),\boldsymbol{\beta}\right)\boldsymbol{\mathbf{X}}_{i}\right]\right\Vert \\
 & =O_{p}\left(p^{p+1/2}\phi_{n}^{-p}\psi_{1}\left(n,p,h_{n}\right)+\sqrt{p^{2}\log\left(pn\right)/n}+p^{3/2}\phi_{n}\right).
\end{align*}
Then if we set $\phi_{n}=p^{\frac{p-1}{p+1}}\psi_{1}^{\frac{1}{p+1}}\left(n,p,h_{n}\right),$
we have that
\begin{align*}
p\phi_{n} & =p^{p}\phi_{n}^{-p}\psi_{1}\left(n,p,h_{n}\right)=p^{\frac{2p}{p+1}}\psi_{1}^{\frac{1}{p+1}}\left(n,p,h_{n}\right)
 \leq p^{\frac{5p+1}{2\left(p+1\right)}}\psi_{1}^{\frac{1}{p+1}}\left(n,p,h_{n}\right)\rightarrow0,
\end{align*}
and
\[
\sqrt{p^{2}\log\left(pn\right)/n}=o\left(p^{\frac{5p+1}{2\left(p+1\right)}}\psi_{1}^{\frac{1}{p+1}}\left(n,p,h_{n}\right)\right),
\]
so 
\[
\sup_{\boldsymbol{\beta}\in\mathcal{B}}\left\Vert \frac{1}{n}\sum_{i=1}^{n}\widehat{G}\left(\left.z\left(\boldsymbol{\mathbf{X}}_{e,i},\boldsymbol{\beta}\right)\right|\boldsymbol{\beta}\right)\boldsymbol{\mathbf{X}}_{i}-\mathbb{E}\left[L\left(z\left(\boldsymbol{\mathbf{X}}_{e,i},\boldsymbol{\beta}\right),\boldsymbol{\beta}\right)\boldsymbol{\mathbf{X}}_{i}\right]\right\Vert =O_{p}\left(p^{\frac{5p+1}{2\left(p+1\right)}}\psi_{1}^{\frac{1}{p+1}}\left(n,p,h_{n}\right)\right).
\]
This finishes the whole proof. 
\end{proof}
\begin{lemma}
\label{lem5}Suppose that $p$ is fixed. If all the assumptions in
\autoref{lem3} hold with $\upsilon_{G}=4$, $\upsilon_{K}=3$,
and $\upsilon_{f}=4$, we have that  \autoref{lem3}(i) holds. Moreover, 

(i) There holds
\[
\sup_{\left(\boldsymbol{\mathbf{X}}_{e},\boldsymbol{\beta}\right)\in\mathcal{X}_{e}\times\in\mathcal{B}}\left|\mathbb{E}_{\mathscr{D}_{n}}A_{n,\cdot}\left(\mathbf{X}_{e},\boldsymbol{\beta}\right)-A_{\cdot}\left(\boldsymbol{\mathbf{X}}_{e},\boldsymbol{\beta}\right)\right|=O_{p}\left(h_{n}^{3}\right);
\]

(ii) There holds
\[
\sup_{\left(\boldsymbol{\mathbf{X}}_{e},\boldsymbol{\beta}\right)\in\mathcal{X}_{e}\times\in\mathcal{B}}\left|A_{n,\cdot}\left(\boldsymbol{\mathbf{X}}_{e},\boldsymbol{\beta}\right)-A_{\cdot}\left(\boldsymbol{\mathbf{X}}_{e},\boldsymbol{\beta}\right)\right|=O_{p}\left(h^{-1}\sqrt{\log\left(nh^{-1}\right)/n}+h^{3}\right).
\]
\end{lemma}
\begin{proof}
The proof is similar to the proof of  \autoref{lem3} so is omitted. 
\end{proof}
\begin{lemma}
\label{lem6}Suppose that $p$ is fixed. For any $\mathbf{X}_{e}\in\mathcal{X}_{e}$
and $\boldsymbol{\beta}\in\mathcal{B}$, define 
\[
A_{n,\cdot}^{\prime}\left(\boldsymbol{\mathbf{X}}_{e},\boldsymbol{\beta}\right)=\frac{1}{nh_{n}^{2}}\sum_{j=1}^{n}K^{\prime}\left(\left(z\left(\boldsymbol{\mathbf{X}}_{e},\boldsymbol{\beta}\right)-z\left(\boldsymbol{\mathbf{X}}_{e,j},\boldsymbol{\beta}\right)\right)/h_{n}\right)\left(\boldsymbol{\mathbf{X}}-\boldsymbol{\mathbf{X}}_{j}\right)\cdot\left(\cdot_{j}\right),
\]
where $\cdot=1$ or $\cdot=y$. If all the assumptions in \autoref{lem3}
hold with $\upsilon_{G}=4$, $\upsilon_{K}=3$, and $\upsilon_{f}=4$,
then 

(i) There holds
\[
\sup_{\left(\boldsymbol{\boldsymbol{\mathbf{X}}}_{e},\boldsymbol{\beta}\right)\in\mathcal{X}_{e}\times\mathcal{B}}\left\Vert A_{n,\cdot}^{\prime}\left(\boldsymbol{\boldsymbol{\mathbf{X}}}_{e},\boldsymbol{\beta}\right)-\mathbb{E}_{\mathscr{D}_{n}}A_{n,\cdot}^{\prime}\left(\boldsymbol{\boldsymbol{\mathbf{X}}}_{e},\boldsymbol{\beta}\right)\right\Vert =O_{p}\left(h_{n}^{-2}\sqrt{\log\left(nh_{n}^{-1}\right)/n}\right);
\]

(ii) Define $A_{y}^{\prime}\left(\boldsymbol{\boldsymbol{\mathbf{X}}}_{e},\boldsymbol{\beta}\right)=\lim_{n\rightarrow\infty}\mathbb{E}_{\mathscr{D}_{n}}A_{n,y}^{\prime}\left(\boldsymbol{\boldsymbol{\mathbf{X}}}_{e},\boldsymbol{\beta}\right)$
and $A_{1}^{\prime}\left(\boldsymbol{\boldsymbol{\mathbf{X}}}_{e},\boldsymbol{\beta}\right)=\lim_{n\rightarrow\infty}\mathbb{E}_{\mathscr{D}_{n}}A_{n,1}^{\prime}\left(\boldsymbol{\boldsymbol{\mathbf{X}}}_{e},\boldsymbol{\beta}\right)$.
We have that $A_{y}^{\prime}\left(\boldsymbol{\boldsymbol{\mathbf{X}}}_{e},\boldsymbol{\beta}\right)=\left.\partial H_{1}\left(\left.z,\boldsymbol{\boldsymbol{\mathbf{X}}}\right|\boldsymbol{\beta}\right)/\partial z\right|_{z=z\left(\boldsymbol{\boldsymbol{\mathbf{X}}}_{e},\boldsymbol{\beta}\right)}$
and $A_{1}^{\prime}\left(\boldsymbol{\boldsymbol{\mathbf{X}}}_{e},\boldsymbol{\beta}\right)=\left.\partial H_{2}\left(\left.z,\boldsymbol{\boldsymbol{\mathbf{X}}}\right|\boldsymbol{\beta}\right)/\partial z\right|_{z=z\left(\boldsymbol{\boldsymbol{\mathbf{X}}}_{e},\boldsymbol{\beta}\right)}$,
where 
\[
H_{1}\left(\left.z,\boldsymbol{\boldsymbol{\mathbf{X}}}\right|\boldsymbol{\beta}\right)=\int_{\mathcal{X}}G\left(z-\widetilde{\boldsymbol{\mathbf{X}}}^{\mathrm{T}}\Delta\boldsymbol{\beta}\right)f_{e}\left(z-\widetilde{\boldsymbol{\mathbf{X}}}^{\mathrm{T}}\boldsymbol{\beta}, \widetilde{\boldsymbol{\mathbf{X}}} \right)\left(\boldsymbol{\boldsymbol{\mathbf{X}}}-\widetilde{\boldsymbol{\mathbf{X}}}\right)d\widetilde{\boldsymbol{\mathbf{X}}},
\]
\[
H_{2}\left(\left.z,\boldsymbol{\boldsymbol{\mathbf{X}}}\right|\boldsymbol{\beta}\right)=\int_{\mathcal{X}}f_{e}\left(z-\widetilde{\boldsymbol{\mathbf{X}}}^{\mathrm{T}}\boldsymbol{\beta}, \widetilde{\boldsymbol{\mathbf{X}}} \right)\left(\boldsymbol{\boldsymbol{\mathbf{X}}}-\widetilde{\boldsymbol{\mathbf{X}}}\right)d\widetilde{\boldsymbol{\mathbf{X}}},
\]
and the differentiation of $H_{1}$ and $H_{2}$ are element-wise.
Moreover, there holds
\[
\sup_{\left(\boldsymbol{\boldsymbol{\mathbf{X}}}_{e},\boldsymbol{\beta}\right)\in\mathcal{X}_{e}\times\mathcal{B}}\left\Vert \mathbb{E}_{\mathscr{D}_{n}}A_{n,\cdot}^{\prime}\left(\boldsymbol{\boldsymbol{\mathbf{X}}}_{e},\boldsymbol{\beta}\right)-A_{\cdot}^{\prime}\left(\boldsymbol{\boldsymbol{\mathbf{X}}}_{e},\boldsymbol{\beta}\right)\right\Vert =O_{p}\left(h_{n}^{3}\right),
\]

(iii) There holds
\[
\sup_{\left(\boldsymbol{\mathbf{X}}_{e},\boldsymbol{\beta}\right)\in\mathcal{X}_{e}\times\mathcal{B}}\left\Vert A_{n,\cdot}^{\prime}\left(\mathbf{X}_{e},\boldsymbol{\beta}\right)-A_{\cdot}^{\prime}\left(\boldsymbol{\mathbf{X}}_{e},\boldsymbol{\beta}\right)\right\Vert =O_{p}\left(h_{n}^{-2}\sqrt{\log\left(nh_{n}^{-1}\right)/n}+h_{n}^{3}\right).
\]
 
\end{lemma}
\begin{proof}
\autoref{lem6}(i) is a direct result of \autoref{lemS.1} if we note that for each $1\leq l \leq p$,  $h_n^{-2}K^{\prime}\left(\left(z\left(\boldsymbol{\mathbf{X}}_{e},\boldsymbol{\beta}\right)-z\left(\boldsymbol{\mathbf{X}}_{e,j},\boldsymbol{\beta}\right)\right)/h_{n}\right)\left(X_l-X_{l,j}\right)\cdot\left(\cdot_{j}\right)$ is bounded by $Ch_n^{-2}$ and its derivatives with respect to $\boldsymbol{\beta}$ and $\mathbf{X}$ are both upper bounded since $p$ is fixed. 

To prove \autoref{lem6}(ii), we note that 
\begin{align*}
 & \mathbb{E}_{\mathscr{D}_{n}}A_{n,y}^{\prime}\left(\boldsymbol{\mathbf{X}}_{e},\boldsymbol{\beta}\right)\\
 & =\frac{1}{h_{n}^{2}}\mathbb{E}_{\mathscr{D}_{n}}\left[K^{\prime}\left(\left(z\left(\boldsymbol{\mathbf{X}}_{e},\boldsymbol{\beta}\right)-z\left(\boldsymbol{\mathbf{X}}_{e,j},\boldsymbol{\beta}\right)\right)/h_{n}\right)\left(\boldsymbol{\mathbf{X}}-\boldsymbol{\mathbf{X}}_{j}\right)\cdot G\left(X_{0,j}+\boldsymbol{\mathbf{X}}_{j}^{\mathrm{T}}\boldsymbol{\beta}^{\star}\right)\right]\\
 & =\frac{1}{h_{n}^{2}}\mathbb{E}_{\mathscr{D}_{n}}\left[K^{\prime}\left(\left(z\left(\boldsymbol{\mathbf{X}}_{e},\boldsymbol{\beta}\right)-z\left(\boldsymbol{\mathbf{X}}_{e,j},\boldsymbol{\beta}\right)\right)/h_{n}\right)\left(\boldsymbol{\mathbf{X}}-\boldsymbol{\mathbf{X}}_{j}\right)\cdot G\left(z\left(\boldsymbol{\mathbf{X}}_{e,j},\boldsymbol{\beta}\right)-\boldsymbol{\mathbf{X}}_{j}^{\mathrm{T}}\Delta\boldsymbol{\beta}\right)\right]\\
 & =\frac{1}{h_{n}^{2}}\int K^{\prime}\left(\left(z\left(\boldsymbol{\mathbf{X}}_{e},\boldsymbol{\beta}\right)-z\right)/h_{n}\right)dz\int_{\mathcal{X}}\left[G\left(z-\widetilde{\mathbf{X}}^{\mathrm{T}}\Delta\boldsymbol{\beta}\right)f_{\mathbf{X},z}\left(\left.\widetilde{\mathbf{X}},z\right|\boldsymbol{\beta}\right)\left(\mathbf{X}-\widetilde{\mathbf{X}}\right)\right]d\widetilde{\mathbf{X}}\\
 & =\frac{1}{h_{n}^{2}}\int\left[K^{\prime}\left(\left(z\left(\boldsymbol{\mathbf{X}}_{e},\boldsymbol{\beta}\right)-z\right)/h_{n}\right)H_{1}\left(\left.z,\mathbf{X}\right|\boldsymbol{\beta}\right)\right]dz\\
 & =\frac{1}{h_{n}}\int\left[K^{\prime}\left(z\right)H_{1}\left(\left.z\left(\boldsymbol{\mathbf{X}}_{e},\boldsymbol{\beta}\right)-h_{n}z,\boldsymbol{\mathbf{X}}\right|\boldsymbol{\beta}\right)\right]dz
\end{align*}
Note that both $G$ and $f_{e}$ have up to fourth bounded derivatives
with respect to $z$, and the upper bounds hold uniformly with respect
to $z$, $\boldsymbol{\mathbf{X}}$ and $\boldsymbol{\beta}$. This
implies that each element of  $H_{1}\left(\left.z,\boldsymbol{\mathbf{X}}\right|\boldsymbol{\beta}\right)$
has up to fourth bounded derivatives with respect to $z$. Also ote
that $\int K^{\prime}\left(v\right)dv=\left.K\left(v\right)\right|_{-\infty}^{\infty}=0$,
$\int vK^{\prime}\left(v\right)dv=\left.K\left(v\right)\right|_{-\infty}^{\infty}-\int K\left(v\right)dv=-1,$
$\int v^{s}K^{\prime}\left(v\right)dv=\left.v^{s}K\left(v\right)\right|_{-\infty}^{\infty}-s\int v^{s-1}K\left(v\right)dv=0$
for $s=2,3$, and $\left|\int v^{4}K^{\prime}\left(v\right)dv\right|<\infty.$
This implies that 
\[
\left\Vert \mathbb{E}_{\mathscr{D}_{n}}A_{n,y}^{\prime}\left(\mathbf{X}_{e},\boldsymbol{\beta}\right)-A_{y}^{\prime}\left(\mathbf{X}_{e},\boldsymbol{\beta}\right)\right\Vert =O_{p}\left(h_{n}^{3 }\right)
\]
uniform with respect to $\mathbf{X}_{e}$ and $\boldsymbol{\beta}$.
The proof of the uniform distance between $\mathbb{E}_{\mathscr{D}_{n}}A_{n,1}^{\prime}\left(\mathbf{X}_{e},\boldsymbol{\beta}\right)$
and $A_{1}^{\prime}\left(\mathbf{X}_{e},\boldsymbol{\beta}\right)$
is similar. So we finish the proof of \autoref{lem6}(ii).

Finally, \autoref{lem6}(iii) is a combination of \autoref{lem6}(i) and \autoref{lem6}(ii). 
\end{proof}
\begin{lemma}
\label{lem8}Suppose that $p$ is fixed. If all the assumptions in
\autoref{lem3} hold with $\upsilon_{G}=4$, $\upsilon_{K}=3$,
and $\upsilon_{f}=4$, we have that 
\begin{align*}
\sup_{\left(\boldsymbol{\mathbf{X}}_{e},\boldsymbol{\beta}\right)\in\mathcal{X}_{e}^{\phi}\times\mathcal{B}} & \left\Vert \frac{\partial\widehat{G}\left(\left.z\left(\mathbf{X}_{e},\boldsymbol{\beta}\right)\right|\boldsymbol{\beta}\right)}{\partial\boldsymbol{\beta}}-\frac{\partial H_{1}\left(z\left(\boldsymbol{\mathbf{X}}_{e},\boldsymbol{\beta}\right),\boldsymbol{\mathbf{X}}_{e}\right)/\partial z}{f_{z}\left(z\left(\boldsymbol{\mathbf{X}}_{e},\boldsymbol{\beta}\right)\right)}\right.\\
 & \left.+L\left(z\left(\boldsymbol{\mathbf{X}}_{e},\boldsymbol{\beta}\right),\boldsymbol{\beta}\right)\frac{\partial H_{2}\left(z\left(\boldsymbol{\mathbf{X}}_{e},\boldsymbol{\beta}\right),\boldsymbol{\mathbf{X}}_{e}\right)/\partial z}{f_{z}\left(z\left(\boldsymbol{\mathbf{X}}_{e},\boldsymbol{\beta}\right)\right)}\right\Vert =O_{p}\left(h_{n}^{-2}\sqrt{\log\left(nh_{n}^{-1}\right)/n}+h_{n}^{3}\right),
\end{align*}
where $\mathcal{X}_e^{\phi}$ is defined in (\ref{eq:truncation_seet}) in the main text.
\end{lemma}
\begin{proof}
Note that 
\begin{align*}
\frac{\partial\widehat{G}\left(\left.z\left(\boldsymbol{\mathbf{X}}_{e},\boldsymbol{\beta}\right)\right|\boldsymbol{\beta}\right)}{\partial\boldsymbol{\beta}} & =\frac{\partial A_{n,y}\left(\boldsymbol{\mathbf{X}}_{e},\boldsymbol{\beta}\right)/\partial\boldsymbol{\beta}}{A_{n,1}\left(\boldsymbol{\mathbf{X}}_{e},\boldsymbol{\beta}\right)}-\frac{A_{n,y}\left(\boldsymbol{\mathbf{X}}_{e},\boldsymbol{\beta}\right)}{A_{n,1}\left(\boldsymbol{\mathbf{X}}_{e},\boldsymbol{\beta}\right)}\cdot\frac{\partial A_{n,1}\left(\boldsymbol{\mathbf{X}}_{e},\boldsymbol{\beta}\right)/\partial\boldsymbol{\beta}}{A_{n,1}\left(\boldsymbol{\mathbf{X}}_{e},\boldsymbol{\beta}\right)}\\
 & =\frac{A_{n,y}^{\prime}\left(\boldsymbol{\mathbf{X}}_{e},\boldsymbol{\beta}\right)}{A_{n,1}\left(\boldsymbol{\mathbf{X}}_{e},\boldsymbol{\beta}\right)}-\frac{A_{n,y}\left(\boldsymbol{\mathbf{X}}_{e},\boldsymbol{\beta}\right)}{A_{n,1}\left(\boldsymbol{\mathbf{X}}_{e},\boldsymbol{\beta}\right)}\frac{A_{n,1}^{\prime}\left(\boldsymbol{\mathbf{X}}_{e},\boldsymbol{\beta}\right)}{A_{n,1}\left(\boldsymbol{\mathbf{X}}_{e},\boldsymbol{\beta}\right)}.
\end{align*}
Then 
\begin{align}
\left\Vert \frac{A_{n,y}^{\prime}\left(\boldsymbol{\mathbf{X}}_{e},\boldsymbol{\beta}\right)}{A_{n,1}\left(\boldsymbol{\mathbf{X}}_{e},\boldsymbol{\beta}\right)}-\frac{\partial H_{1}\left(z\left(\boldsymbol{\mathbf{X}}_{e},\boldsymbol{\beta}\right),\boldsymbol{\mathbf{X}}_{e}\right)/\partial z}{f_{z}\left(z\left(\boldsymbol{\mathbf{X}}_{e},\boldsymbol{\beta}\right)\right)}\right\Vert  & =\left\Vert \frac{A_{n,y}^{\prime}\left(\boldsymbol{\mathbf{X}}_{e},\boldsymbol{\beta}\right)}{A_{n,1}\left(\boldsymbol{\mathbf{X}}_{e},\boldsymbol{\beta}\right)}-\frac{A_{y}^{\prime}\left(\boldsymbol{\mathbf{X}}_{e},\boldsymbol{\beta}\right)}{A_{1}\left(\boldsymbol{\mathbf{X}}_{e},\boldsymbol{\beta}\right)}\right\Vert \nonumber \\
 & \leq\left\Vert \frac{A_{n,y}^{\prime}\left(\boldsymbol{\mathbf{X}}_{e},\boldsymbol{\beta}\right)-A_{y}^{\prime}\left(\boldsymbol{\mathbf{X}}_{e},\boldsymbol{\beta}\right)}{A_{n,1}\left(\boldsymbol{\mathbf{X}}_{e},\boldsymbol{\beta}\right)}\right\Vert \label{Ap25}\\
 & +\left\Vert \frac{A_{y}^{\prime}\left(\boldsymbol{\mathbf{X}}_{e},\boldsymbol{\beta}\right)}{A_{1}\left(\boldsymbol{\mathbf{X}}_{e},\boldsymbol{\beta}\right)}\frac{A_{n,1}\left(\boldsymbol{\mathbf{X}}_{e},\boldsymbol{\beta}\right)-A_{1}\left(\boldsymbol{\mathbf{X}}_{e},\boldsymbol{\beta}\right)}{A_{n,1}\left(\boldsymbol{\mathbf{X}}_{e},\boldsymbol{\beta}\right)}\right\Vert .\label{Ap26}
\end{align}
Now for any $\left(\boldsymbol{\mathbf{X}}_{e},\boldsymbol{\beta}\right)\in\mathcal{X}_{e}^{\phi}\times\mathcal{B}$,
$A_{1}\left(\boldsymbol{\mathbf{X}}_{e},\boldsymbol{\beta}\right)$
is uniformly lower-bounded according to  \autoref{lem4}, so $A_{n,1}^{-1}\left(\boldsymbol{\mathbf{X}}_{e},\boldsymbol{\beta}\right)=O_{p}\left(1\right)$
also uniformly holds. Moreover, $\left\Vert A_{y}^{\prime}\left(\boldsymbol{\mathbf{X}}_{e},\boldsymbol{\beta}\right)\right\Vert $
is upper bounded, so $\left\Vert A_{n,y}^{\prime}\left(\boldsymbol{\mathbf{X}}_{e},\boldsymbol{\beta}\right)\right\Vert =O_{p}\left(1\right)$
also uniformly holds. Then (\ref{Ap25}) is $O_{p}\left(h_{n}^{-2}\sqrt{\log\left(nh_{n}^{-1}\right)/n}+h_{n}^{3}\right)$
and (\ref{Ap26}) is $O_{p}\left(h_{n}^{-1}\sqrt{\log\left(nh_{n}^{-1}\right)/n}+h_{n}^{3}\right)$.
Similar method can be used to show that 
\[
\frac{A_{n,y}\left(\boldsymbol{\mathbf{X}}_{e},\boldsymbol{\beta}\right)}{A_{n,1}\left(\boldsymbol{\mathbf{X}}_{e},\boldsymbol{\beta}\right)}\frac{A_{n,1}^{\prime}\left(\boldsymbol{\mathbf{X}}_{e},\boldsymbol{\beta}\right)}{A_{n,1}\left(\boldsymbol{\mathbf{X}}_{e},\boldsymbol{\beta}\right)}-\frac{L\left(z\left(\boldsymbol{\mathbf{X}}_{e},\boldsymbol{\beta}\right),\boldsymbol{\beta}\right)\partial H_{2}\left(z\left(\boldsymbol{\mathbf{X}}_{e},\boldsymbol{\beta}\right),\boldsymbol{\mathbf{X}}_{e}\right)/\partial z}{f_{z}\left(z\left(\boldsymbol{\mathbf{X}}_{e},\boldsymbol{\beta}\right)\right)}
\]
is also $O_{p}\left(h_{n}^{-2}\sqrt{\log\left(nh_{n}^{-1}\right)/n}+h_{n}^{3}\right)$.
This finishes the proof.
\end{proof}
\begin{lemma}
\label{lem9}Suppose that $p$ is fixed. If all the assumptions in
\autoref{lem3} hold with $\upsilon_{G}=4$, $\upsilon_{K}=3$,
and $\upsilon_{f}=4$, then for any $\overline{\mathcal{B}}\subseteq\mathcal{B}$,
we have that
\begin{align*}
\sup_{\left(\boldsymbol{\mathbf{X}}_{e},\boldsymbol{\beta}\right)\in\mathcal{X}_{e}^{\phi}\times\overline{\mathcal{B}}}\left\Vert \frac{\partial\widehat{G}\left(\left.Z\left(\mathbf{X}_{e},\boldsymbol{\beta}\right)\right|\boldsymbol{\beta}\right)}{\partial\boldsymbol{\beta}}-\int W\left(\boldsymbol{\mathbf{X}}_{e},\widetilde{\boldsymbol{\mathbf{X}}}_{e},\boldsymbol{\beta}\right)\left(\mathbf{X}-\widetilde{\boldsymbol{\mathbf{X}}}_{e}\right)d\widetilde{\boldsymbol{\mathbf{X}}}_{e}\right\Vert  & \leq\alpha_{1,n}+\alpha_{2},
\end{align*}
where $\alpha_{1,n}=O_{p}\left(h_{n}^{-2}\sqrt{\log\left(nh_{n}^{-1}\right)/n}+h_{n}^{3}\right)$
and $\alpha_{2}=O_{p}\left(\sup_{\boldsymbol{\beta}\in\overline{\mathcal{B}}}\left\Vert \Delta\boldsymbol{\beta}\right\Vert \right)$.
\end{lemma}
\begin{proof}
We only need to show that 
\begin{align*}
\sup_{\left(\mathbf{X}_{e},\boldsymbol{\beta}\right)\in\mathcal{\overline{X}}_{e}\times\mathcal{B}} & \left\Vert \frac{\partial H_{1}\left(z\left(\boldsymbol{\mathbf{X}}_{e},\boldsymbol{\beta}\right),\boldsymbol{\mathbf{X}}_{e}\right)/\partial z}{f_{z}\left(z\left(\boldsymbol{\mathbf{X}}_{e},\boldsymbol{\beta}\right)\right)}-L\left(z\left(\boldsymbol{\mathbf{X}}_{e},\boldsymbol{\beta}\right),\boldsymbol{\beta}\right)\frac{\partial H_{2}\left(z\left(\boldsymbol{\mathbf{X}}_{e},\boldsymbol{\beta}\right),\boldsymbol{\mathbf{X}}_{e}\right)/\partial z}{f_{z}\left(z\left(\boldsymbol{\mathbf{X}}_{e},\boldsymbol{\beta}\right)\right)}\right.\\
 & \left.-\int W\left(\boldsymbol{\mathbf{X}}_{e},\widetilde{\mathbf{X}}_{e},\boldsymbol{\beta}\right)\left(\mathbf{X}-\widetilde{\mathbf{X}}\right)d\widetilde{\mathbf{X}}\right\Vert =O\left(\left\Vert \Delta\boldsymbol{\beta}\right\Vert \right).
\end{align*}
Note that 
\begin{align*}
 & \partial H_{1}\left(z\left(\boldsymbol{\mathbf{X}}_{e},\boldsymbol{\beta}\right),\mathbf{X}_{e}\right)/\partial z-L\left(z\left(\boldsymbol{\mathbf{X}}_{e},\boldsymbol{\beta}\right),\boldsymbol{\beta}\right)\partial H_{2}\left(z\left(\boldsymbol{\mathbf{X}}_{e},\boldsymbol{\beta}\right),\boldsymbol{\mathbf{X}}_{e}\right)/\partial z\\
 & =\int G^{\prime}\left(z\left(\mathbf{X}_{e},\boldsymbol{\beta}\right)-\widetilde{\mathbf{X}}\Delta\boldsymbol{\beta}\right)f_{e}\left(z\left(\boldsymbol{\boldsymbol{\mathbf{X}}}_{e},\boldsymbol{\beta}\right)-\widetilde{\boldsymbol{\mathbf{X}}}^{\mathrm{T}}\boldsymbol{\beta},\widetilde{\boldsymbol{\mathbf{X}}}\right)\left(\mathbf{X}-\widetilde{\mathbf{X}}\right)d\widetilde{\mathbf{X}}\\
 & +\int G\left(z\left(\boldsymbol{\mathbf{X}}_{e},\boldsymbol{\beta}\right)-\widetilde{\boldsymbol{\mathbf{X}}}^{\mathrm{T}}\Delta\boldsymbol{\beta}\right)\left(\partial f_{e}\left(z\left(\boldsymbol{\boldsymbol{\mathbf{X}}}_{e},\boldsymbol{\beta}\right)-\widetilde{\boldsymbol{\mathbf{X}}}^{\mathrm{T}}\boldsymbol{\beta},\widetilde{\boldsymbol{\mathbf{X}}}\right)/\partial z\right)\left(\boldsymbol{\mathbf{X}}-\widetilde{\mathbf{X}}\right)d\widetilde{\mathbf{X}}\\
 & -L\left(z\left(\mathbf{X}_{e},\boldsymbol{\beta}\right),\boldsymbol{\beta}\right)\int\left(\partial f_{e}\left(z\left(\boldsymbol{\boldsymbol{\mathbf{X}}}_{e},\boldsymbol{\beta}\right)-\widetilde{\boldsymbol{\mathbf{X}}}^{\mathrm{T}}\boldsymbol{\beta},\widetilde{\boldsymbol{\mathbf{X}}}\right)/\partial z\right)\left(\mathbf{X}-\widetilde{\mathbf{X}}\right)d\widetilde{\mathbf{X}}\\
 & =\int G^{\prime}\left(z\left(\boldsymbol{\mathbf{X}}_{e},\boldsymbol{\beta}\right)-\boldsymbol{\mathbf{X}}^{\mathrm{T}}\Delta\boldsymbol{\beta}\right)f_{e}\left(z\left(\boldsymbol{\boldsymbol{\mathbf{X}}}_{e},\boldsymbol{\beta}\right)-\widetilde{\boldsymbol{\mathbf{X}}}^{\mathrm{T}}\boldsymbol{\beta},\widetilde{\boldsymbol{\mathbf{X}}}\right)\left(\mathbf{X}-\widetilde{\mathbf{X}}\right)d\widetilde{\mathbf{X}}\\
 & +\int\left[G\left(z\left(\boldsymbol{\mathbf{X}}_{e},\boldsymbol{\beta}\right)-\boldsymbol{\mathbf{X}}^{\mathrm{T}}\Delta\boldsymbol{\beta}\right)-G\left(z\left(\boldsymbol{\mathbf{X}}_{e},\boldsymbol{\beta}\right)\right)\right]\left(\partial f_{e}\left(z\left(\boldsymbol{\boldsymbol{\mathbf{X}}}_{e},\boldsymbol{\beta}\right)-\widetilde{\boldsymbol{\mathbf{X}}}^{\mathrm{T}}\boldsymbol{\beta},\widetilde{\boldsymbol{\mathbf{X}}}\right)/\partial z\right)\left(\mathbf{X}-\widetilde{\mathbf{X}}\right)d\widetilde{\mathbf{X}}\\
 & -\left(L\left(z\left(\mathbf{X}_{e},\boldsymbol{\beta}\right),\boldsymbol{\beta}\right)-G\left(z\left(\mathbf{X}_{e},\boldsymbol{\beta}\right)\right)\right)\int\left(\partial f_{e}\left(z\left(\boldsymbol{\boldsymbol{\mathbf{X}}}_{e},\boldsymbol{\beta}\right)-\widetilde{\boldsymbol{\mathbf{X}}}^{\mathrm{T}}\boldsymbol{\beta},\widetilde{\boldsymbol{\mathbf{X}}}\right)/\partial z\right)\left(\mathbf{X}-\widetilde{\mathbf{X}}\right)d\widetilde{\mathbf{X}}.
\end{align*}
Note that 
\begin{align*}
 & \left\Vert \int\left[G\left(z\left(\boldsymbol{\mathbf{X}}_{e},\boldsymbol{\beta}\right)-\widetilde{\boldsymbol{\mathbf{X}}}^{\mathrm{T}}\Delta\boldsymbol{\beta}\right)-G\left(z\left(\boldsymbol{\mathbf{X}}_{e},\boldsymbol{\beta}\right)\right)\right]\left(\partial f_{e}\left(z\left(\boldsymbol{\boldsymbol{\mathbf{X}}}_{e},\boldsymbol{\beta}\right)-\widetilde{\boldsymbol{\mathbf{X}}}^{\mathrm{T}}\boldsymbol{\beta},\widetilde{\boldsymbol{\mathbf{X}}}\right)/\partial z\right)\left(\mathbf{X}-\widetilde{\mathbf{X}}\right)d\widetilde{\mathbf{X}}\right\Vert \\
 & \leq C\cdot\sup_{\widetilde{\boldsymbol{\mathbf{X}}}\in\mathcal{X}}\left|G\left(z\left(\boldsymbol{\mathbf{X}}_{e},\boldsymbol{\beta}\right)-\widetilde{\mathbf{X}}^{\mathrm{T}}\Delta\boldsymbol{\beta}\right)-G\left(z\left(\boldsymbol{\mathbf{X}}_{e},\boldsymbol{\beta}\right)\right)\right|\cdot m\left(\mathbb{X}\left(z\left(\boldsymbol{\mathbf{X}}_{e},\boldsymbol{\beta}\right),\mathbf{X}\right)\right)\\
 & \leq C\cdot\left\Vert \Delta\boldsymbol{\beta}\right\Vert \cdot m\left(\mathbb{X}\left(z\left(\boldsymbol{\mathbf{X}}_{e},\boldsymbol{\beta}\right),\mathbf{X}\right)\right),
\end{align*}
and according to our choice of $\mathcal{X}_{e}^{\phi}$, we know
that $m\left(\mathbb{X}\left(z\left(\boldsymbol{\mathbf{X}}_{e},\boldsymbol{\beta}\right),\mathbf{X}\right)\right)>0$.
On the other side, 
\begin{align*}
 & \left\Vert \left(L\left(z\left(\boldsymbol{\mathbf{X}}_{e},\boldsymbol{\beta}\right),\boldsymbol{\beta}\right)-G\left(z\left(\boldsymbol{\mathbf{X}}_{e},\boldsymbol{\beta}\right)\right)\right)\int\left(\partial f_{e}\left(z\left(\boldsymbol{\boldsymbol{\mathbf{X}}}_{e},\boldsymbol{\beta}\right)-\widetilde{\boldsymbol{\mathbf{X}}}^{\mathrm{T}}\boldsymbol{\beta},\widetilde{\boldsymbol{\mathbf{X}}}\right)/\partial z\right)\left(\boldsymbol{\mathbf{X}}-\widetilde{\boldsymbol{\mathbf{X}}}\right)d\widetilde{\boldsymbol{\mathbf{X}}}\right\Vert \\
 & \leq C\cdot\left|L\left(z\left(\boldsymbol{\boldsymbol{\mathbf{X}}}_{e},\boldsymbol{\beta}\right),\boldsymbol{\beta}\right)-G\left(z\left(\boldsymbol{\boldsymbol{\mathbf{X}}}_{e},\boldsymbol{\beta}\right)\right)\right|\cdot m\left(\mathbb{X}\left(z\left(\boldsymbol{\mathbf{X}}_{e},\boldsymbol{\beta}\right),\mathbf{X}\right)\right)\\
 & =C\cdot\left|L\left(z\left(\boldsymbol{\boldsymbol{\mathbf{X}}}_{e},\boldsymbol{\beta}\right),\boldsymbol{\beta}\right)-L\left(z\left(\boldsymbol{\boldsymbol{\mathbf{X}}}_{e},\boldsymbol{\beta}\right),\boldsymbol{\beta}^{\star}\right)\right|\cdot m\left(\mathbb{X}\left(z\left(\boldsymbol{\mathbf{X}}_{e},\boldsymbol{\beta}\right),\mathbf{X}\right)\right)\\
 & \leq C\cdot\left(\sup_{z,\boldsymbol{\beta}}\left\Vert \partial L\left(z,\boldsymbol{\beta}\right)/\partial\boldsymbol{\beta}\right\Vert \right)\cdot\left\Vert \Delta\boldsymbol{\beta}\right\Vert \cdot m\left(\mathbb{X}\left(z\left(\boldsymbol{\mathbf{X}}_{e},\boldsymbol{\beta}\right),\mathbf{X}\right)\right)\\
 & \leq C\cdot\left\Vert \Delta\boldsymbol{\beta}\right\Vert \cdot m\left(\mathbb{X}\left(z\left(\boldsymbol{\mathbf{X}}_{e},\boldsymbol{\beta}\right),\mathbf{X}\right)\right)
\end{align*}
due to the upper boundedness of $\left\Vert \partial L\left(z,\boldsymbol{\beta}\right)/\partial\boldsymbol{\beta}\right\Vert $
according to  \autoref{lem:2}(viii). Note that 
\[
f_{z}\left(\left.z\left(\mathbf{X}_{e},\boldsymbol{\beta}\right)\right|\boldsymbol{\beta}\right)>C\cdot m\left(\mathbb{X}\left(z\left(\boldsymbol{\mathbf{X}}_{e},\boldsymbol{\beta}\right),\mathbf{X}\right)\right)
\]
for some $C>0$ due to \autoref{assu:4}(i) and the choice
of $\mathcal{X}_{e}^{\phi}$, so we have that 
\begin{align*}
 & \left\Vert \left(\partial H_{1}\left(z\left(\mathbf{X}_{e},\boldsymbol{\beta}\right),\boldsymbol{\mathbf{X}}_{e}\right)/\partial z-L\left(z\left(\boldsymbol{\mathbf{X}}_{e},\boldsymbol{\beta}\right),\boldsymbol{\beta}\right)\partial H_{2}\left(z\left(\boldsymbol{\mathbf{X}}_{e},\boldsymbol{\beta}\right),\boldsymbol{\mathbf{X}}_{e}\right)/\partial z\right)/f_{z}\left(\left.z\left(\boldsymbol{\mathbf{X}}_{e},\boldsymbol{\beta}\right)\right|\boldsymbol{\beta}\right)\right.\\
 & \left.-\int G^{\prime}\left(z\left(\boldsymbol{\mathbf{X}}_{e},\boldsymbol{\beta}\right)-\widetilde{\boldsymbol{\mathbf{X}}}^{\mathrm{T}}\Delta\boldsymbol{\beta}\right)f_{e}\left(z\left(\boldsymbol{\mathbf{X}}_{e},\boldsymbol{\beta}\right)-\widetilde{\mathbf{X}}^{\mathrm{T}}\boldsymbol{\beta},\widetilde{\mathbf{X}}\right)\left(\mathbf{X}-\widetilde{\mathbf{X}}\right)d\widetilde{\mathbf{X}}/f_{z}\left(\left.z\left(\boldsymbol{\mathbf{X}}_{e},\boldsymbol{\beta}\right)\right|\boldsymbol{\beta}\right)\right\Vert \\
 & =\left\Vert \left(\partial_{z}H_{1}\left(z\left(\boldsymbol{\mathbf{X}}_{e},\boldsymbol{\beta}\right),\boldsymbol{\mathbf{X}}_{e}\right)-L\left(z\left(\boldsymbol{\mathbf{X}}_{e},\boldsymbol{\beta}\right),\boldsymbol{\beta}\right)\partial_{z}H_{2}\left(z\left(\boldsymbol{\mathbf{X}}_{e},\boldsymbol{\beta}\right),\boldsymbol{\mathbf{X}}_{e}\right)\right)/f_{z}\left(\left.z\left(\boldsymbol{\mathbf{X}}_{e},\boldsymbol{\beta}\right)\right|\boldsymbol{\beta}\right)\right.\\
 & \left.-\int W\left(\boldsymbol{\mathbf{X}}_{e},\widetilde{\boldsymbol{\mathbf{X}}}_{e},\boldsymbol{\beta}\right)\left(\boldsymbol{\mathbf{X}}-\widetilde{\mathbf{X}}\right)d\widetilde{\mathbf{X}}\right\Vert \leq C\cdot\left\Vert \Delta\boldsymbol{\beta}\right\Vert .
\end{align*}
This proves the results.
\end{proof}
\medskip{}

Now we prove \autoref{lem3.2} in the main text.
\begin{proof}[Proof of \autoref{lem3.2}]
Note that 
\begin{align}
\sup_{\boldsymbol{\beta}\in\mathcal{B}_{n}} & \left\Vert \frac{1}{n}\sum_{i=1}^{n}\boldsymbol{\boldsymbol{\mathbf{X}}}_{i}^{\phi}\frac{\partial\widehat{G}\left(\left.z\left(\boldsymbol{\mathbf{X}}_{e,i},\boldsymbol{\beta}\right)\right|\boldsymbol{\beta}\right)}{\partial\boldsymbol{\beta}}-\varLambda_{\phi}\left(\boldsymbol{\beta}\right)\right\Vert \nonumber \\
\leq\sup_{\boldsymbol{\beta}\in\mathcal{B}_{n}} & \left\Vert \frac{1}{n}\sum_{i=1}^{n}\boldsymbol{\boldsymbol{\mathbf{X}}}_{i}^{\phi}\left(\frac{\partial\widehat{G}\left(\left.z\left(\boldsymbol{\mathbf{X}}_{e,i},\boldsymbol{\beta}\right)\right|\boldsymbol{\beta}\right)}{\partial\boldsymbol{\beta}}-\int W\left(\boldsymbol{\boldsymbol{\mathbf{X}}}_{e,i},\boldsymbol{\boldsymbol{\mathbf{X}}}_{e},\boldsymbol{\beta}\right)\left(\boldsymbol{\boldsymbol{\mathbf{X}}}_{i}-\boldsymbol{\boldsymbol{\mathbf{X}}}\right)d\boldsymbol{\boldsymbol{\mathbf{X}}}\right)\right\Vert \label{Ap14}\\
+\sup_{\boldsymbol{\beta}\in\mathcal{B}_{n}} & \left\Vert \frac{1}{n}\sum_{i=1}^{n}\boldsymbol{\boldsymbol{\mathbf{X}}}_{i}^{\phi}\left(\int W\left(\boldsymbol{\boldsymbol{\mathbf{X}}}_{e,i},\boldsymbol{\boldsymbol{\mathbf{X}}}_{e},\boldsymbol{\beta}\right)\left(\boldsymbol{\boldsymbol{\mathbf{X}}}_{i}-\boldsymbol{\boldsymbol{\mathbf{X}}}\right)d\boldsymbol{\boldsymbol{\mathbf{X}}}\right)-\varLambda_{\phi}\left(\boldsymbol{\beta}\right)\right\Vert .\label{Ap15}
\end{align}
Obviously, (\ref{Ap14}) is of order $O_{p}\left(h_{n}^{-2}\sqrt{\log\left(nh_{n}^{-1}\right)/n}+h_{n}^{3}+\sup_{\boldsymbol{\beta}\in\mathcal{B}_{n}}\left\Vert \Delta\boldsymbol{\beta}\right\Vert \right)$
according to \autoref{lem9}. Using \autoref{lemS.1}, we can
show that (\ref{Ap15}) is $O_{p}\left(\sqrt{\left(\log n\right)/n}\right)$
by noting  that each element of $\int W\left(\boldsymbol{\boldsymbol{\mathbf{X}}}_{e,i},\boldsymbol{\boldsymbol{\mathbf{X}}}_{e},\boldsymbol{\beta}\right)\left(\boldsymbol{\boldsymbol{\mathbf{X}}}_{i}-\boldsymbol{\boldsymbol{\mathbf{X}}}\right)d\boldsymbol{X}$
is bounded and that $\int_{\mathcal{X}}\left\Vert \partial W\left(\boldsymbol{\mathbf{X}}_{e},\widetilde{\boldsymbol{\mathbf{X}}}_{e},\boldsymbol{\beta}\right)/\partial\boldsymbol{\beta}\right\Vert d\widetilde{\boldsymbol{\mathbf{X}}}$
is uniformly upper bounded according to \autoref{lem:2}(x). This
finishes the proof of  \autoref{lem3.2}. 
\end{proof}
\medskip{}

Now we prove  \autoref{lem3.3} in the main text.

\begin{proof}[Proof of \autoref{lem3.3}]
We first show that 
\[
\boldsymbol{\xi}_{n}^{\phi}=\frac{1}{n^{2}}\sum_{i=1}^{n}\sum_{j=1}^{n}K_{h_{n}}\left(z_{j}^{\star}-z_{i}^{\star}\right)\left(\frac{y_{j}-y_{i}}{f_{z}^{\star}\left(z_{i}^{\star}\right)}\right)\mathbf{X}_{i}^{\phi}+o_{p}\left(\frac{1}{\sqrt{n}}\right),
\]
Define $f_{z}^{\star}\left(z_{i}^{\star}\right)=f_{z}\left(\left.z\right|\boldsymbol{\beta}^{\star}\right)$
and $f_{\mathbf{X},z}^{\star}\left(\mathbf{X},z\right)=f_{\mathbf{X},z}\left(\left.\mathbf{X},z\right|\boldsymbol{\beta}^{\star}\right)$.
Recall that $z_{i}^{\star}=z\left(\mathbf{X}_{e,i},\boldsymbol{\beta}^{\star}\right)$,
so
\begin{align*}
 & \boldsymbol{\xi}_{n}^{\phi}-\frac{1}{n^{2}}\sum_{i=1}^{n}\sum_{j=1}^{n}K_{h_{n}}\left(z_{j}^{\star}-z_{i}^{\star}\right)\left(\frac{y_{j}-y_{i}}{f_{z}^{\star}\left(z_{i}^{\star}\right)}\right)\mathbf{X}_{i}^{\phi}\\
 & =\frac{1}{n}\sum_{i=1}^{n}\left[\frac{1}{n}\sum_{j=1}^{n}K_{h_{n}}\left(z_{j}^{\star}-z_{i}^{\star}\right)\left(y_{j}-y_{i}\right)\right]\left[\frac{1}{\frac{1}{n}\sum_{j=1}^{n}K_{h_{n}}\left(z_{j}^{\star}-z_{i}^{\star}\right)}-\frac{1}{f_{z}^{\star}\left(z_{i}^{\star}\right)}\right]\mathbf{X}_{i}^{\phi}\\
 & =\frac{1}{n}\sum_{i=1}^{n}\left[\frac{1}{n}\sum_{j=1}^{n}K_{h_{n}}\left(z_{j}^{\star}-z_{i}^{\star}\right)\left(y_{j}-G\left(z_{i}^{\star}\right)\right)\right]\left[\frac{1}{\frac{1}{n}\sum_{j=1}^{n}K_{h_{n}}\left(z_{j}^{\star}-z_{i}^{\star}\right)}-\frac{1}{f_{z}^{\star}\left(z_{i}^{\star}\right)}\right]\mathbf{X}_{i}^{\phi}(i)\\
 & -\frac{1}{n}\sum_{i=1}^{n}\varepsilon_{i}\left[\frac{1}{n}\sum_{j=1}^{n}K_{h_{n}}\left(z_{j}^{\star}-z_{i}^{\star}\right)\right]\left[\frac{1}{\frac{1}{n}\sum_{j=1}^{n}K_{h_{n}}\left(z_{j}^{\star}-z_{i}^{\star}\right)}-\frac{1}{f_{z}^{\star}\left(z_{i}^{\star}\right)}\right]\mathbf{X}_{i}^{\phi}(ii).
\end{align*}
For term (i), due to truncation, we have that 
\[
\max_{1\leq i\leq n}\left\Vert \left[\frac{1}{\frac{1}{n}\sum_{j=1}^{n}K_{h_{n}}\left(z_{j}^{\star}-z_{i}^{\star}\right)}-\frac{1}{f_{z}\left(z_{i}^{\star}\right)}\right]\mathbf{X}_{i}^{\phi}\right\Vert =O_{p}\left(h_{n}^{-1}\sqrt{\log\left(n\right)/n}+h_{n}^{3}\right).
\]
We further provide a uniform bound for $\frac{1}{n}\sum_{j=1}^{n}K_{h_{n}}\left(z_{j}^{\star}-z_{i}^{\star}\right)\left(y_{j}-G\left(z_{i}^{\star}\right)\right)\mathbf{X}_{i}^{\phi}$
over $i$. We first note that
\[
\mathbb{E}_{\mathscr{D}_{n}}\left[\frac{1}{n}\sum_{j=1}^{n}K_{h_{n}}\left(z_{j}^{\star}-z_{i}^{\star}\right)\left(y_{j}-G\left(z_{i}^{\star}\right)\right)\mathbf{X}_{i}^{\phi}\right]=\mathbb{E}_{\mathscr{D}_{n}}\left[\frac{1}{n}\sum_{j=1}^{n}K_{h_{n}}\left(z_{j}^{\star}-z_{i}^{\star}\right)\left(G\left(z_{j}^{\star}\right)-G\left(z_{i}^{\star}\right)\right)\mathbf{X}_{i}^{\phi}\right],
\]
where the RHS is equivalent to 
\begin{align*}
 & \mathbb{E}\left\{ \mathbb{E}\left[\left.\frac{1}{n}\sum_{j=1}^{n}K_{h_{n}}\left(z_{j}^{\star}-z_{i}^{\star}\right)\left(G\left(z_{j}^{\star}\right)-G\left(z_{i}^{\star}\right)\right)\mathbf{X}_{i}^{\phi}\right|\mathbf{X}_{e,i}\right]\right\} \\
 & =\frac{n-1}{n}\mathbb{E}\left\{ \mathbf{X}_{i}^{\phi}\int\left[K_{h_{n}}\left(z-z_{i}^{\star}\right)\left(G\left(z\right)-G\left(z_{i}^{\star}\right)\right)f_{z}^{\star}\left(z\right)\right]dz\right\} \\
 & =\frac{n-1}{n}\mathbb{E}\left\{\mathbf{X}_{i}^{\phi}0 \int\left[K\left(z\right)\left(G\left(z_{i}^{\star}+zh_{n}\right)-G\left(z_{i}^{\star}\right)\right)f_{z}^{\star}\left(z_{i}+zh_{n}\right)\right]dz\right\} .
\end{align*}
Now note that since $G$ and $f_{z}^{\star}$ both have up to fourth
order bounded derivatives, we have that 
\begin{align*}
 & \left(G\left(z_{i}^{\star}+zh_{n}\right)-G\left(z_{i}^{\star}\right)\right)f_{z}^{\star}\left(z_{i}+zh_{n}\right)\\
 & =\left(G^{\prime}\left(z_{i}^{\star}\right)zh_{n}+\frac{1}{2}G^{\prime\prime}\left(z_{i}^{\star}\right)z^{2}h_{n}^{2}+\frac{1}{6}G^{\prime\prime\prime}\left(z_{i}^{\star}\right)z^{3}h_{n}^{3}+O\left(z^{4}h_{n}^{4}\right)\right)\left(f_{z}^{\star}\left(z_{i}^{\star}\right)+O\left(zh_{n}\right)\right)\\
 & =G^{\prime}\left(z_{i}^{\star}\right)f_{z}^{\star}\left(z_{i}^{\star}\right)zh_{n}+\frac{1}{2}G^{\prime\prime}\left(z_{i}^{\star}\right)f_{z}^{\star}\left(z_{i}^{\star}\right)z^{2}h_{n}^{2}+\frac{1}{6}G^{\prime\prime\prime}\left(z_{i}^{\star}\right)f_{z}^{\star}\left(z_{i}^{\star}\right)z^{3}h_{n}^{3}+O\left(z^{4}h_{n}^{4}\right).
\end{align*}
So 
\[
\int\left[K\left(z\right)\left(G\left(z_{i}^{\star}+zh_{n}\right)-G\left(z_{i}^{\star}\right)\right)f_{z}^{\star}\left(z_{i}+zh_{n}\right)\right]dz=O\left(h_{n}^{3}\right),
\]
where the bound does not depend on $i$. So 
\[
\max_{1\leq i\leq n}\left\Vert \mathbb{E}\left[\frac{1}{n}\sum_{j=1}^{n}K_{h_{n}}\left(z_{j}^{\star}-z_{i}^{\star}\right)\left(G\left(z_{j}^{\star}\right)-G\left(z_{i}^{\star}\right)\right)\boldsymbol{\mathbf{X}}_{i}^{\phi}\right]\right\Vert =O\left(h_{n}^{3}\right).
\]
On the other side, we have that we have that 
\begin{align*}
\max_{1\leq i\leq n} & \left\Vert \frac{1}{n}\sum_{j=1}^{n}K_{h_{n}}\left(z_{j}^{\star}-z_{i}^{\star}\right)\left(y_{j}-G\left(z_{i}^{\star}\right)\right)\mathbf{X}_{i}^{\phi}\right.\\
 & \left.-\mathbb{E}_{\mathscr{D}_{n}}\left[\frac{1}{n}\sum_{j=1}^{n}K_{h_{n}}\left(z_{j}^{\star}-z_{i}^{\star}\right)\left(y_{j}-G\left(z_{i}^{\star}\right)\right)\mathbf{X}_{i}^{\phi}\right]\right\Vert =O_{p}\left(\sqrt{\left(\log n\right)/nh_{n}^{2}}\right).
\end{align*}
Together we have that 
\[
\max_{1\leq i\leq n}\left\Vert \frac{1}{n}\sum_{j=1}^{n}K_{h_{n}}\left(z_{j}^{\star}-z_{i}^{\star}\right)\left(y_{j}-G\left(z_{i}^{\star}\right)\right)\mathbf{X}_{i}^{\phi}\right\Vert =O_{p}\left(h_{n}^{-1}\sqrt{\left(\log n\right)/n}+h_{n}^{3}\right).
\]
So 
\begin{align*}
 & \left\Vert \frac{1}{n}\sum_{i=1}^{n}\left[\frac{1}{n}\sum_{j=1}^{n}K_{h_{n}}\left(z_{j}^{\star}-z_{i}^{\star}\right)\left(y_{j}-G\left(z_{i}^{\star}\right)\right)\right]\left[\frac{1}{\frac{1}{n}\sum_{j=1}^{n}K_{h_{n}}\left(z_{j}^{\star}-z_{i}^{\star}\right)}-\frac{1}{f_{z}\left(z_{i}^{\star}\right)}\right]\mathbf{X}_{i}^{\phi}\right\Vert \\
 & \leq\max_{1\leq i\leq n}\left\Vert \frac{1}{n}\sum_{j=1}^{n}K_{h_{n}}\left(z_{j}^{\star}-z_{i}^{\star}\right)\left(y_{j}-G\left(z_{i}^{\star}\right)\right)\mathbf{X}_{i}^{\phi}\right\Vert \max_{1\leq i\leq n}\left|\frac{1}{\frac{1}{n}\sum_{j=1}^{n}K_{h_{n}}\left(z_{j}^{\star}-z_{i}^{\star}\right)}-\frac{1}{f_{z}\left(z_{i}^{\star}\right)}\right|\\
 & =O_{p}\left(h_{n}^{-2}\left(\log n\right)/n+h_{n}^{6}\right)=o_{p}\left(1/\sqrt{n}\right),
\end{align*}
according to our choice of $h_n$, so term (i) is $o_{p}\left(1/\sqrt{n}\right)$. 

For term (ii), without of loss of generality, we assume that $\mathbf{X}_{i}^{\phi}=X_{i}^{\phi}$
is a scalar; the general case can be proved similarly. We note that
\begin{align*}
 & \mathbb{E}\left[\sum_{i=1}^{n}\varepsilon_{i}\left[\frac{1}{n}\sum_{j=1}^{n}K_{h_{n}}\left(z_{j}^{\star}-z_{i}^{\star}\right)\right]\left[\frac{1}{\frac{1}{n}\sum_{j=1}^{n}K_{h_{n}}\left(z_{j}^{\star}-z_{i}^{\star}\right)}-\frac{1}{f_{z}^{\star}\left(z_{i}^{\star}\right)}\right]X_{i}^{\phi}\right]\\
 & =\mathbb{E}\sum_{i=1}^{n}\mathbb{E}\left\{ \left.\varepsilon_{i}\left[1-\frac{\frac{1}{n}\sum_{j=1}^{n}K_{h_{n}}\left(z_{j}^{\star}-z_{i}^{\star}\right)}{f_{z}^{\star}\left(z_{i}^{\star}\right)}\right]X_{i}^{\phi}\right|X_{i}\right\}=0 
\end{align*}
due to the fact that the data is i.i.d. and that $\mathbb{E}\left(\left.\varepsilon_{i}\right|\mathbf{X}_{e,i}\right)=0$
for all $i$. Moreover, 
\begin{align*}
 & \mathbb{V}\left[\frac{1}{n}\sum_{i=1}^{n}\varepsilon_{i}\left[1-\frac{\frac{1}{n}\sum_{j=1}^{n}K_{h_{n}}\left(z_{j}^{\star}-z_{i}^{\star}\right)}{f_{z}^{\star}\left(z_{i}^{\star}\right)}\right]X_{i}^{\phi2}\right]\\
 & =\frac{1}{n}\mathbb{E}\left\{ G\left(z_{i}^{\star}\right)\left(1-G\left(z_{i}^{\star}\right)\right)\left[1-\frac{\frac{1}{n}\sum_{j=1}^{n}K_{h_{n}}\left(z_{j}^{\star}-z_{i}^{\star}\right)}{f_{z}^{\star}\left(z_{i}^{\star}\right)}\right]^{2}X_{i}^{\phi2}\right\} \\
 & \leq\frac{C}{n}\mathbb{E}\left\{ \left(\frac{1}{n}\sum_{j=1}^{n}K_{h_{n}}\left(z_{j}^{\star}-z_{i}^{\star}\right)-f_{z}^{\star}\left(z_{i}^{\star}\right)\right)^{2}X_{i}^{\phi2}\right\} \\
 & =\frac{C}{n^{3}}\mathbb{E}X_{i}^{\phi2}\left(\sum_{j\neq i, k\neq i, j\neq k}^{n}\mathbb{E}\left[\left.\left(K_{h_{n}}\left(z_{j}^{\star}-Z_{i}^{\star}\right)-f_{z}^{\star}\left(z_{i}^{\star}\right)\right)\left(K_{h_{n}}\left(z_{k}^{\star}-z_{i}^{\star}\right)-f_{z}^{\star}\left(z_{i}^{\star}\right)\right)\right|X_{i}^{\phi}\right]+O\left(nh_{n}^{-1}\right)\right)
\end{align*}
Note that $\mathbb{E}\left[\left.\left(K_{h_{n}}\left(z_{j}^{\star}-Z_{i}^{\star}\right)-f_{z}^{\star}\left(z_{i}^{\star}\right)\right)\left(K_{h_{n}}\left(z_{k}^{\star}-z_{i}^{\star}\right)-f_{z}^{\star}\left(z_{i}^{\star}\right)\right)\right|X_{i}^{\phi}\right]$ is $ O\left(h_{n}^{6}\right)$
for all $k\neq j$, $j\neq i$, and $k\neq i$. So the above term
is of order $O\left(h_{n}^{6}/n+h_{n}^{-1}/n^{2}\right)$, implying
that 
\[
\left\Vert \frac{1}{n}\sum_{i=1}^{n}\varepsilon_{i}\left[1-\frac{\frac{1}{n}\sum_{j=1}^{n}K_{h_{n}}\left(z_{j}^{\star}-z_{i}^{\star}\right)}{f_{z}^{\star}\left(z_{i}^{\star}\right)}\right]\mathbf{X}_{i}^{\phi}\right\Vert =O_{p}\left(h_{n}^{3}/\sqrt{n}+1/\left(n\sqrt{h_{n}}\right)\right)=o_{p}\left(1/\sqrt{n}\right),
\]
according to the choice of $h_n$. This proves the first result. 

Now we obtain the asymptotic distribution of 
\[
\frac{1}{n^{2}}\sum_{i=1}^{n}\sum_{j=1}^{n}K_{h_{n}}\left(z_{j}^{\star}-z_{i}^{\star}\right)\left(\frac{y_{j}-y_{i}}{f_{z}^{\star}\left(z_{i}^{\star}\right)}\right)\mathbf{X}_{i}^{\phi}.
\]
First note that 
\begin{align*}
 & \frac{1}{n^{2}}\sum_{i=1}^{n}\sum_{j=1}^{n}K_{h_{n}}\left(z_{j}^{\star}-z_{i}^{\star}\right)\left(\frac{y_{j}-y_{i}}{f_{z}^{\star}\left(z_{i}^{\star}\right)}\right)\mathbf{X}_{i}^{\phi}\\
 & =\frac{1}{2n^{2}}\sum_{i=1}^{n}\sum_{j=1}^{n}K_{h_{n}}\left(z_{j}^{\star}-z_{i}^{\star}\right)\left(\frac{y_{j}-y_{i}}{f_{z}^{\star}\left(z_{i}^{\star}\right)}\mathbf{X}_{i}^{\phi}+\frac{y_{i}-y_{j}}{f_{z}^{\star}\left(z_{j}^{\star}\right)}\mathbf{X}_{j}^{\phi}\right)\\
 & =\frac{1}{n^{2}}\sum_{i=1}^{n-1}\sum_{j=i+1}^{n}K_{h_{n}}\left(z_{j}^{\star}-z_{i}^{\star}\right)\left(\frac{y_{j}-y_{i}}{f_{z}^{\star}\left(z_{i}^{\star}\right)}\mathbf{X}_{i}^{\phi}+\frac{y_{i}-y_{j}}{f_{z}^{\star}\left(z_{j}^{\star}\right)}\mathbf{X}_{j}^{\phi}\right)\\
 & =\frac{n\left(n-1\right)}{2n^{2}}\begin{pmatrix}n\\
2
\end{pmatrix}^{-1}\sum_{i=1}^{n-1}\sum_{j=i+1}^{n}K_{h_{n}}\left(z_{j}^{\star}-z_{i}^{\star}\right)\left(\frac{y_{j}-y_{i}}{f_{z}^{\star}\left(z_{i}^{\star}\right)}\mathbf{X}_{i}^{\phi}+\frac{y_{i}-y_{j}}{f_{z}^{\star}\left(z_{j}^{\star}\right)}\mathbf{X}_{j}^{\phi}\right).
\end{align*}
Let $\mathbb{E}_{j|i}$ be the expectation with respect to the $j$-th observation conditional on the $i$-th observation. Note that 
\begin{align*}
 & \mathbb{E}_{j|i}\left[K_{h_{n}}\left(z_{j}^{\star}-z_{i}^{\star}\right)\frac{y_{j}-y_{i}}{f_{z}^{\star}\left(z_{i}^{\star}\right)}\mathbf{X}_{i}^{\phi}\right]\\
 & =\frac{\mathbf{X}_{i}^{\phi}}{f_{z}^{\star}\left(z_{i}^{\star}\right)}\mathbb{E}_{j|i}\left[K_{h_{n}}\left(z_{j}^{\star}-z_{i}^{\star}\right)\left(G\left(z_{j}^{\star}\right)-y_{i}\right)\right]\\
 & =\frac{\mathbf{X}_{i}^{\phi}}{f_{z}^{\star}\left(z_{i}^{\star}\right)}\int K\left(z\right)\left(G\left(z_{i}^{\star}+h_{n}z\right)-y_{i}\right)f_{z}^{\star}\left(z_{i}^{\star}+h_{n}z\right)dz\\
 & =\frac{\mathbf{X}_{i}^{\phi}}{f_{z}^{\star}\left(z_{i}^{\star}\right)}\int K\left(z\right)\left(G\left(z_{i}^{\star}\right)+G^{\prime}\left(z_{i}^{\star}\right)zh_{n}+\frac{1}{2}G^{\prime\prime}\left(z_{i}^{\star}\right)z^{2}h_{n}^{2}+O\left(z^{3}h_{n}^{3}\right)-y_{i}\right)\left(f_{z}^{\star}\left(z_{i}^{\star}\right)+O\left(zh_{n}\right)\right)dz\\
 & =\frac{\mathbf{X}_{i}^{\phi}}{f_{z}^{\star}\left(z_{i}^{\star}\right)}\int K\left(z\right)\left(G\left(z_{i}^{\star}\right)-y_{i}\right)f_{z}^{\star}\left(z_{i}^{\star}\right)dz+O\left(h_{n}^{3}\right)=\mathbf{X}_{i}^{\phi}\left(G\left(z_{i}^{\star}\right)-y_{i}\right)+O\left(h_{n}^{3}\right),
\end{align*}
and
\begin{align*}
& \mathbb{E}_{j|i}\left[K_{h_{n}}\left(z_{j}^{\star}-z_{i}^{\star}\right)\left(\frac{y_{i}-y_{j}}{f_{z}^{\star}\left(z_{j}^{\star}\right)}\right)\mathbf{X}_{j}^{\phi}\right]\\ & =\int\frac{1}{h_{n}}K\left(\frac{z-z_{i}^{\star}}{h_{n}}\right)\left(\frac{y_{i}-G\left(z\right)}{f_{z}^{\star}\left(z\right)}\right)\mathbf{X}^{\phi}f_{\mathbf{X},z}^{\star}\left(\mathbf{X},z\right)dzd\mathbf{X}\\
 & =\int K\left(z\right)\frac{y_{i}-G\left(z_{i}^{\star}+h_{n}z\right)}{f_{z}^{\star}\left(z_{i}^{\star}+h_{n}z\right)}\mathbf{X}^{\phi}f_{\mathbf{X},z}^{\star}\left(\mathbf{X},z_{i}^{\star}+h_{n}z\right)dzd\mathbf{X}\\
 & =\left(y_{i}-G\left(z_{i}^{\star}\right)\right)\int\mathbf{X}^{\phi}f_{\mathbf{X}|z}^{\star}\left(\left.\mathbf{X}\right|z_{i}^{\star}\right)d\mathbf{X}+O\left(h_{n}^{2}\right)\\
 & =\left(y_{i}-G\left(z_{i}^{\star}\right)\right)\mathbb{E}\left(\left.\mathbf{X}^{\phi}\right|z_{i}^{\star}\right)+O\left(h_{n}^{3}\right).
\end{align*}
So 
\[
\mathbb{E}_{j|i}\left[K_{h_{n}}\left(z_{j}^{\star}-z_{i}^{\star}\right)\left(\frac{y_{j}-y_{i}}{f_{z}^{\star}\left(z_{i}^{\star}\right)}\mathbf{X}_{i}^{\phi}+\frac{y_{i}-y_{j}}{f_{z}^{\star}\left(z_{j}^{\star}\right)}\mathbf{X}_{j}^{\phi}\right)\right]=-\varepsilon_i\left(\mathbf{X}_{i}^{\phi}-\mathbb{E}\left(\left.\mathbf{X}_{i}^{\phi}\right|z_{i}^{\star}\right)\right)+O\left(h_{n}^{3}\right).
\]
We also note that 
\begin{align*}
\mathbb{E}\left\Vert K_{h_{n}}\left(z_{j}^{\star}-z_{i}^{\star}\right)\left(\frac{y_{i}-y_{j}}{f_{z}^{\star}\left(z_{j}^{\star}\right)}\right)\mathbf{X}_{j}^{\phi}\right\Vert ^{2} & \leq C\mathbb{E}\left(K_{h_{n}}^{2}\left(z_{j}^{\star}-z_{i}^{\star}\right)\right)=O\left(h_{n}^{-2}\right)=o\left(n\right),
\end{align*}
\[
\mathbb{E}_{i}\mathbb{E}_{j|i}\left[K_{h_{n}}\left(z_{j}^{\star}-z_{i}^{\star}\right)\left(\frac{y_{j}-y_{i}}{f_{z}^{\star}\left(z_{i}^{\star}\right)}\mathbf{X}_{i}^{\phi}+\frac{y_{i}-y_{j}}{f_{z}^{\star}\left(z_{j}^{\star}\right)}\mathbf{X}_{j}^{\phi}\right)\right]=O\left(h_{n}^{3}\right)=o\left(\frac{1}{\sqrt{n}}\right),
\]
so according to \citet{powell1989semiparametric}, we have that 
\begin{align*}
 & \sqrt{n}\begin{pmatrix}n\\
2
\end{pmatrix}^{-1}\sum_{i=1}^{n-1}\sum_{j=i+1}^{n}K_{h_{n}}\left(z_{j}^{\star}-z_{i}^{\star}\right)\left(\frac{y_{j}-y_{i}}{f_{z}^{\star}\left(z_{i}^{\star}\right)}\mathbf{X}_{i}^{\phi}+\frac{y_{i}-y_{j}}{f_{z}^{\star}\left(z_{j}^{\star}\right)}\mathbf{X}_{j}^{\phi}\right).\\
 & =-\frac{2}{\sqrt{n}}\sum_{i=1}^{n}\varepsilon_i\left(\mathbf{X}_{i}^{\phi}-\mathbb{E}\left(\left.\mathbf{X}_{i}^{\phi}\right|z_{i}^{\star}\right)\right)+o_{p}\left(1\right).
\end{align*}
This implies that 
\begin{align*}
\sqrt{n}\boldsymbol{\xi}_{n}^{\phi} & =-\frac{1}{\sqrt{n}}\sum_{i=1}^{n}\varepsilon_i\left(\mathbf{X}_{i}^{\phi}-\mathbb{E}\left(\left.\mathbf{X}_{i}^{\phi}\right|z_{i}^{\star}\right)\right)+o_{p}\left(1\right)\rightarrow_{d}N\left(0,\Sigma_{\boldsymbol{\xi}}^{\phi}\right).
\end{align*}
\end{proof}

\begin{lemma}
	\label{lemS.2} Suppose that  \autoref{assu1}, \autoref{assump:2}(i) and (ii),
	and \autoref{assu:6} hold, we have that 
	\[
	\sup_{\boldsymbol{\beta}\in\mathcal{B}}\left\Vert \Gamma_{q,n}\left(\boldsymbol{\beta}\right)-\Gamma_{q}\left(\boldsymbol{\beta}\right)\right\Vert =O_{p}\left(\chi_{1,n}\right).
	\]
\end{lemma}
\begin{proof}[Proof of \autoref{lemS.2}]
	This is a direct result of  \autoref{lemS.1} by noting that  $\left|r_{s}\left(z\right)r_{s}\left(z\right)\right|\leq D_{q,0}^{2}$
	and $\left\Vert \partial\left(r_{s}\left(X_{0}+\mathbf{X}^{\mathrm{T}}\boldsymbol{\beta}\right)r_{s}\left(X_{0}+\mathbf{X}^{\mathrm{T}}\boldsymbol{\beta}\right)\right)/\partial\boldsymbol{\beta}\right\Vert \leq C\sqrt{p}D_{q,0}D_{q,1}$.
\end{proof}
\begin{lemma}
	\label{lemS.3}Suppose that  \autoref{assu1}, \autoref{assump:2}(i) and (ii), 
	and \autoref{assu:6} hold, and $\chi_{1,n}\rightarrow0$ as $n\rightarrow \infty$. We have that
	\[
	\sup_{\boldsymbol{\beta}\in\mathcal{B}}\left\Vert \Gamma_{q,n}^{-1}\left(\boldsymbol{\beta}\right)-\Gamma_{q}^{-1}\left(\boldsymbol{\beta}\right)\right\Vert =O_{p}\left(\chi_{1,n}\right).
	\]
\end{lemma}
\begin{proof}[Proof of \autoref{lemS.3}]
	First note that 
	\[
	\sup_{\boldsymbol{\beta}\in\mathcal{B}}\left|\underline{\lambda}\left(\Gamma_{q,n}\left(\boldsymbol{\beta}\right)\right)-\underline{\lambda}\left(\Gamma_{q}\left(\boldsymbol{\beta}\right)\right)\right|\leq\sup_{\boldsymbol{\beta}\in\mathcal{B}}\left\Vert \Gamma_{q,n}\left(\boldsymbol{\beta}\right)-\Gamma_{q}\left(\boldsymbol{\beta}\right)\right\Vert =O_{p}\left(\chi_{1,n}\right),
	\]
	and 
	\[
	\sup_{\boldsymbol{\beta}\in\mathcal{B}}\left|\overline{\lambda}\left(\Gamma_{q,n}\left(\boldsymbol{\beta}\right)\right)-\overline{\lambda}\left(\Gamma_{q}\left(\boldsymbol{\beta}\right)\right)\right|\leq\sup_{\boldsymbol{\beta}\in\mathcal{B}}\left\Vert \Gamma_{q,n}\left(\boldsymbol{\beta}\right)-\Gamma_{q}\left(\boldsymbol{\beta}\right)\right\Vert =O_{p}\left(\chi_{1,n}\right).
	\]
	Since $\chi_{1,n}\rightarrow0$, we have that with probability going
	to 1, there holds 
	\[
	\sup_{\boldsymbol{\beta}\in\mathcal{B}}\overline{\lambda}\left(\Gamma_{q,n}\left(\boldsymbol{\beta}\right)\right)\leq\frac{3\overline{\lambda}_{\Gamma}}{2},\ \inf_{\boldsymbol{\beta}\in\mathcal{B}}\overline{\lambda}\left(\Gamma_{q,n}\left(\boldsymbol{\beta}\right)\right)\geq\frac{\underline{\lambda}_{\Gamma}}{2},
	\]
	indicating that $\sup_{\boldsymbol{\beta}\in\mathcal{B}}\overline{\lambda}\left(\Gamma_{q,n}^{-1}\left(\boldsymbol{\beta}\right)\right)=O_{p}\left(1\right)$. 
	
	Note that for any positive semi-definite matrices $A$ and $B$,
	there holds 
$
	\min\left\{ \underline{\lambda}_{A}\left\Vert B\right\Vert ,\underline{\lambda}_{B}\left\Vert A\right\Vert \right\} \leq\left\Vert AB\right\Vert \leq\max\left\{ \overline{\lambda}_{A}\left\Vert B\right\Vert ,\overline{\lambda}_{B}\left\Vert A\right\Vert \right\} ,
	$
	so we have that 
	\begin{align*}
		& \sup_{\boldsymbol{\beta}\in\mathcal{B}}\left\Vert \Gamma_{q,n}^{-1}\left(\boldsymbol{\beta}\right)-\Gamma_{q}^{-1}\left(\boldsymbol{\beta}\right)\right\Vert =\sup_{\boldsymbol{\beta}\in\mathcal{B}}\left\Vert \Gamma_{q,n}^{-1}\left(\boldsymbol{\beta}\right)\left(\Gamma_{q,n}\left(\boldsymbol{\beta}\right)-\Gamma_{q}\left(\boldsymbol{\beta}\right)\right)\Gamma_{q}^{-1}\left(\boldsymbol{\beta}\right)\right\Vert \\
		& \leq\left(\sup_{\boldsymbol{\beta}\in\mathcal{B}}\overline{\lambda}\left(\Gamma_{q,n}^{-1}\left(\boldsymbol{\beta}\right)\right)\right)\left(\sup_{\boldsymbol{\beta}\in\mathcal{B}}\overline{\lambda}\left(\Gamma_{q}^{-1}\left(\boldsymbol{\beta}\right)\right)\right)\sup_{\boldsymbol{\beta}\in\mathcal{B}}\left\Vert \Gamma_{q,n}\left(\boldsymbol{\beta}\right)-\Gamma_{q}\left(\boldsymbol{\beta}\right)\right\Vert =O_{p}\left(\chi_{1,n}\right).
	\end{align*}
\end{proof}
\begin{lemma}
	\label{lemS.4}Suppose that  \autoref{assu1}, \autoref{assump:2}(i) and (ii),
	and \autoref{assu:6} hold, and moreover $\chi_{1,n}\rightarrow0$ as $n \rightarrow \infty$. Define
	\[\mathcal{Z}=\left\{ z:z=X_{0}+\mathbf{X}^{\mathrm{T}}\boldsymbol{\beta}\text{ for some }\mathbf{X}_{e}\in\mathcal{X}_{e}\text{ and }\boldsymbol{\beta}\in\mathcal{B}\right\}. \]
	We have that 
	\[
	\sup_{z\in\mathcal{Z}}\sup_{\boldsymbol{\beta}\in\mathcal{B}}\left\Vert \mathfrak{X}_{q,n}\left(z,\boldsymbol{\beta}\right)-\mathfrak{X}_{q}\left(z,\boldsymbol{\beta}\right)\right\Vert =O_{p}\left(\sqrt{p}qD_{q,0}^{2}\chi_{1,n}\right).
	\]
\end{lemma}
\begin{proof}[Proof of \autoref{lemS.4}]
	Note that 
	\begin{align*}
		& \sup_{z\in\mathcal{Z}}\sup_{\boldsymbol{\beta}\in\mathcal{B}}\left\Vert \mathfrak{X}_{q,n}\left(z,\boldsymbol{\beta}\right)-\mathfrak{X}_{q}\left(z,\boldsymbol{\beta}\right)\right\Vert \\
		& \leq\sup_{z\in\mathcal{Z}}\sup_{\boldsymbol{\beta}\in\mathcal{B}}\left\Vert \mathfrak{X}_{q,n}\left(z,\boldsymbol{\beta}\right)-\frac{1}{n}\sum_{i=1}^{n}\left(\boldsymbol{r}_{q}^{\mathrm{T}}\left(X_{0,i}+\mathbf{X}_{i}^{\mathrm{T}}\boldsymbol{\beta}\right)\Gamma_{q}^{-1}\left(\boldsymbol{\beta}\right)\boldsymbol{r}_{q}\left(z\right)\mathbf{X}_{i}\right)\right\Vert \\
		& +\sup_{z\in\mathcal{Z}}\sup_{\boldsymbol{\beta}\in\mathcal{B}}\left\Vert \mathfrak{X}_{q}\left(z,\boldsymbol{\beta}\right)-\frac{1}{n}\sum_{i=1}^{n}\left(\boldsymbol{r}_{q}^{\mathrm{T}}\left(X_{0,i}+\mathbf{X}_{i}^{\mathrm{T}}\boldsymbol{\beta}\right)\Gamma_{q}^{-1}\left(\boldsymbol{\beta}\right)\boldsymbol{r}_{q}\left(z\right)\mathbf{X}_{i}\right)\right\Vert .
	\end{align*}
	For the first term, we have that 
	\begin{align*}
		& \sup_{z\in\mathcal{Z}}\sup_{\boldsymbol{\beta}\in\mathcal{B}}\left\Vert \mathfrak{X}_{q,n}\left(z,\boldsymbol{\beta}\right)-\frac{1}{n}\sum_{i=1}^{n}\left(\boldsymbol{r}_{q}^{\mathrm{T}}\left(X_{0,i}+\mathbf{X}_{i}^{\mathrm{T}}\boldsymbol{\beta}\right)\Gamma_{q}^{-1}\left(\boldsymbol{\beta}\right)\boldsymbol{r}_{q}\left(z\right)\mathbf{X}_{i}\right)\right\Vert \\
		& =\frac{1}{n}\sum_{i=1}^{n}\sup_{z\in\mathcal{Z}}\sup_{\boldsymbol{\beta}\in\mathcal{B}}\left\Vert \boldsymbol{r}_{q}^{\mathrm{T}}\left(X_{0,i}+\mathbf{X}_{i}^{\mathrm{T}}\boldsymbol{\beta}\right)\left(\Gamma_{q,n}^{-1}\left(\boldsymbol{\beta}\right)-\Gamma_{q}^{-1}\left(\boldsymbol{\beta}\right)\right)\boldsymbol{r}_{q}\left(z\right)\mathbf{X}_{i}\right\Vert \\
		& \leq C\sqrt{p}qD_{q,0}^{2}\left\Vert \Gamma_{q,n}^{-1}\left(\boldsymbol{\beta}\right)-\Gamma_{q}^{-1}\left(\boldsymbol{\beta}\right)\right\Vert =O_{p}\left(\sqrt{p}qD_{q,0}^{2}\chi_{1,n}\right).
	\end{align*}
	For the second term, we note that 
	\begin{align*}
		& \sup_{z\in\mathcal{Z}}\sup_{\boldsymbol{\beta}\in\mathcal{B}}\left\Vert \mathfrak{X}_{q}\left(z,\boldsymbol{\beta}\right)-\frac{1}{n}\sum_{i=1}^{n}\left(\boldsymbol{r}_{q}^{\mathrm{T}}\left(X_{0,i}+\mathbf{X}_{i}^{\mathrm{T}}\boldsymbol{\beta}\right)\Gamma_{q}^{-1}\left(\boldsymbol{\beta}\right)\boldsymbol{r}_{q}\left(z\right)\mathbf{X}_{i}\right)\right\Vert \\
		& \leq\sup_{\boldsymbol{\beta}\in\mathcal{B}}\sup_{\widetilde{\boldsymbol{\beta}}\in\mathcal{B}}\sup_{\mathbf{X}_{e}\in\mathcal{X}_{e}}\left\Vert \mathfrak{X}_{q}\left(X_{0}+\mathbf{X}^{\mathrm{T}}\widetilde{\boldsymbol{\beta}},\boldsymbol{\beta}\right)-\frac{1}{n}\sum_{i=1}^{n}\left(\boldsymbol{r}_{q}^{\mathrm{T}}\left(X_{0,i}+\mathbf{X}_{i}^{\mathrm{T}}\boldsymbol{\beta}\right)\Gamma_{q}^{-1}\left(\boldsymbol{\beta}\right)\boldsymbol{r}_{q}\left(X_{0}+\mathbf{X}^{\mathrm{T}}\widetilde{\boldsymbol{\beta}}\right)\mathbf{X}_{i}\right)\right\Vert ,
	\end{align*}
	where uniformly for all $\boldsymbol{\beta},\boldsymbol{\beta}_{1},\boldsymbol{\beta}_{2},\widetilde{\boldsymbol{\beta}}\in\mathcal{B}$,
	$\mathbf{X}_{e}\in\mathcal{X}_{e}$, and $\mathbf{X}_{i}\in\mathcal{X}$,
	there hold
	\[
	\left|\boldsymbol{r}_{q}^{\mathrm{T}}\left(X_{0,i}+\mathbf{X}_{i}^{\mathrm{T}}\boldsymbol{\beta}\right)\Gamma_{q}^{-1}\left(\boldsymbol{\beta}\right)\boldsymbol{r}_{q}\left(X_{0}+\mathbf{X}^{\mathrm{T}}\widetilde{\boldsymbol{\beta}}\right)X_{i,j}\right|\leq CqD_{q,0}^{2},
	\]
	and 
	\[
	\left\Vert \frac{\partial\boldsymbol{r}_{q}^{\mathrm{T}}\left(X_{0,i}+\mathbf{X}_{i}^{\mathrm{T}}\boldsymbol{\beta}\right)\Gamma_{q}^{-1}\left(\boldsymbol{\beta}\right)\boldsymbol{r}_{q}\left(X_{0}+\mathbf{X}^{\mathrm{T}}\widetilde{\boldsymbol{\beta}}\right)X_{i,j}}{\partial\mathbf{X}_{e}}\right\Vert \leq C\sqrt{p}qD_{q,0}D_{q,1},
	\]
	\[
	\left\Vert \frac{\partial\boldsymbol{r}_{q}^{\mathrm{T}}\left(X_{0,i}+\mathbf{X}_{i}^{\mathrm{T}}\boldsymbol{\beta}\right)\Gamma_{q}^{-1}\left(\boldsymbol{\beta}\right)\boldsymbol{r}_{q}\left(X_{0}+\mathbf{X}^{\mathrm{T}}\widetilde{\boldsymbol{\beta}}\right)X_{i,j}}{\partial\widetilde{\boldsymbol{\beta}}}\right\Vert \leq C\sqrt{p}qD_{q,0}D_{q,1},
	\]
	\begin{align*}
		& \left\Vert \boldsymbol{r}_{q}^{\mathrm{T}}\left(X_{0,i}+\mathbf{X}_{i}^{\mathrm{T}}\boldsymbol{\beta}_{1}\right)\Gamma_{q}^{-1}\left(\boldsymbol{\beta}_{1}\right)\boldsymbol{r}_{q}\left(X_{0}+\mathbf{X}^{\mathrm{T}}\widetilde{\boldsymbol{\beta}}\right)-\boldsymbol{r}_{q}^{\mathrm{T}}\left(X_{0,i}+\mathbf{X}_{i}^{\mathrm{T}}\boldsymbol{\beta}_{2}\right)\Gamma_{q}^{-1}\left(\boldsymbol{\beta}_{2}\right)\boldsymbol{r}_{q}\left(X_{0}+\mathbf{X}^{\mathrm{T}}\widetilde{\boldsymbol{\beta}}\right)\right\Vert \\
		& \leq\left\Vert \left(\boldsymbol{r}_{q}^{\mathrm{T}}\left(X_{0,i}+\mathbf{X}_{i}^{\mathrm{T}}\boldsymbol{\beta}_{1}\right)-\boldsymbol{r}_{q}^{\mathrm{T}}\left(X_{0,i}+\mathbf{X}_{i}^{\mathrm{T}}\boldsymbol{\beta}_{2}\right)\right)\Gamma_{q}^{-1}\left(\boldsymbol{\beta}_{1}\right)\boldsymbol{r}_{q}\left(X_{0}+\mathbf{X}^{\mathrm{T}}\widetilde{\boldsymbol{\beta}}\right)\right\Vert \\
		& +\left\Vert \boldsymbol{r}_{q}^{\mathrm{T}}\left(X_{0,i}+\mathbf{X}_{i}^{\mathrm{T}}\boldsymbol{\beta}_{2}\right)\left(\Gamma_{q}^{-1}\left(\boldsymbol{\beta}_{1}\right)-\Gamma_{q}^{-1}\left(\boldsymbol{\beta}_{2}\right)\right)\boldsymbol{r}_{q}\left(X_{0}+\mathbf{X}^{\mathrm{T}}\widetilde{\boldsymbol{\beta}}\right)\right\Vert \\
		& \leq C\sqrt{p}qD_{q,0}D_{q,1}\left\Vert \boldsymbol{\beta}_{1}-\boldsymbol{\beta}_{2}\right\Vert +CqD_{q,0}^{2}\left\Vert \Gamma_{q}\left(\boldsymbol{\beta}_{1}\right)-\Gamma_{q}\left(\boldsymbol{\beta}_{2}\right)\right\Vert \leq C\sqrt{p}q^{2}D_{q,0}^{3}D_{q,1}\left\Vert \boldsymbol{\beta}_{1}-\boldsymbol{\beta}_{2}\right\Vert .
	\end{align*}
	So we have that the second term is of order 
	$
	O_{p}\left(\sqrt{p}\chi_{1,n}\right).
	$
	This finishes the proof.
\end{proof}
\begin{lemma}
	\label{lemS.5}Suppose that  \autoref{assu1}, \autoref{assump:2}(i)-(iii),
	and \autoref{assu:6} hold with $\upsilon_G \geq1$, and that $\chi_{1,n}\rightarrow0$ as $n \rightarrow \infty$, then we have that 
	\begin{align*}
		\sup_{\boldsymbol{\beta}\in\mathcal{B}} & \left\Vert \frac{1}{n}\sum_{i=1}^{n}\left(\mathbf{X}_{i}-\mathfrak{X}_{q}\left(X_{0,i}+\mathbf{X}_{i}^{\mathrm{T}}\boldsymbol{\beta},\boldsymbol{\beta}\right)\right)\left(G\left(X_{0,i}+\mathbf{X}_{i}^{\mathrm{T}}\boldsymbol{\beta}\right)-G\left(X_{0,i}+\mathbf{X}_{i}^{\mathrm{T}}\boldsymbol{\beta}^{\star}\right)\right)\right.\\
		& \left.-\mathbb{E}\left(\left(\mathbf{X}_{i}-\mathfrak{X}_{q}\left(X_{0,i}+\mathbf{X}_{i}^{\mathrm{T}}\boldsymbol{\beta},\boldsymbol{\beta}\right)\right)\left(G\left(X_{0,i}+\mathbf{X}_{i}^{\mathrm{T}}\boldsymbol{\beta}\right)-G\left(X_{0,i}+\mathbf{X}_{i}^{\mathrm{T}}\boldsymbol{\beta}^{\star}\right)\right)\right)\right\Vert =O_{p}\left(\sqrt{p}\chi_{1,n}\right).
	\end{align*}
\end{lemma}
\begin{proof}[Proof of \autoref{lemS.5}]
	
	We only need to note that uniformly for all $\mathbf{X}_{e,i}$, $1\leq j\leq p$,
	and $\boldsymbol{\beta},\boldsymbol{\beta}_{1},\boldsymbol{\beta}_{2}\in\mathcal{B}$,
	there hold
	\begin{align*}
		& \left|\left(X_{i,j}-\mathbb{E}_{\mathbf{X}_e}\left(\boldsymbol{r}_{q}^{\mathrm{T}}\left(X_{0}+\mathbf{X}^{\mathrm{T}}\boldsymbol{\beta}\right)\Gamma_{q}^{-1}\left(\boldsymbol{\beta}\right)\boldsymbol{r}_{q}\left(X_{0,i}+\mathbf{X}_{i}^{\mathrm{T}}\boldsymbol{\beta}\right)X_{j}\right)\right)\left(G\left(X_{0,i}+\mathbf{X}_{i}^{\mathrm{T}}\boldsymbol{\beta}\right)-G\left(X_{0,i}+\mathbf{X}_{i}^{\mathrm{T}}\boldsymbol{\beta}^{\star}\right)\right)\right|\\
		& \leq CqD_{q,0}^{2},
	\end{align*}
	and 
	\begin{align*}
		& \left\Vert G\left(X_{0,i}+\mathbf{X}_{i}^{\mathrm{T}}\boldsymbol{\beta}_{1}\right)\mathbb{E}_{\mathbf{X}_{e}}\left(\boldsymbol{r}_{q}^{\mathrm{T}}\left(X_{0}+\mathbf{X}^{\mathrm{T}}\boldsymbol{\beta}_{1}\right)\Gamma_{q}^{-1}\left(\boldsymbol{\beta}_{1}\right)\boldsymbol{r}_{q}\left(X_{0,i}+\mathbf{X}_{i}^{\mathrm{T}}\boldsymbol{\beta}_1\right)X_{j}\right)\right.\\
		& \left.-G\left(X_{0,i}+\mathbf{X}_{i}^{\mathrm{T}}\boldsymbol{\beta}_{2}\right)\mathbb{E}_{\mathbf{X}_{e}}\left(\boldsymbol{r}_{q}^{\mathrm{T}}\left(X_{0}+\mathbf{X}^{\mathrm{T}}\boldsymbol{\beta}_{2}\right)\Gamma_{q}^{-1}\left(\boldsymbol{\beta}_{2}\right)\boldsymbol{r}_{q}\left(X_{0,i}+\mathbf{X}_{i}^{\mathrm{T}}\boldsymbol{\beta}_2\right)X_{j}\right)\right\Vert \\
		& \leq\left\Vert \left(G\left(X_{0,i}+\mathbf{X}_{i}^{\mathrm{T}}\boldsymbol{\beta}_{1}\right)-G\left(X_{0,i}+\mathbf{X}_{i}^{\mathrm{T}}\boldsymbol{\beta}_{2}\right)\right)\mathbb{E}_{\mathbf{X}_{e}}\left(\boldsymbol{r}_{q}^{\mathrm{T}}\left(X_{0}+\mathbf{X}^{\mathrm{T}}\boldsymbol{\beta}_{1}\right)\Gamma_{q}^{-1}\left(\boldsymbol{\beta}_{1}\right)\boldsymbol{r}_{q}\left(X_{0,i}+\mathbf{X}_{i}^{\mathrm{T}}\boldsymbol{\beta}_1\right)X_{j}\right)\right\Vert \\
		& +\left\Vert G\left(X_{0,i}+\mathbf{X}_{i}^{\mathrm{T}}\boldsymbol{\beta}_{2}\right)\mathbb{E}_{\mathbf{X}_{e}}\left(\left(\boldsymbol{r}_{q}^{\mathrm{T}}\left(X_{0}+\mathbf{X}^{\mathrm{T}}\boldsymbol{\beta}_{1}\right)-\boldsymbol{r}_{q}^{\mathrm{T}}\left(X_{0}+\mathbf{X}^{\mathrm{T}}\boldsymbol{\beta}_{2}\right)\right)\Gamma_{q}^{-1}\left(\boldsymbol{\beta}_{1}\right)\boldsymbol{r}_{q}\left(X_{0,i}+\mathbf{X}_{i}^{\mathrm{T}}\boldsymbol{\beta}_1\right)X_{j}\right)\right\Vert \\
		& +\left\Vert G\left(X_{0,i}+\mathbf{X}_{i}^{\mathrm{T}}\boldsymbol{\beta}_{2}\right)\mathbb{E}_{\mathbf{X}_{e}}\left(\boldsymbol{r}_{q}^{\mathrm{T}}\left(X_{0}+\mathbf{X}^{\mathrm{T}}\boldsymbol{\beta}_{2}\right)\left(\Gamma_{q}^{-1}\left(\boldsymbol{\beta}_{1}\right)-\Gamma_{q}^{-1}\left(\boldsymbol{\beta}_{2}\right)\right)\boldsymbol{r}_{q}\left(X_{0,i}+\mathbf{X}_{i}^{\mathrm{T}}\boldsymbol{\beta}_1\right)X_{j}\right)\right\Vert \\
		& +\left\Vert G\left(X_{0,i}+\mathbf{X}_{i}^{\mathrm{T}}\boldsymbol{\beta}_{2}\right)\mathbb{E}_{\mathbf{X}_{e}}\left(\boldsymbol{r}_{q}^{\mathrm{T}}\left(X_{0}+\mathbf{X}^{\mathrm{T}}\boldsymbol{\beta}_{2}\right)\Gamma_{q}^{-1}\left(\boldsymbol{\beta}_{2}\right)\left(\boldsymbol{r}_{q}\left(X_{0,i}+\mathbf{X}_{i}^{\mathrm{T}}\boldsymbol{\beta}_1\right) - \boldsymbol{r}_{q}\left(X_{0,i}+\mathbf{X}_{i}^{\mathrm{T}}\boldsymbol{\beta}_2\right)\right)X_{j}\right)\right\Vert \\
		& \leq C\sqrt{p}q^{2}D_{q,0}^{3}D_{q,1}\left\Vert \boldsymbol{\beta}_{1}-\boldsymbol{\beta}_{2}\right\Vert .
	\end{align*}
\end{proof}
\begin{lemma}
	\label{lemS.6}Suppose that  \autoref{assu1}, \autoref{assump:2}(i)-(iii),
	and \autoref{assu:6} hold with $\upsilon_G \geq1$, and that $\chi_{1,n}\rightarrow0$ as $n \rightarrow \infty$, then we have that 
	\[
	\sup_{\boldsymbol{\beta}\in\mathcal{B}}\left\Vert \frac{1}{n}\sum_{i=1}^{n}\mathbf{X}_{i}\boldsymbol{r}_{q}^{\mathrm{T}}\left(z\left(\mathbf{X}_{e,i},\boldsymbol{\beta}\right)\right)\Gamma_{q,n}^{-1}\left(\boldsymbol{\beta}\right)\left(\frac{1}{n}\sum_{j=1}^{n}\boldsymbol{r}_{q}^{\mathrm{T}}\left(z\left(\mathbf{X}_{e,j},\boldsymbol{\beta}\right)\right)R_{q}\left(z\left(\mathbf{X}_{e,j},\boldsymbol{\beta}\right)\right)\right)\right\Vert =O_{p}\left(\sqrt{p}qD_{q,0}^{2}\mathcal{E}_{q,0}\right),
	\]
	\[
	\sup_{\boldsymbol{\beta}\in\mathcal{B}}\left\Vert \frac{1}{n}\sum_{i=1}^{n}\mathbf{X}_{i}\boldsymbol{r}_{q}^{\mathrm{T}}\left(z\left(\mathbf{X}_{e,i},\boldsymbol{\beta}\right)\right)\Gamma_{q,n}^{-1}\left(\boldsymbol{\beta}\right)\left(\frac{1}{n}\sum_{j=1}^{n}\boldsymbol{r}_{q}^{\mathrm{T}}\left(z\left(\mathbf{X}_{e,j},\boldsymbol{\beta}\right)\right)\varepsilon_{j}\right)\right\Vert =O_{p}\left(\sqrt{p}\chi_{1,n}\right),
	\]
	and 
	\[
	\sup_{\boldsymbol{\beta}\in\mathcal{B}}\left\Vert \frac{1}{n}\sum_{i=1}^{n}\left(R_{q}\left(z\left(\mathbf{X}_{e,i},\boldsymbol{\beta}\right)\right)\mathbf{X}_{i}+\varepsilon_{i}\mathbf{X}_{i}\right)\right\Vert =O_{p}\left(\sqrt{p}\mathcal{E}_{q,0}+\sqrt{p(\log p)/n}\right).
	\]
\end{lemma}
\begin{proof}[Proof of \autoref{lemS.6}]
	For the first result, we note that 
	\begin{align*}
		& \sup_{\boldsymbol{\beta}\in\mathcal{B}}\left\Vert \mathbf{X}_{i}\boldsymbol{r}_{q}^{\mathrm{T}}\left(z\left(\mathbf{X}_{e,i},\boldsymbol{\beta}\right)\right)\Gamma_{q,n}^{-1}\left(\boldsymbol{\beta}\right)\left(\frac{1}{n}\sum_{j=1}^{n}\boldsymbol{r}_{q}^{\mathrm{T}}\left(z\left(\mathbf{X}_{e,j},\boldsymbol{\beta}\right)\right)R_{q}\left(z\left(\mathbf{X}_{e,j},\boldsymbol{\beta}\right)\right)\right)\right\Vert \\
		& =O_{p}\left(\sqrt{p}\sup_{\boldsymbol{\beta}\in\mathcal{B},\mathbf{X}_{e}\in\mathcal{X}_{e}}\left\Vert \boldsymbol{r}_{q}\left(z\left(\mathbf{X}_{e},\boldsymbol{\beta}\right)\right)\right\Vert \sup_{\boldsymbol{\beta}\in\mathcal{B},\mathbf{X}_{e}\in\mathcal{X}_{e}}\left\Vert \boldsymbol{r}_{q}\left(z\left(\mathbf{X}_{e},\boldsymbol{\beta}\right)\right)R_{q}\left(z\left(\mathbf{X}_{e},\boldsymbol{\beta}\right)\right)\right\Vert \right)\\
		& =O_{p}\left(\sqrt{p}qD_{q,0}^{2}\mathcal{E}_{q,0}\right).
	\end{align*}
	For the second result, we first have that 
	\[
	\sup_{\boldsymbol{\beta}\in\mathcal{B}}\left\Vert \frac{1}{n}\sum_{j=1}^{n}\boldsymbol{r}_{q}^{\mathrm{T}}\left(z\left(\mathbf{X}_{e,j},\boldsymbol{\beta}\right)\right)\varepsilon_{j}\right\Vert =O_{p}\left(\sqrt{pqD_{q,0}^{2}\log\left(pqD_{q,1}n\right)/n}\right),
	\]
	due to the fact that $\left|r_{l}\left(z\left(\mathbf{X}_{e,j},\boldsymbol{\beta}\right)\right)\varepsilon_{j}\right|\leq CD_{q,0}$
	and $\left\Vert \left(\partial r_{l}\left(z\left(\mathbf{X}_{e,j},\boldsymbol{\beta}\right)\right)/\partial\boldsymbol{\beta}\right)\varepsilon_{j}\right\Vert \leq C\sqrt{p}D_{q,1}$
	for all $0\leq l\leq q$. So 
	\begin{align*}
		& \sup_{\boldsymbol{\beta}\in\mathcal{B}}\left\Vert \frac{1}{n}\sum_{i=1}^{n}\mathbf{X}_{i}\boldsymbol{r}_{q}^{\mathrm{T}}\left(z\left(\mathbf{X}_{e,i},\boldsymbol{\beta}\right)\right)\Gamma_{q,n}^{-1}\left(\boldsymbol{\beta}\right)\left(\frac{1}{n}\sum_{j=1}^{n}\boldsymbol{r}_{q}^{\mathrm{T}}\left(z\left(\mathbf{X}_{e,j},\boldsymbol{\beta}\right)\right)\varepsilon_{j}\right)\right\Vert \\
		& =O_{p}\left(\sqrt{pq}D_{q,0}\sqrt{pqD_{q,0}^{2}\log\left(pqD_{q,1}n\right)/n}\right)=O_{p}\left(\sqrt{p}\chi_{1,n}\right).
	\end{align*}
	Finally for the third result, we have that $\left\Vert \frac{1}{n}\sum_{i=1}^{n}R_{q}\left(z\left(\mathbf{X}_{e,i},\boldsymbol{\beta}\right)\right)\mathbf{X}_{i}\right\Vert =O_{p}\left(\sqrt{p}\mathcal{E}_{q,0}\right)$
	and $\left\Vert \frac{1}{n}\sum_{i=1}^{n}\varepsilon_{i}\mathbf{X}_{i}\right\Vert =O_{p}\left(\sqrt{p\left(\log p\right)/n}\right)$. 
	
	Combine the above results, we finish the proof.
\end{proof}
Now we are ready to prove \autoref{lem4.1} in the main text.

\begin{proof}[Proof of \autoref{lem4.1}]
	We note that 
	\begin{align*}
		\boldsymbol{\beta}_{k+1} & =\boldsymbol{\beta}_{k}-\frac{\delta_{k}}{n}\sum_{i=1}^{n}\left(\widehat{G}\left(\left.z_{i,k}\right|\boldsymbol{\beta}_{k}\right)-y_{i}\right)\mathbf{X}_{i}\\
		& =\boldsymbol{\beta}_{k}-\frac{\delta_{k}}{n}\sum_{i=1}^{n}\left(\boldsymbol{r}_{q}^{\mathrm{T}}\left(z_{i,k}\right)\widehat{\boldsymbol{\pi}}_{q,n,k}-\boldsymbol{r}_{q}^{\mathrm{T}}\left(z_{i,k}\right)\boldsymbol{\pi}_{q}^{\star}\right)\mathbf{X}_{i}-\frac{\delta_{k}}{n}\sum_{i=1}^{n}\left(G\left(z_{i,k}\right)-G\left(\boldsymbol{z}_{i}^{\star}\right)\right)\mathbf{X}_{i}\\
		& +\frac{\delta_{k}}{n}\sum_{i=1}^{n}R_{q}\left(z_{i,k}\right)\mathbf{X}_{i}+\frac{\delta}{n}\sum_{i=1}^{n}\varepsilon_{i}\mathbf{X}_{i}.
	\end{align*}
	Now we look at the $\widehat{\boldsymbol{\pi}}_{q,n,k}-\boldsymbol{\pi}_{q}^{\star}$.
	Define $\Gamma_{q,n,k}=\Gamma_{q,n}\left(\boldsymbol{\beta}_{k}\right)$,
	we have that
	\begin{align*}
		\widehat{\boldsymbol{\pi}}_{q,n,k} & =\left(\frac{1}{n}\sum_{i=1}^{n}\boldsymbol{r}_{q}\left(z_{i,k}\right)\boldsymbol{r}_{q}^{\mathrm{T}}\left(z_{i,k}\right)\right)^{-1}\left(\frac{1}{n}\sum_{i=1}^{n}\boldsymbol{r}_{q}\left(z_{i,k}\right)y_{i}\right)\\
		& =\boldsymbol{\pi}_{q}^{\star}-\Gamma_{q,n,k}^{-1}\left(\frac{1}{n}\sum_{i=1}^{n}\boldsymbol{r}_{q}\left(z_{i,k}\right)\left(G\left(z_{i,k}\right)-G\left(\boldsymbol{z}_{i}^{\star}\right)\right)\right)+\Gamma_{q,n,k}^{-1}\left(\frac{1}{n}\sum_{i=1}^{n}\boldsymbol{r}_{q}\left(z_{i,k}\right)R_{q}\left(z_{i,k}\right)\right)\\
		& +\Gamma_{q,n,k}^{-1}\left(\frac{1}{n}\sum_{i=1}^{n}\boldsymbol{r}_{q}\left(z_{i,k}\right)\varepsilon_{i}\right).
	\end{align*}
	Take the above expression of $\widehat{\boldsymbol{\pi}}_{q,n,k}-\boldsymbol{\pi}_{q}^{\star}$
	into the update of $\boldsymbol{\beta}_{k}$, we have that 
	\begin{align*}
		\boldsymbol{\beta}_{k+1} & =\boldsymbol{\beta}_{k}-\frac{\delta_{k}}{n}\sum_{i=1}^{n}\left(\mathbf{X}_{i}-\mathfrak{X}_{q,n}\left(z_{i,k},\boldsymbol{\beta}_{k}\right)\right)\left(G\left(z_{i,k}\right)-G\left(z_{i}^{\star}\right)\right)\\
		& -\frac{\delta_{k}}{n}\sum_{i=1}^{n}\mathbf{X}_{i}\boldsymbol{r}_{q}^{\mathrm{T}}\left(z_{i,k}\right)\Gamma_{q,n,k}^{-1}\left(\frac{1}{n}\sum_{j=1}^{n}\boldsymbol{r}_{q}\left(z_{j,k}\right)R_{q}\left(z_{j,k}\right)+\frac{1}{n}\sum_{i=1}^{n}\boldsymbol{r}_{q}\left(z_{j,k}\right)\varepsilon_{j}\right)\\
		&+\frac{\delta_{k}}{n}\sum_{i=1}^{n}\left(R_{q}\left(z_{i,k}\right)\mathbf{X}_{i}+\varepsilon_{i}\mathbf{X}_{i}\right).
	\end{align*}
	If we define 
	\begin{align*}
		\mathfrak{R}_{n,k} & =\mathbb{E}\left(\mathbf{X}-\mathfrak{X}_{q}\left(z\left(\mathbf{X}_{e},\boldsymbol{\beta}_{k}\right),\boldsymbol{\beta}_{k}\right)\right)\left(G\left(z\left(\mathbf{X}_{e},\boldsymbol{\beta}_{k}\right)\right)-G\left(z\left(\mathbf{X}_{e},\boldsymbol{\beta}^{\star}\right)\right)\right)\\
		&
		-\frac{1}{n}\sum_{i=1}^n \left(\mathbf{X}_i-\mathfrak{X}_{q}\left(z\left(\mathbf{X}_{e,i},\boldsymbol{\beta}_{k}\right),\boldsymbol{\beta}_{k}\right)\right)\left(G\left(z\left(\mathbf{X}_{e,i},\boldsymbol{\beta}_{k}\right)\right)-G\left(z\left(\mathbf{X}_{e,i},\boldsymbol{\beta}^{\star}\right)\right)\right)\\
		&+\frac{1}{n}\sum_{i=1}^n \left(\mathbf{X}_i-\mathfrak{X}_{q}\left(z\left(\mathbf{X}_{e,i},\boldsymbol{\beta}_{k}\right),\boldsymbol{\beta}_{k}\right)\right)\left(G\left(z\left(\mathbf{X}_{e,i},\boldsymbol{\beta}_{k}\right)\right)-G\left(z\left(\mathbf{X}_{e,i},\boldsymbol{\beta}^{\star}\right)\right)\right)\\
		& -\frac{1}{n}\sum_{i=1}^{n}\left(\mathbf{X}_{i}-\mathfrak{X}_{q,n}\left(z\left(\mathbf{X}_{e,i},\boldsymbol{\beta}_{k}\right),\boldsymbol{\beta}_{k}\right)\right)\left(G\left(z\left(\mathbf{X}_{e,i},\boldsymbol{\beta}_{k}\right)\right)-G\left(z\left(\mathbf{X}_{e,i},\boldsymbol{\beta}^{\star}\right)\right)\right)\\
		& -\frac{\delta_{k}}{n}\sum_{i=1}^{n}\mathbf{X}_{i}\boldsymbol{r}_{q}^{\mathrm{T}}\left(z_{i,k}\right)\Gamma_{q,n,k}^{-1}\left(\frac{1}{n}\sum_{j=1}^{n}\boldsymbol{r}_{q}\left(z_{j,k}\right)R_{q}\left(z_{j,k}\right)+\frac{1}{n}\sum_{i=1}^{n}\boldsymbol{r}_{q}\left(z_{j,k}\right)\varepsilon_{j}\right)\\
		& +\frac{\delta_{k}}{n}\sum_{i=1}^{n}\left(R_{q}\left(z_{i,k}\right)\mathbf{X}_{i}+\varepsilon_{i}\mathbf{X}_{i}\right),
	\end{align*}
	we have that 
	\[
	\boldsymbol{\beta}_{k+1}=\boldsymbol{\beta}_{k}-\delta_{k}\mathbb{E}\left[\left(\mathbf{X}-\mathfrak{X}_{q}\left(z\left(\mathbf{X}_{e},\boldsymbol{\beta}_{k}\right),\boldsymbol{\beta}_{k}\right)\right)\left(G\left(z\left(\mathbf{X}_{e},\boldsymbol{\beta}_{k}\right)\right)-G\left(z\left(\mathbf{X}_{e},\boldsymbol{\beta}^{\star}\right)\right)\right)\right]+\delta_{k}\mathfrak{R}_{n,k}.
	\]
	It remains to verify the order of $\sup_{k\geq1}\left\Vert \mathfrak{R}_{n,k}\right\Vert $,
	which is done based on \autoref{lemS.4}, \autoref{lemS.5}, and
     \autoref{lemS.6}. 
\end{proof}

Now we prove \autoref{lem4.2} and \autoref{lem4.3} in the main text.

\begin{proof}[Proof of \autoref{lem4.2}]
	Recall that \[\Psi_{q}\left(t,\boldsymbol{\beta}\right)=\mathbb{E}\left[G^{\prime}\left(z\left(\mathbf{X}_{e},\boldsymbol{\beta}^{\star}\right)+t\mathbf{X}^{\mathrm{T}}\Delta\boldsymbol{\beta}\right)\left(\mathbf{X}\mathbf{X}^{\mathrm{T}}-\mathfrak{X}_{q}\left(z\left(\mathbf{X}_{e},\boldsymbol{\beta}\right),\boldsymbol{\beta}\right)\mathbf{X}^{\mathrm{T}}\right)\right].\]
	We have that 
	\begin{align*}
		& \sup_{0\leq t\leq1,\boldsymbol{\beta}\in\mathcal{B}_{n}}\left\Vert \frac{1}{n}\sum_{i=1}^{n}G^{\prime}\left(z_{i}^{\star}+t\mathbf{X}_{i}^{\mathrm{T}}\Delta\boldsymbol{\beta}\right)\left(\mathbf{X}_{i}\mathbf{X}_{i}^{\mathrm{T}}-\mathfrak{X}_{q,n}\left(z\left(\mathbf{X}_{e,i},\boldsymbol{\beta}\right),\boldsymbol{\beta}\right)\mathbf{X}_{i}^{\mathrm{T}}\right)-\Psi_{q}^{\star}\right\Vert \\
		& \leq\sup_{0\leq t\leq1,\boldsymbol{\beta}\in\mathcal{B}}\left\Vert \frac{1}{n}\sum_{i=1}^{n}G^{\prime}\left(z_{i}^{\star}+t\mathbf{X}_{i}^{\mathrm{T}}\Delta\boldsymbol{\beta}\right)\left(\mathfrak{X}_{q,n}\left(z\left(\mathbf{X}_{e,i},\boldsymbol{\beta}\right),\boldsymbol{\beta}\right)-\mathfrak{X}_{q}\left(z\left(\mathbf{X}_{e,i},\boldsymbol{\beta}\right),\boldsymbol{\beta}\right)\right)\mathbf{X}_{i}^{\mathrm{T}}\right\Vert \\
		& +\sup_{0\leq t\leq1,\boldsymbol{\beta}\in\mathcal{B}}\left\Vert \frac{1}{n}\sum_{i=1}^{n}G^{\prime}\left(z_{i}^{\star}+t\mathbf{X}_{i}^{\mathrm{T}}\Delta\boldsymbol{\beta}\right)\left(\mathbf{X}_{i}\mathbf{X}_{i}^{\mathrm{T}}-\mathfrak{X}_{q}\left(z\left(\mathbf{X}_{e,i},\boldsymbol{\beta}\right),\boldsymbol{\beta}\right)\mathbf{X}_{i}^{\mathrm{T}}\right)-\Psi_{q}\left(t,\boldsymbol{\beta}\right)\right\Vert \\
		& +\sup_{0\leq t\leq1,\boldsymbol{\beta}\in\mathcal{B}_{n}}\left\Vert \Psi_{q}\left(t,\boldsymbol{\beta}\right)-\Psi_{q}^{\star}\right\Vert .
	\end{align*}
	From \autoref{lemS.4}, we know that 
	\[
	\sup_{z\in\mathcal{Z}}\sup_{\boldsymbol{\beta}\in\mathcal{B}}\left\Vert \mathfrak{X}_{q,n}\left(z,\boldsymbol{\beta}\right)-\mathfrak{X}_{q}\left(z,\boldsymbol{\beta}\right)\right\Vert =O_{p}\left(\sqrt{p}qD_{q,0}^{2}\chi_{1,n}\right),
	\]
	and as a result, 
	\begin{align*}
		& \sup_{0\leq t\leq1,\boldsymbol{\beta}\in\mathcal{B}}\left\Vert \frac{1}{n}\sum_{i=1}^{n}G^{\prime}\left(z_{i}^{\star}+t\mathbf{X}_{i}^{\mathrm{T}}\Delta\boldsymbol{\beta}\right)\left(\mathfrak{X}_{q,n}\left(z\left(\mathbf{X}_{e,i},\boldsymbol{\beta}\right),\boldsymbol{\beta}\right)-\mathfrak{X}_{q}\left(z\left(\mathbf{X}_{e,i},\boldsymbol{\beta}\right),\boldsymbol{\beta}\right)\right)\mathbf{X}_{i}^{\mathrm{T}}\right\Vert \\
		& =O_{p}\left(pqD_{q,0}^{2}\chi_{1,n}\right).
	\end{align*}
	For the second term, we have that 
	\begin{align*}
		& \sup_{0\leq t\leq1,\boldsymbol{\beta}\in\mathcal{B}}\left\Vert \frac{1}{n}\sum_{i=1}^{n}G^{\prime}\left(z_{i}^{\star}+t\mathbf{X}_{i}^{\mathrm{T}}\Delta\boldsymbol{\beta}\right)\left(\mathbf{X}_{i}\mathbf{X}_{i}^{\mathrm{T}}-\mathfrak{X}_{q,n}\left(z\left(\mathbf{X}_{e,i},\boldsymbol{\beta}\right),\boldsymbol{\beta}\right)\mathbf{X}_{i}^{\mathrm{T}}\right)-\Psi_{q}\left(t,\boldsymbol{\beta}\right)\right\Vert \\
		& =O_{p}\left(\sqrt{p^{3}q^{2}D_{q,0}^{4}\log\left(pqD_{q,0}D_{q,1}n\right)/n}\right)=O_{p}\left(p\chi_{1,n}\right),
	\end{align*}
	due to the fact that 
	\[
	\left|G^{\prime}\left(z_{i}^{\star}+t\mathbf{X}_{i}^{\mathrm{T}}\Delta\boldsymbol{\beta}\right)\left(X_{i,s}X_{i,t}-\left(\mathfrak{X}_{q}\left(z\left(\mathbf{X}_{e,i},\boldsymbol{\beta}\right),\boldsymbol{\beta}\right)\right)_{s}X_{i,t}\right)\right|\leq CqD_{q.0}^{2},
	\]
	and 
	\begin{align*}
		& \left|G^{\prime}\left(z_{i}^{\star}+t\mathbf{X}_{i}^{\mathrm{T}}\Delta\boldsymbol{\beta}_{1}\right)\left(X_{i,s}X_{i,t}-\left(\mathfrak{X}_{q}\left(z\left(\mathbf{X}_{e,i},\boldsymbol{\beta}_{1}\right),\boldsymbol{\beta}_{1}\right)\right)_{s}X_{i,t}\right)\right.\\
		& \left.-G^{\prime}\left(z_{i}^{\star}+t\mathbf{X}_{i}^{\mathrm{T}}\Delta\boldsymbol{\beta}_{2}\right)\left(X_{i,s}X_{i,t}-\left(\mathfrak{X}_{q}\left(z\left(\mathbf{X}_{e,i},\boldsymbol{\beta}_{2}\right),\boldsymbol{\beta}_{2}\right)\right)_{s}X_{i,t}\right)\right|\\
		& \leq C\sqrt{p}q^{2}D_{q,0}^{3}D_{q,1}\left\Vert \boldsymbol{\beta}_{1}-\boldsymbol{\beta}_{2}\right\Vert .
	\end{align*}
	Finally, 
	\begin{align*}
		& \sup_{0\leq t\leq1,\boldsymbol{\beta}\in\mathcal{B}_{n}}\left\Vert \Psi_{q}\left(t,\boldsymbol{\beta}\right)-\Psi_{q}^{\star}\right\Vert \\
		& \leq\sup_{0\leq t\leq1,\boldsymbol{\beta}\in\mathcal{B}_{n}}\left\Vert \mathbb{E}\left[G^{\prime}\left(z\left(\mathbf{X}_{e},\boldsymbol{\beta}^{\star}\right)+t\mathbf{X}^{\mathrm{T}}\Delta\boldsymbol{\beta}\right)-G^{\prime}\left(z\left(\mathbf{X}_{e},\boldsymbol{\beta}^{\star}\right)\right)\left(\mathbf{X}\mathbf{X}^{\mathrm{T}}-\mathfrak{X}_{q}\left(z\left(\mathbf{X}_{e},\boldsymbol{\beta}\right),\boldsymbol{\beta}\right)\mathbf{X}^{\mathrm{T}}\right)\right]\right\Vert \\
		& +\sup_{\boldsymbol{\beta}\in\mathcal{B}_{n}}\left\Vert \mathbb{E}\left[G^{\prime}\left(z\left(\mathbf{X}_{e},\boldsymbol{\beta}^{\star}\right)\right)\left(\mathfrak{X}_{q}\left(z\left(\mathbf{X}_{e},\boldsymbol{\beta}\right),\boldsymbol{\beta}\right)-\mathfrak{X}_{q}\left(z\left(\mathbf{X}_{e},\boldsymbol{\beta}^{\star}\right),\boldsymbol{\beta}^{\star}\right)\mathbf{X}^{\mathrm{T}}\right)\right]\right\Vert .
	\end{align*}
	Obviously the first term is bounded by $C\sqrt{p^{3}}qD_{q,0}^{2}\sup_{\boldsymbol{\beta}\in\mathcal{B}_{n}}\left\Vert \Delta\boldsymbol{\beta}\right\Vert $,
	while the second term is bounded by 
	\begin{align*}
		& Cp\sup_{\mathbf{X}_{e},\mathbf{\widetilde{X}}_{e}}\left\Vert \left(\boldsymbol{r}_{q}^{\mathrm{T}}\left(z\left(\mathbf{\widetilde{X}}_{e},\boldsymbol{\beta}\right)\right)\Gamma_{q}^{-1}\left(\boldsymbol{\beta}\right)\boldsymbol{r}_{q}\left(z\left(\mathbf{X}_{e},\boldsymbol{\beta}\right)\right)\right)-\left(\boldsymbol{r}_{q}^{\mathrm{T}}\left(z\left(\mathbf{\widetilde{X}}_{e},\boldsymbol{\beta}^{\star}\right)\right)\Gamma_{q}^{-1}\left(\boldsymbol{\beta}^{\star}\right)\boldsymbol{r}_{q}\left(z\left(\mathbf{X}_{e},\boldsymbol{\beta}^{\star}\right)\right)\right)\right\Vert \\
		& \leq Cp\sup_{\mathbf{X}_{e},\mathbf{\widetilde{X}}_{e}}\left\Vert \left(\boldsymbol{r}_{q}\left(z\left(\mathbf{\widetilde{X}}_{e},\boldsymbol{\beta}\right)\right)-\boldsymbol{r}_{q}\left(z\left(\mathbf{\widetilde{X}}_{e},\boldsymbol{\beta}^{\star}\right)\right)\right)^{\mathrm{T}}\Gamma_{q}^{-1}\left(\boldsymbol{\beta}\right)\boldsymbol{r}_{q}\left(z\left(\mathbf{X}_{e},\boldsymbol{\beta}\right)\right)\right\Vert \\
		& +Cp\sup_{\mathbf{X}_{e},\mathbf{\widetilde{X}}_{e}}\left\Vert \boldsymbol{r}_{q}\left(z\left(\mathbf{\widetilde{X}}_{e},\boldsymbol{\beta}^{\star}\right)\right)^{\mathrm{T}}\left(\Gamma_{q}^{-1}\left(\boldsymbol{\beta}\right)-\Gamma_{q}^{-1}\left(\boldsymbol{\beta}^{\star}\right)\right)\boldsymbol{r}_{q}\left(z\left(\mathbf{X}_{e},\boldsymbol{\beta}\right)\right)\right\Vert \\
		& +Cp\sup_{\mathbf{X}_{e},\mathbf{\widetilde{X}}_{e}}\left\Vert \boldsymbol{r}_{q}\left(z\left(\mathbf{\widetilde{X}}_{e},\boldsymbol{\beta}^{\star}\right)\right)^{\mathrm{T}}\Gamma_{q}^{-1}\left(\boldsymbol{\beta}^{\star}\right)\left(\boldsymbol{r}_{q}\left(z\left(\mathbf{X}_{e},\boldsymbol{\beta}\right)\right)-\boldsymbol{r}_{q}\left(z\left(\mathbf{X}_{e},\boldsymbol{\beta}^{\star}\right)\right)\right)\right\Vert \\
		& \leq C\sqrt{p^3}q^{2}D_{q,0}^{3}D_{q,1}\sup_{\boldsymbol{\beta}\in\mathcal{B}_{n}}\left\Vert \Delta\boldsymbol{\beta}\right\Vert .
	\end{align*}
	So 
	\[
	\sup_{0\leq t\leq1,\boldsymbol{\beta}\in\mathcal{B}_{n}}\left\Vert \Psi_{q}\left(t,\boldsymbol{\beta}\right)-\Psi_{q}^{\star}\right\Vert =O_{p}\left(\sqrt{p^3}q^{2}D_{q,0}^{3}D_{q,1}\sup_{\boldsymbol{\beta}\in\mathcal{B}_{n}}\left\Vert \Delta\boldsymbol{\beta}\right\Vert \right).
	\]
	
	Combine the above results, we have that 
	\begin{align*}
		& \sup_{0\leq t\leq1,\boldsymbol{\beta}\in\mathcal{B}_{n}}\left\Vert \frac{1}{n}\sum_{i=1}^{n}G^{\prime}\left(z_{i}^{\star}+t\mathbf{X}_{i}^{\mathrm{T}}\Delta\boldsymbol{\beta}\right)\left(\mathbf{X}_{i}\mathbf{X}_{i}^{\mathrm{T}}-\mathfrak{X}_{q,n}\left(z\left(\mathbf{X}_{e,i},\boldsymbol{\beta}\right),\boldsymbol{\beta}\right)\mathbf{X}_{i}^{\mathrm{T}}\right)-\Psi_{q}^{\star}\right\Vert \\
		&=O_{p}\left(pqD_{q,0}^{2}\chi_{1,n} + \sqrt{p^3}q^{2}D_{q,0}^{3}D_{q,1}\sup_{\boldsymbol{\beta}\in\mathcal{B}_{n}}\left\Vert \Delta\boldsymbol{\beta}\right\Vert \right).
	\end{align*}
\end{proof}

\begin{proof}[Proof of \autoref{lem4.3}]
	According to \autoref{thm4.1}, we have that $\sup_{k\geq k_{1,n}^{SBGD}+1}\left\Vert \Delta\boldsymbol{\beta}_{k}\right\Vert =O_{p}\left(\chi_{2,n}\right)$.
	To prove the lemma, we first show that 
	\begin{align*}
		& \sup_{k\geq k_{1,n}^{SBGD}+1}\left\Vert \frac{1}{n}\sum_{i=1}^{n}\mathbf{X}_{i}\boldsymbol{r}_{q,i,k}^{\mathrm{T}}\Gamma_{q,n,k}^{-1}\left(\frac{1}{n}\sum_{j=1}^{n}\boldsymbol{r}_{q,j,k}R_{q,j,k}+\frac{1}{n}\sum_{i=1}^{n}\boldsymbol{r}_{q,j,k}\varepsilon_{j}-\frac{1}{n}\sum_{i=1}^{n}\boldsymbol{r}_{q,j}^{\star}\varepsilon_{j}\right)\right\Vert \\
		& =O_{p}\left(\sqrt{p}qD_{q,0}^{2}\mathcal{E}_{q,0}+\sqrt{pq}D_{q,0}\chi_{2,n}\chi_{3,n}\right),
	\end{align*}
	where $\chi_{3,n}=\sqrt{p^{2}qD_{q,1}^{2}\log\left(pqD_{q,2}n\right)/n}$.
	Note that 
	\begin{align*}
		\sup_{k\geq k_{1,n}^{SBGD}+1}\left\Vert \frac{1}{n}\sum_{i=1}^{n}\boldsymbol{r}_{q,i,k}\varepsilon_{i}-\frac{1}{n}\sum_{i=1}^{n}\boldsymbol{r}_{q,i}^{\star}\varepsilon_{i}\right\Vert  & =\sup_{k\geq k_{1,n}^{SBGD}+1}\left\Vert \left\{ \int_{0}^{1}\frac{1}{n}\sum_{i=1}^{n}\varepsilon_{i}\boldsymbol{r}_{q}^{\prime}\left(z_{i}^{\star}+t\mathbf{X}_{i}^{\mathrm{T}}\Delta\boldsymbol{\beta}_{k}\right)\mathbf{X}_{i}^{\mathrm{T}}dt\right\} \Delta\boldsymbol{\beta}_{k}\right\Vert \\
		& \leq\sup_{\boldsymbol{\beta}\in\mathcal{B}}\left\Vert \frac{1}{n}\sum_{i=1}^{n}\varepsilon_{i}\boldsymbol{r}_{q}^{\prime}\left(X_{0,i}+\mathrm{\boldsymbol{X}}_{i}^{\mathrm{T}}\boldsymbol{\beta}\right)\mathbf{X}_{i}^{\mathrm{T}}\right\Vert \sup_{k\geq k_{1,n}^{SBGD}+1}\left\Vert \Delta\boldsymbol{\beta}_{k}\right\Vert ,
	\end{align*}
	Obviously, we have that $\sup_{\boldsymbol{\beta}\in\mathcal{B}}\left\Vert \frac{1}{n}\sum_{i=1}^{n}\varepsilon_{i}\boldsymbol{r}_{q}^{\prime}\left(X_{0,j}+\mathbf{X}_{i}^{\mathrm{T}}\boldsymbol{\beta}\right)\mathbf{X}_{i}^{\mathrm{T}}\right\Vert =O_{p}\left(\chi_{3,n}\right)$
	due to the fact that $\left|\varepsilon_{i}r_{s}^{\prime}\left(X_{0,i}+\mathbf{X}_{i}^{\mathrm{T}}\boldsymbol{\beta}\right)X_{t}\right|\leq CD_{q,1}$
	and $\left\Vert \partial\varepsilon_{i}r_{s}^{\prime}\left(X_{0,i}+\mathbf{X}_{i}^{\mathrm{T}}\boldsymbol{\beta}\right)X_{t}/\partial\boldsymbol{\beta}\right\Vert \leq C\sqrt{p}D_{q,2}$,
	so 
	\[
	\sup_{k\geq k_{1,n}^{SBGD}+1}\left\Vert \frac{1}{n}\sum_{i=1}^{n}\boldsymbol{r}_{q,j,k}\varepsilon_{j}-\frac{1}{n}\sum_{i=1}^{n}\boldsymbol{r}_{q,j}^{\star}\varepsilon_{j}\right\Vert =O_{p}\left(\chi_{2,n}\chi_{3,n}\right),
	\]
	which leads to the result if we further note that 
	\begin{align*}
		& \sup_{k\geq k_{1,n}^{SBGD}+1}\left\Vert \frac{1}{n}\sum_{i=1}^{n}\mathbf{X}_{i}\boldsymbol{r}_{q,i,k}^{\mathrm{T}}\Gamma_{q,n,k}^{-1}\left(\frac{1}{n}\sum_{j=1}^{n}\boldsymbol{r}_{q,j,k}R_{q,j,k}+\frac{1}{n}\sum_{i=1}^{n}\boldsymbol{r}_{q,j,k}\varepsilon_{j}-\frac{1}{n}\sum_{i=1}^{n}\boldsymbol{r}_{q,j}^{\star}\varepsilon_{j}\right)\right\Vert \\
		& =O_{p}\left(\sqrt{pq}D_{q,0}\sup_{k\geq k_{1,n}^{SBGD}+1}\left\Vert \frac{1}{n}\sum_{j=1}^{n}\boldsymbol{r}_{q,j,k}R_{q,j,k}+\frac{1}{n}\sum_{i=1}^{n}\boldsymbol{r}_{q,j,k}\varepsilon_{j}-\frac{1}{n}\sum_{i=1}^{n}\boldsymbol{r}_{q,j}^{\star}\varepsilon_{j}\right\Vert \right)\\
		& =O_{p}\left(\sqrt{p}qD_{q,0}^{2}\mathcal{E}_{q,0}+\sqrt{pq}D_{q,0}\chi_{2,n}\chi_{3,n}\right).
	\end{align*}

	Next we show that 
	\[\sup_{k\geq k_{1,n}^{SBGD}+1}\left\Vert \frac{1}{n}\sum_{i=1}^{n}\mathbf{X}_{i}\boldsymbol{r}_{q,i,k}^{\mathrm{T}}\Gamma_{q,n,k}^{-1} - \mathbb{E}\left(\mathbf{X}_{i}\boldsymbol{r}_{q}^{\mathrm{T}}\left(X_{0,i}+\mathbf{X}_{i}^{\mathrm{T}}\boldsymbol{\beta}^{\star}\right)\Gamma_{q}^{-1}\left(\boldsymbol{\beta}^{\star}\right)\right)\right\Vert = O_p\left(p\sqrt{q^3}D_{q,0}^{2}D_{q,1}\chi_{2,n}\right).\]
	
	\begin{align*}
		& \sup_{k\geq k_{1,n}^{SBGD}+1}\left\Vert \frac{1}{n}\sum_{i=1}^{n}\mathbf{X}_{i}\boldsymbol{r}_{q,i,k}^{\mathrm{T}}\Gamma_{q,n,k}^{-1}-\frac{1}{n}\sum_{i=1}^{n}\mathbf{X}_{i}\boldsymbol{r}_{q}^{\mathrm{T}}\left(X_{0,i}+\mathbf{X}_{i}^{\mathrm{T}}\boldsymbol{\beta}^{\star}\right)\Gamma_{q}^{-1}\left(\boldsymbol{\beta}^{\star}\right)\right\Vert \\
		& \leq\sup_{k\geq k_{1,n}^{SBGD}+1}\left\Vert \frac{1}{n}\sum_{i=1}^{n}\mathbf{X}_{i}\boldsymbol{r}_{q}^{\mathrm{T}}\left(X_{0,i}+\mathbf{X}_{i}^{\mathrm{T}}\boldsymbol{\beta}_{k}\right)\Gamma_{q,n,k}^{-1}-\frac{1}{n}\sum_{i=1}^{n}\mathbf{X}_{i}\boldsymbol{r}_{q}^{\mathrm{T}}\left(X_{0,i}+\mathbf{X}_{i}^{\mathrm{T}}\boldsymbol{\beta}^{\star}\right)\Gamma_{q,n,k}^{-1}\right\Vert \\
		& +\sup_{k\geq k_{1,n}^{SBGD}+1}\left\Vert \frac{1}{n}\sum_{i=1}^{n}\mathbf{X}_{i}\boldsymbol{r}_{q}^{\mathrm{T}}\left(X_{0,i}+\mathbf{X}_{i}^{\mathrm{T}}\boldsymbol{\beta}^{\star}\right)\Gamma_{q,n,k}^{-1}-\frac{1}{n}\sum_{i=1}^{n}\mathbf{X}_{i}\boldsymbol{r}_{q}^{\mathrm{T}}\left(X_{0,i}+\mathbf{X}_{i}^{\mathrm{T}}\boldsymbol{\beta}^{\star}\right)\Gamma_{q}^{-1}\left(\boldsymbol{\beta}^{\star}\right)\right\Vert .
	\end{align*}
	The first term is obviously bounded in probability by 
	\begin{align*}
		& C\sup_{k\geq k_{1,n}^{SBGD}+1}\left\Vert \frac{1}{n}\sum_{i=1}^{n}\mathbf{X}_{i}\left(\boldsymbol{r}_{q}\left(X_{0,i}+\mathbf{X}_{i}^{\mathrm{T}}\boldsymbol{\beta}_{k}\right)-\boldsymbol{r}_{q}\left(X_{0,i}+\mathbf{X}_{i}^{\mathrm{T}}\boldsymbol{\beta}^{\star}\right)\right)^{\mathrm{T}}\right\Vert \\
		& \leq Cp\sqrt{q}D_{q,1}\left\Vert \boldsymbol{\beta}_{k}-\boldsymbol{\beta}^{\star}\right\Vert =Cp\sqrt{q}D_{q,1}\chi_{2,n}.
	\end{align*}
	The second term is bounded by 
	\begin{align*}
		& \sup_{k\geq k_{1,n}^{*}}\left\Vert \frac{1}{n}\sum_{i=1}^{n}\mathbf{X}_{i}\boldsymbol{r}_{q}^{\mathrm{T}}\left(X_{0,i}+\mathbf{X}_{i}^{\mathrm{T}}\boldsymbol{\beta}^{\star}\right)\right\Vert \sup_{k\geq k_{1,n}^{*}}\left\Vert \Gamma_{q,n,k}^{-1}-\Gamma_{q}^{-1}\left(\boldsymbol{\beta}^{\star}\right)\right\Vert \\
		& \leq C\sqrt{pq}D_{q,0}\sup_{k\geq k_{1,n}^{SBGD}+1}\left\Vert \Gamma_{q,n,k}^{-1}-\Gamma_{q}^{-1}\left(\boldsymbol{\beta}^{\star}\right)\right\Vert .
	\end{align*}
	Now we provide an upper bound for $\sup_{k\geq k_{1,n}^{SBGD}+1}\left\Vert \Gamma_{q,n,k}^{-1}-\Gamma_{q}^{-1}\left(\boldsymbol{\beta}^{\star}\right)\right\Vert $.
	Note that 
	\begin{align*}
		\sup_{k\geq k_{1,n}^{SBGD}+1}\left\Vert \Gamma_{q,n,k}^{-1}-\Gamma_{q}^{-1}\left(\boldsymbol{\beta}^{\star}\right)\right\Vert  & =O_{p}\left(\sup_{k\geq k_{1,n}^{SBGD}+1}\left\Vert \Gamma_{q,n,k}-\Gamma_{q}\left(\boldsymbol{\beta}^{\star}\right)\right\Vert \right)\\
		& =O_{p}\left(\sup_{k\geq k_{1,n}^{SBGD}+1}\left\Vert \Gamma_{q,n,k}-\Gamma_{q,n}\left(\boldsymbol{\beta}^{\star}\right)\right\Vert +\left\Vert \Gamma_{q,n}\left(\boldsymbol{\beta}^{\star}\right)-\Gamma_{q}\left(\boldsymbol{\beta}^{\star}\right)\right\Vert \right)\\
		& =O_{p}\left(\sqrt{p}qD_{q,0}D_{q,1}\chi_{2,n}+\chi_{1,n}\right)=O_{p}\left(\sqrt{p}qD_{q,0}D_{q,1}\chi_{2,n}\right).
	\end{align*}
	So 
	\begin{align*}
		& \sup_{k\geq k_{1,n}^{*}}\left\Vert \frac{1}{n}\sum_{i=1}^{n}\mathbf{X}_{i}\boldsymbol{r}_{q}^{\mathrm{T}}\left(X_{0,i}+\mathbf{X}_{i}^{\mathrm{T}}\boldsymbol{\beta}^{\star}\right)\Gamma_{q,n,k}^{-1}-\frac{1}{n}\sum_{i=1}^{n}\mathbf{X}_{i}\boldsymbol{r}_{q}^{\mathrm{T}}\left(X_{0,i}+\mathbf{X}_{i}^{\mathrm{T}}\boldsymbol{\beta}^{\star}\right)\Gamma_{q}^{-1}\left(\boldsymbol{\beta}^{\star}\right)\right\Vert \\
		& =O_{p}\left(p\sqrt{q^3}D_{q,0}^{2}D_{q,1}\chi_{2,n}\right),
	\end{align*}
	and together 
	\[
	\sup_{k\geq k_{1,n}^{SBGD}+1}\left\Vert \frac{1}{n}\sum_{i=1}^{n}\mathbf{X}_{i}\boldsymbol{r}_{q,i,k}^{\mathrm{T}}\Gamma_{q,n,k}^{-1}-\frac{1}{n}\sum_{i=1}^{n}\mathbf{X}_{i}\boldsymbol{r}_{q}^{\mathrm{T}}\left(X_{0,i}+\mathbf{X}_{i}^{\mathrm{T}}\boldsymbol{\beta}^{\star}\right)\Gamma_{q}^{-1}\left(\boldsymbol{\beta}^{\star}\right)\right\Vert =O_{p}\left(p\sqrt{q^3}D_{q,0}^{2}D_{q,1}\chi_{2,n}\right).
	\]
Moreover, note that $\left\Vert\frac{1}{n}\sum_{i=1}^{n}\mathbf{X}_{i}\boldsymbol{r}_{q}^{\mathrm{T}}\left(X_{0,i}+\mathbf{X}_{i}^{\mathrm{T}}\boldsymbol{\beta}^{\star}\right)\Gamma_{q}^{-1}\left(\boldsymbol{\beta}^{\star}\right) - \mathbb{E}\left(\mathbf{X}_{i}\boldsymbol{r}_{q}^{\mathrm{T}}\left(X_{0,i}+\mathbf{X}_{i}^{\mathrm{T}}\boldsymbol{\beta}^{\star}\right)\Gamma_{q}^{-1}\left(\boldsymbol{\beta}^{\star}\right)\right)\right\Vert = O_p\left(\sqrt{p^3D_{q,0}^2\log\left(pn\right)/n}\right)$, so we have shown the results.
	
	Based on the above results, we have that 
	\begin{align*}
		& \sup_{k\geq k_{1,n}^{SBGD}+1}\left\Vert \frac{1}{n}\sum_{i=1}^{n}\mathbf{X}_{i}\boldsymbol{r}_{q,i,k}^{\mathrm{T}}\Gamma_{q,n,k}^{-1}\left(\frac{1}{n}\sum_{j=1}^{n}\boldsymbol{r}_{q,j,k}R_{q,j,k}+\frac{1}{n}\sum_{i=1}^{n}\boldsymbol{r}_{q,j,k}\varepsilon_{j}\right)+\frac{1}{n}\sum_{i=1}^{n}R_{q}\left(z_{i,k}\right)\mathbf{X}_{i}-\frac{1}{n}\sum_{i=1}^{n}\mathfrak{X}\left(z_i^{\star},\boldsymbol{\beta}^{\star}\right)\varepsilon_{j}\right\Vert \\
		& \leq\sup_{k\geq k_{1,n}^{SBGD}+1}\left\Vert \frac{1}{n}\sum_{i=1}^{n}\mathbf{X}_{i}\boldsymbol{r}_{q,i,k}^{\mathrm{T}}\Gamma_{q,n,k}^{-1}\left(\frac{1}{n}\sum_{j=1}^{n}\boldsymbol{r}_{q,j,k}R_{q,j,k}+\frac{1}{n}\sum_{i=1}^{n}\boldsymbol{r}_{q,j,k}\varepsilon_{j}-\frac{1}{n}\sum_{i=1}^{n}\boldsymbol{r}_{q,j}^{\star}\varepsilon_{j}\right)\right\Vert \\
		& +\sup_{k\geq k_{1,n}^{SBGD}+1}\left\Vert \left(\frac{1}{n}\sum_{i=1}^{n}\mathbf{X}_{i}\boldsymbol{r}_{q,i,k}^{\mathrm{T}}\Gamma_{q,n,k}^{-1}-\mathbb{E}\left(\mathbf{X}_{i}\boldsymbol{r}_{q}^{\mathrm{T}}\left(X_{0,i}+\mathbf{X}_{i}^{\mathrm{T}}\boldsymbol{\beta}^{\star}\right)\Gamma_{q}^{-1}\left(\boldsymbol{\beta}^{\star}\right)\right)\right)\left(\frac{1}{n}\sum_{i=1}^{n}\boldsymbol{r}_{q,j}^{\star}\varepsilon_{j}\right)\right\Vert \\
		&  +\sup_{k\geq k_{1,n}^{SBGD}+1}\left\Vert \frac{1}{n}\sum_{i=1}^{n}R_{q}\left(z_{i,k}\right)\mathbf{X}_{i}\right\Vert \\
		& =O_{p}\left(\sqrt{p}qD_{q,0}^{2}\mathcal{E}_{q,0}+\sqrt{pq}D_{q,0}\chi_{2,n}\chi_{3,n}+p\sqrt{q^3}D_{q,0}^{2}D_{q,1}\chi_{2,n}\sqrt{(qD_{q,0}^{2}\log q)/n}\right)
	\end{align*}
\end{proof}

\newpage{}
\section{Proofs of Theorems}\label{appendixB}

\subsection*{Proof of  \autoref{prop:Known_G}}
\begin{proof}
We first prove \autoref{prop:Known_G}(i). Recall that $\Delta\boldsymbol{\beta}_{e,k}=\boldsymbol{\beta}_{e,k}-\boldsymbol{\beta}_{e}^{\star}$
and $\varepsilon_{i}=y_{i}-G\left(\boldsymbol{\mathbf{X}}_{e,i}^{\mathrm{T}}\boldsymbol{\beta}_{e}^{\star}\right)$.
We have that 
\[
\Delta\boldsymbol{\beta}_{e,k+1}=\Delta\boldsymbol{\beta}_{e,k}-\frac{\delta}{n}\sum_{i=1}^{n}\left(G\left(\boldsymbol{\mathbf{X}}_{e,i}^{\mathrm{T}}\boldsymbol{\beta}_{e,k}\right)-G\left(\boldsymbol{\mathbf{X}}_{e,i}^{\mathrm{T}}\boldsymbol{\beta}_{e}^{\star}\right)-\varepsilon_{i}\right)\boldsymbol{\mathbf{X}}_{e,i},
\]
so 
\[
\left\Vert \Delta\boldsymbol{\beta}_{e,k+1}\right\Vert \leq\left\Vert \Delta\boldsymbol{\beta}_{e,k}-\frac{\delta}{n}\sum_{i=1}^{n}\left(G\left(\boldsymbol{\mathbf{X}}_{e,i}^{\mathrm{T}}\boldsymbol{\beta}_{e,k}\right)-G\left(\boldsymbol{\mathbf{X}}_{e,i}^{\mathrm{T}}\boldsymbol{\beta}_{e}^{\star}\right)\right)\mathbf{X}_{e,i}\right\Vert +\left\Vert \frac{\delta}{n}\sum_{i=1}^{n}\varepsilon_{i}\mathbf{X}_{e,i}\right\Vert .
\]
Note that mean value theorem leads to 
\begin{align*}
 & \Delta\boldsymbol{\beta}_{e,k}-\frac{\delta}{n}\sum_{i=1}^{n}\left(G\left(\boldsymbol{\mathbf{X}}_{e,i}^{\mathrm{T}}\boldsymbol{\beta}_{e,k}\right)-G\left(\boldsymbol{\mathbf{X}}_{e,i}^{\mathrm{T}}\boldsymbol{\beta}_{e}^{\star}\right)\right)\boldsymbol{\mathbf{X}}_{e,i}\\
 & =\Delta\boldsymbol{\beta}_{e,k}-\int_{0}^{1}\left\{ \frac{\delta}{n}\sum_{i=1}^{n}G^{\prime}\left(\boldsymbol{\mathbf{X}}_{e,i}^{\mathrm{T}}\boldsymbol{\beta}_{e}^{\star}+t\boldsymbol{\mathbf{X}}_{e,i}^{\mathrm{T}}\Delta\boldsymbol{\beta}_{e,k}\right)\boldsymbol{\mathbf{X}}_{e,i}\boldsymbol{\mathbf{X}}_{e,i}^{\mathrm{T}}\Delta\boldsymbol{\beta}_{e,k}\right\} dt\\
 & =\int_{0}^{1}\left\{ \left(I_{p+1}-\delta M_{n}\left(\boldsymbol{\beta}_{e}^{\star}+t\Delta\boldsymbol{\beta}_{e,k}\right)\right)\Delta\boldsymbol{\beta}_{e,k}\right\} dt,
\end{align*}
where the integration is understood to be element-wise, and $\boldsymbol{\beta}_{e}^{\star}+t\Delta\boldsymbol{\beta}_{e,k}\in\mathcal{B}_{e}$
due to convexity of $\mathcal{B}_{e}$. 

We next provide a uniform
upper bound for $\overline{\lambda}\left(I_{p+1}-\delta M_{n}\left(\boldsymbol{\beta}_{e}\right)\right)$
and lower bound for $\underline{\lambda}\left(I_{p+1}-\delta M_{n} \left(\boldsymbol{\beta}_{e}\right)\right)$
with respect to $\boldsymbol{\beta}_{e}\in\mathcal{B}_{e}$ in probability.
Since \autoref{assump:2} holds, we have that $G\left(\mathbf{X}^{\mathrm{T}}_{e,i}\boldsymbol{\beta}\right)X_{i,t}X_{i,s}$ is bounded by $\Vert G\Vert_{\infty}$ and $\Vert\partial G\left(\mathbf{X}^{\mathrm{T}}_{e,i}\boldsymbol{\beta}\right)X_{i,t}X_{i,s}/\partial{\boldsymbol{\beta}}\Vert\leq C\sqrt{p}$. Then according to \autoref{lemS.1}, we have that 
\[
\sup_{\boldsymbol{\beta}_{e}\in\mathcal{B}}\left\Vert M_{n}\left(\boldsymbol{\beta}_{e}\right)-M\left(\boldsymbol{\beta}_{e}\right)\right\Vert =O_{p}\left(\sqrt{\frac{p^{3}\log n}{n}}\right).
\]
Since $p^5(\log p)^2n^{-1}\rightarrow0$ holds, $\sqrt{p^{3}\left(\log n\right)/n}\rightarrow0$
holds, so 
\begin{align*}
\sup_{\boldsymbol{\beta}_{e}\in\mathcal{B}}\left|\overline{\lambda}\left(M_{n}\left(\boldsymbol{\beta}_{e}\right)\right)-\overline{\lambda}\left(M\left(\boldsymbol{\beta}_{e}\right)\right)\right| & =o_{p}\left(1\right),
\end{align*}
and 
\begin{align*}
\sup_{\boldsymbol{\beta}_{e}\in\mathcal{B}}\left|\underline{\lambda}\left(M_{n}\left(\boldsymbol{\beta}_{e}\right)\right)-\underline{\lambda}\left(M\left(\boldsymbol{\beta}_{e}\right)\right)\right| & =o_{p}\left(1\right).
\end{align*}
Due to \autoref{assump:2}(iv), with probability going to 1,
there holds, 
\[
\underline{\lambda}_{e}/2\leq\inf_{\boldsymbol{\beta}_{e}\in\mathcal{B}}\underline{\lambda}\left(M_{n}\left(\boldsymbol{\beta}_{e}\right)\right)\leq\sup_{\boldsymbol{\beta}_{e}\in\mathcal{B}}\overline{\lambda}\left(M_{n}\left(\boldsymbol{\beta}_{e}\right)\right)\leq3\overline{\lambda}_{e}/2.
\]
Since $\delta < 2/(3\overline{\lambda}_e)$, we have that with probability going to 1, there holds
\[
0\leq\inf_{\boldsymbol{\beta}_{e}\in\mathcal{B}}\overline{\lambda}\left(I_{p+1}-\delta M_{n}\left(\boldsymbol{\beta}_{e}\right)\right)\leq\sup_{\boldsymbol{\beta}_{e}\in\mathcal{B}}\overline{\lambda}\left(I_{p+1}-\delta M_{n} \left(\boldsymbol{\beta}_{e}\right)\right)\leq1-\underline{\lambda}_{e}\delta/2.
\]
Based on the above inequality, we have that with probability going
to 1, there holds
\begin{align*}
 & \left\Vert \int_{0}^{1}\left\{ \left(I_{p+1}-\delta M_{n}\left(\boldsymbol{\beta}_{e}^{\star}+t\Delta\boldsymbol{\beta}_{e,k}\right)\right)\Delta\boldsymbol{\beta}_{e,k}\right\} dt\right\Vert \\
 & \leq\int_{0}^{1}\left\{ \sup_{\boldsymbol{\beta}_{e}\in\mathcal{B}}\overline{\lambda}\left(I_{p+1}-\delta M_{n} \left(\boldsymbol{\beta}_{e}\right)\right)\right\} dt\cdot\left\Vert \Delta\boldsymbol{\beta}_{e,k}\right\Vert \leq\left(1-\underline{\lambda}_{e}\delta/2\right)\cdot\left\Vert \Delta\boldsymbol{\beta}_{e,k}\right\Vert .
\end{align*}
So with probability going to 1, for all $k$ there holds
\begin{align*}
\left\Vert \Delta\boldsymbol{\beta}_{e,k+1}\right\Vert  & \leq\left(1-\underline{\lambda}_{e}\delta/2\right)\left\Vert \Delta\boldsymbol{\beta}_{e,k}\right\Vert +\delta\left\Vert \frac{1}{n}\sum_{i=1}^{n}\varepsilon_{i}\mathbf{X}_{e,i}\right\Vert \\
 & \leq\cdots\leq\left(1-\underline{\lambda}_{e}\delta/2\right)^{k}\left\Vert \Delta\boldsymbol{\beta}_{e,1}\right\Vert +\delta\sum_{j=1}^{k}\left(1-\underline{\lambda}_{e}\delta/2\right)^{j-1}\left\Vert \frac{1}{n}\sum_{i=1}^{n}\varepsilon_{i}\mathbf{X}_{e,i}\right\Vert \\
 & \leq\left(1-\underline{\lambda}_{e}\delta/2\right)^{k}\left\Vert \Delta\boldsymbol{\beta}_{e,1}\right\Vert +2\underline{\lambda}_{e}^{-1}\left\Vert \frac{1}{n}\sum_{i=1}^{n}\varepsilon_{i}\mathbf{X}_{e,i}\right\Vert .
\end{align*}
Note that for any $\tau>0$, 
\begin{align*}
P\left(\left\Vert \frac{1}{n}\sum_{i=1}^{n}\varepsilon_{i}\mathbf{X}_{e,i}\right\Vert >\tau\right) & \leq\sum_{j=0}^{p}P\left(\left|\frac{1}{n}\sum_{i=1}^{n}\varepsilon_{i}X_{e,j,i}\right|>\frac{\tau}{\sqrt{p+1}}\right)\\
 & \leq\sum_{j=0}^{p}2\exp\left(Cn\tau^{2}/p\right)=2\exp\left(C_{1}\log p-C_{2}n\tau^{2}/p\right),
\end{align*}
so 
\[
\left\Vert \frac{1}{n}\sum_{i=1}^{n}\varepsilon_{i}\mathbf{X}_{e,i}\right\Vert =O_{p}\left(\sqrt{p\left(\log p\right)/n}\right).
\]
Then for $k$ such that 
\[
\left(1-\underline{\lambda}_{e}\delta/2\right)^{k}\left\Vert \Delta\boldsymbol{\beta}_{e,1}\right\Vert \leq\sqrt{p\left(\log p\right)/n},
\]
or equivalently, 
\[
k\geq k_{1,n}^{BGD}=\frac{\log\left\Vert \Delta\boldsymbol{\beta}_{e,1}\right\Vert +\frac{1}{2}\log\left(n/\left(p\log p\right)\right)}{-\log\left(1-\underline{\lambda}_{e}\delta/2\right)},
\]
we have that 
\[
\left\Vert \Delta\boldsymbol{\beta}_{e,k+1}\right\Vert =O_{p}\left(\sqrt{p\left(\log p\right)/n}\right).
\]
This proves \autoref{prop:Known_G}(i).

Next we prove \autoref{prop:Known_G}(ii). For any $k\geq k_{1,n}^{BGD}+1$, there holds
\begin{align*}
\Delta\boldsymbol{\beta}_{e,k+1} & =\Delta\boldsymbol{\beta}_{e,k}-\frac{\delta}{n}\sum_{i=1}^{n}\left(G\left(\mathbf{X}_{e,i}^{\mathrm{T}}\boldsymbol{\beta}_{e,k}\right)-G\left(\boldsymbol{X}_{e,i}^{\mathrm{T}}\boldsymbol{\beta}_{e}^{\star}\right)-\varepsilon_{i}\right)\mathbf{X}_{e,i},\\
 & =\left(I_{p+1}-\delta M_{n}\left(\boldsymbol{\overline{\beta}}_{e,k}\right)\right)\Delta\boldsymbol{\beta}_{e,k}+\frac{\delta}{n}\sum_{i=1}^{n}\varepsilon_{i}\mathbf{X}_{e,i},
\end{align*}
where $\boldsymbol{\overline{\beta}}_{e,k}$ is element-wise and lies
between $\boldsymbol{\beta}_{e,k}$ and $\boldsymbol{\beta}_{e}^{\star}$.
Since $\left\Vert \Delta\boldsymbol{\beta}_{e,k}\right\Vert =O_{p}\left(\sqrt{p\left(\log p\right)/n}\right)$
for $k\geq k_{1,n}^{BGD}+1$, $\left\Vert \Delta\boldsymbol{\overline{\beta}}_{e,k}\right\Vert =O_{p}\left(\sqrt{p\left(\log p\right)/n}\right)$
also holds. Note that
\[
\left\Vert M_{n}\left(\overline{\boldsymbol{\beta}}_{e,k}\right)-M\left(\boldsymbol{\beta}_{e}^{\star}\right)\right\Vert \leq\left\Vert M_{n}\left(\boldsymbol{\overline{\beta}}_{e,k}\right)-M_{n}\left(\boldsymbol{\beta}_{e}^{\star}\right)\right\Vert +\left\Vert M_{n}\left(\boldsymbol{\beta}_{e}^{\star}\right)-M\left(\boldsymbol{\beta}_{e}^{\star}\right)\right\Vert .
\]
For the second term, $\left\Vert M_{n}\left(\boldsymbol{\beta}_{e}^{\star}\right)-M\left(\boldsymbol{\beta}_{e}^{\star}\right)\right\Vert =O_{p}\left(\sqrt{p^{2}\left(\log p\right)/n}\right)$
obviously holds. For the first term, since $G$ is twice differentiable
with bounded derivatives, we have that 
\begin{align*}
 \sup_{k\geq k_{1,n}^{BGD}+1}\left\Vert M_{n}\left(\overline{\boldsymbol{\beta}}_{e,k}\right)-M_{n}\left(\boldsymbol{\beta}_{e}^{\star}\right)\right\Vert
 & \leq\sup_{k\geq k_{1,n}^{BGD}+1}\frac{1}{n}\sum_{i=1}^{n}\left\Vert \mathbf{X}_{e,i}\mathbf{X}_{e,i}^{\mathrm{T}}\right\Vert \left|G^{\prime\prime}\left(\mathbf{X}_{e,i}^{\mathrm{T}}\check{\boldsymbol{\beta}}_{e,k}\right)\right|\left|\mathbf{X}_{e,i}^{\mathrm{T}}\Delta\boldsymbol{\overline{\beta}}_{e,k}\right|.\\
 & \leq C\sqrt{p^3}\sup_{k\geq k_{1,n}^{BGD}+1}\left\Vert \boldsymbol{\overline{\beta}}_{e,k}-\boldsymbol{\beta}_{e}^{\star}\right\Vert =O_{p}\left(\sqrt{p^{4}\left(\log p\right)/n}\right),
\end{align*}
where $\check{\boldsymbol{\beta}}_{e,k}$ lies somewhere between $\overline{\boldsymbol{\beta}}_{e,k}$
and $\boldsymbol{\beta}_{e}^{\star}$ and is also element-wise, and
the second last inequality comes from the fact that $\left\Vert \mathbf{X}_{e,i}\mathbf{X}_{e,i}^{\mathrm{T}}\right\Vert \leq p$
and $\left|\mathbf{X}_{e,i}^{\mathrm{T}}\Delta\boldsymbol{\overline{\beta}}_{e,k}\right|\leq\left\Vert \mathbf{X}_{e,i}\right\Vert \left\Vert \Delta\boldsymbol{\overline{\beta}}_{e,k}\right\Vert $.
This implies that 
\[
\sup_{k\geq k_{1,n}+1}\left\Vert M_{n}\left(\overline{\boldsymbol{\beta}}_{e,k}\right)-M\left(\boldsymbol{\beta}_{e}^{\star}\right)\right\Vert =O_{p}\left(\sqrt{p^{4}\left(\log p\right)/n}\right).
\]
Define $\omega_{k}=\left(M_{n}\left(\overline{\boldsymbol{\beta}}_{e,k}\right)-M\left(\boldsymbol{\beta}_{e}^{\star}\right)\right)\Delta\boldsymbol{\beta}_{e,k}$.
Obviously, there holds 
\begin{align*}
\sup_{k\geq k_{1,n}^{BGD}+1}\left\Vert \omega_{k}\right\Vert  & \leq\left(\sup_{k\geq k_{1,n}^{BGD}+1}\left\Vert M_{n}\left(\overline{\boldsymbol{\beta}}_{e,k}\right)-M\left(\boldsymbol{\beta}_{e}^{\star}\right)\right\Vert \right)\left(\sup_{k\geq k_{1,n}^{BGD}+1}\left\Vert \Delta\boldsymbol{\beta}_{e,k}\right\Vert \right)\\
 & =O_{p}\left(\sqrt{p^{5}\left(\log p\right)^{2}/n^{2}}\right),
\end{align*}
which is $o_{p}\left(n^{-1/2}\right)$ according to Assumption 2. 

Based on the above result, we have that for any $k\geq1$, 
\begin{align*}
\Delta\boldsymbol{\beta}_{e,k+k_{1,n}^{BGD}+1} & =\left(I_{p+1}-\delta M_{n}\left(\boldsymbol{\overline{\beta}}_{e,k+k_{1,n}^{BGD}}\right)\right)\Delta\boldsymbol{\beta}_{e,k+k_{1,n}^{BGD}}-\frac{\delta}{n}\sum_{i=1}^{n}\varepsilon_{i}\mathbf{X}_{e,i}\\
 & =\left(I_{p+1}-\delta M\left(\boldsymbol{\beta}_{e}^{\star}\right)\right)\Delta\boldsymbol{\beta}_{e,k+k_{1,n}^{BGD}}-\delta\omega_{k+k_{1,n}^{BGD}}-\frac{\delta}{n}\sum_{i=1}^{n}\varepsilon_{i}\mathbf{X}_{e,i}\\
 & =\left(I_{p+1}-\delta M\left(\boldsymbol{\beta}_{e}^{\star}\right)\right)^{k}\Delta\boldsymbol{\beta}_{e,k_{1,n}^{BGD}+1}-\delta\sum_{j=0}^{k-1}\left(I_{p+1}-\delta M\left(\boldsymbol{\beta}_{e}^{\star}\right)\right)^{j}\omega_{k+k_{1,n}^{BGD}-j}\\
 & -\delta\left(\sum_{j=0}^{k-1}\left(I_{p+1}-\delta M\left(\boldsymbol{\beta}_{e}^{\star}\right)\right)^{j}\right)\left(\frac{1}{n}\sum_{i=1}^{n}\varepsilon_{i}\mathbf{X}_{e,i}\right).
\end{align*}
For the first part on the RHS of the last equality, we have that 
\begin{align*}
\left\Vert \left(I_{p+1}-\delta M\left(\boldsymbol{\beta}_{e}^{\star}\right)\right)^{k}\Delta\boldsymbol{\beta}_{e,k_{1,n}^{BGD}+1}\right\Vert  & \leq\left(1-\underline{\lambda}_{e}\delta\right)^{k}\left\Vert \Delta\boldsymbol{\beta}_{e,k_{1,n}^{BGD}+1}\right\Vert \\
 & =\left(1-\underline{\lambda}_{e}\delta\right)^{k}O_{p}\left(\sqrt{p\left(\log p\right)/n}\right).
\end{align*}
For the second part, we have that 
\begin{align*}
\left\Vert \delta\sum_{j=0}^{k-1}\left(I_{p+1}-\delta M\left(\boldsymbol{\beta}_{e}^{\star}\right)\right)^{j}\omega_{k+k_{1,n}^{BGD}-j}\right\Vert  & \leq\delta\sum_{j=0}^{\infty}\left(1-\underline{\lambda}_{e}\delta\right)^{j}\left\Vert \omega_{k+k_{1,n}^{BGD}-j}\right\Vert \\
 & \leq\underline{\lambda}_{e}^{-1}\sup_{k\geq1}\left\Vert \omega_{k+k_{1,n}^{BGD}}\right\Vert =O_{p}\left(\sqrt{p^{5}\left(\log p\right)^{2}/n^{2}}\right)\\
 & =o_{p}\left(n^{-1/2}\right).
\end{align*}
For the third part, we have that 
\begin{align*}
 & \left\Vert \left(\sum_{j=0}^{k-1}\delta\left(I_{p+1}-\delta M\left(\boldsymbol{\beta}_{e}^{\star}\right)\right)^{j}\right)\left(\frac{1}{n}\sum_{i=1}^{n}\varepsilon_{i}\mathbf{X}_{e,i}\right)-M_{n}^{-1}\left(\boldsymbol{\beta}_{e}^{\star}\right)\left(\frac{1}{n}\sum_{i=1}^{n}\varepsilon_{i}\mathbf{X}_{e,i}\right)\right\Vert \\
 & \leq\sum_{j=k}^{\infty}\delta\left(1-\underline{\lambda}_{e}\delta\right)^{j}\left\Vert \frac{1}{n}\sum_{i=1}^{n}\varepsilon_{i}\mathbf{X}_{e,i}\right\Vert =\left(1-\underline{\lambda}_{e}\delta\right)^{k}\left\Vert \frac{1}{n}\sum_{i=1}^{n}\varepsilon_{i}\mathbf{X}_{e,i}\right\Vert \\
 & =\left(1-\underline{\lambda}_{e}\delta\right)^{k}O_{p}\left(\sqrt{p\left(\log p\right)/n}\right).
\end{align*}
This implies that when $\left(1-\underline{\lambda}_e\delta\right)^{k_{2,n}^{BGD}}\sqrt{p\log p}\rightarrow 0$, we have that 
\[
\sup_{k\geq k_{2,n}^{BGD}+1}\left\Vert \sqrt{n}\Delta\boldsymbol{\beta}_{e,k+k_{1,n}^{BGD}}-M^{-1}\left(\boldsymbol{\beta}_{e}^{\star}\right)\frac{1}{\sqrt{n}}\sum_{i=1}^{n}\varepsilon_{i}\mathbf{X}_{e,i}\right\Vert =o_{p}\left(1\right).
\]
 This proves \autoref{prop:Known_G}(ii)

Now we prove \autoref{prop:Known_G}(iii). We first note that for any square matrices $A$, $B$, and $C$, there hold $\left\Vert AB\right\Vert \leq\overline{\sigma}\left(A\right)\left\Vert B\right\Vert $ and  $\left\Vert ABC\right\Vert \leq\overline{\sigma}\left(A\right)\left\Vert BC\right\Vert \leq\overline{\sigma}\left(A\right)\overline{\sigma}\left(B\right)\left\Vert C\right\Vert $. 
So 
\begin{align*}
\left\Vert M^{-1}\left(\boldsymbol{\beta}_{e}^{\star}\right)-M_{n}^{-1}\left(\widehat{\boldsymbol{\beta}}_{e}\right)\right\Vert  & =\left\Vert M^{-1}\left(\boldsymbol{\beta}_{e}^{\star}\right)\left(M_{n}\left(\widehat{\boldsymbol{\beta}}_{e}\right)-M\left(\boldsymbol{\beta}_{e}^{\star}\right)\right)M_{n}^{-1}\left(\widehat{\boldsymbol{\beta}}_{e}\right)\right\Vert \\
 & \leq\overline{\sigma}\left(M^{-1}\left(\boldsymbol{\beta}_{e}^{\star}\right)\right)\cdot\overline{\sigma}\left(M_{n}^{-1}\left(\widehat{\boldsymbol{\beta}}_{e}\right)\right)\cdot\left\Vert M_{n}\left(\widehat{\boldsymbol{\beta}}_{e}\right)-M\left(\boldsymbol{\beta}_{e}^{\star}\right)\right\Vert ,
\end{align*}
due to the fact that $M_{n}^{-1}\left(\widehat{\boldsymbol{\beta}}_{e}\right)$
and $M_{n}\left(\widehat{\boldsymbol{\beta}}_{e}\right)-M\left(\boldsymbol{\beta}_{e}^{\star}\right)$
are both symmetric. Due to \autoref{assump:2}(iv), we have that $\overline{\sigma}\left(M^{-1}\left(\boldsymbol{\beta}_{e}^{\star}\right)\right)=\overline{\lambda}\left(M^{-1}\left(\boldsymbol{\beta}_{e}^{\star}\right)\right)\leq\underline{\lambda}_{e}^{-1}$.
Since $\left\Vert M_n\left(\widehat{\boldsymbol{\beta}}_{e}\right)-M\left(\boldsymbol{\beta}_{e}^{\star}\right)\right\Vert =o_{p}\left(1\right)$
holds according to the previous proof, we have that with
probability going to 1, $\overline{\sigma}\left(M_n^{-1}\left(\widehat{\boldsymbol{\beta}}_{e}\right)\right)=\overline{\lambda}\left(M_n^{-1}\left(\widehat{\boldsymbol{\beta}}_{e}\right)\right)\leq2\underline{\lambda}_{e}^{-1}$.
Then with probability going to 1, we have that 
\[
\left\Vert M^{-1}\left(\boldsymbol{\beta}_{e}^{\star}\right)-M_n^{-1}\left(\widehat{\boldsymbol{\beta}}_{e}\right)\right\Vert \leq2\underline{\lambda}_{e}^{-2}\left\Vert M_n\left(\widehat{\boldsymbol{\beta}}_{e}\right)-M\left(\boldsymbol{\beta}_{e}^{\star}\right)\right\Vert =O_{p}\left(\sqrt{p^{4}\left(\log p\right)/n}\right).
\]
On the other side, we have that 
\begin{align*}
 & \left\Vert \frac{1}{n}\sum_{i=1}^{n}\widehat{G}_{i}\left(1-\widehat{G}_{i}\right)\boldsymbol{\mathbf{X}}_{e,i}\boldsymbol{\mathbf{X}}_{e,i}^{\mathrm{T}}-\mathbb{E}\left[G_{i}^{\star}\left(1-G_{i}^{\star}\right)\boldsymbol{\mathbf{X}}_{e,i}\boldsymbol{\mathbf{X}}_{e,i}^{\mathrm{T}}\right]\right\Vert \\
 & \leq\left\Vert \frac{1}{n}\sum_{i=1}^{n}\widehat{G}_{i}\left(1-\widehat{G}_{i}\right)\boldsymbol{\mathbf{X}}_{e,i}\boldsymbol{\mathbf{X}}_{e,i}^{\mathrm{T}}-\frac{1}{n}\sum_{i=1}^{n}G_{i}^{\star}\left(1-G_{i}^{\star}\right)\boldsymbol{\mathbf{X}}_{e,i}\boldsymbol{\mathbf{X}}_{e,i}^{\mathrm{T}}\right\Vert \\
 & +\left\Vert \frac{1}{n}\sum_{i=1}^{n}G_{i}^{\star}\left(1-G_{i}^{\star}\right)\boldsymbol{\mathbf{X}}_{e,i}\boldsymbol{\mathbf{X}}_{e,i}^{\mathrm{T}}-\mathbb{E}\left[G_{i}^{\star}\left(1-G_{i}^{\star}\right)\boldsymbol{\mathbf{X}}_{e,i}\boldsymbol{\mathbf{X}}_{e,i}^{\mathrm{T}}\right]\right\Vert \\
 & \leq C\sqrt{p^3}\left\Vert \widehat{\boldsymbol{\beta}}_{e}-\boldsymbol{\beta}_{e}^{\star}\right\Vert +O_{p}\left(\sqrt{p^{2}\left(\log p\right)/n}\right)=O_{p}\left(\sqrt{p^{4}(\log p)/n}\right).
\end{align*}
Together, we have that 
\begin{align*}
\left\Vert \widehat{\Sigma}_{1}-\Sigma_{1}^{\star}\right\Vert  & \leq\left\Vert M^{-1}\left(\boldsymbol{\beta}_{e}^{\star}\right)\mathbb{E}\left[G_{i}^{\star}\left(1-G_{i}^{\star}\right)\boldsymbol{\mathbf{X}}_{e,i}\boldsymbol{\mathbf{X}}_{e,i}^{\mathrm{T}}\right]\left(M^{-1}\left(\boldsymbol{\beta}_{e}^{\star}\right)-M_n^{-1}\left(\widehat{\boldsymbol{\beta}}_{e}\right)\right)\right\Vert \\
 & +\left\Vert M^{-1}\left(\boldsymbol{\beta}_{e}^{\star}\right)\left(\frac{1}{n}\sum_{i=1}^{n}\widehat{G}_{i}\left(1-\widehat{G}_{i}\right)\boldsymbol{\mathbf{X}}_{e,i}\boldsymbol{\mathbf{X}}_{e,i}^{\mathrm{T}}-\mathbb{E}\left[G_{i}^{\star}\left(1-G_{i}^{\star}\right)\boldsymbol{\mathbf{X}}_{e,i}\boldsymbol{\mathbf{X}}_{e,i}^{\mathrm{T}}\right]\right)M_n^{-1}\left(\widehat{\boldsymbol{\beta}}_{e}\right)\right\Vert \\
 & +\left\Vert \left(M^{-1}\left(\boldsymbol{\beta}_{e}^{\star}\right)-M_n^{-1}\left(\widehat{\boldsymbol{\beta}}_{e}\right)\right)\left(\frac{1}{n}\sum_{i=1}^{n}\widehat{G}_{i}\left(1-\widehat{G}_{i}\right)\boldsymbol{\mathbf{X}}_{e,i}\boldsymbol{\mathbf{X}}_{e,i}^{\mathrm{T}}\right)M_n^{-1}\left(\widehat{\boldsymbol{\beta}}_{e}\right)\right\Vert \\
 & \leq\overline{\lambda}\left(M^{-1}\left(\boldsymbol{\beta}_{e}^{\star}\right)\right)\overline{\lambda}\left(\mathbb{E}\left[G_{i}^{\star}\left(1-G_{i}^{\star}\right)\boldsymbol{\mathbf{X}}_{e,i}\boldsymbol{\mathbf{X}}_{e,i}^{\mathrm{T}}\right]\right)\left\Vert M^{-1}\left(\boldsymbol{\beta}_{e}^{\star}\right)-M_n^{-1}\left(\widehat{\boldsymbol{\beta}}_{e}\right)\right\Vert \\
 & +\overline{\lambda}\left(M^{-1}\left(\boldsymbol{\beta}_{e}^{\star}\right)\right)\overline{\lambda}\left(M_n^{-1}\left(\widehat{\boldsymbol{\beta}}_{e}\right)\right)\left\Vert \frac{1}{n}\sum_{i=1}^{n}\widehat{G}_{i}\left(1-\widehat{G}_{i}\right)\boldsymbol{\mathbf{X}}_{e,i}\boldsymbol{\mathbf{X}}_{e,i}^{\mathrm{T}}-\mathbb{E}\left[G_{i}^{\star}\left(1-G_{i}^{\star}\right)\boldsymbol{\mathbf{X}}_{e,i}\boldsymbol{\mathbf{X}}_{e,i}^{\mathrm{T}}\right]\right\Vert \\
 & +\overline{\lambda}\left(M_n^{-1}\left(\widehat{\boldsymbol{\beta}}_{e}\right)\right)\overline{\lambda}\left(\frac{1}{n}\sum_{i=1}^{n}\widehat{G}_{i}\left(1-\widehat{G}_{i}\right)\boldsymbol{\mathbf{X}}_{e,i}\boldsymbol{\mathbf{X}}_{e,i}^{\mathrm{T}}\right)\left\Vert M^{-1}\left(\boldsymbol{\beta}_{e}^{\star}\right)-M_n^{-1}\left(\widehat{\boldsymbol{\beta}}_{e}\right)\right\Vert .
\end{align*}
Note that $\overline{\lambda}\left(M^{-1}\left(\boldsymbol{\beta}_{e}^{\star}\right)\right)\leq\underline{\lambda}_{e}^{-1}$,
$\overline{\lambda}\left(\mathbb{E}\left[G_{i}^{\star}\left(1-G_{i}^{\star}\right)\boldsymbol{\mathbf{X}}_{e,i}\boldsymbol{\mathbf{X}}_{e,i}^{\mathrm{T}}\right]\right)\leq\frac{1}{4}\overline{\lambda}\left(\mathbb{E}\left[\boldsymbol{\mathbf{X}}_{e,i}\boldsymbol{\mathbf{X}}_{e,i}^{\mathrm{T}}\right]\right)\leq C$,
$\overline{\lambda}\left(M_n^{-1}\left(\widehat{\boldsymbol{\beta}}_{e}\right)\right)\leq2\underline{\lambda}_{e}^{-1}$
with probability going to 1, and $\overline{\lambda}\left(\frac{1}{n}\sum_{i=1}^{n}\widehat{G}_{i}\left(1-\widehat{G}_{i}\right)\boldsymbol{\mathbf{X}}_{e,i}\boldsymbol{\mathbf{X}}_{e,i}^{\mathrm{T}}\right)\leq C$
with probability going to 1, we have that 
\begin{align*}
\left\Vert \widehat{\Sigma}_{1}-\Sigma_{1}^{\star}\right\Vert  & \leq C\left\Vert M^{-1}\left(\boldsymbol{\beta}_{e}^{\star}\right)-M_n^{-1}\left(\widehat{\boldsymbol{\beta}}_{e}\right)\right\Vert \\
 & +C\left\Vert \frac{1}{n}\sum_{i=1}^{n}\widehat{G}_{i}\left(1-\widehat{G}_{i}\right)\boldsymbol{\mathbf{X}}_{e,i}\boldsymbol{\mathbf{X}}_{e,i}^{\mathrm{T}}-\mathbb{E}\left[G_{i}^{\star}\left(1-G_{i}^{\star}\right)\boldsymbol{\mathbf{X}}_{e,i}\boldsymbol{\mathbf{X}}_{e,i}^{\mathrm{T}}\right]\right\Vert \\
 & =O_{p}\left(\sqrt{p^{4}\left(\log p\right)/n}\right)=o_{p}\left(1\right),
\end{align*}
which validates the result. 

To prove (iv), we only need to show that $\widehat{\sigma}^2\left(\rho\right)-\sigma^2\left(\rho\right)=o_{p}\left(1\right)$.
Note that 
\begin{align*}
\left|\widehat{\sigma}_n^2(\rho)-\sigma^2(\rho)\right| & =\left|\rho^{\mathrm{T}}\left(\widehat{\Sigma}_{1}-\Sigma_{1}^{\star}\right)\rho\right|\leq\left\Vert \rho\right\Vert \left\Vert \left(\widehat{\Sigma}_{1}-\Sigma_{1}^{\star}\right)\rho\right\Vert \leq\left\Vert \rho\right\Vert ^{2}\left\Vert \widehat{\Sigma}_{1}-\Sigma_{1}^{\star}\right\Vert \rightarrow_{p}0
\end{align*}
given that $\left\Vert \rho\right\Vert <\infty$ for all $n$, which
validates the result.
\end{proof}

\subsection*{Proof of  \autoref{prop2}}
\begin{proof}
We first show \autoref{prop2}(i). Note that from the proof in \autoref{prop:Known_G}, we know
that with probability going to 1, we have that 
\begin{align}
\left\Vert \Delta\boldsymbol{\beta}_{e,k+1}\right\Vert  & \leq\sup_{\boldsymbol{\beta}_{e}\in\mathcal{B}}\overline{\lambda}\left(I_{p+1}-\delta_{k}M_{n}\left(\boldsymbol{\beta}_{e}\right)\right)\left\Vert \Delta\boldsymbol{\beta}_{e,k}\right\Vert +\delta_{k}\left\Vert \frac{1}{n}\sum_{i=1}^{n}\varepsilon_{i}\boldsymbol{\mathbf{X}}_{e,i}\right\Vert \nonumber \\
 & \leq\left(1-\underline{\lambda}_{e}\delta_{k}/2\right)\left\Vert \Delta\boldsymbol{\beta}_{e,k}\right\Vert +\delta_{k}\left\Vert \frac{1}{n}\sum_{i=1}^{n}\varepsilon_{i}\boldsymbol{\mathbf{X}}_{e,i}\right\Vert \leq\cdots\nonumber \\
 & \leq\left(\prod_{j=1}^{k}\left(1-\underline{\lambda}_{e}\delta_{j}/2\right)\right)\left\Vert \Delta\boldsymbol{\beta}_{e,1}\right\Vert +\left\{ \sum_{j=0}^{k-1}\delta_{k-j}\left(\prod_{l=0}^{j-1}\left(1-\underline{\lambda}_{e}\delta_{k-l}/2\right)\right)\right\} \left\Vert \frac{1}{n}\sum_{i=1}^{n}\varepsilon_{i}\boldsymbol{\mathbf{X}}_{e,i}\right\Vert ,\label{Ap31}
\end{align}
where $\prod_{l=0}^{j-1}\left(1-\underline{\lambda}_{e}\delta_{k-l}/2\right)=1$
if $j=0$. 

For the first term on the RHS of (\ref{Ap31}), since $e^{x}\geq1+x$
for all $x$, we have $1-\underline{\lambda}_{e}\delta_{j}/2\leq\exp\left(-\underline{\lambda}_{e}\delta_{j}/2\right)$
for all $j$. Define $S_{0}=0$ and $S_{j}=\sum_{l=1}^{j}\delta_{l}$
for $j\geq1$, we have that
\[
\left(\prod_{j=1}^{k}\left(1-\underline{\lambda}_{e}\delta_{j}/2\right)\right)\left\Vert \Delta\boldsymbol{\beta}_{e,1}\right\Vert \leq\exp\left(-\frac{\underline{\lambda}_{e}}{2}\sum_{j=1}^{k}\delta_{j}\right)\left\Vert \Delta\boldsymbol{\beta}_{e,1}\right\Vert =\exp\left(-\frac{\underline{\lambda}_{e}S_{k}}{2}\right)\left\Vert \Delta\boldsymbol{\beta}_{e,1}\right\Vert .
\]

Next we show that $\sum_{j=0}^{k-1}\delta_{k-j}\left(\prod_{l=0}^{j-1}\left(1-\underline{\lambda}_{e}\delta_{k-l}/2\right)\right)$
is upper bounded by $\exp\left(\underline{\lambda}_{e}\delta_{k+1}/2\right)$
up to some constant scale that is independent of $k$. Since $\limsup_{k}\delta_{k-1}/\delta_{k}<\infty$,
we have that
\begin{align*}
 & \sum_{j=0}^{k-1}\delta_{k-j}\left(\prod_{l=0}^{j-1}\left(1-\underline{\lambda}_{e}\delta_{k-l}/2\right)\right)\leq\sum_{j=0}^{k-1}\delta_{k-j}\exp\left(-\frac{\underline{\lambda}_{e}}{2}\sum_{l=0}^{j-1}\delta_{k-l}\right)\\
 & \leq C\sum_{j=0}^{k-1}\delta_{k-j+1}\exp\left(-\frac{\underline{\lambda}_{e}\left(S_{k}-S_{k-j}\right)}{2}\right)\\
 & =C\exp\left(-\frac{\underline{\lambda}_{e}S_{k}}{2}\right)\sum_{j=0}^{k-1}\left(S_{k-j+1}-S_{k-j}\right)\exp\left(\frac{\underline{\lambda}_{e}S_{k-j}}{2}\right)\\
 & \leq 2C\underline{\lambda}_{e}^{-1}\exp\left(-\frac{\underline{\lambda}_{e}S_{k}}{2}\right)\sum_{j=0}^{k-1}\left\{ \exp\left(\frac{\underline{\lambda}_{e}S_{k-j+1}}{2}\right)-\exp\left(\frac{\underline{\lambda}_{e}S_{k-j}}{2}\right)\right\} \leq C\exp\left(\frac{\underline{\lambda}_{e}\delta_{k+1}}{2}\right).
\end{align*}
Then we have that 
\[
\left\Vert \Delta\boldsymbol{\beta}_{e,k+1}\right\Vert =O_{p}\left(\exp\left(-\frac{\underline{\lambda}_{e}S_{k}}{2}\right)\left\Vert \Delta\boldsymbol{\beta}_{e,1}\right\Vert \right)+O_{p}\left(\exp\left(\frac{\underline{\lambda}_{e}\delta_{k+1}}{2}\right)\sqrt{p\left(\log p\right)/n}\right).
\]
When $k\geq\widetilde{k}_{1,n}^{BGD} + 1$, we have that 
\[
\exp\left(-\frac{\underline{\lambda}_{e}S_{k}}{2}\right)\left\Vert \Delta\boldsymbol{\beta}_{e,1}\right\Vert \leq\sqrt{p\left(\log p\right)/n},
\]
and 
\[
\exp\left(\frac{\underline{\lambda}_{e}\delta_{k+1}}{2}\right)\leq e,
\]
so $\left\Vert \Delta\boldsymbol{\beta}_{e,k+1}\right\Vert =O_{p}\left(\sqrt{p\left(\log p\right)/n}\right)$.
This validates \autoref{prop2}(i). 

For \autoref{prop2}(ii), we know that for $k\geq\widetilde{k}_{1,n}^{BGD}+1$, $\left\Vert \Delta\boldsymbol{\beta}_{e,k}\right\Vert =O_{p}\left(\sqrt{p\left(\log p\right)/n}\right)$
holds, so we have that 
\begin{align*}
\Delta\boldsymbol{\beta}_{e,k+1} & =\left(I_{p+1}-\delta_{k}M_{n}\left(\overline{\boldsymbol{\beta}}_{e,k}\right)\right)\Delta\boldsymbol{\beta}_{e,k}-\frac{\delta_{k}}{n}\sum_{i=1}^{n}\varepsilon_{i}\mathbf{X}_{e,i}\\
 & =\left(I_{p+1}-\delta_{k}M\left(\boldsymbol{\beta}_{e}^{\star}\right)\right)\Delta\boldsymbol{\beta}_{e,k}-\delta_{k}\left(M_{n}\left(\overline{\boldsymbol{\beta}}_{e,k}\right)-M\left(\boldsymbol{\beta}_{e}^{\star}\right)\right)\Delta\boldsymbol{\beta}_{e,k}-\frac{\delta_{k}}{n}\sum_{i=1}^{n}\varepsilon_{i}\mathbf{X}_{e,i},
\end{align*}
where $\overline{\boldsymbol{\beta}}_{e,k}$ lies between $\boldsymbol{\beta}_{e,k}$
and $\boldsymbol{\beta}_{e}^{\star}$ and is element-wise. Following
the proof of \autoref{prop:Known_G}, we can easily show that 
\[
\sup_{k\geq\widetilde{k}_{1,n}^{BGD}+1}\left\Vert M_{n}\left(\overline{\boldsymbol{\beta}}_{e,k}\right)-M\left(\boldsymbol{\beta}_{e}^{\star}\right)\right\Vert =O_{p}\left(\sqrt{p^{4}\left(\log p\right)/n}\right).
\]
Recall that $\omega_{k}=\left(M_{n}\left(\overline{\boldsymbol{\beta}}_{e,k}\right)-M\left(\boldsymbol{\beta}_{e}^{\star}\right)\right)\Delta\boldsymbol{\beta}_{e,k}$,
so 
\[
\sup_{k\geq\widetilde{k}_{1,n}^{BGD}+1}\left\Vert \omega_{k}\right\Vert =O_{p}\left(\sqrt{p^{5}\left(\log p\right)^{2}/n^{2}}\right)=o_{p}\left(n^{-1/2}\right).
\]
We have that 
\begin{align*}
\Delta\boldsymbol{\beta}_{e,k+\widetilde{k}_{1,n}^{BGD}+1} & =\left(I_{p+1}-\delta_{\widetilde{k}_{1,n}^{BGD}+k}M\left(\boldsymbol{\beta}_{e}^{\star}\right)\right)\Delta\boldsymbol{\beta}_{e,k+\widetilde{k}_{1,n}^{BGD}}-\delta_{\widetilde{k}_{1,n}^{BGD}+k}\omega_{\widetilde{k}_{1,n}^{BGD}+k}-\delta_{\widetilde{k}_{1,n}^{BGD}+k}\frac{1}{n}\sum_{i=1}^{n}\varepsilon_{i}\mathbf{X}_{e,i}\\
 & =\prod_{j=0}^{k-1}\left(I_{p+1}-\delta_{\widetilde{k}_{1,n}^{BGD}+k-j}M\left(\boldsymbol{\beta}_{e}^{\star}\right)\right)\Delta\boldsymbol{\beta}_{e,\widetilde{k}_{1,n}^{BGD}+1}\\
 & -\sum_{j=0}^{k-1}\left\{ \delta_{\widetilde{k}_{1,n}^{BGD}+k-j}\prod_{l=0}^{j-1}\left(I_{p+1}-\delta_{\widetilde{k}_{1,n}^{BGD}+k-l}M\left(\boldsymbol{\beta}_{e}^{\star}\right)\right)\right\} \omega_{\widetilde{k}_{1,n}^{BGD}+k-j}\\
 & -\sum_{j=0}^{k-1}\left\{ \delta_{\widetilde{k}_{1,n}^{BGD}+k-j}\prod_{l=0}^{j-1}\left(I_{p+1}-\delta_{\widetilde{k}_{1,n}^{BGD}+k-l}M\left(\boldsymbol{\beta}_{e}^{\star}\right)\right)\right\} \frac{1}{n}\sum_{i=1}^{n}\varepsilon_{i}\mathbf{X}_{e,i},
\end{align*}
where $\prod_{l=0}^{j-1}\left(I_{p+1}-\delta_{\widetilde{k}_{1,n}^{BGD}+k-l}M\left(\boldsymbol{\beta}_{e}^{\star}\right)\right)=1$
if $j=0$. For the first part, define $S_{\widetilde{k}_{1,n}^{BGD},k}=\sum_{j=\widetilde{k}_{1,n}^{BGD}+1}^{\widetilde{k}_{1,n}^{BGD}+k}\delta_{j}$,
we have that 
\begin{align*}
\left\Vert \prod_{j=0}^{k-1}\left(I_{p+1}-\delta_{\widetilde{k}_{1,n}^{BGD}+k-j}M\left(\boldsymbol{\beta}_{e}^{\star}\right)\right)\Delta\boldsymbol{\beta}_{e,\widetilde{k}_{1,n}^{BGD}+1}\right\Vert  & \leq\prod_{j=0}^{k-1}\left(1-\underline{\lambda}_{e}\delta_{\widetilde{k}_{1,n}^{BGD}+k-j}/2\right)\left\Vert \Delta\boldsymbol{\beta}_{e,\widetilde{k}_{1,n}^{BGD}+1}\right\Vert \\
 & \leq\exp\left(-\underline{\lambda}_{e}S_{\widetilde{k}_{1,n}^{BGD},k}/2\right)\left\Vert \Delta\boldsymbol{\beta}_{e,\widetilde{k}_{1,n}^{BGD}+1}\right\Vert \\
 & =O_{p}\left(\exp\left(-\underline{\lambda}_{e}S_{\widetilde{k}_{1,n}^{BGD},k}/2\right)\sqrt{p\left(\log p\right)/n}\right).
\end{align*}
For the second term, we have that
\begin{align*}
 & \left\Vert \sum_{j=0}^{k-1}\left\{ \delta_{\widetilde{k}_{1,n}^{BGD}+k-j}\prod_{l=0}^{j-1}\left(I_{p+1}-\delta_{\widetilde{k}_{1,n}^{BGD}+k-l}M\left(\boldsymbol{\beta}_{e}^{\star}\right)\right)\right\} \omega_{\widetilde{k}_{1,n}^{BGD}+k-j}\right\Vert \\
 & \leq\left\{ \sum_{j=0}^{k-1}\delta_{\widetilde{k}_{1,n}^{BGD}+k-j}\prod_{l=0}^{j-1}\left(1-\underline{\lambda}_{e}\delta_{\widetilde{k}_{1,n}^{BGD}+k-l}/2\right)\right\} \left\{ \sup_{k\geq1}\left\Vert \omega_{\widetilde{k}_{1,n}^{BGD}+k}\right\Vert \right\} \\
 & \leq\exp\left(-\underline{\lambda}_{e}S_{\widetilde{k}_{1,n}^{BGD}+k}/2\right)\left\{ \sum_{j=0}^{k-1}\delta_{\widetilde{k}_{1,n}^{BGD}+k-j}\exp\left(\underline{\lambda}_{e}S_{\widetilde{k}_{1,n}^{BGD}+k-j}/2\right)\right\} \left\{ \sup_{k\geq1}\left\Vert \omega_{\widetilde{k}_{1,n}^{BGD}+k}\right\Vert \right\} \\
 & =O_{p}\left(\sqrt{p^{5}\left(\log p\right)^{2}/n^{2}}\right)
\end{align*}
according to the proof of \autoref{prop2}(i). Now we look at the last term. Note
that 
\begin{align*}
\mathcal{M}_{k,n}\equiv: & \sum_{j=0}^{k-1}\left\{ \delta_{\widetilde{k}_{1,n}^{BGD}+k-j}\prod_{l=0}^{j-1}\left(I_{p+1}-\delta_{\widetilde{k}_{1,n}^{BGD}+k-l}M\left(\boldsymbol{\beta}_{e}^{\star}\right)\right)\right\} \\
 & =\delta_{\widetilde{k}_{1,n}^{BGD}+k}I_{p+1}+\delta_{\widetilde{k}_{1,n}^{BGD}+k-1}\left(I_{p+1}-\delta_{\widetilde{k}_{1,n}^{BGD}+k}M\left(\boldsymbol{\beta}_{e}^{\star}\right)\right)+\cdots\\
 & +\delta_{\widetilde{k}_{1,n}^{BGD}+1}\left(I_{p+1}-\delta_{\widetilde{k}_{1,n}^{BGD}+k}M\left(\boldsymbol{\beta}_{e}^{\star}\right)\right)\left(I_{p+1}-\delta_{\widetilde{k}_{1,n}^{BGD}+k-1}M\left(\boldsymbol{\beta}_{e}^{\star}\right)\right)\cdots\left(I_{p+1}-\delta_{\widetilde{k}_{1,n}^{BGD}+2}M\left(\boldsymbol{\beta}_{e}^{\star}\right)\right),
\end{align*}
so 
\[
\mathcal{M}_{k+1,n}=\delta_{\widetilde{k}_{1,n}^{BGD}+k+1}I_{p+1}+\left(I_{p+1}-\delta_{\widetilde{k}_{1,n}^{BGD}+k}M\left(\boldsymbol{\beta}_{e}^{\star}\right)\right)\mathcal{M}_{k,n}.
\]
Note that 
\begin{align*}
 & \mathcal{M}_{k+1,n}-M^{-1}\left(\boldsymbol{\beta}_{e}^{\star}\right)\\
 & =\mathcal{M}_{k,n}-M^{-1}\left(\boldsymbol{\beta}_{e}^{\star}\right)+\delta_{\widetilde{k}_{1,n}^{BGD}+k+1}M\left(\boldsymbol{\beta}_{e}^{\star}\right)\left(M^{-1}\left(\boldsymbol{\beta}_{e}^{\star}\right)-\mathcal{M}_{k,n}\right)\\
 & =\left(I_{p+1}-\delta_{\widetilde{k}_{1,n}^{BGD}+k}M\left(\boldsymbol{\beta}_{e}^{\star}\right)\right)\left(\mathcal{M}_{k,n}-M^{-1}\left(\boldsymbol{\beta}_{e}^{\star}\right)\right),
\end{align*}
so
\begin{align*}
\left\Vert \mathcal{M}_{k+1,n}-M^{-1}\left(\boldsymbol{\beta}_{e}^{\star}\right)\right\Vert  & \leq\overline{\lambda}\left(I_{p+1}-\delta_{\widetilde{k}_{1,n}^{BGD}+k}M_{n}\left(\boldsymbol{\beta}_{e}^{\star}\right)\right)\left\Vert \mathcal{M}_{k,n}-M^{-1}\left(\boldsymbol{\beta}_{e}^{\star}\right)\right\Vert \\
 & \leq\left(1-\delta_{\widetilde{k}_{1,n}^{BGD}+k}\underline{\lambda}_{e}\right)\left\Vert \mathcal{M}_{k,n}-M^{-1}\left(\boldsymbol{\beta}_{e}^{\star}\right)\right\Vert \\
 & \leq\exp\left(-\underline{\lambda}_{e}S_{\widetilde{k}_{1,n}^{BGD},k}\right)\left\Vert \mathcal{M}_{1,n}-M^{-1}\left(\boldsymbol{\beta}_{e}^{\star}\right)\right\Vert .
\end{align*}
Then 
\begin{align*}
 & \sum_{j=0}^{k-1}\left\{ \delta_{\widetilde{k}_{1,n}^{BGD}+k-1-j}\prod_{l=0}^{j-1}\left(I-\delta_{\widetilde{k}_{1,n}^{BGD}+k-1}M\left(\boldsymbol{\beta}_{e}^{\star}\right)\right)\right\} \frac{1}{n}\sum_{i=1}^{n}\varepsilon_{i}\mathbf{X}_{e,i}\\
 & =M^{-1}\left(\boldsymbol{\beta}_{e}^{\star}\right)\frac{1}{n}\sum_{i=1}^{n}\varepsilon_{i}\mathbf{X}_{e,i}+O_{p}\left(\exp\left(-\underline{\lambda}_{e}S_{\widetilde{k}_{1,n}^{BGD},k}\right)\sqrt{p\left(\log p\right)/n}\right).
\end{align*}
So we have 
\begin{align*}
\left\Vert \sqrt{n}\Delta\boldsymbol{\beta}_{e,k+\widetilde{k}_{1,n}^{BGD}}-M^{-1}\left(\boldsymbol{\beta}_{e}^{\star}\right)\frac{1}{\sqrt{n}}\sum_{i=1}^{n}\varepsilon_{i}\mathbf{X}_{e,i}\right\Vert  & =O_{p}\left(\exp\left(-\underline{\lambda}_{e}S_{\widetilde{k}_{1,n}^{BGD},k}/2\right)\sqrt{p\left(\log p\right)/n}\right)\\
 & +O_{p}\left(\sqrt{p^{5}\left(\log p\right)^{2}/n^{2}}\right)\\
 & +O_{p}\left(\exp\left(-S_{\widetilde{k}_{1,n}^{BGD},k}\right)\sqrt{p\left(\log p\right)/n}\right).
\end{align*}
According to the definition of $\widetilde{k}_{2,n}^{BGD}$, we have that for $k\geq\widetilde{k}_{2,n}^{BGD}$, there holds $
S_{\widetilde{k}_{1,n}^{BGD},k}/\log p \rightarrow \infty$,  this proves \autoref{prop2}(ii).

The proof of \autoref{prop2}(iii) and \autoref{prop2}(iv) is the same as that in the proof of \autoref{prop:Known_G}, so is left out.
\end{proof}

\subsection*{Proof of \autoref{thm:3.1}}
\begin{proof}
	Define 
	\[
	\eta_{1,n}\left(\boldsymbol{\beta}\right)=\frac{1}{n}\sum_{i=1}^{n}\widehat{G}\left(\left.z\left(\mathbf{X}_{e,i},\boldsymbol{\beta}\right)\right|\boldsymbol{\beta}\right)\boldsymbol{\mathbf{X}}_{i}-\mathbb{E}\left[L\left(z\left(\boldsymbol{\mathbf{X}}_{e,i},\boldsymbol{\beta}\right),\boldsymbol{\beta}\right)\boldsymbol{\mathbf{X}}_{i}\right],
	\]
	\[
	\eta_{2,n}=\left(\frac{1}{n}\sum_{i=1}^{n}G\left(z_{i}^{\star}\right)\boldsymbol{\mathbf{X}}_{i}-\mathbb{E}\left[G\left(z_{i}^{\star}\right)\boldsymbol{\mathbf{X}}_{i}\right]\right)+\frac{1}{n}\sum_{i=1}^{n}\varepsilon_{i}\cdot\boldsymbol{\mathbf{X}}_{i}.
	\]
	Note that when $\boldsymbol{\beta}^{\star}\in\mathcal{B}$ and $\boldsymbol{\beta}_{k}\in\mathcal{B}$,
	we have that $\boldsymbol{\beta}^{\star}+t\Delta\boldsymbol{\beta}_{k}\in\mathcal{B}$
	for all $0\leq t\leq1$, so 
	\begin{align*}
		\left\Vert \Delta\boldsymbol{\beta}_{k+1}\right\Vert  & \leq\left\Vert \int_{0}^{1}\left(I_{p}-\delta\varLambda\left(\boldsymbol{\beta}^{\star}+t\Delta\boldsymbol{\beta}_{k}\right)\right)dt\Delta\boldsymbol{\beta}_{k}\right\Vert +\delta \left\Vert \eta_{1,n}\left(\boldsymbol{\beta}_{k}\right)\right \Vert+\delta\left\Vert\eta_{2,n}\right\Vert \\
		& \leq\left\{ \sup_{\boldsymbol{\beta}\in\mathcal{B}}\overline{\sigma}\left(I_{p}-\delta\varLambda\left(\boldsymbol{\beta}\right)\right)\right\} \left\Vert \Delta\boldsymbol{\beta}_{k}\right\Vert +\delta \left\Vert \eta_{1,n}\left(\boldsymbol{\beta}_{k}\right)\right \Vert+\delta\left\Vert\eta_{2,n}\right\Vert .
	\end{align*}
	Note that for any $1\leq s,t\leq p$, 
	\begin{align*}
		\left|\left(\varLambda\left(\boldsymbol{\beta}\right)\right)_{s,t}\right| & =\left|\mathbb{E}\left[\int_{\mathcal{X}}\left(X_{s,i}X_{t,i}-X_{s,i}X_{t}\right)W\left(\boldsymbol{\mathbf{X}}_{e,i},\boldsymbol{\mathbf{X}}_{e},\boldsymbol{\beta}\right)d\boldsymbol{\mathbf{X}}\right]\right|\\
		& \leq2\left\Vert G^{\prime}\right\Vert _{\infty}\mathbb{E}\left[\int_{\mathcal{X}}f_{\boldsymbol{\mathbf{X}}|z}\left(\left.\mathbf{X}\right|z\left(\boldsymbol{\mathbf{X}}_{e,i},\boldsymbol{\beta}\right),\boldsymbol{\beta}\right)d\boldsymbol{\mathbf{X}}\right]=2\left\Vert G^{\prime}\right\Vert _{\infty},
	\end{align*}
	so each element of $\varLambda^{\mathrm{T}}\left(\boldsymbol{\beta}\right)\varLambda\left(\boldsymbol{\beta}\right)$ is bounded by $2p\Vert G^{\prime}\Vert_{\infty} $, and we have that
	\begin{align*}
		& \sup_{\boldsymbol{\beta}\in\mathcal{B}}\left|\overline{\sigma}^{2}\left(I_{p}-\delta\varLambda\left(\boldsymbol{\beta}\right)\right)-\overline{\lambda}\left(I_{p}-\delta\left(\varLambda\left(\boldsymbol{\beta}\right)+\varLambda^{\mathrm{T}}\left(\boldsymbol{\beta}\right)\right)\right)\right|\\
		& \leq\sup_{\boldsymbol{\beta}\in\mathcal{B}}\delta^{2}\left\Vert \varLambda^{\mathrm{T}}\left(\boldsymbol{\beta}\right)\varLambda\left(\boldsymbol{\beta}\right)\right\Vert \leq2\left\Vert G^{\prime}\right\Vert _{\infty}p^{2}\delta^{2}.
	\end{align*}
	Then according to \autoref{assu:5}, we have that 
	\[
	\sup_{\boldsymbol{\beta}\in\mathcal{B}}\overline{\sigma}^{2}\left(I_{p}-\delta\varLambda\left(\boldsymbol{\beta}\right)\right)\leq1-\delta\underline{\lambda}_{\varLambda}+2\left\Vert G^{\prime}\right\Vert _{\infty}p^{2}\delta^{2}.
	\]
	When $\delta<\min\left\{ 1/\left(2\underline{\lambda}_{\varLambda}\right),1/\left(4\left\Vert G^{\prime}\right\Vert _{\infty}p^{2}\right)\right\} $,
	we have that 
	\[
	0\leq1-\delta\underline{\lambda}_{\varLambda}+2\left\Vert G^{\prime}\right\Vert _{\infty}p^{2}\delta^{2}\leq1-\delta\underline{\lambda}_{\varLambda}/2<1.
	\]
	So 
	\[
	\sup_{\boldsymbol{\beta}\in\mathcal{B}}\overline{\sigma}\left(I_{p}-\delta\varLambda\left(\boldsymbol{\beta}\right)\right)\leq\sqrt{1-\delta\underline{\lambda}_{\varLambda}/2}\leq1-\delta\underline{\lambda}_{\varLambda}/4,
	\]
	and
	\begin{align*}
		\left\Vert \Delta\boldsymbol{\beta}_{k+1}\right\Vert  & \leq\left(1-\delta\underline{\lambda}_{\varLambda}/4\right)\left\Vert \Delta\boldsymbol{\beta}_{k}\right\Vert +\delta \left\Vert \eta_{1,n}\left(\boldsymbol{\beta}_{k}\right)\right \Vert+\delta\left\Vert\eta_{2,n}\right\Vert \\
		& \leq\cdots\leq\left(1-\delta\underline{\lambda}_{\varLambda}/4\right)^{k}\left\Vert \Delta\boldsymbol{\beta}_{1}\right\Vert +\delta\cdot\sum_{j=0}^{k-1}\left(1-\delta\underline{\lambda}_{\varLambda}/4\right)^{j}\left(\left\Vert \eta_{1,n}\left(\boldsymbol{\beta}_{j}\right)\right\Vert+\left\Vert\eta_{2,n}\right\Vert\right)\\
		& \leq\left(1-\delta\underline{\lambda}_{\varLambda}/4\right)^{k}\left\Vert \Delta\boldsymbol{\beta}_{1}\right\Vert +\delta\cdot\sum_{j=0}^{\infty}\left(1-\delta\underline{\lambda}_{\varLambda}/4\right)^{j}\left(\sup_{\boldsymbol{\beta}\in\mathcal{B}}\left\Vert \eta_{1,n}\left(\boldsymbol{\beta}\right)\right\Vert +\left\Vert \eta_{2,n}\right\Vert \right)\\
		& =\left(1-\delta\underline{\lambda}_{\varLambda}/4\right)^{k}\left\Vert \Delta\boldsymbol{\beta}_{1}\right\Vert +4\underline{\lambda}_{\varLambda}^{-1}\left(\sup_{\boldsymbol{\beta}\in\mathcal{B}}\left\Vert \eta_{1,n}\left(\boldsymbol{\beta}\right)\right\Vert +\left\Vert \eta_{2,n}\right\Vert \right).
	\end{align*}
	Note that 
	\[
	\sup_{\boldsymbol{\beta}\in\mathcal{B}}\left\Vert \eta_{1,n}\left(\boldsymbol{\beta}\right)\right\Vert =p^{\frac{5p+1}{2\left(p+1\right)}}\psi^{\frac{1}{p+1}}\left(n,p,h_{n}\right)
	\]
	according to  \autoref{lem:3.1}, and 
	\[
	\left\Vert \eta_{2,n}\right\Vert =O_{p}\left(\sqrt{p\left(\log p\right)/n}\right)=o_{p}\left(p^{\frac{5p+1}{2\left(p+1\right)}}\left(\psi\left(n,p,h_{n}\right)\right)^{\frac{1}{3p+3}}\right)
	\]
	under any choices of $h_{n}\rightarrow0$. This implies that when
	\[
	\left(1-\delta\underline{\lambda}_{\varLambda}/4\right)^{k}\left\Vert \Delta\boldsymbol{\beta}_{1}\right\Vert \leq p^{\frac{5p+1}{2\left(p+1\right)}}\left(\psi\left(n,p,h_{n}\right)\right)^{\frac{1}{p+1}},
	\]
	or equivalently, 
	\[
	k\geq k_{1,n}^{KBGD}=\frac{\log\left(\left\Vert \Delta\boldsymbol{\beta}_{1}\right\Vert \right)-\frac{5p+1}{2\left(p+1\right)}\log p-\frac{1}{p+1}\log\psi\left(n,p,h_{n}\right)}{-\log\left(1-\delta\underline{\lambda}_{\varLambda}/4\right)},
	\]
	we have that $\sup_{k\geq k_{1,n}^{KBGD}+1}\left\Vert \Delta\boldsymbol{\beta}_{k}\right\Vert =O_{p}\left(p^{\frac{5p+1}{2\left(p+1\right)}}\psi^{\frac{1}{p+1}}\left(n,p,h_{n}\right)\right).$
\end{proof}

\subsection*{Proof of \autoref{thm:3.2}}
\begin{proof}
	We first note that 
	\begin{align*}
		\left\Vert \int_{\mathcal{X}}V\left(\mathbf{X}_{e,i},\mathbf{X}_{e},\boldsymbol{\beta}\right)d\mathbf{X}\right\Vert  & \leq2p\left\Vert G^{\prime}\right\Vert _{\infty}\int_{\mathcal{X}}f_{\boldsymbol{\mathbf{X}}|z}\left(\left.\mathbf{X}\right|z\left(\mathbf{X}_{e,i},\boldsymbol{\beta}\right),\boldsymbol{\beta}\right)d\boldsymbol{\mathbf{X}}=2p\left\Vert G^{\prime}\right\Vert _{\infty},
	\end{align*}
	for all $\mathbf{X}_{e,i}$, so 
	\begin{align*}
		\sup_{\boldsymbol{\beta}\in\mathcal{B}}\left\Vert \varLambda_{\phi}\left(\boldsymbol{\beta}\right)-\varLambda\left(\boldsymbol{\beta}\right)\right\Vert  & \leq2p\left\Vert G^{\prime}\right\Vert _{\infty}\mathbb{E}\left(1-I_{i}^{\phi}\right)\leq2\zeta p^{2}\left\Vert G^{\prime}\right\Vert _{\infty}\phi,
	\end{align*}
	where the last inequality comes from the fact that $m\left(\mathcal{X}_{e}^{\phi}\right)=1-\left(1-\phi\right)^{p}\leq p\phi$.
	So 
	\begin{equation}
		\sup_{\boldsymbol{\beta}\in\mathcal{B}}\left\Vert \varLambda_{\phi}\left(\boldsymbol{\beta}\right)-\varLambda\left(\boldsymbol{\beta}\right)\right\Vert \leq\delta\underline{\lambda}_{\varLambda}/8\label{Ap21}
	\end{equation}
	holds under the choice of $\phi$.
	
	Based on (\ref{Ap21}), the following proof is similar to the proof
	of \autoref{thm:3.1}. Define 
	\[
	\eta_{1,n}^{\phi}\left(\boldsymbol{\beta}\right)=\frac{1}{n}\sum_{i=1}^{n}\widehat{G}\left(\left.z\left(\boldsymbol{\mathbf{X}}_{e,i},\boldsymbol{\beta}\right)\right|\boldsymbol{\beta}\right)\boldsymbol{\mathbf{X}}_{i}^{\phi}-\mathbb{E}\left[L\left(z\left(\boldsymbol{\mathbf{X}}_{e,i},\boldsymbol{\beta}\right),\boldsymbol{\beta}\right)\boldsymbol{\mathbf{X}}_{i}^{\phi}\right],
	\]
	\[
	\eta_{2,n}^{\phi}=\frac{1}{n}\sum_{i=1}^{n}G\left(z_{i}^{\star}\right)\boldsymbol{\mathbf{X}}_{i}^{\phi}-\mathbb{E}\left[G\left(z_{i}^{\star}\right)\boldsymbol{\mathbf{X}}_{i}^{\phi}\right]+\frac{1}{n}\sum_{i=1}^{n}\varepsilon_{i}\boldsymbol{\mathbf{X}}_{i}^{\phi}.
	\]
	We have that 
	\begin{align*}
		\Delta\boldsymbol{\beta}_{k+1} & =\Delta\boldsymbol{\beta}_{k}-\frac{\delta}{n}\sum_{i=1}^{n}\left(\widehat{G}\left(\left.z_{i,k}\right|\boldsymbol{\beta}_{k}\right)-Y_{i}\right)\boldsymbol{\mathbf{X}}_{i}^{\phi}\\
		& =\Delta\boldsymbol{\beta}_{k}-\delta\mathbb{E}\left[\left(L\left(z_{i,k},\boldsymbol{\beta}_{k}\right)-G\left(Z_{i}^{\star}\right)\right)\boldsymbol{\mathbf{X}}_{i}^{\phi}\right]+\delta\left(\eta_{1,n}^{\phi}\left(\boldsymbol{\beta}_{k}\right)+\eta_{2,n}^{\phi}\right)\\
		& =\int_{0}^{1}\left\{ I_{p}-\delta\varLambda_{\phi}\left(\boldsymbol{\beta}^{\star}+t\Delta\boldsymbol{\beta}_{k}\right)\right\} \Delta\boldsymbol{\beta}_{k}dt+\delta\left(\eta_{1,n}^{\phi}\left(\boldsymbol{\beta}_{k}\right)+\eta_{2,n}^{\phi}\right),
	\end{align*}
	so 
	\begin{align*}
		\left\Vert \boldsymbol{\beta}_{k+1}\right\Vert  & \leq\sup_{\boldsymbol{\beta}\in\mathcal{B}}\overline{\sigma}\left(I_{p}-\delta\varLambda_{\phi}\left(\boldsymbol{\beta}\right)\right)\left\Vert \boldsymbol{\beta}_{k}\right\Vert +\delta\left(\sup_{\boldsymbol{\beta}\in\mathcal{B}}\left\Vert \eta_{1,n}^{\phi}\left(\boldsymbol{\beta}\right)\right\Vert +\left\Vert \eta_{2,n}^{\phi}\right\Vert \right).
	\end{align*}
	Obviously, since $p$ is fixed, we have that $\left\Vert \eta_{2,n}^{\phi}\right\Vert =O_{p}\left(n^{-1/2}\right)$.
	Due to trimming, we also have that $\sup_{\boldsymbol{\beta}\in\mathcal{B}}\left\Vert \eta_{1,n}^{\phi}\left(\boldsymbol{\beta}\right)\right\Vert =O_{p}\left(\psi\left(n,p,h_{n}\right)\right)$.
	Note that (\ref{Ap21}) holds, so we have that 
	\begin{align*}
		\sup_{\boldsymbol{\beta}\in\mathcal{B}} & \left\Vert \left\{ I_{p}-\delta\varLambda_{\phi}\left(\boldsymbol{\beta}\right)\right\} -\left\{ I_{p}-\delta\varLambda\left(\boldsymbol{\beta}\right)\right\} \right\Vert \leq\delta\underline{\lambda}_{\varLambda}/8.
	\end{align*}
	According to the proof of \autoref{thm:3.1}, there holds $\sup_{\boldsymbol{\beta}\in\mathcal{B}}\overline{\sigma}\left(I_{p}-\delta\varLambda\left(\boldsymbol{\beta}\right)\right)\leq1-\delta\underline{\lambda}_{\varLambda}/4$
	under the choice of $\delta$, so we have that 
	\[
	\sup_{\boldsymbol{\beta}\in\mathcal{B}}\overline{\sigma}\left(I_{p}-\delta\varLambda_{\phi}\left(\boldsymbol{\beta}\right)\right)\leq1-\delta\underline{\lambda}_{\varLambda}/8.
	\]
	Then based on the proof of \autoref{thm:3.1}, it remains to note
	that 
	\[
	\sup_{\boldsymbol{\beta}\in\mathcal{B}}\left(\left\Vert \eta_{1,n}^{\phi}\left(\boldsymbol{\beta}\right)\right\Vert +\left\Vert \eta_{2,n}^{\phi}\right\Vert \right)=O_{p}\left(\psi\left(n,p,h_{n}\right)\right)
	\]
	holds under any fixed trimming parameter $\phi$.
\end{proof}

\subsection*{Proof of \autoref{thm:3.3}}
\begin{proof}
	Note that under the choice of $\delta$ and $\phi$  , $\sup_{k\geq\widetilde{k}_{1,n}^{KBGD}+1}\left\Vert \boldsymbol{\beta}_{k}-\boldsymbol{\beta}^{\star}\right\Vert =O_{p}\left(\psi\left(n,p,h_{n}\right)\right)$
	according to  \autoref{thm:3.2}. According to (\ref{eq:gradient_update}),
	we have that
	\begin{align*}
		& \left\Vert \Delta\boldsymbol{\beta}_{k+\widetilde{k}_{1,n}^{KBGD}+1}\right\Vert\\
		&  \leq\sup_{k\geq\widetilde{k}_{1,n}^{KBGD}+1,,t\in\left[0,1\right]}\overline{\sigma}\left\{ I_{p}-\frac{\delta}{n}\sum_{i=1}^{n}\left[\left.\mathbf{X}_{i}^{\phi}\frac{\partial\widehat{G}\left(\left.z\left(\mathbf{X}_{e,i},\boldsymbol{\beta}\right)\right|\boldsymbol{\beta}\right)}{\partial\boldsymbol{\beta}^{\mathrm{T}}}\right|_{\boldsymbol{\beta}=\boldsymbol{\beta}^{\star}+t\Delta\boldsymbol{\beta}_{k}}\right]\right\} \left\Vert \Delta\boldsymbol{\beta}_{k}\right\Vert +\delta\left\Vert \boldsymbol{\xi}_{n}^{\phi}\right\Vert .
	\end{align*}
	According to  \autoref{lem3.2}, we have that 
	\begin{align}
		& \sup_{k\geq\widetilde{k}_{1,n}^{KBGD},t\in\left[0,1\right]}\left\Vert \left\{ I_{p}-\frac{\delta}{n}\sum_{i=1}^{n}\left[\left.\mathbf{X}_{i}^{\phi}\frac{\partial\widehat{G}\left(\left.z\left(\mathbf{X}_{e,i},\boldsymbol{\beta}\right)\right|\boldsymbol{\beta}\right)}{\partial\boldsymbol{\beta}^{\mathrm{T}}}\right|_{\boldsymbol{\beta}=\boldsymbol{\beta}^{\star}+t\Delta\boldsymbol{\beta}_{k}}\right]\right\} \right.\nonumber \\
		& \left.-\left\{ I_{p}-\delta\varLambda_{\phi}\left(\boldsymbol{\beta}^{\star}+t\Delta\boldsymbol{\beta}_{k}\right)\right\} \right\Vert =\delta O_{p}\left(h_{n}^{-2}\sqrt{\left(\log\left(nh_{n}^{-1}\right)\right)/n}+h_{n}^{3}\right),\label{14}
	\end{align}
	due to the fact that 
	\[
	\sup_{k\geq\widetilde{k}_{1,n}^{KBGD}+1}\left\Vert \Delta\boldsymbol{\beta}_{k}\right\Vert =O_{p}\left(\psi_{1}\left(n,p,h_{n}\right)\right)=o_{p}\left(h_{n}^{-2}\sqrt{\left(\log\left(nh_{n}^{-1}\right)\right)/n}+h_{n}^{3}\right),
	\]
	when $p$ is fixed and $h_{n}\rightarrow0$. 
	
	When $nh_{n}^{6}\rightarrow0$ and $h_{n}^{4}n/\left(\log n\right)^{2}\rightarrow\infty$,
	we have that $h_{n}^{-2}\sqrt{\left(\log\left(nh_{n}^{-1}\right)\right)/n}+h_{n}^{3}\rightarrow0$.
	So we have that  (\ref{14}) is smaller than $\delta\underline{\lambda}_{\varLambda}/16$
	with probability going to 1. According to the choice of $\phi$ and
	$\delta$, we have that $\sup_{\boldsymbol{\beta}\in\mathcal{B}}\overline{\sigma}\left(I_{p}-\delta\varLambda_{\phi}\left(\boldsymbol{\beta}\right)\right)\leq1-\delta\underline{\lambda}_{\varLambda}/8$
	according to the proof of \autoref{thm:3.2}. So as $n$ increases,
	with probability going to 1, there holds
	\[
	\sup_{k\geq\widetilde{k}_{1,n}^{KBGD}+1,t\in\left[0,1\right]}\overline{\sigma}\left(I_{p}-\frac{\delta}{n}\sum_{i=1}^{n}\left[\left.\boldsymbol{\mathbf{X}}_{i}^{\phi}\frac{\partial\widehat{G}\left(\left.z\left(\mathbf{X}_{e,i},\boldsymbol{\beta}\right)\right|\boldsymbol{\beta}\right)}{\partial\boldsymbol{\beta}^{\mathrm{T}}}\right|_{\boldsymbol{\beta}=\boldsymbol{\beta}^{\star}+t\Delta\boldsymbol{\beta}_{k}}\right]\right)\leq1-\delta\underline{\lambda}_{\varLambda}/16,
	\]
	Then as $n$ increases, with probability going to 1 there holds
	\begin{align*}
		\left\Vert \Delta\boldsymbol{\beta}_{k+\widetilde{k}_{1,n}^{KBGD}+1}\right\Vert  & \leq\left(1-\delta\underline{\lambda}_{\varLambda}/16\right)\left\Vert \Delta\boldsymbol{\beta}_{k+\widetilde{k}_{1,n}^{KBGD}}\right\Vert +\delta\left\Vert \boldsymbol{\xi}_{n}^{\phi}\right\Vert \\
		& \leq\cdots\leq\left(1-\delta\underline{\lambda}_{\varLambda}/16\right)^{k}\left\Vert \Delta\boldsymbol{\beta}_{\widetilde{k}_{1,n}^{KBGD}+1}\right\Vert +16\underline{\lambda}_{\varLambda}^{-1}\left\Vert \boldsymbol{\xi}_{n}^{\phi}\right\Vert .
	\end{align*}
	According to \autoref{lem3.3}, $\left\Vert \boldsymbol{\xi}_{n}^{\phi}\right\Vert =O_{p}\left(n^{-1/2}\right)$.
	Also note that $\left\Vert \Delta\boldsymbol{\beta}_{\widetilde{k}_{1,n}^{KBGD}+1}\right\Vert =O_{p}\left(\psi\left(n,p,h_{n}\right)\right)$,
	then if we choose $k_{2,n}^{KBGD}$ such that $\left(1-\delta\underline{\lambda}_{\varLambda}/16\right)^{k_{2,n}^{KBGD}-1}\leq n^{-1/2}\psi^{-1}\left(n,p,h_{n}\right)$,
	or equivalently,
	\[
	k_{2,n}^{KBGD}\geq-\frac{\log\left(n^{1/2}\right)+\log\left(\psi\left(n,p,h_{n}\right)\right)}{\log\left(1-\delta\underline{\lambda}_{\varLambda}/16\right)}+1,
	\]
	we have that $\sup_{k\geq k_{2,n}^{KBGD}+1}\left\Vert \Delta\boldsymbol{\beta}_{k+\widetilde{k}_{1,n}^{KBGD}}\right\Vert =O_{p}\left(n^{-1/2}\right)$.
	This proves (i).
	
	To prove (ii), we consider the following decomposition, 
	\[
	\Delta\boldsymbol{\beta}_{k+1}=\left(I_{p}-\delta\varLambda_{\phi}\left(\boldsymbol{\beta}^{\star}\right)\right)\Delta\boldsymbol{\beta}_{k}+\delta\overline{\omega}_{1}\left(\boldsymbol{\beta}_{k}\right)+\delta\overline{\omega}_{2}\left(\boldsymbol{\beta}_{k}\right)-\delta\boldsymbol{\xi}_{n}^{\phi},
	\]
	where 
	\begin{align*}
		\overline{\omega}_{1}\left(\boldsymbol{\beta}_{k}\right) & =\int_{0}^{1}\left\{ \varLambda_{\phi}\left(\boldsymbol{\beta}^{\star}+t\Delta\boldsymbol{\beta}_{k}\right)-\frac{1}{n}\sum_{i=1}^{n}\left[\mathbf{X}_{i}^{\phi}\left.\frac{\partial\widehat{G}\left(\left.z\left(\boldsymbol{\mathbf{X}}_{e,i},\boldsymbol{\beta}\right)\right|\boldsymbol{\beta}\right)}{\partial\boldsymbol{\beta}^{\mathrm{T}}}\right|_{\boldsymbol{\beta}=\boldsymbol{\beta}^{\star}+t\Delta\boldsymbol{\beta}_{k}}\right]\right\} dt\Delta\boldsymbol{\beta}_{k},
	\end{align*}
and
	\begin{align*}
		\overline{\omega}_{2}\left(\boldsymbol{\beta}_{k}\right) & =\int_{0}^{1}\left\{ \varLambda_{\phi}\left(\boldsymbol{\beta}^{\star}\right)-\varLambda_{\phi}\left(\boldsymbol{\beta}^{\star}+t\Delta\boldsymbol{\beta}_{k}\right)\right\} dt\Delta\boldsymbol{\beta}_{k}.
	\end{align*}
	Obviously, according to  \autoref{lem3.2}, 
	\[
	\sup_{k\geq\widetilde{k}_{1,n}^{KBGD}+k_{2,n}^{KBGD}+1}\left\Vert \overline{\omega}_{1}\left(\boldsymbol{\beta}_{k}\right)\right\Vert =O_{p}\left(h_{n}^{-2}\sqrt{\left(\log\left(nh_{n}^{-1}\right)\right)/n}+h_{n}^{3}\right)O_{p}\left(n^{-\frac{1}{2}}\right)=o_{p}\left(n^{-\frac{1}{2}}\right).
	\]
	We also note that each element of matrix $I_{i}^{\phi}\cdot\int_{\mathcal{X}}V\left(\boldsymbol{\mathbf{X}}_{e,i},\boldsymbol{\mathbf{X}}_{e},\boldsymbol{\beta}\right)d\boldsymbol{\mathbf{X}}$
	has bounded derivative with respect to $\boldsymbol{\beta}$ for any
	$\boldsymbol{\mathbf{X}}_{e,i}$. This is because, if $\boldsymbol{\mathbf{X}}_{e,i}\notin \mathcal{X}_{e}^{\phi}$,
	$I_{i}^{\phi}=0$ so each element will be zero and the results hold;
	if $\mathbf{X}_{e,i}\in\mathcal{X}_{e}^{\phi}$, then $f_{z}\left(\left.z\left(\mathbf{X}_{e,i},\boldsymbol{\beta}\right)\right|\boldsymbol{\beta}\right)>0$,
	so $\int_{\mathcal{X}}\left\Vert \partial W\left(\mathbf{X}_{e,i},\mathbf{X}_{e},\boldsymbol{\beta}\right)/\partial\boldsymbol{\beta}\right\Vert d\boldsymbol{\mathbf{X}}$
	is bounded according to  \autoref{lem:2}(x). This implies that
	
	\begin{align*}
		\sup_{k\geq\widetilde{k}_{1,n}^{KBGD}+k_{2,n}^{KBGD}+1}\left\Vert \overline{\omega}_{2}\left(\boldsymbol{\beta}_{k}\right)\right\Vert  & \leq C \left\Vert \Delta\boldsymbol{\beta}_{k}\right\Vert ^{2}=o_{p}\left(n^{-\frac{1}{2}}\right).
	\end{align*}
	Then 
	\begin{align*}
		&\Delta\boldsymbol{\beta}_{k+\widetilde{k}_{1,n}^{KBGD}+k_{2,n}^{KBGD}+1}\\
		& =\left(I_{p}-\delta\varLambda_{\phi}\left(\boldsymbol{\beta}^{\star}\right)\right)^{k}\Delta\boldsymbol{\beta}_{\widetilde{k}_{1,n}^{KBGD}+k_{2,n}^{KBGD}+1}+\delta\sum_{j=1}^{k}\left(I_{p}-\delta\varLambda_{\phi}\left(\boldsymbol{\beta}^{\star}\right)\right)^{k-j}\overline{\omega}_{1}\left(\boldsymbol{\beta}_{\widetilde{k}_{1,n}^{KBGD}+k_{2,n}^{KBGD}+j}\right)\\
		& +\sum_{j=1}^{k}\left(I_{p}-\delta\varLambda_{\phi}\left(\boldsymbol{\beta}^{\star}\right)\right)^{k-j}\overline{\omega}_{2}\left(\boldsymbol{\beta}_{\widetilde{k}_{1,n}^{KBGD}+k_{2,n}^{KBGD}+j}\right)-\delta\sum_{j=1}^{k}\left(I_{p}-\delta\varLambda_{\phi}\left(\boldsymbol{\beta}^{\star}\right)\right)^{k-j}\boldsymbol{\xi}_{n}^{\phi}.
	\end{align*}
	Note that $\sup_{\boldsymbol{\beta}\in\mathcal{B}}\overline{\sigma}\left(I_{p}-\delta\varLambda_{\phi}\left(\boldsymbol{\beta}\right)\right)\leq1-\delta\underline{\lambda}_{\varLambda}/8,$
	so 
	\begin{align*}
		\left\Vert \left(I_{p}-\delta\varLambda_{\phi}\left(\boldsymbol{\beta}^{\star}\right)\right)^{k}\Delta\boldsymbol{\beta}_{\widetilde{k}_{1,n}^{KBGD}+k_{2,n}^{KBGD}+1}\right\Vert & \leq\left(1-\delta\underline{\lambda}_{\varLambda}/8\right)^{k}\left\Vert \Delta\boldsymbol{\beta}_{\widetilde{k}_{1,n}^{KBGD}+k_{2,n}^{KBGD}+1}\right\Vert ,\\
		\delta\left\Vert \sum_{j=1}^{k}\left(I_{p}-\delta\varLambda_{\phi}\left(\boldsymbol{\beta}^{\star}\right)\right)^{k-j}\omega_{1}\left(\boldsymbol{\beta}_{\widetilde{k}_{1,n}^{KBGD}+k_{2,n}^{KBGD}+j}\right)\right\Vert & \leq\delta\sum_{j=0}^{\infty}\left(1-\delta\underline{\lambda}_{\varLambda}/8\right)^{j}\sup_{k\geq\widetilde{k}_{1,n}^{KBGD}+k_{2,n}^{KBGD}+1}\left\Vert \overline{\omega}_{1}\left(\boldsymbol{\beta}_{k}\right)\right\Vert\\
		& =o_{p}\left(n^{-1/2}\right),\\
		\delta\left\Vert \sum_{j=1}^{k}\left(I_{p}-\delta\varLambda_{\phi}\left(\boldsymbol{\beta}^{\star}\right)\right)^{k-j}\omega_{2}\left(\boldsymbol{\beta}_{\widetilde{k}_{1,n}^{KBGD}+k_{2,n}^{KBGD}+j}\right)\right\Vert & \leq\delta\sum_{j=0}^{\infty}\left(1-\delta\underline{\lambda}_{\varLambda}/8\right)^{j}\sup_{k\geq\widetilde{k}_{1,n}^{KBGD}+k_{2,n}^{KBGD}}\left\Vert \overline{\omega}_{2}\left(\boldsymbol{\beta}_{k}\right)\right\Vert \\
		& =o_{p}\left(n^{-1/2}\right),\\
		\left\Vert \varLambda_{\phi}^{-1}\left(\boldsymbol{\beta}^{\star}\right)\boldsymbol{\xi}_{n}^{\phi}-\delta\sum_{j=1}^{k}\left(I_{p}-\delta\varLambda_{\phi}\left(\boldsymbol{\beta}^{\star}\right)\right)^{k-j}\boldsymbol{\xi}_{n}^{\phi}\right\Vert &  \leq8\lambda_{\varLambda}^{-1}\left(1-\delta\underline{\lambda}_{\varLambda}/8\right)^{k+1}\left\Vert \boldsymbol{\xi}_{n}^{\phi}\right\Vert .
	\end{align*}
	As $k\rightarrow\infty$, we have that $\lambda_{\varLambda}^{-1}\left(1-\delta\underline{\lambda}_{\varLambda}/8\right)^{k+1}\left\Vert \boldsymbol{\xi}_{n}^{\phi}\right\Vert =o_{p}\left(n^{-1/2}\right)$,
	so 
	\[
	\Delta\boldsymbol{\beta}_{k+\widetilde{k}_{1,n}^{KBGD}+k_{2,n}^{KBGD}}=\varLambda_{\phi}^{-1}\left(\boldsymbol{\beta}^{\star}\right)\boldsymbol{\xi}_{n}^{\phi}+o_{p}\left(n^{-1/2}\right).
	\]
	According to  \autoref{lem3.3}, we have that $\sqrt{n}\boldsymbol{\xi}_{n}^{\phi}\rightarrow N\left(0,\Sigma_{\boldsymbol{\xi}}^{\phi}\right)$,
	so we have that 
	\[
	\sqrt{n}\Delta\boldsymbol{\beta}_{k+\widetilde{k}_{1,n}^{KBGD}+k_{2,n}^{KBGD}}=\varLambda_{\phi}^{-1}\left(\boldsymbol{\beta}^{\star}\right)\sqrt{n}\boldsymbol{\xi}_{n}^{\phi}+o_{p}\left(1\right)\rightarrow_{d}N\left(0,\varLambda_{\phi}^{-1}\left(\boldsymbol{\beta}^{\star}\right)\Sigma_{\boldsymbol{\xi}}^{\phi}\left(\varLambda_{\phi}^{-1}\left(\boldsymbol{\beta}^{\star}\right)\right)^{\mathrm{T}}\right).
	\]
\end{proof}

\subsection*{Proof of \autoref{thm:3.4}}
\begin{proof}
	We only need to show that $\left\Vert \widehat{\varLambda}_{\phi,n}^{-1}\left(\widehat{\boldsymbol{\beta}}\right)-\varLambda_{\phi}^{-1}\left(\boldsymbol{\beta}^{\star}\right)\right\Vert \rightarrow_{p}0$
	and $\left\Vert \widehat{\Sigma}_{\boldsymbol{\xi}}^{\phi}-\Sigma_{\boldsymbol{\xi}}^{\phi}\right\Vert \rightarrow_{p}0$
	both hold. Note that \autoref{lem3.2} indicates that $\left\Vert \widehat{\varLambda}_{\phi,n}\left(\widehat{\boldsymbol{\beta}}\right)-\varLambda_{\phi}\left(\boldsymbol{\beta}^{\star}\right)\right\Vert \rightarrow_{p}0$,
	which implies that $\left\Vert \widehat{\varLambda}_{\phi,n}^{-1}\left(\widehat{\boldsymbol{\beta}}\right)-\varLambda_{\phi}^{-1}\left(\boldsymbol{\beta}^{\star}\right)\right\Vert \rightarrow_{p}0$
	also holds. 
	
	Now we show that $\left\Vert \widehat{\Sigma}_{\boldsymbol{\xi}}^{\phi}-\Sigma_{\boldsymbol{\xi}}^{\phi}\right\Vert \rightarrow_{p}0$
	holds. Our basic proof method is similar to that of \autoref{lem:3.1}.
	In particular, let $\phi_{n}\downarrow0$ and $\mathcal{X}_{e,n}$
	be as defined as in the proof of  \autoref{lem:3.1}. Then we have
	that $f_{z}^{\star}\left(z_{i}^{\star}\right)\geq C\phi_{n}^{p}$
	as long as $\mathbf{X}_{e,i}\in\mathcal{X}_{e,n}$. Denote $G_{i}^{\star}=G\left(z_{i}^{\star}\right)$,
	we have
	\begin{align}
		\left\Vert \widehat{\Sigma}_{\boldsymbol{\xi}}^{\phi}-\Sigma_{\boldsymbol{\xi}}^{\phi}\right\Vert \leq & \left\Vert \frac{1}{n}\sum_{i=1}^{n}\left(I_{n,i}\cdot\widehat{G}_{i}\left(1-\widehat{G}_{i}\right)\left(\mathbf{X}_{i}^{\phi}-\widehat{\mathbb{E}}\left(\left.\mathbf{X}_{i}^{\phi}\right|\widehat{z}_{i}\right)\right)\left(\mathbf{X}_{i}^{\phi}-\widehat{\mathbb{E}}\left(\left.\mathbf{X}_{i}^{\phi}\right|\widehat{z}_{i}\right)\right)^{\mathrm{T}}\right)\right.\nonumber \\
		& \left.-\mathbb{E}\left(I_{n,i}\cdot G_{i}^{\star}\left(1-G_{i}^{\star}\right)\left(\mathbf{X}_{i}^{\phi}-\mathbb{E}\left(\left.\mathbf{X}_{i}^{\phi}\right|z_{i}^{\star}\right)\right)\left(\mathbf{X}_{i}^{\phi}-\widehat{\mathbb{E}}\left(\left.\mathbf{X}_{i}^{\phi}\right|z_{i}^{\star}\right)\right)^{\mathrm{T}}\right)\right\Vert \label{Ap32}\\
		+ & \left\Vert \frac{1}{n}\sum_{i=1}^{n}\left(\left(1-I_{n,i}\right)\cdot\widehat{G}_{i}\left(1-\widehat{G}_{i}\right)\left(\mathbf{X}_{i}^{\phi}-\widehat{\mathbb{E}}\left(\left.\mathbf{X}_{i}^{\phi}\right|\widehat{z}_{i}\right)\right)\left(\mathbf{X}_{i}^{\phi}-\widehat{\mathbb{E}}\left(\left.\mathbf{X}_{i}^{\phi}\right|\widehat{z}_{i}\right)\right)^{\mathrm{T}}\right)\right.\nonumber \\
		& \left.-\mathbb{E}\left(\left(1-I_{n,i}\right)\cdot G_{i}^{\star}\left(1-G_{i}^{\star}\right)\left(\mathbf{X}_{i}^{\phi}-\mathbb{E}\left(\left.\mathbf{X}_{i}^{\phi}\right|z_{i}^{\star}\right)\right)\left(\mathbf{X}_{i}^{\phi}-\widehat{\mathbb{E}}\left(\left.\mathbf{X}_{i}^{\phi}\right|z_{i}^{\star}\right)\right)^{\mathrm{T}}\right)\right\Vert .\label{Ap33}
	\end{align}
	Note that $\widehat{G}_{i}$, $G_{i}^{\star}$, $\mathbf{X}_{i}^{\phi}$,
	$\widehat{\mathbb{E}}\left(\left.\mathbf{X}_{i}^{\phi}\right|\widehat{z}_{i}\right)$,
	and $\mathbb{E}\left(\left.\mathbf{X}_{i}^{\phi}\right|z_{i}^{\star}\right)$
	are all upper bounded, so (\ref{Ap33}) is $O_{p}\left(\phi_{n}\right)$.
	Now we look at (\ref{Ap32}). Note that
	\[
	\widehat{G}_{i}-\frac{\sum_{j=1}^{n}K_{h_{n}}\left(z_{i}^{\star}-z_{j}^{\star}\right)y_{j}}{\sum_{j=1}^{n}K_{h_{n}}\left(z_{i}^{\star}-z_{j}^{\star}\right)}=\frac{\partial\widehat{G}\left(\left.z\left(\mathbf{X}_{e,i},\widetilde{\boldsymbol{\beta}}\right)\right|\widetilde{\boldsymbol{\beta}}\right)}{\partial\boldsymbol{\beta}^{\mathrm{T}}}\Delta\widehat{\boldsymbol{\beta}},
	\]
	where $\widetilde{\boldsymbol{\beta}}$ lies somewhere between $\widehat{\boldsymbol{\beta}}$
	and $\boldsymbol{\beta}^{\star}$. According to the proof of 
	\autoref{lem8}, we have that 
	\[
	\sup_{\left(\mathbf{X}_{e},\boldsymbol{\beta}\right)\in\mathcal{X}_{e,n}\times\mathcal{B}}\left\Vert \frac{\partial\widehat{G}\left(\left.z\left(\mathbf{X}_{e},\boldsymbol{\beta}\right)\right|\boldsymbol{\beta}\right)}{\partial\boldsymbol{\beta}^{\mathrm{T}}}\right\Vert =O_{p}\left(1\right)
	\]
	if $\phi_{n}^{-p}\left(h_{n}^{-2}\sqrt{\log\left(nh_{n}^{-1}\right)/n}+h_{n}^{3}\right)\rightarrow0$, since $\left\Vert f_{z}^{-1}\left(z\left(\boldsymbol{\mathbf{X}}_{e},\boldsymbol{\beta}\right)\right)\partial H_{1}\left(z\left(\boldsymbol{\mathbf{X}}_{e},\boldsymbol{\beta}\right),\boldsymbol{\mathbf{X}}_{e}\right)/\partial z\right\Vert$ and $\left\Vert L\left(z\left(\boldsymbol{\mathbf{X}}_{e},\boldsymbol{\beta}\right),\boldsymbol{\beta}\right)f_{z}^{-1}\left(z\left(\boldsymbol{\mathbf{X}}_{e},\boldsymbol{\beta}\right)\right)\partial H_{2}\left(z\left(\boldsymbol{\mathbf{X}}_{e},\boldsymbol{\beta}\right),\boldsymbol{\mathbf{X}}_{e}\right)/\partial z\right\Vert$ are both bounded for all $\boldsymbol{\beta}\in \mathcal{B}$ and $\mathbf{X}_e\in\mathcal{X}_{e,n}$.
	So 
	\[
	\max_{1\leq i\leq n}\left|\left(\widehat{G}_{i}-\frac{\sum_{j=1}^{n}K_{h_{n}}\left(z_{i}^{\star}-z_{j}^{\star}\right)y_{j}}{\sum_{j=1}^{n}K_{h_{n}}\left(z_{i}^{\star}-z_{j}^{\star}\right)}\right)\cdot I_{n,i}\right|= O_{p}\left(n^{-1/2}\right).
	\]
	Also note that when $\phi_{n}^{-p}\left(h_{n}^{-2}\sqrt{\log\left(nh_{n}^{-1}\right)/n}+h_{n}^{3}\right)\rightarrow0$,
	\[
	\max_{1\leq i\leq n}\left|\left(\frac{\sum_{j=1}^{n}K_{h_{n}}\left(z_{i}^{\star}-z_{j}^{\star}\right)y_{j}}{\sum_{j=1}^{n}K_{h_{n}}\left(z_{i}^{\star}-z_{j}^{\star}\right)}-G\left(z_{i}^{\star}\right)\right)\cdot I_{n,i}\right|=O_{p}\left(\phi_{n}^{-p}\left(h_{n}^{-1}\sqrt{\log\left(nh_{n}^{-1}\right)/n}+h_{n}^{3}\right)\right),
	\]
	this indicates that
	\[
	\max_{1\leq i\leq n}I_{n,i}\cdot\left|\widehat{G}_{i}-G\left(z_{i}^{\star}\right)\right|=O_{p}\left(\phi_{n}^{-p}\left(h_{n}^{-1}\sqrt{\log\left(nh_{n}^{-1}\right)/n}+h_{n}^{3}\right)\right),
	\]
	due to $n^{1/2}\left(h_{n}^{-1}\sqrt{\log\left(nh_{n}^{-1}\right)/n}+h_{n}^{3}\right)\rightarrow\infty$ under the choice of $h_{n}$.
	Using similar argument, we can also show that 
	\[
	\max_{1\leq i\leq n}\left\Vert \left(\widehat{\mathbb{E}}\left(\left.\mathbf{X}_{i}^{\phi}\right|\widehat{z}_{i}\right)-\mathbb{E}\left(\left.\mathbf{X}_{i}^{\phi}\right|z_{i}^{\star}\right)\right)\cdot I_{n,i}\right\Vert =O_{p}\left(\phi_{n}^{-p}\left(h_{n}^{-1}\sqrt{\log\left(nh_{n}^{-1}\right)/n}+h_{n}^{3}\right)\right).
	\]
	So we have that (\ref{Ap32}) is of order $O_{p}\left(\phi_{n}^{-p}\left(h_{n}^{-1}\sqrt{\log\left(nh_{n}^{-1}\right)/n}+h_{n}^{3}\right)+n^{-1/2}\right)$.
	It remains to choose
	\[
	\phi_{n}=O\left(\left(h_{n}^{-1}\sqrt{\log\left(nh_{n}^{-1}\right)/n}+h_{n}^{3}\right)^{\frac{1}{p+1}}\right)
	\]
	to conclude the proof. 
\end{proof}

\subsection*{Proof of \autoref{thm4.1}}
\begin{proof}
	The proof is similar to that of  \autoref{thm:3.1}. Note
	that 
	\begin{align*}
		& \sup_{0\leq t\leq1,\boldsymbol{\beta}\in\mathcal{B}}\left|\overline{\sigma}^{2}\left(I_{p}-\delta\Psi_{q}\left(t,\boldsymbol{\beta}\right)\right)-\overline{\lambda}\left(I_{p}-\delta\left(\Psi_{q}\left(t,\boldsymbol{\beta}\right)+\Psi_{q}^{\mathrm{T}}\left(t,\boldsymbol{\beta}\right)\right)\right)\right|\\
		&  \leq\delta^{2}\sup_{0\leq t\leq1,\boldsymbol{\beta}\in\mathcal{B}}\left\Vert \Psi_{q}\left(t,\boldsymbol{\beta}\right)\right\Vert ^{2}\leq\delta^{2}\left\Vert G^{\prime}\right\Vert _{\infty}^{2}p^{2}\left\{ 1+\underline{\lambda}_{\Gamma}^{-1}qD_{q,0}^{2}\right\} ^{2}.
	\end{align*}
	So if $\delta^{2}\left\Vert G^{\prime}\right\Vert _{\infty}^{2}p^{2}\left\{ 1+\underline{\lambda}_{\Gamma}^{-1}qD_{q,0}^{2}\right\} ^{2}\leq\frac{1}{2}\underline{\lambda}_{\Psi}\delta$,
	or equivalently, 
	$
	\delta\leq \underline{\lambda}_{\Psi}/\left(2\left\Vert G^{\prime}\right\Vert _{\infty}^{2}p^{2}\left\{ 1+\underline{\lambda}_{\Gamma}^{-1}qD_{q,0}^{2}\right\} ^{2}\right),
	$
	we have that 
	\[
	\sup_{0\leq t\leq1,\boldsymbol{\beta}\in\mathcal{B}}\left|\overline{\sigma}^{2}\left(I_{p}-\delta\Psi_{q}\left(t,\boldsymbol{\beta}\right)\right)-\overline{\lambda}\left(I_{p}-\delta\left(\Psi_{q}\left(t,\boldsymbol{\beta}\right)+\Psi_{q}^{\mathrm{T}}\left(t,\boldsymbol{\beta}\right)\right)\right)\right|\leq\underline{\lambda}_{\Psi}\delta/2,
	\]
	so
	\[
	\sup_{0\leq t\leq1,\boldsymbol{\beta}\in\mathcal{B}}\overline{\sigma}^{2}\left(I_{p}-\delta\Psi_{q}\left(t,\boldsymbol{\beta}\right)\right)\leq1-\underline{\lambda}_{\Psi}\delta/2<1,
	\]
	and 
	\[
	\sup_{0\leq t\leq1,\boldsymbol{\beta}\in\mathcal{B}}\overline{\sigma}\left(I_{p}-\delta\Psi_{q}\left(t,\boldsymbol{\beta}\right)\right)\leq1-\underline{\lambda}_{\Psi}\delta/4.
	\]
	Then we have that 
	\begin{align*}
		\left\Vert \Delta\boldsymbol{\beta}_{k+1}\right\Vert  & \leq\left\Vert \int_{0}^{1}\left(I_{p}-\delta\Psi_{q}\left(t,\boldsymbol{\beta}_{k}\right)\right)\Delta\boldsymbol{\beta}dt+\delta\mathfrak{R}_{n,k}\right\Vert \\
		& \leq\sup_{0\leq t\leq1,\boldsymbol{\beta}\in\mathcal{B}}\overline{\sigma}\left(I_{p}-\delta\Psi_{q}\left(t,\boldsymbol{\beta}\right)\right)\left\Vert \Delta\boldsymbol{\beta}_{k}\right\Vert +\delta_{k}\left\Vert \mathfrak{R}_{n,k}\right\Vert  \leq\left(1-\underline{\lambda}_{\Psi}\delta/4\right)\left\Vert \Delta\boldsymbol{\beta}_{k}\right\Vert +\delta\left\Vert \mathfrak{R}_{n,k}\right\Vert \leq\cdots\\
		& \leq\left(1-\underline{\lambda}_{\Psi}\delta/4\right)^{k}\left\Vert \Delta\boldsymbol{\beta}_{1}\right\Vert +\delta\sum_{j=1}^{k}\left(1-\underline{\lambda}_{\Psi}\delta/4\right)^{k-j}\left\Vert \mathfrak{R}_{n,j}\right\Vert \\
		& \leq\left(1-\underline{\lambda}_{\Psi}\delta/4\right)^{k}\left\Vert \Delta\boldsymbol{\beta}_{1}\right\Vert +4/\underline{\lambda}_{\Psi}O_{p}\left(\sup_{k\geq1}\left\Vert \mathfrak{R}_{n,k}\right\Vert \right).
	\end{align*}
	When $\left(1-\underline{\lambda}_{\Psi}\delta\right/4)^{k}\left\Vert \Delta\boldsymbol{\beta}_{1}\right\Vert \leq\chi_{2,n}$,
	or equivalently, 
	$
	k\geq\frac{\log\left(\left\Vert \Delta\boldsymbol{\beta}_{1}\right\Vert \right)-\log\left(\chi_{2,n}\right)}{-\log\left(1-\underline{\lambda}_{\Psi}\delta/4\right)}=k_{1,n}^{SBGD},
	$
	there holds $\left\Vert \Delta\boldsymbol{\beta}_{k+1}\right\Vert =O_{p}\left(\chi_{2,n}\right)$. 
\end{proof}
\subsection*{Proof of \autoref{thm4.2}}
\begin{proof}
	We first prove \autoref{thm4.2} (i). Note that 
	\begin{align*}
		\Delta\boldsymbol{\beta}_{k+1} & =\left\{ \int_{0}^{1}\left(I_{p}-\delta\Psi_{q}^{\star}\right)dt\right\} \Delta\boldsymbol{\beta}_{k}+\delta\mathfrak{R}_{n,k}\\
		& =\left(I_{p}-\delta\Psi_{q}^{\star}\right)\Delta\boldsymbol{\beta}_{k}+\frac{\delta}{n}\sum_{i=1}^{n}\left(\mathbf{X}_{i} - \mathfrak{X}_{q,i}\right)\varepsilon_{i}  +\delta\left\{\mathfrak{R}_{n,k}-\frac{1}{n}\sum_{i=1}^{n}\left(\mathbf{X}_{i} - \mathfrak{X}_{q,i}\right)\varepsilon_{i} \right.\\
		& \left. +\int_{0}^{1}\left(\Psi_{q}^{\star}-\frac{1}{n}\sum_{i=1}^{n}G^{\prime}\left(z_{i}^{\star}+t\mathbf{X}_{i}^{\mathrm{T}}\Delta\boldsymbol{\beta}\right)\left(\mathbf{X}_{i}\mathbf{X}_{i}^{\mathrm{T}}-\mathfrak{X}_{q,n}\left(z\left(\mathbf{X}_{e,i},\boldsymbol{\beta}\right),\boldsymbol{\beta}\right)\mathbf{X}_{i}^{\mathrm{T}}\right)\right)dt\Delta\boldsymbol{\beta}_{k}\right\}.
	\end{align*}
	Define 
	\begin{align*}
	\widetilde{\mathfrak{R}}_{n,k} & =\mathfrak{R}_{n,k}-\frac{\delta}{n}\sum_{i=1}^{n}\left(\mathbf{X}_{i} - \mathfrak{X}_{q,i}\right)\varepsilon_{i}+\\
	&  \int_{0}^{1}\left(\Psi_{q}^{\star}-\frac{1}{n}\sum_{i=1}^{n}G^{\prime}\left(z_{i}^{\star}+t\mathbf{X}_{i}^{\mathrm{T}}\Delta\boldsymbol{\beta}\right)\left(\mathbf{X}_{i}\mathbf{X}_{i}^{\mathrm{T}}-\mathfrak{X}_{q,n}\left(z\left(\mathbf{X}_{e,i},\boldsymbol{\beta}\right),\boldsymbol{\beta}\right)\mathbf{X}_{i}^{\mathrm{T}}\right)\right)dt\Delta\boldsymbol{\beta}_{k}.
\end{align*}
	According to \autoref{lem4.2}, we have that 
	\begin{align*}
		& \sup_{k\geq k_{1,n}^{SBGD}+1}\left\Vert \int_{0}^{1}\left(\Psi_{q}^{\star}-\frac{1}{n}\sum_{i=1}^{n}G^{\prime}\left(z_{i}^{\star}+t\mathbf{X}_{i}^{\mathrm{T}}\Delta\boldsymbol{\beta}\right)\left(\mathbf{X}_{i}\mathbf{X}_{i}^{\mathrm{T}}-\mathfrak{X}_{q,n}\left(z\left(\mathbf{X}_{e,i},\boldsymbol{\beta}\right),\boldsymbol{\beta}\right)\mathbf{X}_{i}^{\mathrm{T}}\right)\right)dt\Delta\boldsymbol{\beta}_{k}\right\Vert \\
		& \leq\sup_{k\geq k_{1,n}^{SBGD}+1,0\leq t\leq1}\left\Vert \Psi_{q}^{\star}-\frac{1}{n}\sum_{i=1}^{n}G^{\prime}\left(z_{i}^{\star}+t\mathbf{X}_{i}^{\mathrm{T}}\Delta\boldsymbol{\beta}\right)\left(\mathbf{X}_{i}\mathbf{X}_{i}^{\mathrm{T}}-\mathfrak{X}_{q,n}\left(z\left(\mathbf{X}_{e,i},\boldsymbol{\beta}\right),\boldsymbol{\beta}\right)\mathbf{X}_{i}^{\mathrm{T}}\right)\right\Vert \sup_{k\geq k_{1,n}^{SBGD}+1}\left\Vert \Delta\boldsymbol{\beta}_{k}\right\Vert \\
		& =O_{p}\left(\sqrt{p}qD_{q,0}^{2}\left(p+qD_{q,0}D_{q,1}\right)\sup_{k\geq k_{1,n}^{SBGD}+1}\left\Vert \Delta\boldsymbol{\beta}\right\Vert ^{2}\right)\\
		& =O_{p}\left(\sqrt{p}qD_{q,0}^{2}\left(p+qD_{q,0}D_{q,1}\right)\chi_{2,n}^{2}\right).
	\end{align*}
	According to \autoref{lem4.3}, we have that 
	\[
	\sup_{k\geq k_{1,n}^{SBGD}+1}\left\Vert \mathfrak{R}_{n,k}-\frac{\delta}{n}\sum_{i=1}^{n}\left(\mathbf{X}_{i} -\mathfrak{X}_{q,i}\right)\varepsilon_{i}\right\Vert =O_{p}\left(\chi_{4,n}\right).
	\]
	This shows the result.
	
	To prove \autoref{thm4.2}(ii), we note that 
	\begin{align*}
		\Delta\boldsymbol{\beta}_{k+k_{1,n}^{SBGD}+1} & =\left(I_{p}-\delta\Psi_{q}^{\star}\right)\Delta\boldsymbol{\beta}_{k+k_{1,n}^{SBGD}}+\frac{\delta}{n}\sum_{i=1}^{n}\left(\mathbf{X}_{i} -\mathfrak{X}_{q,i}\right)\varepsilon_{i}+\widetilde{\mathfrak{R}}_{n,k+k_{1,n}^{SBGD}},\\
		& =\left(I_{p}-\delta\Psi_{q}^{\star}\right)^{k}\Delta\boldsymbol{\beta}_{k_{1,n}^{SBGD}+1}+\sum_{j=1}^{k}\left(I_{p}-\delta\Psi_{q}^{\star}\right)^{j-1}\left(\frac{\delta}{n}\sum_{i=1}^{n}\left(\mathbf{X}_{i} -\mathfrak{X}_{q,i}\right)\varepsilon_{i}\right)\\
		& +\sum_{j=1}^{k}\left(I_{p}-\delta\Psi_{q}^{\star}\right)^{j-1}\widetilde{\mathfrak{R}}_{n,k+k_{1,n}^{SBGD}+1-j}\\
		& =\Psi_{q}^{\star-1}\frac{1}{n}\sum_{i=1}^{n}\left(\mathbf{X}_{i} -\mathfrak{X}_{q,i}\right)\varepsilon_{i}+\left(I_{p}-\delta\Psi_{q}^{\star}\right)^{k}\Delta\boldsymbol{\beta}_{k_{1,n}^{SBGD}+1} +\sum_{j=1}^{k}\left(I_{p}-\delta\Psi_{q}^{\star}\right)^{j-1}\widetilde{\mathfrak{R}}_{n,k+k_{1,n}^{SBGD}+1-j}\\
		& +\sum_{j=k+1}^{\infty}\left(I_{p}-\delta\Psi_{q}^{\star}\right)^{j-1}\left(\frac{\delta}{n}\sum_{i=1}^{n}\left(\mathbf{X}_{i} -\mathfrak{X}_{q,i}\right)\varepsilon_{i}\right).
	\end{align*}
	Then since 
	\[
	\left\Vert \left(I_{p}-\delta\Psi_{q}^{\star}\right)^{k}\Delta\boldsymbol{\beta}_{k_{1,n}^{SBGD}+1}\right\Vert =O_{p}\left(\left(1-\underline{\lambda}_{\Psi}\delta/4\right)^{k}\chi_{2,n}\right),
	\]
	\[
	\left\Vert \sum_{j=1}^{k}\left(I_{p}-\delta\Psi_{q}^{\star}\right)^{j-1}\widetilde{\mathfrak{R}}_{n,k+k_{1,n}^{SBGD}+1-j}\right\Vert \leq\sum_{j=1}^{\infty}\left(1-\underline{\lambda}_{\Psi}\delta/4\right)^{j-1}\sup_{k\geq k_{1,n}^{SBGD}+1}\left\Vert \widetilde{\mathfrak{R}}_{n,k}\right\Vert =O_{p}\left(\chi_{5,n}\right),
	\]
	and 
	\begin{align*}
		\left\Vert \sum_{j=k+1}^{\infty}\left(I_{p}-\delta\Psi_{q}^{\star}\right)^{j-1}\left(\frac{\delta}{n}\sum_{i=1}^{n}\left(\mathfrak{V}_{q}\boldsymbol{r}_{q}\left(z_{i}^{\star}\right)+\mathbf{X}_{i}\right)\varepsilon_{i}\right)\right\Vert  & \leq\left(1-\underline{\lambda}_{\Psi}\delta/4\right)^{k}\left\Vert \frac{4}{\underline{\lambda}_{\Psi}n}\sum_{i=1}^{n}\left(\mathbf{X}_{i} -\mathfrak{X}_{q,i}\right)\varepsilon_{i}\right\Vert \\
		& =O_{p}\left(\left(1-\underline{\lambda}_{\Psi}\delta/4\right)^{k}\sqrt{\frac{pqD_{q,0}^{2}\left(\log p\right)}{n}}\right)\\
		& =O_{p}\left(\left(1-\underline{\lambda}_{\Psi}\delta/4\right)^{k}\chi_{2,n}\right).
	\end{align*}
	So as long as $\left(1-\underline{\lambda}_{\Psi}\delta/4\right)^{k}\chi_{2,n}\leq n^{-1/2}$,
	or equivalently, 
	$
	k\geq k_{2,n}^{SBGD}=\frac{-\log\chi_{2,n}+\log\sqrt{n}}{-\log\left(1-\underline{\lambda}_{\Psi}\delta/4\right)},
	$
	we have that 
	\[
	\sup_{k\geq k_{2,n}^{SBGD}+1}\left\Vert \Delta\boldsymbol{\beta}_{k+k_{1,n}^{SBGD}+1}-\Psi_{q}^{\star-1}\frac{1}{n}\sum_{i=1}^{n}\left(\mathfrak{V}_{q}\boldsymbol{r}_{q}\left(z_{i}^{\star}\right)+\mathbf{X}_{i}\right)\varepsilon_{i}\right\Vert =o_{p}\left(n^{-\frac{1}{2}}\right).
	\]
	The following results hold trivially. 
\end{proof}

\subsection*{Proof of \autoref{thm4.3}}
\begin{proof}
	Note that under all the conditions imposed in \autoref{thm4.2}, we have that
	\[
	\left\Vert \widehat{\boldsymbol{\beta}}-\boldsymbol{\beta}^{\star}\right\Vert =O_{p}\left(\sqrt{pq^{2}D_{q,0}^{4}\left(\log p\right)/n}\right),
	\]
	due to the fact that each element of $\left(\mathbf{X}_{i} -\mathfrak{X}_{q,i}\right)\varepsilon_{i}$
	is bounded by $CqD_{q,0}^{2}$ and \autoref{assu:7} holds.
	
	To prove the theorem, we first show that 
	\[
	\sup_{1\leq i\leq n}\left|\widehat{G}_{i}-G_{i}\left(z_i^{\star}\right)\right|=O_{p}\left(\sqrt{p^2q^{4}D_{q,0}^{8}\left(\log p\right)/n}+qD_{q,0}^{2}\mathcal{E}_{q,0}\right).
	\]
	Define $\widehat{z}_{i}=z\left(\mathbf{X}_{e,i},\widehat{\boldsymbol{\beta}}\right)$.
	To show the above result, note that 
	\begin{align*}
		& \sup_{1\leq i\leq n}\left|\widehat{G}_{i}-G\left(z_i^{\star}\right)\right|\leq\sup_{1\leq i\leq n}\left|\widehat{\boldsymbol{r}}_{q,i}^{\mathrm{T}}\left(\widehat{\boldsymbol{\pi}}_{q}-\boldsymbol{\pi}_{q}^{\star}\right)\right|\\
		& +\sup_{1\leq i\leq n}\left|\widehat{\boldsymbol{r}}_{q,i}^{\mathrm{T}}\boldsymbol{\pi}_{q}^{\star}-G\left(\widehat{z}_{i}\right)\right|+\sup_{1\leq i\leq n}\left|G\left(\widehat{z}_{i}\right)-G\left(z_{i}^{\star}\right)\right|.
	\end{align*}
	Obviously, the second and third terms on RHS are of order $O_{p}\left(\mathcal{E}_{q,0}\right)$
	and $O_{p}\left(\sqrt{p^{2}q^{2}D_{q,0}^{4}\left(\log p\right)/n}\right)$,
	while the first term is bounded by $\sqrt{q}D_{q,0}\left\Vert \widehat{\boldsymbol{\pi}}_{q}-\boldsymbol{\pi}_{q}^{\star}\right\Vert $.
	Note that 
	\begin{align*}
		\widehat{\boldsymbol{\pi}}_{q}-\boldsymbol{\pi}_{q}^{\star} & =\Gamma_{q,n}^{-1}\left(\widehat{\boldsymbol{\beta}}\right)\left(\frac{1}{n}\sum_{i=1}^{n}\widehat{\boldsymbol{r}}_{q,i}\left(G\left(\widehat{z}_{i}\right)-G\left(z_{i}^{\star}\right)\right)\right)+\Gamma_{q,n}^{-1}\left(\widehat{\boldsymbol{\beta}}\right)\left(\frac{1}{n}\sum_{i=1}^{n}\widehat{\boldsymbol{r}}_{q,i}R_{q}\left(\widehat{z}_{i}\right)\right)\\
		& +\Gamma_{q,n}^{-1}\left(\widehat{\boldsymbol{\beta}}\right)\left(\frac{1}{n}\sum_{i=1}^{n}\boldsymbol{r}_{q}\left(\widehat{z}_{i}\right)\varepsilon_{i}\right).
	\end{align*}
	So we have that $\left\Vert \widehat{\boldsymbol{\pi}}_{q}-\boldsymbol{\pi}_{q}^{\star}\right\Vert =O_{p}\left(\sqrt{p^{2}q^{3}D_{q,0}^{6}\left(\log p\right)/n}+\sqrt{q}D_{q,0}\mathcal{E}_{q,0}\right)$
	and the third term is of order $O_{p}\left(\sqrt{p^{2}q^{4}D_{q,0}^{8}\left(\log p\right)/n}+qD_{q,0}^{2}\mathcal{E}_{q,0}\right).$
	This proves the first result.

	We also note that according to the proof of \autoref{lemS.4}, we have that 
	\[
	\sup_{1\leq i\leq n}\left\Vert \mathfrak{X}_{q,n}\left(\widehat{z}_{i},\widehat{\boldsymbol{\beta}}\right)-\mathfrak{X}_{q}\left(z_{i}^{\star},\boldsymbol{\beta}^{\star}\right)\right\Vert =O_{p}\left(\sqrt{p^{3}q^{6}D_{q,0}^{10}D_{q,1}^2\log\left(pn\right)/n}\right).
	\] Then we show that 
	\begin{align*}
		& \max_{1\leq i\leq n}\left\Vert \widehat{G}_{i}\left(1-\widehat{G}_{i}\right)\left(\mathbf{X}_{i} - \mathfrak{X}_{q,n}\left(\widehat{z}, \widehat{\boldsymbol{\beta}}\right)\right)\left(\mathbf{X}_{i} -  \mathfrak{X}_{q,n}\left(\widehat{z}, \widehat{\boldsymbol{\beta}}\right) \right)^{\mathrm{T}}-G_{i}\left(1-G_{i}\right)\left(\mathbf{X}_{i}-\mathfrak{X}_{q}\left(z_{i}^{\star},\boldsymbol{\beta}^{\star}\right)\right)\left( \mathbf{X}_{i} - \mathfrak{X}_{q}\left(z_{i}^{\star},\boldsymbol{\beta}^{\star}\right)\right)^{\mathrm{T}}\right\Vert \\
		& =O_{p}\left(\sqrt{p^{4}q^{8}D_{q,0}^{14}\left(\log pn\right)/n}\left(D_{q,0}+D_{q,1}\right)+pq^{3}D_{q,0}^{6}\mathcal{E}_{q,0}\right).
	\end{align*}
	Note that the above is bounded by 
	\begin{align*}
		& \max_{1\leq i\leq n}\left\Vert \left(\widehat{G}_{i}\left(1-\widehat{G}_{i}\right)-G_{i}\left(1-G_{i}\right)\right)\left(\mathbf{X}_{i} - \mathfrak{X}_{q,n}\left(\widehat{z}, \widehat{\boldsymbol{\beta}}\right)\right)\left(\mathbf{X}_{i} - \mathfrak{X}_{q,n}\left(\widehat{z}, \widehat{\boldsymbol{\beta}}\right)\right)^{\mathrm{T}}\right\Vert \\
		& +\max_{1\leq i\leq n}\left\Vert G_{i}\left(1-G_{i}\right)\left(\left(\mathbf{X}_{i} - \mathfrak{X}_{q,n}\left(\widehat{z}, \widehat{\boldsymbol{\beta}}\right)\right)\left(\mathbf{X}_{i} - \mathfrak{X}_{q,n}\left(\widehat{z}, \widehat{\boldsymbol{\beta}}\right)\right)^{\mathrm{T}}-\left(\mathbf{X}_{i}-\mathfrak{X}_{q}\left(z_{i}^{\star},\boldsymbol{\beta}^{\star}\right)\right)\left( \mathbf{X}_{i} - \mathfrak{X}_{q}\left(z_{i}^{\star},\boldsymbol{\beta}^{\star}\right)\right)^{\mathrm{T}}\right)\right\Vert ,
	\end{align*}
	where the first term is of order $O_{p}\left(\sqrt{p^{4}q^{8}D_{q,0}^{16}\left(\log p\right)/n}+pq^{3}D_{q,0}^{6}\mathcal{E}_{q,0}\right)$,
	while the second term is of order $O_{p}\left(\sqrt{p^{4}q^{8}D_{q,0}^{14}D_{q,1}^{2}\left(\log pn\right)/n}\right)$.
	Together we show the result. 
	
	Next we show that 
	\[
	\left\Vert \widehat{\Psi}_{q}^{\star}-\Psi_{q}^{\star}\right\Vert =O_{p}\left(\sqrt{p^{4}q^{4}D_{q,0}^{4}\log\left(pqD_{q,0}D_{q,1}n\right)/n}\right).
	\]
Since $\upsilon_{G}\geq2$, we have that 
	\begin{align*}
		& \sup_{1\leq i\leq n}\left|\widehat{G}_{i}^{\prime}-G^{\prime}\left(z\left(\mathbf{X}_{e,i},\boldsymbol{\beta}^{\star}\right)\right)\right|\leq\sup_{1\leq i\leq n}\left|\widehat{\boldsymbol{r}}_{q,i}^{\prime\mathrm{T}}\left(\widehat{\boldsymbol{\pi}}_{q}-\boldsymbol{\pi}_{q}^{\star}\right)\right|\\
		& +\sup_{1\leq i\leq n}\left|\widehat{\boldsymbol{r}}_{q,i}^{\prime\mathrm{T}}\boldsymbol{\pi}_{q}^{\star}-G^{\prime}\left(\widehat{z}_{i}\right)\right|+\sup_{1\leq i\leq n}\left|G^{\prime}\left(\widehat{z}_{i}\right)-G^{\prime}\left(z_{i}^{\star}\right)\right|\\
		& =O_{p}\left(\sqrt{p^2q^{4}D_{q,0}^{8}D_{q,1}^2\left(\log p\right)/n}+qD_{q,0}D_{q,1}\mathcal{E}_{q,0}+\mathcal{E}_{q,1}\right).
	\end{align*}
	So 
	\begin{align*}
		\left\Vert \widehat{\Psi}_{q}^{\star}-\Psi_{q}^{\star}\right\Vert  & \leq\left\Vert \frac{1}{n}\sum_{i=1}^{n}\left(\widehat{G}_{i}^{\prime}-G^{\prime}\left(z_{i}^{\star}\right)\right)\cdot\left(\mathbf{X}_{i}\mathbf{X}_{i}^{\mathrm{T}}-\mathfrak{X}_{q,n}\left(\widehat{z}_{i},\widehat{\boldsymbol{\beta}}\right)\mathbf{X}_{i}^{\mathrm{T}}\right)\right\Vert ,\\
		& +\left\Vert \frac{1}{n}\sum_{i=1}^{n}G^{\prime}\left(z_{i}^{\star}\right)\cdot\left(\left(\mathfrak{X}_{q,n}\left(\widehat{z}_{i},\widehat{\boldsymbol{\beta}}\right)-\mathfrak{X}_{q}\left(z_{i}^{\star},\boldsymbol{\beta}^{\star}\right)\right)\mathbf{X}_{i}^{\mathrm{T}}\right)\right\Vert \\
		& +\left\Vert \frac{1}{n}\sum_{i=1}^{n}G^{\prime}\left(z_{i}^{\star}\right)\cdot\mathfrak{X}_{q}\left(z_{i}^{\star},\boldsymbol{\beta}^{\star}\right)\mathbf{X}_{i}^{\mathrm{T}}-\Psi_{q}^{\star}\right\Vert \\
		& =O_{p}\left(\sqrt{p^{4}q^{6}D_{q,0}^{12}D_{q,1}^{2}\log \left(pn\right)/n}+pq^{2}D_{q,0}^{3}D_{q,1}\mathcal{E}_{q,0}+pqD_{q,0}^{2}\mathcal{E}_{q,1}\right),
	\end{align*}
	which also implies that $\overline{\sigma}\left(\widehat{\Psi}_{q}^{\star-1}\right)=O_{p}\left(1\right)$,
	and 
	\[
	\left\Vert \widehat{\Psi}_{q}^{\star-1}-\Psi_{q}^{\star-1}\right\Vert =O_{p}\left(\sqrt{p^{4}q^{6}D_{q,0}^{12}D_{q,1}^{2}\left(\log pn\right)/n}+pq^{2}D_{q,0}^{3}D_{q,1}\mathcal{E}_{q,0}+qD_{q,0}^{2}\mathcal{E}_{q,1}\right)
	\]
	
	Now we are ready to demonstrate the consistency of the variance estimator.
	Note that 
	\begin{align*}
		& \left|\widehat{\sigma}_{S}^{2}\left(\rho\right)-\sigma_{S}^{2}\left(\rho\right)\right| \\
		& \leq\left\Vert \rho\right\Vert ^{2}\left\Vert \widehat{\Psi}_{q}^{\star-1}\frac{1}{n}\sum_{i=1}^{n}\left\{ \widehat{G}_{i}\left(1-\widehat{G}_{i}\right)\left(\mathbf{X}_{i} - \mathfrak{X}_{q,n}\left(\widehat{z}, \widehat{\boldsymbol{\beta}}\right)\right)\left(\mathbf{X}_{i} - \mathfrak{X}_{q,n}\left(\widehat{z}, \widehat{\boldsymbol{\beta}}\right)\right)^{\mathrm{T}}\right\} \left(\widehat{\Psi}_{q}^{\star-1}\right)^{\mathrm{T}}\right.\\
		& -\left.\Psi_{q}^{\star-1}\mathbb{E}\left\{ G\left(z_{i}^{\star}\right)\left(1-G\left(z_{i}^{\star}\right)\right)\left( \mathbf{X}_{i} - \mathfrak{X}_{q}\left(z_{i}^{\star},\boldsymbol{\beta}^{\star}\right)\right)\left( \mathbf{X}_{i} - \mathfrak{X}_{q}\left(z_{i}^{\star},\boldsymbol{\beta}^{\star}\right)\right)^{\mathrm{T}}\right\} \left(\Psi_{q}^{\star-1}\right)^{\mathrm{T}}\right\Vert \\
		& \leq\left\Vert \rho\right\Vert ^{2}  \left\Vert \widehat{\Psi}_{q}^{\star-1}-\Psi_{q}^{\star-1}\right\Vert \left\Vert \frac{1}{n}\sum_{i=1}^{n}\left\{ \widehat{G}_{i}\left(1-\widehat{G}_{i}\right)\left(\mathbf{X}_{i} - \mathfrak{X}_{q,n}\left(\widehat{z}, \widehat{\boldsymbol{\beta}}\right)\right)\left(\mathbf{X}_{i} - \mathfrak{X}_{q,n}\left(\widehat{z}, \widehat{\boldsymbol{\beta}}\right)\right)^{\mathrm{T}}\right\} \left(\Psi_{q}^{\star-1}\right)^{\mathrm{T}}\right\Vert \\
		& +\left\Vert \rho\right\Vert ^{2} \left\Vert \Psi_{q}^{\star-1}\left(\frac{1}{n}\sum_{i=1}^{n}\left\{ \widehat{G}_{i}\left(1-\widehat{G}_{i}\right)\left(\mathbf{X}_{i} - \mathfrak{X}_{q,n}\left(\widehat{z}, \widehat{\boldsymbol{\beta}}\right)\right)\left(\mathbf{X}_{i} - \mathfrak{X}_{q,n}\left(\widehat{z}, \widehat{\boldsymbol{\beta}}\right)\right)^{\mathrm{T}}\right\} \right.\right.\\
		& \left.\left.-\mathbb{E}\left\{ G\left(z_{i}^{\star}\right)\left(1-G\left(z_{i}^{\star}\right)\right)\left(\mathbf{X}_{i} - \mathfrak{X}_{q,n}\left(\widehat{z}, \widehat{\boldsymbol{\beta}}\right)\right)\left(\mathbf{X}_{i} - \mathfrak{X}_{q,n}\left(\widehat{z}, \widehat{\boldsymbol{\beta}}\right)\right)^{\mathrm{T}}\right\} \right)\left(\widehat{\Psi}_{q}^{\star-1}\right)^{\mathrm{T}}\right\Vert \\
		& +\left\Vert \rho\right\Vert ^{2}  \left\Vert \Psi_{q}^{\star-1}\mathbb{E}\left\{ G\left(z_{i}^{\star}\right)\left(1-G\left(z_{i}^{\star}\right)\right)\left( \mathbf{X}_{i} - \mathfrak{X}_{q}\left(z_{i}^{\star},\boldsymbol{\beta}^{\star}\right)\right)\left( \mathbf{X}_{i} - \mathfrak{X}_{q}\left(z_{i}^{\star},\boldsymbol{\beta}^{\star}\right)\right)^{\mathrm{T}}\right\} \left(\widehat{\Psi}_{q}^{\star-1}-\Psi_{q}^{\star-1}\right)\right\Vert .
	\end{align*}
	The first and the third terms are of order $O_{p}\left(\sqrt{p^{6}q^{8}D_{q,0}^{16}D_{q,1}^{2}\left(\log pn\right)/n}+p^{2}q^{3}D_{q,0}^{4}D_{q,1}\mathcal{E}_{q,0}+pq^{2}D_{q,0}^{4}\mathcal{E}_{q,1}\right)$,
	and the second term is of order $O_{p}\left(\sqrt{p^4q^8D_{q,0}^{14}\left(\log pn\right)/n}\left(D_{q,0}+D_{q,1}\right)+pq^{3}D_{q,0}^{6}\mathcal{E}_{q,0}\right)$.
	Together, we have that 
	\[
	\left|\widehat{\sigma}_{S}^{2}\left(\rho\right)-\sigma_{S}^{2}\left(\rho\right)\right|=O_{p}\left(\sqrt{p^{6}q^{8}D_{q,0}^{16}D_{q,1}^{2}\left(\log pn\right)/n}+pq^{3}D_{q,0}^{4}\left(pD_{q,1} + D_{q,0}^2\right)\mathcal{E}_{q,0}+pq^{2}D_{q,0}^{4}\mathcal{E}_{q,1}\right),
	\]
	which implies that $\left|\widehat{\sigma}_{S}^{2}\left(\rho\right)-\sigma_{S}^{2}\left(\rho\right)\right|\rightarrow_{p}0$
	under all the conditions.
\end{proof}

\newpage
\bibliographystyle{plainnat}
\bibliography{overall_1}

\end{document}